\documentclass[
10pt, 
french,
singlespacing, 
parskip, 
headsepline, 
chapterinoneline, 
]{MastersDoctoralThesis} 

\usepackage[utf8]{inputenc} 
\usepackage[T1]{fontenc} 

\usepackage{mathpazo} 

\usepackage[backend=bibtex, style=numeric, natbib=true, backref=true, refsegment = chapter, defernumbers = true]{biblatex} 
\defbibheading{bibliography}[Bibliographie]{%
  \chapter*{#1}%
  \addcontentsline{toc}{part}{Bibliographie}
  \markboth{#1}{#1}
}

\usepackage[autostyle=true]{csquotes} 

\def\signed #1{{\leavevmode\unskip\nobreak\hfil\penalty50\hskip2em
  \hbox{}\nobreak\hfil(#1)%
  \parfillskip=0pt \finalhyphendemerits=0 \endgraf}}

\newsavebox\mybox
\newenvironment{aquote}[1]
  {\savebox\mybox{#1}\begin{quote}}
  {\signed{\usebox\mybox}\end{quote}}

\usepackage{imakeidx}
\makeindex
\usepackage{idxlayout}

\usepackage{xparse}
\DeclareDocumentCommand\firstocc{ m g }{%
    {\textcolor{red}{\emph{#1}}%
        \IfNoValueF{#2} {\index{#2}}%
    }%
}

\usepackage{amsopn}

\DeclareMathOperator{\ucirc}{\underline{\circ}}

\usepackage{bbold}

\usepackage{stmaryrd}

\usepackage{float}

\usepackage{amssymb, amsthm, amsmath}

\usepackage{thmtools}
\declaretheorem[thmbox=S, title=Exemple]{example}
\declaretheorem[thmbox=S, title=Théorème]{theorem}
\declaretheorem[thmbox=S, title=Définition]{definition}
\declaretheorem[thmbox=S, title=Corollaire]{corollary}
\declaretheorem[thmbox=S, title=Proposition]{proposition}
\declaretheorem[thmbox=S, title=Lemme]{lemma}

\usepackage{tikz}
  \usetikzlibrary{automata}
  \usetikzlibrary{shadows}
  \usetikzlibrary{arrows}
  \usetikzlibrary{shapes}
  \usetikzlibrary{decorations.pathmorphing}
  \usetikzlibrary{fit}
  \usetikzlibrary{matrix}

	\tikzstyle{every picture}=[>=stealth',shorten >=1pt,node distance=1.44cm,bend angle=45,initial text=,every state/.style={inner sep=0.75mm, minimum size=1mm},font=\scriptsize]

\makeatletter
\@namedef{ver@framed.sty}{9999/12/31}
\@namedef{opt@framed.sty}{}
\makeatother
\usepackage[frozencache]{minted}

\usepackage{tcolorbox}
\tcbuselibrary{skins, breakable}
\newtcolorbox[auto counter, list inside=implementationBox]{implementationBox}[2][]{%
   enhanced, breakable, colback=white, colframe=black, fonttitle=\bfseries, title=Code Pseudo-Haskell.~\thetcbcounter: #2, #1}

\newtcolorbox[auto counter]{remarqueBox}[2][]{%
  enhanced, breakable,
  title=Remarque~\thetcbcounter: #2, #1,
  colframe=blue!50!black,
  colback=blue!10!white,
  colbacktitle=blue!5!yellow!10!white,
  fonttitle=\bfseries,
  coltitle=black,
  attach boxed title to top center={yshift=-0.25mm-\tcboxedtitleheight/2,yshifttext=2mm-\tcboxedtitleheight/2},
  boxed title style={boxrule=0.5mm, frame code={ \path[tcb fill frame] ([xshift=-4mm]frame.west)-- (frame.north west) -- (frame.north east) -- ([xshift=4mm]frame.east)-- (frame.south east) -- (frame.south west) -- cycle; },
  interior code={ \path[tcb fill interior] ([xshift=-2mm]interior.west)-- (interior.north west) -- (interior.north east)-- ([xshift=2mm]interior.east) -- (interior.south east) -- (interior.south west)-- cycle;}  }
}

\allowdisplaybreaks%

\thesistitle{Une proposition d'implantation des structures d'automates, d'expressions et de leurs algorithmes associés utilisant les catégories enrichies} 
\supervisor{Pr. Pascal \textsc{Caron}} 
\examiner{} 
\degree{Habilitation à diriger des recherches} 
\author{Ludovic \textsc{Mignot}} 
\addresses{} 

\subject{Informatique} 
\keywords{} 
\university{\href{www.univ-rouen.fr}{Université de Rouen Normandie}} 
\department{\href{dpt-info-sciences.univ-rouen.fr}{Département d'informatique}} 
\group{\href{www.gr2if.fr}{Groupe de Recherche Rouennais en Informatique Fondamentale}} 
\faculty{\href{sciences-techniques.univ-rouen.fr}{U.F.R. Sciences et Techniques}} 

\AtBeginDocument{
\hypersetup{pdftitle=\ttitle} 
\hypersetup{pdfauthor=\authorname} 
\hypersetup{pdfkeywords=\keywordnames} 
}

\begin{document}

\frontmatter 

\pagestyle{plain} 


\begin{titlepage}
  \begin{center}

    \vspace*{.06\textheight}
    {\scshape\LARGE \univname\par}\vspace{1.5cm} 
    \textsc{\Large Habilitation à diriger des recherches}\\[0.5cm] 

    {\HRule}\\[0.4cm] 
    {\huge \bfseries \ttitle\par}\vspace{0.4cm} 
    {\HRule}\\[1.5cm] 

    \begin{minipage}[t]{0.4\textwidth}
      \begin{flushleft} \large
        \emph{Auteur~:}\\
        {\authorname} 
      \end{flushleft}
    \end{minipage}
    \begin{minipage}[t]{0.4\textwidth}
      \begin{flushright} \large
        \emph{Directeur~:} \\
        {\supname} 
      \end{flushright}
    \end{minipage}\\[3cm]

    \vfill

    \large
    Soutenue publiquement le 15 décembre 2020 devant le jury composé de\\
      \begin{tabular}{lccr}
        Pascal CARON       & PU & université de Rouen Normandie  & Garant    \\
        Olivier CARTON     & PU & université Paris Diderot & Rapporteur          \\
        Thomas COLCOMBET   & DR & université Paris Diderot      & Rapporteur        \\
        Samuele GIRAUDO    & MCF HDR & université Gustave Eiffel & Examinateur          \\
        Sylvain LOMBARDY   & PU & institut polytechnique de Bordeaux & Rapporteur \\
        Jean-Gabriel LUQUE & PU & université de Rouen Normandie    & Examinateur  \\
        Djelloul ZIADI     & PU & université de Rouen Normandie    & Examinateur  \\
      \end{tabular}\\
    {\groupname}\\{\deptname}\\[2cm] 

    \vfill


    \vfill
  \end{center}
\end{titlepage}


%
%


\begin{abstract}
  \addchaptertocentry{\abstractname} 
  Dans ce document, nous proposons une description, \emph{via} une implantation Haskell, d'une généralisation de la notion d'expressions rationnelles permettant de factoriser les définitions et les méthodes de construction d'automates de mots ou d'arbres sur une seule structure, basée sur des outils de théorie des catégories enrichies.

  Pour cela, nous rappellerons tout d'abord les différentes méthodes, communes et existantes, de conversion d'expressions en automates, en mettant en lumière les similitudes entre le cas des mots et celui des arbres.

  Nous produirons ensuite une étude originale de l'apport de la théorie des catégories enrichies à l'implantation des automates, des expressions et des algorithmes associés, en utilisant des concepts avancés de programmation fonctionnelle, tout en proposant simultanément une implantation Haskell des notions de théorie des catégories enrichies et des automates associés.

  Nous aboutirons alors à l'ébauche d'une définition d'expressions généralisées basée sur la notion de produit tensoriel, laissant entrevoir les perspectives d'un projet de recherche pour lequel je souhaite être habilité.
\end{abstract}




\tableofcontents 
\addcontentsline{toc}{chapter}{Table des matières}

\listoffigures 

\listoftables 

\tcblistof[\chapter]{implementationBox}{Pseudo-codes}


%
%




\mainmatter

\pagestyle{thesis} 



\chapter*{Introduction}
\addcontentsline{toc}{chapter}{Introduction}
\markboth{}{}

\begin{flushright}
  \emph{
    Personne n'ignore qu'il y a deux entrées par où les opinions sont reçues dans l'âme,\\
    qui sont ses deux principales puissances, l'entendement et la volonté.}
  \\
  Blaise Pascal
\end{flushright}

Les expressions rationnelles sont des objets mathématiques inductifs permettant de représenter d'une façon concise les langages rationnels, de mots comme d'arbres.
Depuis le début de mes recherches, j'ai travaillé notamment sur de multiples modélisations d'opérations sur les langages par la définition d'opérateurs permettant de les représenter d'un point de vue syntaxique.

Le sujet de ma thèse de doctorat~\cite{Mig10}, les opérateurs de multi-tildes-barres~\cite{CCM12b, CCM10}, m'a permis d'étendre des constructions d'automates connues (comme la dérivation d'expressions ou la méthode des positions) à de nouveaux opérateurs en conservant les bases de ces méthodes.

J'ai ensuite participé à ce travail d'extension à de nombreuses reprises par la suite, comme:
\begin{itemize}
  \item pour des expressions rationnelles permettant de représenter des langages approchés~\cite{CJM13}, c'est-à-dire des langages obtenus à partir des mots à une certaine distance (Hamming ou Levenshtein par exemple) d'un mot d'un langage initial;
  \item pour des expressions linéaires permettant de représenter les langages obtenus par \emph{hairpin completion}~\cite{CDJM13}, une opération basée sur un mécanisme biochimique de l'ADN;
  \item pour des expressions à contraintes~\cite{CMN16}, paramétrées par des prédicats de logique d'ordre 0 (c'est-à-dire des prédicats avec des variables mais sans quantificateurs).
\end{itemize}

Les processus mis en {\oe}uvre lors de ces travaux, c'est-à-dire généralement définition d'un nouvel opérateur, de sa sémantique, puis d'une façon de résoudre le problème d'appartenance d'un mot, ont mis en lumières de nombreux rapprochements possibles entre ces diverses méthodes.

J'ai participé également à des extensions de ces méthodes sur d'autres structures d'entrées, mettant une nouvelle fois en lumières les liens entre toutes ces méthodes:
\begin{itemize}
  \item l'extension des méthodes classiques de construction d'automates aux automates d'arbres depuis des expressions d'arbres~\cite{MOZ14b, AMZ18a, MOZ17} est une application presque directe des méthodes définies sur les mots (déjà remarquée pour des algorithmes classiques comme la déterminisation par construction de la partie accessible de l'automate des parties~\cite{CDGLJLTT07});
  \item la définition d'une méthode de quotient de langage d'arbres non-nullaires~\cite{CMOZ17} et de dérivation d'expressions d'arbres basées sur la méthode de Brzozowski~\cite{AMZ19} est quant à elle une méthode semblant, à première vue, assez différente de celle des expressions de mots.
\end{itemize}

En parallèle, j'ai travaillé sur les liens entre opérades et opérateurs de multi-tildes-barres~\cite{GLMN16, LMN13} et également sur la représentation multi-linéaire des PROs libres~\cite{LLMN19}.
Les opérades sont des structures algébriques permettant de représenter, par exemple, la composition des arbres.
Les PROs sont des catégories monoïdales strictes dont les objets sont les entiers et le produit tensoriel la somme.

À l'aide de la manipulation de ces outils, il m'a paru possible d'essayer de construire des expressions utilisant des notions de théorie des catégories afin de généraliser les expressions d'arbres et de mots (voire plus) ainsi que différentes méthodes de construction,
le préalable étant alors de définir une structure compatible d'automates pouvant elle-aussi supporter cette généricité, ou plutôt polymorphisme.
Cette structure devrait en effet pouvoir être assez générique pour unifier automates déterministes et non-déterministes (pour pouvoir par exemple unifier les méthodes de dérivation et dérivation partielle), voire d'autres types d'automates connus (alternants, à multiplicités, \emph{etc.}).

Ce document est ainsi la présentation de la démarche de formalisation de ces notions, qui est un projet de recherche et un travail original réalisé pour ce document, se découpant en deux parties et d'une partie préliminaire.

D'une façon classique, les préliminaires rappellent les définitions générales des structures d'automates et des notions de base sur des structures algébriques.
Il s'agit d'un premier pas vers la généralisation catégorique par l'exhibition de parallèles entre différents types d'automates.

La Partie~\ref{part:autArbres} est un résumé de certains travaux d'extension des méthodes de constructions d'automates d'arbres auxquels j'ai pu participer;
cette partie a pour but de mettre en lumières les différents points communs entre les constructions d'automates de mots et d'arbres.

Nous aborderons dans cette partie les constructions d'automates d'arbres Top-Down (Chapitre~\ref{chapTopDownCons}) puis les constructions Bottom-Up (Chapitre~\ref{chapBotUp}).

La Partie~\ref{partAutCat} est une description formelle d'une méthode de représentation d'expressions généralisées au niveau des catégories monoïdales.
Les notions manipulées tout au long de cette partie sont de plusieurs ordres.
De nombreuses notions de théories des catégories seront présentées en utilisant des diagrammes commutatifs.
Au lieu de rappeler les nombreuses preuves classiques existant au niveau des automates de mots et d'arbres et de les ré-écrire au niveau des automates catégoriques, une autre approche sera choisie:
afin de conserver une approche pratique, toutes les notions de la seconde partie seront implantées en Haskell, et le code servira de base à la compréhension et aux exemples.

Cette seconde partie se découpe de la façon suivante.
Dans un premier temps (Chapitre~\ref{chapAutCatMots}), une structure d'automate de mots sur une catégorie sera implantée d'une façon à unifier différents types d'automates (déterministes, non-déterministes, \emph{etc.}), suivant la définition de Colcombet et Petrisan~\cite{CP17}.
Ce chapitre permettra de se familiariser avec les notions de théorie des catégories de base (catégories, foncteurs, monades, \emph{etc.}) et leurs liens avec la programmation fonctionnelle, qui seront à la base de la section suivante.
Les différents types d'automates seront décrits comme des cas particuliers de la construction unifiée, et illustrés par des exemples également manipulables dans un interpréteur Haskell.

Dans un deuxième temps (Chapitre~\ref{chapAutCatMonoidaux}), le modèle précédent sera alors étendu en utilisant des catégories enrichies, permettant d'inclure des entrées plus structurées.
Comme précédemment, les structures existantes seront définies comme des cas particuliers et illustrés par des exemples exécutables.

La dernière partie (Chapitre~\ref{chapRegExpCat}) sera alors une ébauche d'implantation d'expression, ouvrant la voie vers des perspectives de recherche vers lesquelles je souhaiterais m'orienter pour la suite de ma carrière et pour lesquelles je demande à être habilité.


\newcommand{\arite}{\mathrm{arit}\acute{\mathrm{e}}}
\newcommand{\racine}{\mathrm{racine}}
\newcommand{\Racine}{\mathrm{Racine}}
\newcommand{\parent}{\mathrm{parent}}
\newcommand{\Parent}{\mathrm{Parent}}
\mathchardef\mhyphen="2D 
\newcommand{\fils}[1]{\mathrm{#1}^{\mathrm{e}}\mhyphen\mathrm{fils}}
\newcommand{\feuilles}{\mathrm{feuilles}}

\newenvironment{retirer}{\color{green}}{\color{black}}

\chapter*{Préliminaires}
\addcontentsline{toc}{chapter}{Préliminaires}

\begin{flushright}
  \emph{Que l'on me donne six heures pour couper un arbre,\\
  j'en passerai quatre à préparer ma hache.}\\
  Abraham Lincoln (apocryphe)
\end{flushright}

  Dans cette section, nous rappelons des définitions mathématiques classiques ainsi que différentes notations utilisées tout au long de ce document.
  Certaines constructions rappelées peuvent sembler alambiquées; mais les abstractions proposées prendront tout leur sens une fois les diverses extensions possibles abordées.
  Cette section a donc pour but de servir de point de départ aux différentes généralisations mises en {\oe}uvres tout au long de ce document, d'où la concision et le style condensé de cette partie.

  Nous commencerons par rappeler des notations générales, puis présenterons les différentes définitions d'automates que nous manipulerons tout au long du document.
  La présentation essaiera de mettre en avant les différents points communs entre ces différentes structures, afin de préparer la réflexion pour la seconde partie de ce document.
  Cette section rappellera également les différentes structures d'expressions rationnelles simples\footnote{des structures plus complexes apparaîtront naturellement dans la seconde partie de ce document.} sur lesquelles s'appuieront les différentes constructions rappelées dans la première partie de ce document: la proximité des constructions entre mots et arbres servira de point d'appui, encore une fois, à la réflexion de leur unification.

\section*{Notations générales}
Ainsi, commençons par rappeler des définitions d'ordre général.
Dans la suite, on note pour tous trois ensembles \( P \), \( Q \), et \( R \):
  \begin{itemize}
    \item \( P \rightarrow Q \) l'\firstocc{ensemble des fonctions}{ensemble!des fonctions}\ (totales) de \( P \) vers \( Q \);
    \item \( P\times Q \) l'\firstocc{ensemble des couples}{ensemble!des couples} \( (p, q) \) avec \( p \) dans \( P \) et \( q \) dans \( Q \);
    \item \( P+Q \) l'\firstocc{ensemble somme}{ensemble!sommes} isomorphe à \( P \times \{0\} \cup Q \times \{1\} \), égal à \( P\cup Q \) si \( P \) et \( Q \) sont disjoints;
    \item \( \circ \) la \firstocc{composition de fonctions}{composition!de fonctions}, élément de
    \begin{equation*}
      (Q\rightarrow R)\times(P\rightarrow Q) \rightarrow P\rightarrow R,
    \end{equation*}
    définie par
    \begin{equation*}
      (f\circ g) (x) = f(g(x));
    \end{equation*}
    \item \( \mathrm{Id} \) la \firstocc{fonction identité}{fonction!identité}, élément de \( P \rightarrow P \), définie par
    \begin{equation*}
      \mathrm{Id}(x) = x;
    \end{equation*}
    \item \( \bot \) l'élément \firstocc{indéfini}{élément!indéfini} permettant de définir
    toute fonction partielle de \( P \) vers \( Q \) par une fonction totale de \( P\rightarrow Q + \{\bot \} \).
  \end{itemize}

  Si \( \sim \) est une relation d'équivalence sur \( P \), on notera \( P_\sim \) l'ensemble de ses classes d'équivalence et \( {[p]}_\sim \) la classe d'équivalence d'un élément \( p \) de \( P \), voire \( [p] \) s'il n'y a pas d'ambiguïté.

\section*{Les monoïdes, semi-anneaux et semimodules}
Dans la seconde partie de ce document, nous manipulerons des notions de théorie des catégories\footnote{monoïde objet de catégorie monoïdale} permettant d'abstraire et de généraliser des réflexions sur des structures algébriques particulières, notamment en unifiant les réflexions sur les structures algébriques associées respectivement aux mots et aux arbres.
Ainsi, commençons par rappeler les structures algébriques classiquement liées aux automates de mots.

Un \firstocc{monoïde}{monoïde} \( \mathcal{M} \) est un triplet \( (M, \cdot, \varepsilon) \) avec \( M \) un ensemble, \( \cdot \) une loi de composition interne à \( M \) associative et unitaire pour l'élément \( \varepsilon \) de \( M \).
Un monoïde est dit \firstocc{commutatif}{monoïde!commutatif} si sa loi de composition l'est également.
\newline
Un \firstocc{morphisme}{morphisme!de monoïde} d'un monoïde
\((M, \cdot, \varepsilon)\) en un monoïde \((M', \cdot', \varepsilon')\)
est une fonction \(\phi \)
de \(M \rightarrow M'\) satisfaisant
\begin{align*}
  \phi(\varepsilon) &= \varepsilon', & \phi(m_1\cdot m_2) &= \phi(m_1) \cdot' \phi(m_2).
\end{align*}
Le \firstocc{monoïde libre}{monoïde!libre} engendré par un ensemble \( S \) est le triplet \((S^*, \cdot, \varepsilon) \) avec:
\begin{itemize}
  \item \( S^* \) l'ensemble des mots écrits sur l'ensemble \( S \),
  \item \( \cdot \) la concaténation des mots,
  \item \( \varepsilon \) le mot vide.
\end{itemize}
Pour ne pas alourdir l'écriture, nous assimilerons dans la suite les symboles de \( S \) aux mots de longueur \( 1 \) de \( S^* \) lorsqu'il n'y aura pas d'ambiguïté.

Un \firstocc{semianneau}{semianneau} est un quintuplet \((K, +, \times, 1, 0)\) tel que \((K, +, 0)\) est un monoïde commutatif, \((K, \times, 1)\) est un monoïde et \(\times \) se distribue sur \(+\).
Dans la littérature, en fonction des références, l'élément \(0\) peut aussi être considéré comme un absorbant pour la multiplication.

Un \firstocc{semimodule}{semimodule} sur un semianneau \((K, +, \times, 1, 0)\) est un couple \((\mathbb{M}, :)\) tel que \(\mathbb{M} = (M, \pm, \underline{0})\) est un monoïde commutatif et \( : \) est une fonction de \(K\times M\rightarrow M\) telle que:
\begin{itemize}
  \item \(k : (m_1 \pm m_2) = (k : m_1) \pm (k : m_2) \),
  \item \((k \times k') : m = k : (k' : m)\),
  \item \(1 : m = m\),
  \item \((k + k') : m = (k : m) \pm (k' : m)\).
\end{itemize}
En fonction du contexte, un axiome supplémentaire peut être inséré:
\begin{itemize}
  \item \(0 : m = k : \underline{0} = \underline{0}\).
\end{itemize}
Un \firstocc{morphisme}{morphisme!de semimodule} d'un semimodule \((\mathbb{M}, :)\) en un semimodule \((\mathbb{M}', :')\) est un morphisme \(\phi \) du monoïde \(\mathbb{M}\) en \(\mathbb{M}'\) satisfaisant
\begin{equation*}
  \phi(k : m) = k :' \phi(m).
\end{equation*}
Le \firstocc{semimodule libre}{semimodule!libre} sur un semianneau \(\mathbb{K}=(K, +, \times, 1, 0)\) engendré par un ensemble \(S\) est la structure \((\mathbb{K}:S, :)\) définie comme suit:
\begin{itemize}
  \item \(\mathbb{K}:S = (K:S, \pm, \underline{0})\) est le monoïde commutatif défini par
    \begin{itemize}
      \item \(K:S\) l'ensemble des fonctions de \(S \rightarrow K\) telles que seul un nombre fini d'éléments de \(S\) ont une image non-nulle,
      \item \((f \pm f')(x) = f(x) + f'(x)\),
      \item \(\underline{0}(x) = 0\).
    \end{itemize}
  \item \((k : f)(x) = k \times (f (x))\).
\end{itemize}
De façons équivalentes, \(K:S\) peut être vu comme une somme (finie, formelle) d'éléments de \(K\times S\), ou comme un sous-ensemble fini de \(K\times S\).

\section*{Les automates de mots}
Intéressons-nous maintenant à la première structure d'automates que nous allons aborder dans ce document, les automates de mots.
Dans cette section, nous verrons comment rapprocher dans leurs mécanismes d'utilisation les automates déterministes, non-déterministes et à multiplicités.
Dans la seconde partie de ce document, nous aborderons également les automates alternants (et leur généralisation) ainsi que les automates à pile.
Toutes ces structures seront factorisables en une unique structure polymorphe; pour cela, commençons par mettre en lumière les similitudes de fonctionnement, et plus particulièrement au sein de leurs fonctions de transition.

Un \firstocc{automate non-déterministe}{automate!de mots!non-déterministe} est un quintuplet \( A=(\Sigma,Q,I,F,\delta) \) avec
\begin{itemize}
  \item \( \Sigma \) un ensemble, l'\firstocc{alphabet}\ de l'automate,
  \item \( Q \) l'\firstocc{ensemble des états}\ de l'automate,
  \item \( I\subset Q \) l'ensemble des états \firstocc{initiaux}\ de l'automate,
  \item \( F\subset Q \) l'ensemble des états \firstocc{finaux}\ de l'automate,
  \item \( \delta \) la \firstocc{fonction de transition}\ de l'automate, élément de \( Q\times \Sigma \rightarrow 2^Q \).
\end{itemize}
La fonction \( \delta \) pourra être assimilée naturellement à l'ensemble de triplets défini par \( (p,a,q)\in\delta \Leftrightarrow q\in\delta(p,a) \).
Cette fonction peut aussi être étendue comme la fonction \(\delta'\) de \(2^Q \times \Sigma^* \rightarrow 2^Q\) par récurrence sur les mots de la façon suivante:
\begin{align*}
  \delta'(Q',\varepsilon) &= Q', & \delta'(Q', a\cdot w) &= \delta'(\delta''(Q', a), w'),
\end{align*}
où \(\delta''\) est la fonction de \(2^Q\times\Sigma \rightarrow 2^Q\) définie par
\begin{equation*}
  \delta(Q', a) = \bigcup_{q'\in Q'} \delta(q', a).
\end{equation*}
Dans la suite, nous assimilerons les fonctions \(\delta \), \(\delta'\) et \(\delta''\).
\newline
Le \firstocc{langage reconnu}{langage!reconnu par un automate!de mots} par un automate \( A=(\Sigma,\_,I,F,\delta) \) est l'ensemble des mots envoyant un état initial sur un état final, plus formellement défini comme
\begin{equation*}
  L(A) = \{w\in\Sigma^* \mid \delta(I,w)\cap F\neq\emptyset \}.
\end{equation*}

Un \firstocc{automate déterministe}{automate!de mots!déterministe} est un quintuplet \( A=(\Sigma,Q,i,F,\delta) \) avec
\begin{itemize}
  \item \( \Sigma \) un ensemble, l'\firstocc{alphabet}\ de l'automate,
  \item \( Q \) l'\firstocc{ensemble des états}\ de l'automate,
  \item \( i\in Q + \{\bot \} \) l'état \firstocc{initial}\ de l'automate,
  \item \( F\subset Q \) l'ensemble des états \firstocc{finaux}\ de l'automate,
  \item \( \delta \) la \firstocc{fonction de transition}\ de l'automate, élément de \( Q\times \Sigma \rightarrow Q + \{\bot \} \).
\end{itemize}
La fonction \( \delta \) peut être assimilée naturellement à une fonction de \( (Q + \{\bot \})\times \Sigma^*\rightarrow Q + \{\bot \} \) par récurrence sur la longueur des mots comme suit:
\begin{align*}
  \delta'(p,\varepsilon) &= p, & \delta'(p, a\cdot w) &= \delta'(\delta''(p, a), w'),
\end{align*}
où \(\delta''\) est la fonction de \((Q + \{\bot \})\times\Sigma \rightarrow Q + \{\bot \} \) définie par
\begin{equation*}
  \delta''(p, a) =
    \begin{cases}
      \bot & \text{ si } p = \bot,\\
      \delta(p, a) & \text{ sinon.}
    \end{cases}
\end{equation*}
Dans la suite, comme précédemment, nous assimilerons les fonctions \(\delta \), \(\delta'\) et \(\delta''\).

\begin{remarqueBox}[label={remExtdelta}]{Extension des fonctions de transition}
  L'extension de la fonction
  \(\delta \),
  que ce soit dans le cas des automates déterministes comme non-déterministes, suit le même processus.

  La fonction \(\delta'' \) peut être vue comme une \emph{promotion}
  de la fonction \( \delta \) de \(Q \times \Sigma \rightarrow F(Q)\)
  en une fonction de signature \(F(Q) \times \Sigma \rightarrow F(Q)\)
  où \(F\) est une \emph{transformation}\footnote{spoiler alert: c'est un foncteur, et même une monade.} de l'ensemble \(Q\) et des fonctions associées: le passage à l'ensemble des parties dans le cas des automates non-déterministes, l'adjonction d'un élément dans le cas des automates déterministes.

  Ensuite, la fonction \(\delta'\) peut être vue comme la \emph{promotion} de la fonction \(\delta''\) de \(F(Q) \times \Sigma \rightarrow F(Q)\) en une fonction de signature \(F(Q) \times \Sigma^* \rightarrow F(Q)\); remarquons qu'en cas d'application partielle d'un symbole, \(\delta''_a\) est un élément du monoïde \((F(Q) \rightarrow F(Q), \circ,\mathrm{Id})\).
  D'une façon équivalente\footnote{à une transformation naturelle près pour être précis.}, \(\delta'\) promeut\footnote{spoiler alert: il s'agit adjonction fonctorielle} \(\delta''\) de fonction de \(\Sigma \rightarrow (F(Q)\rightarrow F(Q))\) en morphisme du monoïde libre \((\Sigma^*,\cdot,\varepsilon)\) vers \((F(Q) \rightarrow F(Q), \circ,\mathrm{Id})\).

  Cette extension, que nous retrouverons dans le cas des arbres, sera à la base des constructions unifiées et de la définition de langage dans la seconde partie de ce document.
\end{remarqueBox}
\noindent Le \firstocc{langage reconnu}{langage!reconnu par un automate!déterministe de mots} par un automate déterministe \( A=(\Sigma,\_,i,F,\delta) \) est l'ensemble
\begin{equation*}
  L(A) = \{w\in\Sigma^* \mid \delta(i,w)\in F\}.
\end{equation*}
Un automate non-déterministe peut être transformé en un automate déterministe en construisant la partie accessible de l'automate des parties~\cite{RS59}.
De plus, nous assimilerons également les automates déterministes et les automates non-déterministes dont l'ensemble des états initiaux \(I\) est vide ou un singleton, et tel que pour tout mot \(w\), \(|\delta(I,w)|\leq 1\).
Ainsi, ces deux structures admettent des pouvoirs de représentation identiques.

Si le processus d'extension des fonctions de transition est généralisable (Remarque~\ref{remExtdelta}), nous verrons que c'est également le cas de la définition des langages, par l'intermédiaire de la notion de poids d'un mot (poids booléen reflétant l'appartenance d'un mot au langage reconnu dans le cas présent).
Ce sera également le cas de la structure générale des automates, en considérant que le comportement d'un automate (quelque soit son type) est d'envoyer, par l'action d'une entrée, une configuration initiale sur une configuration qui sera à terme évaluée pour l'obtention d'un poids.
Nous verrons que cette description est valable dans le cas des automates, de mots ou d'arbres, déterministes, non déterministes, à multiplicités, alternants, et même à pile.
Dernière illustration avec la fonction de transition des automates à multiplicités, définis comme suit.

Un \firstocc{automate à multiplicités}{automate!de mots!à multiplicités} sur un semianneau \((K, +, \times, 1, 0)\) est un quintuplet \( A=(\Sigma, Q, I, F, \delta) \) avec
\begin{itemize}
  \item \( \Sigma \) un ensemble, l'\firstocc{alphabet}\ de l'automate,
  \item \( Q \) l'\firstocc{ensemble des états}\ de l'automate,
  \item \( I \in K:Q \) la \firstocc{configuration  initiale}\ de l'automate,
  \item \( F \in K:Q\) l'ensemble des états \firstocc{finaux}\ de l'automate,
  \item \( \delta \) la \firstocc{fonction de transition}\ de l'automate, élément de \( Q\times \Sigma \rightarrow K:Q \).
\end{itemize}
Comme précédemment, la fonction \( \delta \) peut être étendue comme une fonction de \( (K:Q)\times \Sigma^*\rightarrow (K:Q) \) par récurrence sur la longueur des mots comme suit:
\begin{align*}
  \delta'(f,\varepsilon) &= p, & \delta'(f, a\cdot w) &= \delta'(\delta''(f, a), w'),
\end{align*}
où \(\delta''\) est la fonction de \((K:Q)\times\Sigma \rightarrow (K:Q)\) définie par
\begin{equation*}
  \delta''(\{(k_1, q_1), \ldots, (k_n, q_n)\}, a) = q_1 : \delta(q_1, a) \pm \cdots \pm k_n : \delta(q_n, a).
\end{equation*}
Dans la suite, nous assimilerons les fonctions \(\delta \), \(\delta'\) et \(\delta''\).
\newline
Le \firstocc{poids d'un mot}{poids!d'un mot!dans un automate} \(w\) dans un automate à multiplicités \( A=(\Sigma, \_, I, F, \delta) \) sur un semianneau \((K, +, \times, 1, 0)\) est le scalaire
\begin{equation*}
  \mathrm{weight}_A(w) = k_1 \times F(q_1) + \cdots + k_n \times F(q_n)
\end{equation*}
où
\begin{equation*}
  \delta(I, w) = \{(k_1, q_1), \ldots, (k_n, q_n)\}.
\end{equation*}
Un automate non-déterministe peut être vu, d'une façon équivalente, comme un automate à multiplicité sur le semianneau booléen, et réciproquement.


\section*{Les expressions rationnelles}
Les expressions rationnelles sont une façon compacte de représenter un ensemble de mots.
Dans cette section, nous présenterons uniquement des expressions rationnelles simples et sans multiplicités; cependant, nous manipulerons d'autres types d'expressions dans la seconde partie de ce document.
Nous montrerons ensuite comment transformer une expression rationnelle en un automate équivalent dans la première partie de ce document en présentant différentes méthodes de construction dans le cas des expressions d'arbres, généralisations des expressions de mots, dont la proximité nous permettra de raisonner pour unifier ces deux types d'expressions.

Une \firstocc{expression rationnelle}{expression rationnelle} \( E \) sur un alphabet \( \Sigma \) est définie inductivement comme suit:
\begin{align*}
  E &= a, & E &= \emptyset, & E &= \varepsilon,\\
  E &= (F + G), & E &= (F \cdot G), & E &= (F^*),
\end{align*}
avec \( a \) un symbole de \( \Sigma \), et \( F \) et \( G \) deux expressions rationnelles sur \( \Sigma \).
Les parenthèses pourront être omises en considérant la concaténation prioritaire sur la somme, l'étoile comme l'opération la plus prioritaire et les opérations binaires associatives à gauche.
\newline
Le \firstocc{langage dénoté par une expression rationnelle}{langage!dénoté par une expression rationnelle!de mots} \( E \) est l'ensemble \( L(E) \) défini inductivement comme suit:
\begin{align*}
  L(a) &= \{a\}, & L(\emptyset) &= \emptyset, & L(\varepsilon) &= \{\varepsilon \},\\
  L(F+G) &= L(F) \cup L(G), & L(F\cdot G) &= L(F)\cdot L(G), & L(F^*) &= {(L(F))}^*,
\end{align*}
avec \( a \) un symbole de \( \Sigma \), \( F \) et \( G \) deux expressions rationnelles sur \( \Sigma \), et \( ^* \) l'opération définie pour tout sous-ensemble \( L \) de \( \Sigma^* \) par
\begin{align*}
  L^* &= \bigcup_{n\geq 0} L^n,  &
  \text{avec } L^n &=
    \begin{cases}
      \{\varepsilon \} & \text{ si } n = 0,\\
      L\cdot L^{n - 1} & \text{ sinon.}
    \end{cases}
\end{align*}
Tout langage dénoté par une expression rationnelle est appelé \firstocc{langage rationnel}{langage!rationnel}.

Le lien entre expressions rationnelles et automates, c'est-à-dire celui entre langages rationnels et reconnaissables, s'énonce \emph{via} le \firstocc{théorème de Kleene}{théorème!de Kleene}~\cite{Kle56}: tout langage est reconnaissable par un automate si et seulement s'il est dénoté par une expression rationnelle.
En d'autres termes, les langages rationnels sont exactement les langages reconnaissables.
Une preuve constructive de ce théorème peut être faite en considérant des algorithmes de conversion préservant le langage.

Un automate peut être converti en une expression rationnelle équivalente par des méthodes telles que l'élimination d'états~\cite{BMC63}, la construction inductive~\cite{MY60}, ou la résolution de systèmes en utilisant le lemme d'Arden~\cite{Ard61}.

Concernant la transformation inverse, nous présenterons diverses méthodes la réalisant dans la suite de ce document en nous intéressant aux langages d'arbres, généralisations des automates de mots, comme présenté dans les sous-sections suivantes.

\section*{Les arbres}
Les arbres sont une extension plus structurée de la notion de mots; en effet, chaque symbole n'est plus potentiellement suivi par un mot, mais par une liste d'arbres, dont la longueur est fixé par la racine (son arité).
Nous nous efforcerons, dans la suite de ce document, à illustrer les mécanismes communs entre les structures de mots et d'arbres pour finalement proposer des reconnaisseurs factorisés.
Pour cela, commençons donc par introduire des définitions préliminaires.

Un ensemble \( S \) est \firstocc{gradué}{ensemble!gradué} s'il existe une famille \( {(S_k)}_{k\in\mathbb{N}} \) d'ensembles disjoints telle que \( S=\bigcup_{k\in\mathbb{N}} S_k \).
Une \firstocc{fonction graduée}{fonction!graduée} entre deux ensembles gradués \( S \) et \( \underline{S} \) est une fonction de \( S \) dans \( \underline{S} \) qui envoie, pour tout entier \( k \), \( S_k \) dans \( \underline{S}_k \).

Un arbre sur un alphabet gradué \( \Sigma \) est soit l'arbre vide \( \varepsilon \) d'arité \( 1 \), soit un couple constitué d'un symbole \( f \) de \( \Sigma \) d'arité \( k \) (sa \firstocc{racine}) et d'un \( k \)-uplet d'arbres \( (t_1,\ldots,t_k) \) (ses \firstocc{fils}) sur \( \Sigma \).
L'arité d'un arbre \( t \), notée \( \arite(t) \) est de \( 1 \) si \( t=\varepsilon \), \( \sum_{1\leq i\leq k} \arite(t_i) \) si \( t=\_(t_1,\ldots,t_k) \).
Un \firstocc{sous-arbre}\ d'un arbre \( t \) est soit \( t \), soit un sous-arbre d'un de ses fils.
Nous noterons \( s\preccurlyeq t \) la relation ``\( s \) est un sous-arbre de \( t \)''.
La racine d'un arbre non-vide \( t \) est notée \( \racine(t) \), son \( k \)\up{e} fils (s'il existe) \( \fils{k}(t) \), l'ensemble de ses \firstocc{feuilles}\ (c'est à dire l'ensemble des symboles d'arité \( 0 \) présents) \( \feuilles(t) \).

  Les \firstocc{prédécesseurs}\ d'un symbole \( f \) dans un arbre \( t \) sont les symboles qui apparaissent directement ``au-dessus de lui'', c'est à dire les racines des sous-arbres de \( t \) dont un des fils a pour racine \( f \).
  On note \( \parent(t,f) \), pour un arbre \( t \) et un symbole \( f \) les couples
      \begin{equation*}
        \parent(t,f) = \{(g,i)\in\Sigma_l\times\mathbb{N} \mid \exists g(s_1,\ldots,s_l)\preccurlyeq t, \racine(s_i)=f\}.
      \end{equation*}
  Ces couples relient les prédécesseurs de \( f \) et les indices des sous-arbres de \( t \) dont \( f \) est la racine.
  Pour un arbre \( t=g(t_1,\ldots,t_k) \) et un symbole \( f \), et selon la définition de la structure d'un arbre, un prédécesseur de \( f \) dans \( t \) est un prédécesseur de \( f \) dans un sous-arbre \( t_i \) de \( t \), ou \( g \) si \( f \) est la racine d'un sous-arbre \( t_i \) de \( t \).
  Ainsi:
    \begin{equation*}
      \parent(t,f) = \bigcup_{i\leq n} \parent(t_i,f) \cup \{(g,i)\mid f\in\racine(t_i)\}.
    \end{equation*}
  L'ensemble des arbres d'arité \( 0 \) sur l'alphabet gradué \( \Sigma \), noté \( \Sigma^*_0 \), sera noté \( \Sigma^* \) lorsqu'il n'y aura pas d'ambiguïté.

\section*{Les opérades}\label{pageOperades}

Si le monoïde est la structure algébrique de prédilection pour les mots, l'opérade est celle pour les arbres.

Une \firstocc{opérade}{opérade}~\cite{Laz55,May06} est une structure algébrique qui simule la composition des fonctions.
Plus formellement, une opérade \( \mathcal{O}=(O,{(\circ_{j})}_{j\in\mathbb{N}},\mathrm{Id}) \) est constituée d'un ensemble gradué \( O \) et d'un élément remarquable \( \mathrm{Id} \) dans \( O_1 \), munis des opérations de composition \( \circ_{j} \) définies pour tout triplet d'entiers \( (i, j, k) \)\footnote{tout couple \( (i,k) \) définit une fonction \( \circ_j \) sans ambiguïté sur le domaine ni sur le codomaine.}
avec \( 0<j\leq k \) dans \( O_k\times O_{i} \rightarrow O_{k+i-1} \) satisfaisant les trois conditions suivantes:
pour tous trois éléments \( p_1 \) dans \( O_m \), \( p_2 \) dans \( O_n \) et \( p_3 \) dans \( O_p \):
  \begin{enumerate}
    \item pour tout entier \( 0<j\leq m \):
      \begin{equation*}
        \mathrm{Id} \circ_1 p_1 = p_1 \circ_j \mathrm{Id} =p_1,
      \end{equation*}
    \item (\firstocc{associativité verticale}{associativité!verticale}, voir Figure~\ref{fig vert assoc}) pour tous deux entiers \( 0<j\leq m \) et \( 0<j'\leq n \):
      \begin{equation*}
        p_1 \circ_j (p_2 \circ_{j'} p_3) = (p_1 \circ_j p_2) \circ_{j+{j'}-1} p_3,
      \end{equation*}
    \item (\firstocc{associativité horizontale}{associativité!horizontale}, voir Figure~\ref{fig hori assoc}) pour tous deux entiers \( 0<{j'}\leq j\leq m \):
      \begin{equation*}
        (p_1 \circ_j p_2) \circ_{j'} p_3 = (p_1 \circ_{j'} p_3) \circ_{j+p-1} p_2.
      \end{equation*}
  \end{enumerate}
		\begin{figure}[H]
		 \begin{minipage}{0.45\linewidth}
		  \centerline{
			  \begin{tikzpicture}
			    \foreach \x/\y/\z in {0/0/1,2/-4.5/3,1/-2.5/2}{
		        \draw (\x,\y) rectangle (\x+1,\y+0.5);
		        \draw (\x+0.5,\y+0.25) node {\( p_{\z} \)};
		        \foreach \n in {0.1,0.3,...,0.9}{
		          \draw (\x+\n,\y) -- (\x+\n,\y-0.3);
		        };
		      };
		      \node[shape=circle,fill=black,inner sep=0.5pt] at (0.7,-0.25)  {};
		      \node at (0.7,-0.4)  {\tiny{\( j \)}};
		      \draw (0.7,-0.5) -- (1.5,-2);
		      \node[shape=circle,fill=black,inner sep=0.5pt] at (1.7,-2.75)  {};
		      \node at (1.7,-2.9)  {\tiny{\( {j'} \)}};
		      \draw (0.5,-1.5) rectangle (3.5,-5.5);
		      \node at (3,-1.8)  {\tiny{\( p_2\circ_{j'} p_3 \)}};
		      \foreach \n in {0.7,0.9,1.1, 1.6,1.8,2,2.2,2.4, 2.7}{
		        \draw (0.5+\n,-5.5) -- (0.5+\n,-5.8);
		      };
		      \draw (1.7,-3) -- (2.5,-4);
		    \end{tikzpicture}
		  }
		 \end{minipage}
		 \hfill
		 \begin{minipage}{0.05\linewidth}
		   \vfill
		   \( = \)
		   \vfill
		 \end{minipage}
		 \hfill
		 \begin{minipage}{0.45\linewidth}
		  \centerline{
			  \begin{tikzpicture}
			    \foreach \x/\y/\z in {0/0/1,2/-4.5/3,1/-1.5/2}{
		        \draw (\x,\y) rectangle (\x+1,\y+0.5);
		        \draw (\x+0.5,\y+0.25) node {\( p_{\z} \)};
		        \foreach \n in {0.1,0.3,...,0.9}{
		          \draw (\x+\n,\y) -- (\x+\n,\y-0.3);
		        };
		      };
		      \node[shape=circle,fill=black,inner sep=0.5pt] at (0.7,-0.25)  {};
		      \node at (0.7,-0.4)  {\tiny{\( j \)}};
		      \draw (0.7,-0.5) -- (1.5,-1);
		      \node[shape=circle,fill=black,inner sep=0.5pt] at (1.7,-1.75)  {};
		      \node at (1.7,-1.9)  {\tiny{\( {j'} \)}};
		      \draw (-0.5,1) rectangle (2.5,-2.5);
		      \node at (2,0.8)  {\tiny{\( p_1\circ_j p_2 \)}};
		      \foreach \n in {0.5,0.7,0.9,1.3,1.5,1.7,1.9,2.1,2.7}{
		        \draw (-0.5+\n,-2.5) -- (-0.5+\n,-2.8);
		      };
		      \draw (1.4,-3) -- (2.5,-4);
		      \node[shape=circle,fill=black,inner sep=0.5pt] at (1.4,-2.75)  {};
		      \node at (1.4,-2.9) {\tiny{\( j+{j'}-1 \)}};
		    \end{tikzpicture}
		  }
		 \end{minipage}
		  \caption{Associativité verticale.}%
		  \label{fig vert assoc}
		\end{figure}
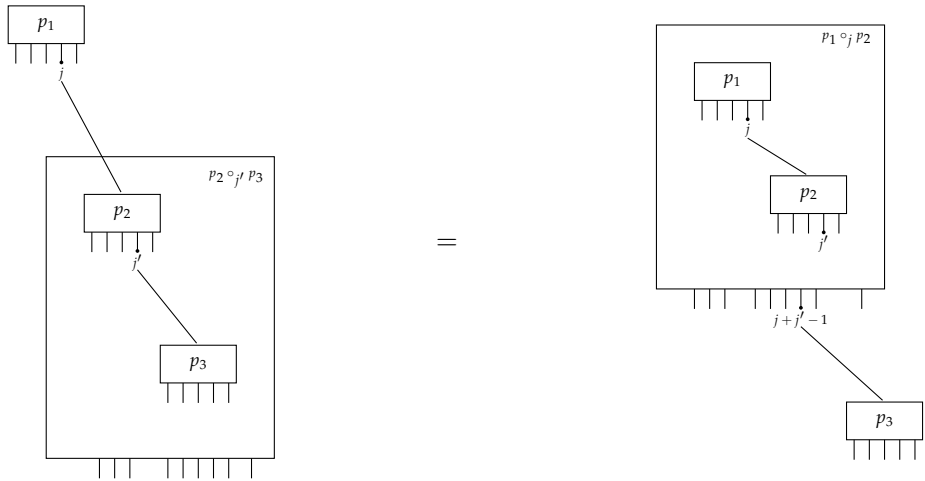
		\begin{figure}[H]
		 \begin{minipage}{0.45\linewidth}
		  \centerline{
			  \begin{tikzpicture}
			    \foreach \x/\y/\z in {0/0/1,-1/-4.5/3,1/-1.5/2}{
		        \draw (\x,\y) rectangle (\x+1,\y+0.5);
		        \draw (\x+0.5,\y+0.25) node {\( p_{\z} \)};
		        \foreach \n in {0.1,0.3,...,0.9}{
		          \draw (\x+\n,\y) -- (\x+\n,\y-0.3);
		        };
		      };
		      \node[shape=circle,fill=black,inner sep=0.5pt] at (0.3,-0.25)  {};
		      \node at (0.3,-0.4)  {\tiny{\( {j'} \)}};
		      \node[shape=circle,fill=black,inner sep=0.5pt] at (0.7,-0.25)  {};
		      \node at (0.7,-0.4)  {\tiny{\( j \)}};
		      \draw (0.7,-0.5) -- (1.5,-1);
		      \node[shape=circle,fill=black,inner sep=0.5pt] at (0.9,-0.25)  {};
		      \node at (0.9,-0.4)  {\tiny{\( m \)}};
		      \node[shape=circle,fill=black,inner sep=0.5pt] at (1.9,-1.75)  {};
		      \node at (1.9,-1.9)  {\tiny{\( n \)}};
		      \draw (-0.5,1) rectangle (2.5,-2.5);
		      \node at (2,0.8)  {\tiny{\( p_1\circ_j p_2 \)}};
		      \foreach \n in {0.5,0.7,0.9,1.3,1.5,1.7,1.9,2.1,2.7}{
		        \draw (-0.5+\n,-2.5) -- (-0.5+\n,-2.8);
		      };
		      \node[shape=circle,fill=black,inner sep=0.5pt] at (0.2,-2.75)  {};
		      \node at (0.2,-2.9) {\tiny{\( {j'} \)}};
		      \draw (0.2,-3) -- (-.5,-4);
		      \node[shape=circle,fill=black,inner sep=0.5pt] at (2.2,-2.75)  {};
		      \node at (2.2,-2.9) {\tiny{\( m+n-1 \)}};
		    \end{tikzpicture}
		  }
		 \end{minipage}
		 \hfill
		 \begin{minipage}{0.05\linewidth}
		   \vfill
		   \( = \)
		   \vfill
		 \end{minipage}
		 \hfill
		 \begin{minipage}{0.45\linewidth}
		  \centerline{
			  \begin{tikzpicture}
			    \foreach \x/\y/\z in {0/0/1,-1/-1.5/3,1/-4.5/2}{
		        \draw (\x,\y) rectangle (\x+1,\y+0.5);
		        \draw (\x+0.5,\y+0.25) node {\( p_{\z} \)};
		        \foreach \n in {0.1,0.3,...,0.9}{
		          \draw (\x+\n,\y) -- (\x+\n,\y-0.3);
		        };
		      };
		      \node[shape=circle,fill=black,inner sep=0.5pt] at (0.3,-0.25)  {};
		      \node at (0.3,-0.4)  {\tiny{\( {j'} \)}};
		      \node[shape=circle,fill=black,inner sep=0.5pt] at (0.7,-0.25)  {};
		      \node at (0.7,-0.4)  {\tiny{\( j \)}};
		      \node[shape=circle,fill=black,inner sep=0.5pt] at (-0.1,-1.75)  {};
		      \node at (-0.1,-1.9)  {\tiny{\( p \)}};
		      \draw (-1.5,-2.5) rectangle (1.5,1);
		      \node at (-1,0.8)  {\tiny{\( p_1\circ_{j'} p_3 \)}};
		      \draw (0.3,-0.5) -- (-0.5,-1);
		      \foreach \n in {0.5, 0.9,1.1,1.3,1.5,1.7, 2.3,2.5,2.7}{
		        \draw (-1.5+\n,-2.5) -- (-1.5+\n,-2.8);
		      };
		      \node[shape=circle,fill=black,inner sep=0.5pt] at (1,-2.75)  {};
		      \node at (1,-2.9)  {\tiny{\( j+p-1 \)}};
		      \draw (1,-3) -- (1.5,-4);
		    \end{tikzpicture}
		  }
		 \end{minipage}
		  \caption{Associativité horizontale.}%
		  \label{fig hori assoc}
		\end{figure}
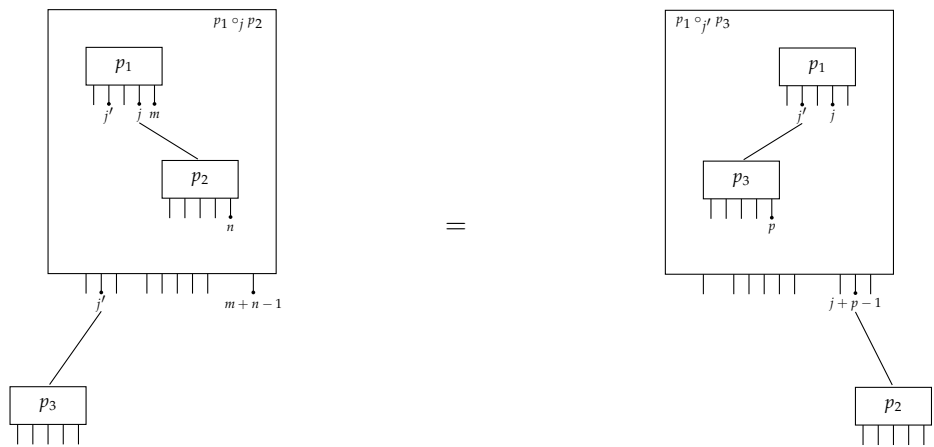
En combinant ces compositions \( \circ_j \), il est possible de définir une composition \( \circ \) envoyant \( O_k \times O_{i_1}\times \cdots\times O_{i_k} \) dans \( O_{i_1+\cdots+i_k} \):
pour tout élément \( (p, q_{1}, \ldots, q_{k}) \) de \( O_k \times O^k \),
\begin{equation*}
  p \circ (q_{1},\ldots,q_{k})= (\cdots((p \circ_k q_{k})\circ_{k-1} q_{{k-1}}\cdots)\cdots)\circ_1 q_{1}.
\end{equation*}
Inversement, la composition \( \circ \) permet de définir les compositions \( \circ_j \) en complétant la composition par l'élément neutre de l'opérade:
pour tout élément \( (p,q) \) de \( O_k\times O_i \), pour tout entier \( 0<j\leq k \):
\begin{equation*}
  p\circ_j q = p\circ (\underbrace{\mathrm{Id},\ldots,\mathrm{Id}}_{j-1\text{ fois}},q,\underbrace{\mathrm{Id},\ldots,\mathrm{Id}}_{k-j\text{ fois}}).
\end{equation*}

  Dans la suite de ce document, nous nous intéresserons à plusieurs opérades remarquables.
  Par exemple, l'ensemble des fonctions envoyant une puissance cartésienne d'un ensemble dans lui-même, ou dualement cet ensemble sur une de ses puissances cartésiennes, où la puissance 0 d'un ensemble est l'ensemble \( \mathbb{1} \) ne possédant qu'un seul élément noté \( () \), sont tous les deux des ensembles que l'on peut munir d'une structure d'opérade.
  Notons d'ailleurs qu'un élément \(x\) d'un ensemble \(S\) est naturellement équivalent à une fonction \( x() =x \) de \( \mathbb{1} \) vers \( S \).


Un \firstocc{morphisme}{opérade!morphisme d'} \( \phi \) d'une opérade \( \mathcal{O}=(O,\circ,\mathrm{Id}) \) en une opérade \( \underline{\mathcal{O}}=(\underline{O},\ucirc,\underline{\mathrm{Id}}) \) est une fonction graduée de \( O \) dans \( \underline{O} \) satisfaisant pour tout couple d'arbres \( (p,q) \) de \( O_m\times O \) et pour tout entier \( 0<j\leq m \):
\begin{align*}
  \phi(p \circ_j q) &= \phi(p) \ucirc_j \phi(q), &
  \phi(\mathrm{Id}) &= \underline{\mathrm{Id}}.
\end{align*}

L'\firstocc{opérade libre}{opérade!libre} générée par un ensemble gradué \( S \) est l'opérade \( (S^*,\circ_j,\varepsilon) \) avec:
\begin{itemize}
  \item \( S^* \) l'ensemble gradué des arbres étiquetés par des éléments de \( S \),
  \item \( \circ_j \) la composition des arbres,
  \item \( \varepsilon \) l'arbre vide.
\end{itemize}
Pour ne pas alourdir l'écriture, nous assimilerons dans la suite les symboles de \( S \) aux arbres de hauteur \( 1 \) de \( S^* \) lorsqu'il n'y aura pas d'ambiguïté.

\section*{Les automates d'arbres ascendants}\label{section Def Aut Arbres}
Comme pour les mots, il est possible de décrire des reconnaisseurs pour les arbres.
Cependant, contrairement aux mots, les arbres ne sont pas des structures \emph{symétriques}: si un mot peut se lire d'une façon similaire de gauche à droite ou de droite à gauche, un arbre n'est pas interprété de la même façon s'il est lu de bas en haut ou de haut en bas.
Cette dissymétrie inhérente se traduit par deux types de reconnaisseurs dont le fonctionnement diffère\footnote{mais que nous pourrons uniformiser par la suite}: les automates ascendants et les automates descendants, interprétant les arbres de bas en haut ou de haut en bas.

Un \firstocc{automate d'arbres ascendant non-déterministe}{automate!d'arbres!ascendant!non-déterministe} (ascendant ou Bottom-Up, ici sans variables\footnote{les variables, apparaissant uniquement comme feuilles, seront prises en compte dans la seconde partie de ce document}) est un quadruplet \( A=(\Sigma,Q,F,\delta) \) avec:
\begin{itemize}
  \item \( \Sigma \) un ensemble gradué, l'\firstocc{alphabet}\ de l'automate,
  \item \( Q \) l'ensemble des \firstocc{états}\ de l'automate,
  \item \( F\subset Q \) l'ensemble des états \firstocc{finaux}\ de l'automate,
  \item \( \delta \) la fonction de \firstocc{transition}\ de l'automate, appartenant à \( Q^n\times \Sigma_n\rightarrow 2^Q \).
\end{itemize}
Comme dans le cas des mots, la fonction \(\delta \) peut être étendue en une fonction \(\delta'\) de signature \({(2^{Q})}^n \rightarrow \Sigma_n^*\rightarrow 2^Q\) d'une façon relativement proche de l'extension de la fonction de transition des automates de mots:
\begin{align*}
  \delta'(Q', \varepsilon) &= Q', &
  \delta'((Q_1,\ldots,Q_n), t \circ_j f) &=
    \delta'((Q_1, \ldots, Q_{j - 1}, \delta''((Q_j, \ldots, Q_{j + \arite(f)}), f),\ldots,Q_n ), t)
\end{align*}
où \(\delta''\) est la fonction de \({(2^{Q})}^n \times \Sigma_n\rightarrow 2^Q\) définie par
\begin{equation*}
  \delta''(Q_1, \ldots, Q_n, f) = \bigcup_{(q_1,\ldots,q_n)\in Q_1\times\cdots\times Q_n} \delta(q_1,\ldots,q_n,f).
\end{equation*}
On remarquera que cette extension est équivalente à la suivante, plus classique:
\begin{align*}
  \delta'(Q', \varepsilon) &= Q', \\
  \delta'((Q_1,\ldots,Q_n), f(t_1,\ldots,t_k)) &=
    \delta((\delta'((Q_1,\ldots,Q_{\arite(t_1)}), t_1),\ldots,\delta'((Q_{n - \arite(t_k) + 1}, \ldots, Q_n), t_k)), f).
\end{align*}
Nous assimilerons naturellement par la suite toutes ces fonctions à la fonction \( \delta \) et nous les distinguerons en fonction de leurs signatures.
De même, nous assimilerons la fonction \( \delta \) à l'ensemble de triplets défini par
\begin{equation*}
  ((p_1,\ldots,p_n),f,q)\in\delta \Leftrightarrow q\in\delta((p_1,\ldots,p_n),f).
\end{equation*}
Enfin, pour un arbre \( t \) de \( \Sigma^*_0 \), nous représenterons l'ensemble \( \delta((),t) \) par \( \delta(t) \).
Ainsi, le \firstocc{langage reconnu}{langage!reconnu par un automate!d'arbres}  par un automate d'arbres \( A=(\Sigma,\_,F,\delta) \) est l'ensemble
\begin{equation*}
  L(A)=\{t\in\Sigma^*_0\mid \delta(t)\cap F\neq\emptyset \}.
\end{equation*}

Un \firstocc{automate d'arbres ascendant déterministe}{automate!d'arbres!ascendants!déterministe}\label{defAutoArbreBUDet} est un quadruplet \( A=(\Sigma, Q, F, \delta) \) avec:
\begin{itemize}
  \item \( \Sigma \) un ensemble gradué, l'\firstocc{alphabet}\ de l'automate,
  \item \( Q \) l'ensemble des \firstocc{états}\ de l'automate,
  \item \( F\subset Q \) l'ensemble des états \firstocc{finaux}\ de l'automate,
  \item \( \delta \) la fonction de \firstocc{transition}, élément de \( Q^n\times \Sigma^n \rightarrow Q + \{\bot \} \).
\end{itemize}
Comme dans le cas des automates d'arbres ascendants non-déterministes, la fonction \( \delta \) peut être étendue comme une fonction de \( {(Q + \{\bot \})}^n\times \Sigma^*_n\rightarrow (Q + \{\bot \}) \), en ne modifiant que la formule de la fonction \(\delta''\), d'une façon similaire à la fonction \(\delta''\) associée aux automates de mots déterministes, renvoyant \(\bot \) si une des composantes du \(n\)-uplet de départ vaut \(\bot \).
De plus, nous assimilerons naturellement dans la suite de ce document les automates déterministes et les automates non-déterministes dont la fonction de transition \( \delta \) vérifie pour tout arbre \( t \) \( |\delta(t)| \leq 1 \).
\newline
Le \firstocc{langage reconnu}{langage!reconnu par un automate!déterministe d'arbres} par un automate déterministe \( A=(\Sigma,\_,F,\delta) \) est l'ensemble
\begin{equation*}
  L(A) = \{t\in\Sigma^*_0 \mid \delta(t)\in F\}.
\end{equation*}

Notoirement, un langage est reconnaissable par un automate d'arbres déterministe si et seulement s'il est reconnaissable par un automate d'arbres (par construction de la partie accessible de l'automate des parties~\cite{CDGLJLTT07}).

Dans la seconde partie de ce document, l'extension de la fonction de transition sera elle aussi utilisée dans le cas d'autres types d'automates d'arbres ascendants (déterministes, à multiplicités, sur des monoïdes multi-opérateurs) et descendants.
L'unification sera encore une fois basée sur le même principe: celui de l'envoi d'une configuration initiale sur une configuration évaluée \emph{a posteriori}.
Pour les automates d'arbres ascendants, la configuration initiale sera au départ associée la configuration constante \( () \)\footnote{ce ne sera plus le cas lors de la prise en compte des variables, permettant de définir la configuration initiale depuis une affectation de configuration aux variables}.

\section*{Les morphismes d'automates}
Comme pour les structures algébriques présentées précédemment, il est possible de considérer des fonctions particulières compatibles avec la structure d'automate.
Ces morphismes seront utilisés lors de la présentation de certains algorithmes de transformation d'expression en automate.

Soit \( \phi \) une fonction graduée entre deux alphabets \( \Sigma \) et \( \Sigma' \).
Cette fonction peut-être étendue en un \firstocc{morphisme alphabétique}{morphisme alphabétique} (il s'agit alors d'un morphisme d'opérades\footnote{spoiler alert: encore une adjonction fonctorielle.}) \(\phi'\) de \( \Sigma^* \) vers \( \Sigma'^* \) en posant
\begin{equation*}
  \phi'(f(t_1,\ldots,t_n)) = \phi(f)(\phi'(t_1),\ldots,\phi'(t_n)).
\end{equation*}
Par abus, nous assimilerons les fonctions \(\phi \) et \(\phi'\).
Par exemple, le \firstocc{morphisme de délinéarisation}{morphisme de délinéarisation} \( \mathrm{h} \) envoyant un alphabet indicé vers sa version non-indicée est un morphisme alphabétique.
Pour un langage \( L \), on note \( \phi(L) \) l'ensemble \( \{\phi(t)\mid t\in L\} \).
L'\firstocc{image par}{image d'un automate par un morphisme} \( \phi \) d'un automate \( A=(\Sigma,Q,Q_F,\delta) \) est l'automate \( \phi(A)=(\Sigma',Q,Q_F,\delta') \) où
\begin{equation*}
  \delta' = \{(q_1,\ldots,q_n,\phi(f),q) \mid (q_1,\ldots,q_n,f,q) \in\delta \}.
\end{equation*}
Par une induction triviale sur la structure des arbres, on peut montrer que
  \begin{equation*}
    \phi(L(A)) = L(\phi(A)).
  \end{equation*}
Un morphisme alphabétique est un cas particulier de \firstocc{morphisme entre deux automates}{morphisme d'automates} \( A=(\Sigma,Q,F,\delta) \) et \( B=(\Sigma',Q',F',\delta') \), qui est une fonction \( \phi \) envoyant \( \Sigma_n \) vers \( \Sigma'_n \) pour tout entier \( n \), \( Q \) vers \( Q' \), \( F \) vers \( F' \) et \( \delta \) vers \( \delta' \) telle que
\begin{equation*}
  \delta'((\phi(q_1),\ldots,\phi(q_n)),\phi(f)) = \{\phi(q) \mid q\in \delta((q_1,\ldots,q_n),f)\}.
\end{equation*}
Dans ce cas, on note \( \phi(A)=(\phi(\Sigma),\phi(Q),\phi(F),\phi(\delta)) \).
\newline
Deux automates \( A \) et \( B \) sont dits \firstocc{isomorphes}{isomorphisme d'automates} s'il existe deux morphismes \( \phi \) et \( \phi' \) satisfaisant
\begin{align*}
  A &= \phi'(\phi(A)), &
  B &= \phi(\phi'(B)).
\end{align*}

On peut montrer que certains morphismes particuliers préservent le langage d'un automate.
Nous nous intéresserons à deux types de morphismes basés sur des relations d'équivalence, et plus précisément des congruences: leur différence sera, encore une fois, la direction de lecture des transitions.

Soient \( A=(\Sigma, Q, F, \delta) \) un automate d'arbre ascendant déterministe et \( \sim \) une relation d'équivalence sur \( Q + \{\bot \} \) telle que pour tous deux éléments équivalents \( p \) et \( p' \),
\begin{equation*}
  p\in F \Leftrightarrow p'\in F.
\end{equation*}
La relation \( \sim \) est dite \firstocc{congruence Bottom-Up}{congruence!Bottom-Up} pour \( \delta \) si et seulement si pour tous deux états \( p \) et \( p' \) de \( Q \), pour tout symbole \( f \) de \( \Sigma_m \), pour tout entier \( n\leq m \), pour tout \( m-1 \) états \( q_1 \), \( \ldots \), \( q_{n-1} \), \( q_{n+1} \), \( \ldots \), \( q_m \) de \( Q \),
\begin{equation}\label{eq def bot up cong}
  p\sim p' \Rightarrow \delta(c,f) \sim \delta(c',f),
\end{equation}
où
\begin{align*}
  c &= (q_1,\ldots,q_{n-1},p,q_{n+1},\ldots,q_m), & c' &= (q_1,\ldots,q_{n-1},p',q_{n+1},\ldots,q_m),
\end{align*}
Deux états Bottom-Up-congruents peuvent être appelés \firstocc{interchangeables}~\cite{AT90} et \( \sim \) une \emph{forward bisimulation}~\cite{HMM09}.
L'\firstocc{automate quotient}{automate!quotient!congruence Bottom-Up} de \( A \) par la congruence \( \sim \) est l'automate \( A_\sim=(\Sigma,Q_\sim,F_\sim,\delta') \) avec
\begin{equation*}
  \delta'(([q_1],\ldots,[q_m]),f) = [\delta((q_1,\ldots,q_m),f)].
\end{equation*}
Ce morphisme préserve le langage, c'est-à-dire que \( L(A) = L(A_\sim) \).

Soient \( A=(\Sigma, Q, F, \delta) \) un automate d'arbre non-déterministe et \( \sim \) une relation d'équivalence sur \( Q \).
La relation \( \sim \) est une \firstocc{congruence Top-Down}{congruence!Top-Down} si et seulement si pour tous deux états équivalents \( p \) et \( p' \) de \( Q \), pour toute transition \( ((q_1,\ldots,q_m),f,p) \) de \( \delta \), il existe \( ((q'_1,\ldots,q'_m),f,p') \) dans \( \delta \) satisfaisant \( q_i \sim q'_i \) pour \( 1\leq i\leq m \).
Remarquons alors que, pour tous deux états équivalents \( p \) et \( p' \) de \( Q \), pour tout arbre \( t \) de \( \Sigma^*_0 \),
  \begin{equation*}
    p\in \delta(t) \Leftrightarrow p'\in \delta(t).
  \end{equation*}
Notons que la réciproque n'est pas vraie.
L'\firstocc{automate quotient}{automate!quotient!congruence Top-Down} de \( A \) par \( \sim \) est l'automate \( A_\sim=(\Sigma,Q_\sim,F',\delta') \) avec
\begin{align*}
  F' &= \{[q]\mid q\in F\},\\
  \delta'(([q_1],\ldots,[q_m]),f) &= \delta(([q_1],\ldots,[q_m]),f).
\end{align*}
Remarquons que \( L(A) = L(A_\sim) \), que si \( A \) est déterministe, alors \( A_\sim \) ne l'est pas nécessairement, et que des états finaux et non-finaux peuvent être fusionnés pendant le calcul du quotient.

\section*{Les expressions rationnelles d'arbres}

Classiquement, les expressions rationnelles d'arbres, comme celles de mots, peuvent être définies par induction.
Il est également possible de considérer, d'une façon équivalente, une définition algébrique depuis les opérades\footnote{Ce genre de définition sera utilisée dans la seconde partie de ce document pour définir, par exemple, des expressions booléennes.}.

Ainsi, une \firstocc{expression rationnelle d'arbres}{expression rationnelle!d'arbres} sur un alphabet gradué \( \Sigma \) est un arbre d'arité \( 0 \) de l'opérade libre générée par \( \Sigma\cup \{ \emptyset,+\}\cup \{ \cdot_a,^{*_a} \mid a\in\Sigma_0\} \) où \( \emptyset \) est d'arité \( 0 \), \( + \) d'arité \( 2 \), et où pour tout symbole \( a \) de \( \Sigma_0 \), \( \cdot_a \) est d'arité \( 2 \) et \( ^{*_a} \) d'arité \( 1 \).
Comme dans le cas des expressions rationnelles de mots, ces expressions seront possiblement parenthésées selon la convention usuelle, à savoir les opérateurs binaires en notation infixe associatifs à gauche en considérant le parenthésage fonction de l'ordre considérant la concaténation \( \cdot_a \) prioritaire sur la somme \( + \), et l'étoile \( ^{*_a} \) comme l'opération la plus prioritaire.

Pour interpréter inductivement le langage dénoté par une expression rationnelle d'arbres, l'action des symboles sur les langages doit être explicitée.
Ici, pour un symbole \( f \) d'arité \( k \) et \( k \) langages \( (L_1,\ldots,L_k) \), nous noterons
\begin{equation*}
  f(L_1,\ldots,L_k) = \{f(t_1,\ldots,t_k) \mid (t_1,\ldots,t_k) \in L_1\times\cdots\times L_k\}.
\end{equation*}
Il faut également fixer une sémantique pour les opérateurs de concaténation et d'étoile.
Pour un arbre \( t \) de \( \Sigma^*_0 \), pour deux symboles \( a \) et \( b \) distincts de \( \Sigma_0 \) et un langage \( L\subset \Sigma^*_0 \), on note \( t \cdot_a L \) l'ensemble défini inductivement par
\begin{align*}
  b \cdot_a L &= \{b\}, &
  a \cdot_a L &= L, &
  f(t_1,\ldots,t_k) \cdot_a L &= f(t_1 \cdot_a L,\ldots,t_k \cdot_a L).
\end{align*}
Par extension, pour tout langage \( L'\subset \Sigma^*_0 \),
\begin{equation*}
  L \cdot_a L' = \bigcup_{t\in L} t\cdot_a L'.
\end{equation*}
L'étoile est alors l'itération infinie de cette opération de substitution.
En effet, pour un langage d'arbres \( L\subset \Sigma^*_0 \) donné,
\begin{align*}
  L^{*_a} &= \bigcup_{n\in\mathbb{N}} L^{a,n} &
  \text{avec }L^{a,n} &=
    \begin{cases}
      \{a\} & \text{ si }n=0,\\
      L^{a,n-1} \cdot_a L & \text{ sinon.}
    \end{cases}
\end{align*}

Le \firstocc{langage dénoté}{langage!dénoté par une expression rationnelle!d'arbres} par une expression rationnelle d'arbres \( E \) sur un alphabet \( \Sigma \) est le sous-langage \( L(E) \) de \( \Sigma^*_0 \) défini inductivement par:
\begin{gather*}
  L(f(E_1,\ldots,E_n)) = f(L(E_1),\ldots,L(E_n)),\\
  \begin{aligned}
    L(\emptyset) &= \emptyset, & L(E_1+E_2) &= L(E_1) \cup L(E_2),\\
    L(E_1\cdot_a E_2) &= L(E_1) \cdot_a L(E_2), & L(E_1^{*_a}) &= {L(E_1)}^{*_a}
  \end{aligned}
\end{gather*}
avec \( f \) un symbole de \( \Sigma_n \), \( k \) expressions rationnelles d'arbres \( E_1,\ldots,E_k \)  sur \( \Sigma \) et \( a \) un symbole de \( \Sigma_0 \).
\newline
Cette définition inductive du langage peut être vue, d'une façon équivalente, comme un morphisme d'opérade, envoyant l'opérade libre des expressions rationnelles sur les fonctions \(n\)-aires de \(\Sigma^*\) sur lui-même: par exemple, serait associée à la somme l'union de langage, ou à un symbole \(n\)-aire son action sur un \(n\)-uplet de langages, \emph{etc.}
Nous reviendrons sur cette interprétation dans la seconde partie de ce document.


\section*{Les automates d'arbres compressés}
Un des inconvénients de la structure d'automate d'arbres est la profusion de transitions dans le pire des cas, due au fait que les transitions relient \(n\)-uplets d'états à état; cependant, nous verrons que certaines constructions d'automates d'arbres peuvent tirer profit des structures condensées d'expressions rationnelles.
L'idée de base de la compression est de considérer le produit cartésien d'ensembles.
Imaginons qu'un automate d'arbres contiennent les 4 transitions binaires \( ((q_1,q_1),f,q_3) \), \( ((q_1,q_2),f,q_3) \), \( ((q_2,q_1),f,q_3) \) et \( ((q_2,q_2),f,q_3) \).
Sans perdre d'information, ces 4 transitions peuvent être \emph{factorisées} en une
\firstocc{transition compressée}\  \(  (\{q_1,q_2\}, \{q_1,q_2\}, f, q_3) \) en
utilisant des ensembles d'états plutôt que des états.
Le comportement de cet automate peut être simulé en considérant le produit cartésien des états d'origine de la transition.

Un \firstocc{automate d'arbres compressé}{automate!d'arbres!compressé} sur un alphabet gradué \( \Sigma \) est un \( 4 \)-uplet \( (\Sigma,Q,Q_F,\delta) \) où
\begin{itemize}
  \item \( Q \) est un ensemble d'\firstocc{états},
  \item \( Q_F\subset Q \) est l'ensemble des \firstocc{états finaux},
  \item \( \delta \subset {(2^Q)}^n\times \Sigma_n\times 2^Q \) est l'ensemble des \firstocc{transitions compressées}\ pouvant être vu comme une fonction de \( {(2^{Q})}^k\times\Sigma_k \) vers \( 2^Q \) définie par
      \begin{equation*}
        (Q_1,\ldots,Q_k,f,q)\in\delta \Leftrightarrow q\in \delta(Q_1,\ldots,Q_k,f).
      \end{equation*}
\end{itemize}

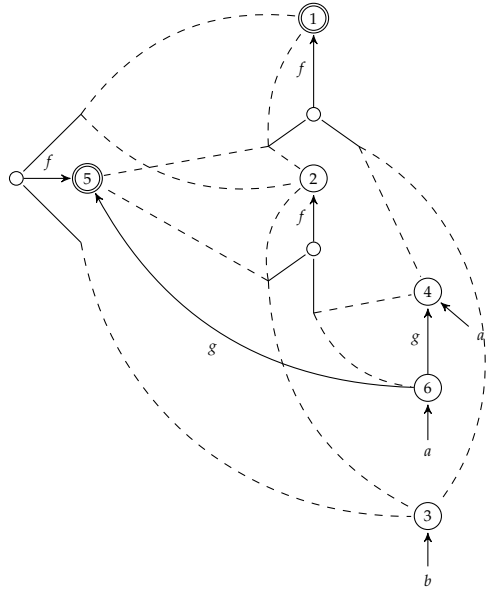
\begin{figure}[H]
    \centerline{
    \begin{tikzpicture}[node distance=2.5cm,bend angle=30,transform shape,scale=0.85]
      \node[state] (2)  {\( 2 \)} ;
     \node[state, double,above of=2, node distance = 2.5cm] (1) {\( 1 \)};
     \node[state,double,  left  of=2,node distance=3.5cm] (5) {\( 5 \)};
      \node[state, below right of=2,node distance = 2.5cm] (4) {\( 4 \)};
      \node[state, below of=4, node distance = 3.5cm] (3) {\( 3 \)};
      \node[state,below  of=4,node distance=1.5cm] (6) {\( 6 \)};
       \node[state, below of=1,node distance=1.5cm] (cerc) {};
       \node[state,below of=2,node distance=1.1cm] (cerc1) {};
       \node[,state,left of=5,node distance=1.1cm] (cerc2) {};
        \draw (3) ++(0cm,-1cm) node {\( b \)}  edge[->] (3);
        \draw (6) ++(0cm,-1cm) node {\( a \)}  edge[->] (6);
        \draw[left] (4) ++(1cm,-0.7cm) node {\( a \)}  edge[above,->] (4);
         \path[->]
       (cerc) edge[->, left] node {\( f \)} (1)
       (cerc1) edge[->, left] node {\( f \)} (2)
       (6) edge[->, left] node {\( g \)} (4)
       (6) edge[->, below left,bend left] node {\( g \)} (5)
       (cerc2) edge[->, above] node {\( f \)} (5);
       \draw (cerc) ++(-0.7 cm,-0.5cm)edge node [above,pos=0.5] {} (cerc)  edge[dashed]node[left,above,pos=0.6]{} (5)  edge[dashed] node[left,above,pos=0.5]{}(2) edge[dashed,bend left=20] node[bend left=20]{}(1);
       \draw (cerc) ++(0.7 cm,-0.5cm)edge node [above,pos=0.5] {} (cerc)  edge[dashed,bend left=50 ]node[left,above,pos=0.6]{} (3)  edge[dashed] node[left,above,pos=0.5]{}(4);
      \draw (cerc1) ++(-0.7 cm,-0.5cm)edge node [above,pos=0.5] {} (cerc1)  edge[dashed,bend left=30 ]node[left,above,pos=0.6]{} (2)  edge[dashed] node[left,above,pos=0.5]{}(5) edge[dashed,bend right=30 ]node[left,above,pos=0.6]{} (3)  ;
      \draw (cerc1) ++(0cm,-1cm)edge node [right,pos=0.5] {} (cerc1)  edge[dashed,bend right=30 ]node[left,above,pos=0.6]{} (6)
       edge[dashed] node[left,above,pos=0.5]{}(4);
       \draw (cerc2) ++(1cm,1cm)edge node [left,pos=0.5] {} (cerc2)  edge[dashed,bend left=30 ]node[left,above,pos=0.6]{} (1)  edge[dashed,bend right=30 ]node[left,above,pos=0.6]{} (2);
       \draw (cerc2) ++(1cm,-1cm)edge node [left,pos=0.5] {} (cerc2)  edge[dashed,bend right=40 ]node[left,above,pos=0.6]{} (3);
    \end{tikzpicture}
    }
    \caption{L'automate compressé \( A \).}%
    \label{exp comp aut}
\end{figure}

\begin{example}\label{ex:aut compr}
Considérons l'automate compressé \( A=(\Sigma,Q,Q_F,\delta) \) de la Figure~\ref{exp comp aut}.
Ses transitions (compressées) sont
\begin{multline*}
  \delta = \{ (\{1,2,5\},\{3,4\},f,1), (\{2,3,5\},\{4,6\},f,2),\\
    (\{1,2\},\{3\},f,5),(\{6\},g,4), (\{6\},g,5), (a,6), (a,4), (b,3)\}.
\end{multline*}
\end{example}

La fonction de transition \( \delta \) peut être restreinte en une fonction de \( Q^n\times \Sigma_n \) vers \( 2^Q \) pour simuler le comportement d'un automate (non compressé) en considérant pour un \( k \)-uplet \( (q_1,\ldots,q_k) \) d'états et un symbole \( f \) de \( \Sigma_k \) toutes les transitions \emph{actives} \( ((Q_1,\ldots,Q_k),f,q) \), qui sont les transitions où \( q_i \) est dans \( Q_i \) pour \( i\leq k \).
Plus formellement, pour tout \( k \) états \( (q_1,\ldots,q_k) \) de \( Q^k \), pour tout symbole \( f \) de \( \Sigma_k \),
    \begin{equation}\label{eq:extdeltaEnsComp}
      \delta((q_1,\ldots,q_k),f) = \bigcup_{\substack{(Q_1,\ldots, Q_k, f,q)\in\delta,\\ \forall i\leq k, q_i\in Q_i}} \{q\}.
    \end{equation}
  On dira alors qu'un automate compressé est \firstocc{déterministe}{automate!d'arbres!compressé!déterministe} si pour tout \( k \) états \( (q_1,\ldots,q_k) \) de \( Q^k \), pour tout symbole \( f \) de \( \Sigma_k \),
      \begin{equation*}
        | \delta((q_1,\ldots,q_k),f)| \leq 1.
      \end{equation*}
L'ensemble de transition \( \delta \) peut être étendu en une fonction \( \delta \) de \( \Sigma^*_0 \) vers \( 2^Q \) en considérant inductivement, pour un arbre \( f(t_1,\ldots,t_k) \) les transitions \emph{actives} \( ((Q_1,\ldots,Q_k), f,q) \) une fois l'arbre lu, c'est-à-dire quand \( \delta(q_i) \) et \( Q_i \) ont un état commun pour tout \( i\leq k \).
Plus formellement, pour tout arbre \( t=f(t_1,\ldots,t_k) \) de \( \Sigma^*_0 \),
  \begin{equation*}
    \delta(t) = \bigcup_{\substack{(Q_1,\ldots,Q_k,f,q)\in\delta,\\ \forall i\leq k, \delta(t_i)\cap Q_i\neq\emptyset}} \{q\}.
  \end{equation*}
Ainsi,
  \begin{equation}\label{eq:lienExtsTransComp}
    \delta(f(t_1,\ldots,t_n)) = \bigcup_{c \in\delta(t_1)\times\cdots\times\delta(t_n)} \delta(c,f).
  \end{equation}
Le \firstocc{langage reconnu}{langage!reconnu par un automate!d'arbres!compressé} par un automate compressé \( A=(\Sigma,\_,Q_F,\delta) \) est le sous-ensemble \( L(A) \) de \( \Sigma^* \) défini par
  \begin{equation*}
    L(A) = \{t\in \Sigma^* \mid \delta(t)\cap Q_F\neq\emptyset \}.
  \end{equation*}
\begin{example}
  Considérons l'automate de la Figure~\ref{exp comp aut} et montrons alors que l'arbre \( t=f(f(b,a),g(a)) \) appartient à \( L(A) \).
  Pour cela, calculons \( \delta(t') \) pour tout sous-arbre \( t' \) de \( t \).
  Tout d'abord, par définition,
  \begin{align*}
    \delta(a)&=\{4,6\}, & \delta(b)&=\{3\}.
  \end{align*}
  Puisque la seule transition de \( \delta \) étiquetée par \( f \) contenant \( 3 \) dans son premier ensemble d'origine et \( 4 \) ou \( 6 \) dans son second est la transition \( (\{2,3,5\},\{4,6\},f,2) \), alors
  \begin{equation*}
    \delta(f(b,a)) = \{2\}.
  \end{equation*}
  Puisque les deux transitions étiquetées par \( g \) sont \( (\{6\},g,4) \) et \( (\{6\},g,5) \),
  \begin{equation*}
    \delta(g(a)) = \{4,5\}.
  \end{equation*}
  Enfin, il y a deux transitions étiquetées par \( f \) contenant \( 2 \) dans leur première origine et \( 4 \) ou \( 5 \) dans leur seconde:
  \begin{align*}
    (\{2,3,5\},&\{4,6\},f,2), & (\{1,2,5\},&\{3,4\},f,1).
  \end{align*}
  Ainsi
  \begin{equation*}
    \delta(f(f(b,a),g(a))) = \{1,2\}.
  \end{equation*}
  En conclusion, puisque \( 1 \) est final, \( t\in L(A) \).
\end{example}
Soit \( \phi \) un morphisme alphabétique entre deux alphabets \( \Sigma \) et \( \Sigma' \).
L'\firstocc{image par}\  \( \phi \) d'un automate compressé \( A=(\Sigma,Q,Q_F,\delta) \) est l'automate compressé \( \phi(A)=(\Sigma',Q,Q_F,\delta') \) où
  \begin{equation*}
    \delta'=\{(Q_1,\ldots,Q_n,\phi(f),q)\mid (Q_1,\ldots,Q_n,f,q) \in \delta \}.
  \end{equation*}
Par une induction triviale, on peut montrer que
  \begin{equation}\label{eq:lien morph lang compress}
    L(\phi(A)) = \phi(L(A)).
  \end{equation}

\section*{Les automates d'arbres descendants}
Dernière structure d'automates que nous allons considérer dans cette section, les automates descendants, permettant d'interpréter les arbres selon une forme semblant plus naturelle, de la racine aux feuilles, plus proche de leur définition inductive.

Un \firstocc{automate d'arbres descendant non-déterministe}{automate!d'arbres!descendant!non-déterministe} (descendant ou Top-Down, ici sans variables\footnote{comme dans le cas des automates ascendants, les variables seront prises en compte dans la seconde partie de ce document.}) est un quadruplet \( A=(\Sigma, Q, I,  \delta) \) avec:
\begin{itemize}
  \item \( \Sigma \) un ensemble gradué, l'\firstocc{alphabet}\ de l'automate,
  \item \( Q \) l'ensemble des \firstocc{états}\ de l'automate,
  \item \( I \subset Q \) l'ensemble des états \firstocc{initiaux}\ de l'automate,
  \item \( \delta \) la fonction de \firstocc{transition}\ de l'automate, appartenant à \( Q \times \Sigma_n \rightarrow 2^{Q^n} \).
\end{itemize}
Comme dans les cas précédents, la fonction \(\delta \) peut être étendue en une fonction \(\delta'\) de signature \( 2^Q \times \Sigma_n \rightarrow 2^{Q^n} \) d'une façon relativement proche de la précédente:
\begin{align*}
  \delta'(Q', \varepsilon) &= Q', &
  \delta'(Q', f(t_1, \ldots,t_k)) &=
    \bigcup_{(q_1,\ldots,q_n) \in \delta''(Q',f)} \delta'(q_1,t_1) \times \cdots\times \delta'(q_n, t_n)
\end{align*}
où \(\delta''\) est la fonction de \(2^{Q} \times \Sigma_n \rightarrow 2^{Q^n}\) définie par
\begin{equation*}
  \delta''(Q', f) = \bigcup_{q'\in Q'} \delta(q', f).
\end{equation*}
Un arbre nullaire enverra alors un état dans un ensemble soit vide, soit réduit à un seul élément (en effet, \(2^{Q^0} = 2^{\mathbb{1}}= \{\emptyset,\{()\} \} \)).

Nous assimilerons naturellement par la suite toutes ces fonctions à la fonction \( \delta \) et nous les distinguerons en fonction de leurs signatures.
De même, nous assimilerons la fonction \( \delta \) à l'ensemble de triplets défini par
\begin{equation*}
  ((p_1,\ldots,p_n),f,q)\in\delta \Leftrightarrow (p_1,\ldots,p_n)\in\delta(q,f).
\end{equation*}
Ainsi, le \firstocc{langage reconnu}{langage!reconnu par un automate!d'arbres! descendant}  par un automate d'arbres \( A=(\Sigma,\_,I,\delta) \) est l'ensemble
\begin{equation*}
  L(A)=\{t\in\Sigma^*_0\mid \delta(I,t) \neq\emptyset \}.
\end{equation*}

Un \firstocc{automate d'arbres descendant déterministe}{automate!d'arbres!descendant!déterministe}\label{defAutoArbreTDDet} est un quadruplet \( A=(\Sigma, Q, I, \delta) \) avec:
\begin{itemize}
  \item \( \Sigma \) un ensemble gradué, l'\firstocc{alphabet}\ de l'automate,
  \item \( Q \) l'ensemble des \firstocc{états}\ de l'automate,
  \item \( I\subset Q \) l'ensemble des états \firstocc{finaux}\ de l'automate,
  \item \( \delta \) la fonction de \firstocc{transition}, élément de \( Q\times \Sigma^n \rightarrow Q^n + \{\bot \} \).
\end{itemize}
Comme dans les cas précédents, la fonction \(\delta \) peut être étendue en une fonction \(\delta'\) de signature \( (Q + \bot) \times \Sigma_n \rightarrow Q^n + \bot \) d'une façon relativement proche de la précédente:
\begin{align*}
  \delta'(p, \varepsilon) &= p, &
  \delta'(p, f(t_1, \ldots,t_k)) &=
    \begin{cases}
      \bot & \text{si } \delta''(p, f) = \bot,\\
      \delta'(q_1,t_1) \times \cdots\times \delta'(q_n, t_n) & \text{si }   \delta''(p, f) = (q_1,\ldots,q_n),
    \end{cases}
\end{align*}
où \(\delta''\) est la fonction de \((Q + \bot) \times \Sigma_n \rightarrow Q^n + \bot \) définie par
\begin{equation*}
  \delta''(p, f) =
  \begin{cases}
    \bot & \text{ si } p = \bot,\\
    \delta(p, f) & \text{ sinon.}
  \end{cases}
\end{equation*}
Un arbre nullaire enverra alors un état soit dans \(\bot \), soit dans \(()\), unique élément de l'ensemble \(Q^0\).
\newline
De plus, nous assimilerons naturellement dans la suite de ce document les automates déterministes et les automates non-déterministes ayant aucun ou un unique état initial et dont la fonction de transition \( \delta \) vérifie pour tout arbre \( t \) \( |\delta(I, t)| \leq 1 \).
\newline
Le \firstocc{langage reconnu}{langage!reconnu par un automate!déterministe d'arbres! descendant} par un automate déterministe \( A=(\Sigma,\_,I,\delta) \) est l'ensemble
\begin{equation*}
  L(A) = \{t\in\Sigma^*_0 \mid \delta(I,t)=()\}.
\end{equation*}

Un des inconvénients des automates descendants est qu'il n'y a pas équivalence entre automates déterministes et non-déterministes\footnote{prendre par exemple le langage \( \{ f(a,b), f(b,a) \} \).}.
Cependant, automates non-déterministes ascendants et descendants sont quant à eux équivalents en terme d'expressivité, la conversion étant directe en considérant leurs fonctions de transition comme des ensembles de transition.
Ainsi, dans la suite de ce document, les constructions proposées seront définies à partir d'ensemble de transitions dans la plupart des cas afin de conserver la généralité d'implantation.

\newpage

\part{Des expressions rationnelles d'arbres aux automates d'arbres}\label{part:autArbres}

\chapter*{Présentation}
\addcontentsline{toc}{chapter}{Présentation}

\begin{flushright}
  \emph{Pour sauver un arbre, mangez un castor!}\\
  Henri Prades
\end{flushright}

Dans cette partie, nous nous intéressons à différentes méthodes de conversion entre expressions rationnelles et automates.
Les différentes méthodes présentées ici considèrent des expressions et des automates d'arbres, et sont essentiellement des extensions de méthodes classiques pour construire des automates de mots.

Ainsi, les constructions classiques pourront être retrouvées dans la plus grande partie des cas en considérant la transformation inverse de l'inclusion des automates de mots en automate d'arbres: toute lettre de mot peut être en effet considérée comme une lettre d'arbre d'arité \(1\); les états initiaux peuvent être modélisés comme des transitions par des lettres d'arité \(0\).

Afin d'aider à la compréhension de ces différentes méthodes, des applications Web sont accessibles, implantées en Haskell suivant le paradigme de programmation réactive fonctionnelle, utilisant la bibliothèque \href{https://hackage.haskell.org/package/reflex}{Reflex}, puis converties en Javascript.
Les constructions d'automates de mots décrites dans cette partie sont disponibles \href{http://ludovicmignot.free.fr/programmes/WordAutomataConstructions/}{à cette adresse}~\cite{WordAutCons}.
Les constructions d'automates d'arbres, exceptée la méthode Bottom-Up par dérivation, sont disponibles \href{http://ludovicmignot.free.fr/programmes/treeAutCompar/}{à cette adresse}~\cite{TreeAutCons}.
La méthode de dérivation Bottom-Up, valables pour les expressions d'arbres étendues, est disponible \href{http://ludovicmignot.free.fr/programmes/BottomUpDerivatives/}{à cette adresse}~\cite{BotUpDerCons}.

Afin de présenter d'une façon simple ces différentes méthodes de conversion, nous suivrons une méthodologie simple.
Nous aborderons tout d'abord dans cette introduction la méthode en la présentant sur les langages et expressions de mots pour introduire les concepts mis en {\oe}uvres, puis nous formaliserons dans les sections suivantes ces mêmes méthodes et leurs extensions aux expressions et langages d'arbres.

Une des premières méthodes de construction d'automate depuis une expression rationnelle est la méthode des positions, due indépendamment à Glushkov~\cite{Glu61} et Mc~Naughton et Yamada~\cite{MY60}.
Cette méthode consiste à calculer à partir de l'expression cinq fonctions dites \emph{de positions},
  permettant de déterminer le rôle de chacune des occurrences des symboles dans la construction du langage.
Ces cinq fonctions sont les suivantes:
\begin{itemize}
  \item \( \mathrm{Pos} \) calcule les différentes positions de l'expression, à savoir les différentes occurrences des symboles~;
  \item \( \mathrm{Null} \) est une fonction indicatrice de la présence du mot vide dans le langage dénoté par l'expression~;
  \item \( \mathrm{First} \) est une fonction renvoyant les positions pouvant commencer un mot du langage~;
  \item \( \mathrm{Last} \) est la fonction duale de \( \mathrm{First} \), calculant les positions pouvant terminer les mots du langage;
  \item \( \mathrm{Follow} \) calcule, pour une position \( x \) donnée, les positions succédant directement à \( x \) dans un mot du langage.
\end{itemize}
Une fois ces fonctions calculées, un automate, dit \emph{de positions}, peut être calculé en considérant les positions comme des états dont la finalité est déterminé par les fonctions de positions, et en définissant les différentes transitions en fonction de la présence des positions dans chacune des fonctions de positions.

Un automate plus petit peut ensuite être calculé, en remarquant que si certaines positions admettent les mêmes successeurs, alors les fusionner préserve le langage reconnu par l'automate.
Cette équivalence est à la base de la définition de l'automate des Follows de Ilie et Yu~\cite{IY03}: les états ne sont alors plus des positions de l'expression, mais des ensembles de positions dites \emph{similaires}, c'est-à-dire dont les ensembles Follows sont égaux.
Cet automate est ainsi un quotient de l'automate des positions.

Une alternative à la prise en compte des positions de l'expression peut être considérée par une opération particulière sur les expressions: la dérivation.
On peut par exemple se demander s'il existe une expression \( E' \) dénotant l'ensemble des mots \( w \) tels que \( a\cdot w \) soit un mot du langage dénoté par une expression \( E \), pour un symbole \( a \) donné.
L'expression \( E' \) est alors appelée \firstocc{dérivée de} \( E \) par rapport à \( a \).
Cette opération, due à Brzozowski~\cite{Brz64} et notée \( \frac{d}{d_a} \), peut être étendue inductivement en l'opération \( \frac{d}{d_w} \) par son application successive en considérant chaque lettre constituant un mot \( w \), et en définissant la dérivation par le mot vide comme étant la fonction identité\footnote{Nous verrons qu'il s'agit, comme dans le cas de l'extension de la fonction de transition, d'une adjonction fonctorielle.}.
Cette opération permet tout d'abord de résoudre le test d'appartenance d'une façon directe: déterminer si un mot \( w \) appartient au langage dénoté par une expression \( E \) revient à déterminer si le mot vide appartient au langage dénoté l'expression \( \frac{d}{d_w}(E) \).
Mais cette opération permet également de construire un automate déterministe, l'\firstocc{automate des dérivées}: l'état initial est l'expression \( E \), les états sont les dérivées de \( E \), les transitions sont de la forme \( (F, a, \frac{d}{d_a}(F)) \), et la finalité d'un état \( F \) est déterminée par la présence du mot vide dans \( L(F) \).
Cependant, rien n'indique ici que l'ensemble des états est fini.
Et effectivement, il existe des expressions dont l'ensemble des dérivées est infini.
Cependant, en considérant une relation d'équivalence particulière sur les expressions au lieu de l'égalité, il est alors possible d'obtenir un ensemble fini et par conséquent un automate fini déterministe.
Il est important de noter que l'automate peut avoir un nombre exponentiel d'états, par rapport au nombre de symboles de l'expression de départ et que cette méthode permet de construire un automate depuis une expression rationnelle étendue, c'est-à-dire où l'intersection et le complémentaire sont des opérateurs autorisés.

Il est également possible de concevoir un automate non-déterministe en utilisant une opération très proche de la dérivation: la \firstocc{dérivation partielle}.
Pour cela, il suffit de changer le codomaine de l'opération: au lieu de considérer une opération transformant une expression en une autre expression, on peut considérer une opération transformant une expression en un ensemble d'expressions (ou ensemble de \firstocc{termes dérivés}) dont la somme serait équivalente à la dérivée classique\footnote{Nous verrons dans la seconde partie qu'il s'agit ici de ce qu'on appelle un morphisme de Kleisli.}.
Cette opération est appelée dérivation partielle (due à Antimirov~\cite{Ant96}) et produit alors l'\firstocc{automate des termes dérivés}, ou \firstocc{automate des équations}.

Cet automate, dont le nombre d'états est linéaire par rapport au nombre d'états de l'expression de départ est également un quotient de l'automate des positions.
Pour démontrer ce résultat~\cite{CZ02}, Champarnaud et Ziadi construisent les \firstocc{c-continuations}\ d'une expression, représentants canoniques des continuations de Berry et Sethi~\cite{BS86}.
Ces expressions, si produites depuis une expression linéarisée (c'est-à dire où les symboles sont indicés par leur position respective dans l'expression), permettent de construire un automate isomorphe à l'automate des positions mais dont les états ne sont pas des positions mais des expressions.
Cette particularité permet d'établir alors un lien morphique entre l'automate des positions et l'automate des termes dérivés, en délinéarisant les états de l'automate.

Dans la suite de cette partie, nous montrerons comment étendre ces méthodes aux expressions et automates d'arbres.
Pour cela, nous considérerons les deux sens classique de lecture des arbres: l'approche Top-Down (de la racine aux feuilles) et l'approche Bottom-Up (des feuilles à la racine).
Ces approches sont équivalentes lorsque l'on considère des automates non-déterministes.
Cependant, en se restreignant aux automates déterministes, le pouvoir de représentativité de l'approche Top-Down est moins puissante.
Par exemple, le langage \( \{f(a,a),f(b,b)\} \) ne peut être reconnu par un automate d'arbres Top-Down.

Ainsi, les méthodes de positions pourront être étendues dans les deux sens.
Afin d'étendre le calcul des c-dérivées, nous rappellerons la construction de l'automate des termes dérivés de Kuske et Meinecke~\cite{KM11}, afin de montrer que cet automate est bien un quotient de l'automate des positions.

La méthode par dérivation, quant à elle, ne pourra être réalisée qu'en partant des feuilles vers la racine: au lieu d'éliminer des n{\oe}uds de la racine aux feuilles pour obtenir récursivement des \( n \)-uplets de longueur croissante comme le font Kuske et Meinecke pour l'automate des termes dérivés, il est possible d'éliminer des feuilles puis des n{\oe}uds jusqu'à arriver à la racine de l'arbre.
L'inconvénient de cette méthode est que, contrairement aux méthodes Top-Down, l'arité des arbres n'est pas conservée.
Une fois une feuille éliminée, l'arbre voit son arité augmenter.
Et une fois un n{\oe}ud (c'est-à-dire étiqueté par un symbole d'arité différente de \(0\)) éliminé, l'arité peut évoluer à la baisse.
De plus, les expressions classiques ne permettent pas d'exprimer de telles variations d'arité, ni d'exprimer des arbres d'arités non-nulles.
Nous commencerons ainsi par définir de nouveaux opérateurs, puis de nouvelles formules de quotients permettant d'exprimer ces opérations.
Enfin, nous présenterons un algorithme purement syntaxique résolvant le test d'appartenance.

\chapter{Les constructions Top-Down}\label{chapTopDownCons}

Les constructions Top-Down ont l'avantage de suivre la définition \emph{naturelle} des arbres, de la racine aux feuilles.
Cependant, comme nous l'avons déjà rappelé, le déterminisme réduit le pouvoir de représentativité des automates.

Ainsi, dans cette section, nous nous intéresserons aux méthodes utilisant la notion de position, à la dérivation partielle, et aux liens (déjà connus dans le cas des mots) entre ces différentes constructions.

La dérivation classique sera quant à elle abordée dans la section suivante, qui traitera des automates ascendants.


\section{L'automate des positions Top-Down}\label{sec aut pos}

Les résultats de cette section sont tirés de~\cite{MOZ14, MOZ17}, écrits avec la collaboration de Nadia Ouali-Sebti et de Djelloul Ziadi.

Comme rappelé dans la section précédente, la méthode dite de positions est basée sur le calcul de 5 fonctions particulières dans le cas des mots.
Ces fonctions permettent de déterminer les positions commençant ou finissant un mot, et pouvant succéder à d'autres positions.
Il y a également une fonction particulière indicatrice de la présence du mot vide dans le langage dénoté par l'expression de départ.
Plus formellement, pour une expression (de mots) \( E \) \firstocc{linéaire}\ (c'est-à-dire où chaque symbole n'apparaît qu'une fois) sur un alphabet \( \Sigma \),
\begin{align*}
  \mathrm{Pos}(E) &= \{a\in\Sigma\mid \exists u, v\in \Sigma^*, uav\in L(E)\},\\
  \mathrm{Null}(E) &= (\varepsilon \in L(E)),\\
  \mathrm{First}(E) &= \{a\in\Sigma\mid \exists u \in \Sigma^*, au\in L(E)\},\\
  \mathrm{Last}(E) &= \{a\in\Sigma\mid \exists u \in \Sigma^*, ua\in L(E)\},\\
  \mathrm{Follow}(a, E) &= \{b\in\Sigma\mid \exists u,v \in \Sigma^*, uabv\in L(E)\}.
\end{align*}

Si l'expression n'est pas linéaire, il est possible de la linéariser~: afin de distinguer chacune des occurrences des symboles, ces derniers peuvent être indicés par leur position respective dans l'expression.
Cette opération est appelée \firstocc{linéarisation}{opération!linéarisation}, transformant une expression \( E \) en sa linéarisée \( \overline{E} \).
Par définition, une expression linéarisée est linéaire.
L'opération inverse de la linéarisation, la \firstocc{délinéarisation}{opération!délinéarisation}, est le morphisme \( \mathrm{h} \) obtenu depuis de la fonction associant à chaque symbole indicé \( f_k \) le symbole \( f \).

Une fois ces fonctions calculées, l'automate des positions peut être construit comme le quintuplet \( (\Sigma, \mathrm{Pos}(E), \{0\}, F,\delta) \), \( 0 \) étant un nouvel état, où
\begin{align*}
  F &=
    \begin{cases}
      \mathrm{Last}(E) \cup \{0\} & \text{ si } \mathrm{Null}(E),\\
      \mathrm{Last}(E) & \text{ sinon,}
    \end{cases}\\
  \delta &= \{(0,x,x) \mid x \in \mathrm{First}(E)\} \cup \{(x, y, y) \mid y \in \mathrm{Follow}(E, x)\}.
\end{align*}

L'extension au cas des arbres, dans une approche Top-Down, est relativement proche de la méthode classique.
On peut en effet s'intéresser aux liens de succession des différentes positions d'une expression linéaire, en considérant un arbre depuis sa racine vers ses feuilles.
La linéarisation, tout comme la délinéarisation, se définissent d'une manière identique dans le cas des expressions d'arbres.

Cependant, les fonctions de position sont légèrement différentes~: si la fonction \( \mathrm{First} \) (\emph{première position}) peut être interprétée comme renvoyant la racine d'un arbre, les fonctions \( \mathrm{Last} \) ou \( \mathrm{Follow} \) ne peuvent s'appliquer avec la même sémantique que celle des mots.
En effet, un symbole d'arbre d'arité \( n \) n'admet pas qu'un seul successeur, mais un \( n \)-uplet~: il nous faut donc spécifier un indice entre \( 1 \) et \( n \) afin de réaliser la projection de ce \( n \)-uplet.
De même, un arbre n'admet pas qu'une seule position de fin, mais un ensemble de feuilles

Ainsi, une fois la linéarisation opérée si nécessaire, les \emph{rôles} des différents symboles d'une expression linéaire \( E \) sur un alphabet \( \Sigma \) peuvent être explicités \emph{via} le calcul des fonctions de positions suivantes~:
\begin{itemize}
  \item \( \mathrm{First}(E) \) est le sous-ensemble des racines des arbres du langage \( L(E) \);
  \item \( \mathrm{Last}(E) \) est le sous-ensemble des feuilles des arbres du langage \( L(E) \);
  \item pour un symbole \( f \) d'arité \( m \) et un entier \( k\leq m \), \( \mathrm{Follow}(E,f,k) \) est le sous-ensemble des racines des \( k \)\up{e} fils des sous-arbres de racine \( f \) des arbres du langage \( L(E) \).
\end{itemize}

Plus formellement,
\begin{align*}
  \mathrm{First}(E) &= \{\racine(t) \in \Sigma \mid t\in L(E)\},\\
  \mathrm{Last}(E) &= \displaystyle\bigcup_{t \in L(E)}\feuilles(t),\\
  \mathrm{Follow}(E,f,k) &= \{g\in \Sigma \mid \exists t\in L(E), \exists s\preccurlyeq t, \racine(s)=f, \fils{k}(s)=g\}.
\end{align*}

\begin{example}\label{Pos Automat}
  Considérons l'alphabet \( \Sigma=\Sigma_0\cup\Sigma_1\cup\Sigma_2 \) défini par
  \begin{align*}
    \Sigma_0 &= \{a,b,c\}, &
    \Sigma_1 &= \{f,h\}, &
    \Sigma_2 &= \{g\}.
  \end{align*}
  Considérons l'expression rationnelle \( E \) et sa linéarisée définies par~:
  \begin{align*}
    E&={({f(a)}^{*_a}\cdot_a b+ h(b))}^{*_b}+{g(c,a)}^{*_c}\cdot_c {({f(a)}^{*_a}\cdot_a b+ h(b))}^{*_b},\\
    \overline{E} &={({f_1(a)}^{*_a}\cdot_a b+ h_2(b))}^{*_b}+{g_3(c,a)}^{*_c}\cdot_c {({f_4(a)}^{*_a}\cdot_a b+ h_5(b))}^{*_b}.
  \end{align*}
  Le langage dénoté par \( \overline{E} \) est l'ensemble
  \begin{align*}
    L(\overline{E}) =&  \{b, f_1(b),f_1(f_1(b)),f_1(h_2(b)),h_2(b),h_2(f_1(b)),h_2(h_2(b)), \ldots,\\
  & g_3(b,a),
g_3(g_3(b,a),a),
g_3(f_4(b),a),g_3(h_5(b),a),\\
& f_4(f_4(b)),f_4(h_5(b),h_5(f_4(b))),h_5(h_5(b)),\ldots \}.
  \end{align*}
  Par conséquent,
  \begin{align*}
    \mathrm{First}(\overline{E})&=\{b,f_1,h_2,g_3,f_4,h_5\}, &
    \mathrm{Last}(\overline{E})&=\{a,b\},\\
    \mathrm{Follow}(\overline{E},f_1,1)&=\{b,f_1,h_2\},&
    \mathrm{Follow}(\overline{E},h_2,1)&=\{b,f_1,h_2\}, \\
    \mathrm{Follow}(\overline{E},g_3,1)&=\{b,g_3,f_4,h_5\},&
    \mathrm{Follow}(\overline{E},g_3,2)&=\{a\}, \\
    \mathrm{Follow}(\overline{E},f_4,1)&=\{b,f_4,h_5\},&
    \mathrm{Follow}(\overline{E},h_5,1)&=\{b,f_4,h_5\}.
  \end{align*}
\end{example}

Les fonctions de position pour les expressions d'arbres, tout comme celles des expressions de mots, sont récursivement calculables~:

\begin{table}[H]
  \begin{align*}
    \mathrm{First}(0)&=\emptyset, \\
    \mathrm{First}(f(\_, \ldots,\_))&=\{ f\}, \\
    \mathrm{First}(E_1+E_2)&=\mathrm{First}(E_1)\cup\mathrm{First}(E_2), \\
    \mathrm{First}({E_1}^{*_c})&=\mathrm{First}(E_1)\cup \{ c\}, \\
    \mathrm{First}(E_1\cdot_c E_2)&=
      \begin{cases}
        (\mathrm{First}(E_1)\setminus \{c\}) \cup \mathrm{First}(E_2) & \text{ si } c\in\mathrm{First}(E_1),\\
        \mathrm{First}(E_1) & \text{ sinon.}
      \end{cases}
    \end{align*}
  \caption{Définition inductive de \( \mathrm{First} \).}\label{table:first}
\end{table}

\begin{table}[H]
  \begin{align*}
    \mathrm{Last}(0)&=\emptyset,\\
    \mathrm{Last}(f(E_1, \ldots,E_n))&=
      \begin{cases}
        \{ f\} & \text{ si }n=0,\\
        \bigcup_{j\leq n} \mathrm{Last}(E_j) & \text{sinon,}
      \end{cases}\\
    \mathrm{Last}(E_1+E_2)&=\mathrm{Last}(E_1)\cup\mathrm{Last}(E_2),\\
    \mathrm{Last}({E_1}^{*_c})&=\mathrm{Last}(E_1)\cup \{ c\},\\
    \mathrm{Last}(E_1\cdot_c E_2) &=
      \begin{cases}
        (\mathrm{Last}(E_1)\setminus \{c\}) \cup \mathrm{Last}(E_2) & \text{ si } c\in\mathrm{Last}(E_1),\\
        \mathrm{Last}(E_1) & \text{ sinon.}
      \end{cases}
  \end{align*}
  \caption{Définition inductive de \( \mathrm{Last} \).}
\end{table}

\begin{table}[H]
  \begin{align*}
    \mathrm{Follow}(0,f,k) &= \emptyset,\\
    \mathrm{Follow}(g(E_1,\ldots,E_n),f,k)&=
      \begin{cases}
        \mathrm{First}(E_k) & \text{ si } f=g,\\
        \bigcup_{j\leq n} \mathrm{Follow}(E_j,f,k) & \text{ sinon,}
      \end{cases}\\
    \mathrm{Follow}(E_1+E_2,f,k)&= \mathrm{Follow}(E_1,f,k) \cup \mathrm{Follow}(E_2,f,k),\\
    \mathrm{Follow}(E_1 \cdot_c E_2,f,k)&=
      \begin{cases}
          (\mathrm{Follow}(E_1,f,k)\setminus \{c\}) \cup \mathrm{First}(E_2) & \text{ si } c\in\mathrm{Follow}(E_1,f,k),\\
          \mathrm{Follow}(E_1,f,k) \cup \mathrm{Follow}(E_2,f,k) & \text{ si }
          c\in\mathrm{Last}(E_1),\\
          \mathrm{Follow}(E_1,f,k) & \text{ sinon,}
      \end{cases}\\
    \mathrm{Follow}(E_1^{*_c},f,k)&=
      \begin{cases}
          \mathrm{Follow}(E_1,f,k) \cup \mathrm{First}(E_1) & \text{ si } c\in\mathrm{Follow}(E_1,f,k),\\
          \mathrm{Follow}(E_1,f,k) & \text{ sinon.}\\
      \end{cases}
    \end{align*}
  \caption{Définition inductive de \( \mathrm{Follow} \).}
\end{table}

À partir de ces fonctions de position, il est très aisé de définir un automate reconnaissant le langage dénoté par une expression rationnelle.
Il suffit pour cela de construire les transitions à partir des fonctions \( \mathrm{First} \) et \( \mathrm{Follow} \) comme une application directe de leurs définitions respectives.
Les points de départ et d'arrivée des transitions étant relatifs à la fois aux symboles et à l'indice des fils considérés, il suffit de considérer des couples états-entiers leur correspondant (nous noterons les entiers comme des exposants pour diminuer le nombre de parenthèses nécessaires).
L'adjonction d'un état terminal unique, étiqueté par l'élément \( \varepsilon \) ne pouvant appartenir à l'alphabet, est utilisé pour recueillir les transitions étiquetées par les symboles de racine.

Plus formellement, la méthode se décrit comme suit.

\begin{definition}\label{def aut pos}
  L'\firstocc{automate des positions}{Construction d'automates!Automates des positions} \( \mathcal{P}_E \) d'une expression linéarisée \( E \) est l'automate \( (Q,\Sigma,F,\delta) \) défini par
  \begin{align*}
    Q &=  \{f^k \mid f\in \Sigma_m, 1\leq k\leq m\} \cup \{\varepsilon \},\\
    F &= \{\varepsilon \},\\
    \delta &=
      \{((g^1,\ldots,g^n),g,f^k)\mid f\in\Sigma_m, k\leq m, g\in\Sigma_n \cap \mathrm{Follow}(E,f,k)\} \\
      & \qquad \cup \{((f^1,\ldots,f^m),f,\varepsilon)\mid f \in \Sigma_m \cap \mathrm{First}(E)\}.
  \end{align*}
\end{definition}

Cet automate reconnaît le langage dénoté par l'expression de départ et possède un nombre d'états égal à la somme des arités de chacune de ses positions, plus un pour l'état initial/final\footnote{en fonction du sens de lecture, ascendant ou descendant} \( \varepsilon \).

Cette construction peut être étendue à toute expression non nécessairement linéaire en utilisant les opérations de linéarisation et de délinéarisation.
En effet, l'\emph{automate des positions} \( \mathcal{P}_E \) associé à une expression (non linéaire) \( E \) est obtenu en remplaçant toute transition  \( ((g^1,\dots,g^n),g,f^k) \) de l'automate \( \mathcal{P}_{\overline{E}} \) par \( ((g^1,\dots,g^n),\mathrm{h}(g),f^k) \).

\begin{example}
  Considérons les deux expressions de l'Exemple~\ref{Pos Automat}:
  \begin{align*}
    E &={({f(a)}^{*_a}\cdot_a b+ h(b))}^{*_b}+{g(c,a)}^{*_c}\cdot_c {({f(a)}^{*_a}\cdot_a b+ h(b))}^{*_b},\\
    \overline{E}&={({f_1(a)}^{*_a}\cdot_a b+ h_2(b))}^{*_b}+{g_3(c,a)}^{*_c}\cdot_c {({f_4(a)}^{*_a}\cdot_a b+ h_5(b))}^{*_b}.
  \end{align*}
  L'automate des positions \( \mathcal{P}_{\overline{E}} \) est représenté Figure~\ref{fig aut pos}.
  Les ensembles d'états sont
  \begin{align*}
    Q &=\{\varepsilon,f^1_1,h^1_2,g^1_3,g^2_3,f^1_4,h^1_5\},\\
    F &=\{\varepsilon \}.
  \end{align*}
  L'ensemble des transitions est constitué des transitions suivantes~:

  \centerline{\( \begin{array}{ccc}
    (f_1^1, f_1,  \varepsilon) &
    (f_1^1, f_1,  f_1^1) &
    (f_1^1, f_1,  h_2^1)\\
    (h_2^1, h_2,  \varepsilon) &
    (h_2^1, h_2,  f_1^1) &
    (h_2^1, h_2,  h_2^1)\\
    ((g_3^1,g_3^2), g_3,  g_3^1) &
    ((g_3^1,g_3^2), g_3,  \varepsilon) &
    (f_4^1, f_4,  \varepsilon)\\
    (f_4^1, f_4,  g_3^1) &
    (f_4^1, f_4,  f_4^1) &
    (f_4^1, f_4,  h_5^1)\\
    (h_5^1, h_5,  \varepsilon) &
    (h_5^1, h_5,  g_3^1) &
    (h_5^1, h_5,  f_4^1)\\
    (h_5^1, h_5,  h_5^1) &
    (a, g_3^2) &
    (b,  \varepsilon)\\
    (b,  f_1^1) &
    (b,  h_2^1) &
    (b,  g_3^1)\\
    (b,  f_4^1) &
    (b,  h_5^1)
  \end{array} \)}

  Le nombre d'états est \( |Q|=7 \) et le nombre de transitions est \( |\delta|=23 \).
  L'automate des positions \( \mathcal{P}_E \), obtenu en délinéarisant les étiquettes de \( \mathcal{P}_{\overline{E}} \) est représenté Figure~\ref{fig aut pos delin}.
\end{example}

\begin{minipage}{0.45\linewidth}
  \begin{figure}[H]
    \centerline{
    	\begin{tikzpicture}[node distance=2.5cm,bend angle=30,transform shape,scale=1]
    	  \node[accepting,state] (eps) {\( \varepsilon \)};
    	  \node[state, above left of=eps] (f11) {\( f^1_1 \)};
    	  \node[state, above right of=eps] (h12) {\( h^1_2 \)};
        \node[state, below of=eps] (g13) {\( g^1_3 \)};
        \node[state, right of=g13,node distance=3.5cm] (g23) {\( g^2_3 \)};
    	  \node[state, below left of=g13,node distance=3.5cm] (h15) {\( h^1_5 \)};
        \node[state, below right of=g13,node distance=3.5cm] (f14) {\( f^1_4 \)};
    	  \draw (eps) ++(-1cm,0cm) node {\( b \)}  edge[->] (eps);
    	  \draw (f11) ++(-1cm,0cm) node {\( b \)}  edge[->] (f11);
    	  \draw (h12) ++(1cm,0cm) node {\( b \)}  edge[->] (h12);
    	  \draw (h15) ++(0cm,-1cm) node {\( b \)}  edge[->] (h15);
    	  \draw (g23) ++(1cm,0cm) node {\( a \)}  edge[->] (g23);
    	  \draw (g13) ++(-1cm,0cm) node {\( b \)}  edge[->] (g13);
    	  \draw (f14) ++(0cm,-1cm) node {\( b \)}  edge[->] (f14);
        \path[->]
          (f11) edge[->,below left] node {\( f_1 \)} (eps)
      		(f11) edge[->,loop,above] node {\( f_1 \)} ()
      		(h12) edge[->,bend right,above] node {\( h_2 \)} (f11)
      		%
      		(h12) edge[->,loop,above] node {\( h_2 \)} ()
      		(h12) edge[->,below right] node {\( h_2 \)} (eps)
      		(f11) edge[->,bend right,above] node {\( f_1 \)} (h12)
      	    %
      		(h15) edge[->, in=135,out=-135,loop,left] node {\( h_5 \)} ()
      		(h15) edge[->,above left] node {\( h_5 \)} (eps)
      		(h15) edge[->,above left] node {\( h_5 \)} (g13)
      		(h15) edge[->,bend right,above] node {\( h_5 \)} (f14)
      		(f14) edge[->,in=45,out=-45,loop,right] node {\( f_4 \)} ()
      		(f14) edge[->,bend right,above] node {\( f_4 \)} (h15)
      		(f14) edge[->,above right] node {\( f_4 \)} (eps)
      		(f14) edge[->,above right] node {\( f_4 \)} (g13)
      	  ;
        \draw (eps) ++(1.75cm,-0.75cm)  edge[->,in=0,out=90] node[above right,pos=0] {\( g_3 \)} (eps) edge (g13) edge (g23);
        \draw (g13) ++(1.5cm,1cm)  edge[->,in=90,out=145] node[above right,pos=0] {\( g_3 \)} (g13) edge (g13) edge (g23);
      \end{tikzpicture}
    }
    \caption{L'automate des positions \( \mathcal{P}_{\overline{E}} \).}%
    \label{fig aut pos}
  \end{figure}
\end{minipage}
\hfill
\begin{minipage}{0.45\linewidth}
  \begin{figure}[H]
    \centerline{
    	\begin{tikzpicture}[node distance=2.5cm,bend angle=30,transform shape,scale=1]
    	  \node[accepting,state] (eps) {\( \varepsilon \)};
    	  \node[state, above left of=eps] (f11) {\( f^1_1 \)};
    	  \node[state, above right of=eps] (h12) {\( h^1_2 \)};
        \node[state, below of=eps] (g13) {\( g^1_3 \)};
        \node[state, right of=g13,node distance=3.5cm] (g23) {\( g^2_3 \)};
    	  \node[state, below left of=g13,node distance=3.5cm] (h15) {\( h^1_5 \)};
        \node[state, below right of=g13,node distance=3.5cm] (f14) {\( f^1_4 \)};
    	  \draw (eps) ++(-1cm,0cm) node {\( b \)}  edge[->] (eps);
    	  \draw (f11) ++(-1cm,0cm) node {\( b \)}  edge[->] (f11);
    	  \draw (h12) ++(1cm,0cm) node {\( b \)}  edge[->] (h12);
    	  \draw (h15) ++(0cm,-1cm) node {\( b \)}  edge[->] (h15);
    	  \draw (g23) ++(1cm,0cm) node {\( a \)}  edge[->] (g23);
    	  \draw (g13) ++(-1cm,0cm) node {\( b \)}  edge[->] (g13);
    	  \draw (f14) ++(0cm,-1cm) node {\( b \)}  edge[->] (f14);
        \path[->]
          (f11) edge[->,below left] node {\( f \)} (eps)
      		(f11) edge[->,loop,above] node {\( f \)} ()
      		(h12) edge[->,bend right,above] node {\( h \)} (f11)
      		%
      		(h12) edge[->,loop,above] node {\( h \)} ()
      		(h12) edge[->,below right] node {\( h \)} (eps)
      		(f11) edge[->,bend right,above] node {\( f \)} (h12)
      	    %
      		(h15) edge[->, in=135,out=-135,loop,left] node {\( h \)} ()
      		(h15) edge[->,above left] node {\( h \)} (eps)
      		(h15) edge[->,above left] node {\( h \)} (g13)
      		(h15) edge[->,bend right,above] node {\( h \)} (f14)
      		(f14) edge[->,in=45,out=-45,loop,right] node {\( f \)} ()
      		(f14) edge[->,bend right,above] node {\( f \)} (h15)
      		(f14) edge[->,above right] node {\( f \)} (eps)
      		(f14) edge[->,above right] node {\( f \)} (g13)
      	  ;
        \draw (eps) ++(1.75cm,-0.75cm)  edge[->,in=0,out=90] node[above right,pos=0] {\( g \)} (eps) edge (g13) edge (g23);
        \draw (g13) ++(1.5cm,1cm)  edge[->,in=90,out=145] node[above right,pos=0] {\( g \)} (g13) edge (g13) edge (g23);
      \end{tikzpicture}
    }
    \caption{L'automate des positions \( \mathcal{P}_{E} \).}%
    \label{fig aut pos delin}
  \end{figure}
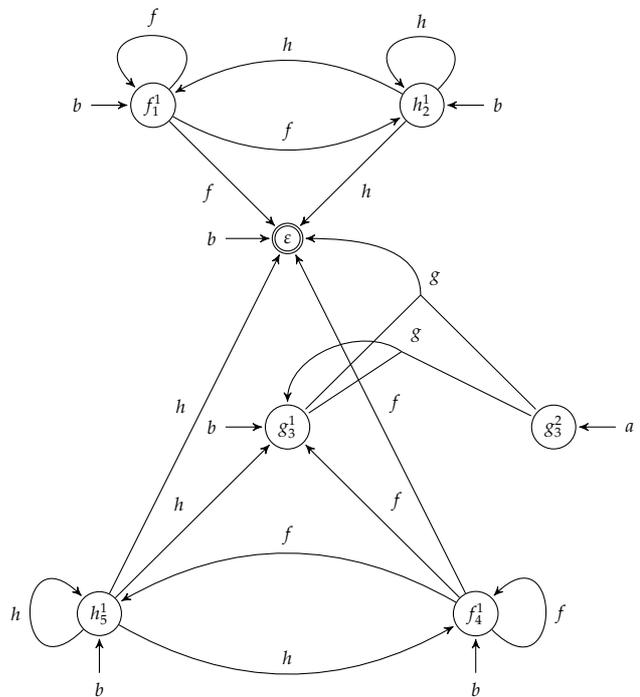
\end{minipage}


\section{L'automate des follows}\label{sec:follows}

Les résultats de cette section sont tirés de~\cite{MOZ14, MOZ17}, écrits avec la collaboration de Nadia Ouali-Sebti et de Djelloul Ziadi.

Dans la section précédente, nous avons montré comment construire l'automate des positions d'une expression donnée, chaque état correspondant à une position.
Cependant, il est possible que certaines positions partagent des mêmes propriétés, telles que d'avoir des ensembles \( \mathrm{Follow} \) égaux.
Ainsi, Ilie et Yu~\cite{IY03} ont proposé une relation d'équivalence sur les positions qui est également une congruence pour la fonction de transition de l'automate des positions.
Il s'agit de la \firstocc{congruence des \( \mathrm{Follow} \)}, définie de la façon suivante pour tout couple de position \( (f,g) \) d'une expression linéaire \( E \)~:
\begin{equation*}
  f \sim_{\mathcal{F}} g \Leftrightarrow \mathrm{Follow}(f, E) = \mathrm{Follow}(g, E).
\end{equation*}
Cette congruence permet ainsi de calculer un automate quotient de celui des positions, qui est ainsi plus petit en nombre d'états.
Pour cela, la congruence est étendue en considérant que l'ensemble \(\mathrm{Follow}\) associé à l'état initial de l'automate de Glushkov est l'ensemble \(\mathrm{First}\).

Pour construire l'automate des \( \mathrm{Follow} \) d'une expression d'arbres, nous suivrons donc le même principe.
Pour commencer, il suffit de définir un automate où les états ne sont plus les couples positions-entiers de l'automate des positions, mais des ensembles de ces couples \( f^k \) partageant les mêmes ensembles de \emph{successeurs}, en considérant que les successeurs de l'état \(\varepsilon \) sont l'ensemble des racines, l'ensemble \(\mathrm{First}(E)\).

\begin{definition}
  L'\firstocc{automate des Follows}{Construction d'automates!Automates des Follows} d'une expression linéaire \( E \) sur un alphabet \( \Sigma \) est l'automate d'arbres \( \mathcal{F}_E=(Q,\Sigma,\mathrm{First}({E}),\delta) \) défini par
  \begin{align*}
    Q &= \{ \mathrm{First}({E})\}\cup
      \{\mathrm{Follow}({E},f,k) \mid f \in\Sigma_m \wedge 1\leq k\leq m \},\\
    \delta &=
      \{(\mathrm{Follow}({E},f,1) , \ldots, \mathrm{Follow}({E},f,m), f, S) \mid S\in Q \wedge f\in S\cap \Sigma_m \}.
  \end{align*}
\end{definition}

L'automate des \( \mathrm{Follow} \) d'une expression \( E \) est par définition même plus petit que l'automate des positions de \( E \), puisque l'on peut considérer la surjection envoyant les couples positions-entiers vers leurs ensembles \( \mathrm{Follow} \).
Afin de montrer que cet automate reconnaît \( L(E) \), on peut montrer que \( \mathcal{F}_E \) est un quotient de \( \mathcal{P}_E \) selon une congruence Top-Down particulière, la relation \( \sim_{\mathcal{F}} \), définie pour tous deux états \( f^k \) et \( g^l \) de \( Q \) par
\begin{equation*}
  f^k \sim_{\mathcal{F}} g^l \Leftrightarrow \mathrm{Follow}({E},f,k)=\mathrm{Follow}({E},g,l).
\end{equation*}
Cette relation est une congruence très particulière.
Il s'agit en effet d'une \firstocc{relation de similarité}{relation!similarité}, c'est-à-dire une relation d'équivalence \( \sim \) sur \( Q \) telle que pour tous deux états équivalents \( q \) et \( q' \) dans \( Q \), la transition \( ((q_1,\dots,q_n),f,q) \) est dans \( \delta \) si et seulement si la transition \( ((q_1,\dots,q_n),f,q') \) l'est également.
En d'autres termes, deux états similaires admettent exactement les mêmes prédécesseurs pour chacun des symboles, y compris les symboles nullaires.
Toute relation de similarité \( \sim \) pour un automate \( A \) est trivialement une congruence Top-Down et ainsi,
\begin{equation*}
  L(A) = L(A_\sim).
\end{equation*}
Cette relation de similarité est la plus large pour \( \mathcal{P}_E \).
De plus, l'appliquer produit l'automate des follows.
\begin{proposition}\label{prop quot eq}
  Le quotient de l'automate des positions d'une expression linéaire \( E \) par \( \sim_{\mathcal{F}} \) est isomorphe à l'automate des \( \mathrm{Follow} \) de \( E \).
\end{proposition}
Cela permet ainsi de montrer aisément la correction de la construction.
\begin{theorem}\label{thm LFE eq LE}
  L'automate des \( \mathrm{Follow} \) d'une expression linéaire \( E \) reconnaît \( L(E) \).
\end{theorem}
Cette construction peut être étendue à des expressions non nécessairement linéaires comme suit.
L'\firstocc{automate des follows} \( \mathcal{F}_E \) associé à une expression \( E \) est obtenu en remplaçant chaque transition
\begin{equation*}
  ((\mathrm{Follow}(E,f,1) , \ldots,\mathrm{Follow}(E,f,m)), f, S)
\end{equation*}
de \( \mathcal{F}_{\overline{E}} \) par sa version délinéarisée, c'est-à-dire par
\begin{equation*}
  ((\mathrm{Follow}({E},f,1) , \ldots, \mathrm{Follow}({E},f,m)),h(f),S).
\end{equation*}
\begin{corollary}
	L'automate des follows d'une expression \( E \) reconnaît \( L(E) \).
\end{corollary}

\begin{figure}[H]
  \centerline{
	\begin{tikzpicture}[node distance=2.5cm,bend angle=30,transform shape,scale=1]
	  \node[accepting,state,rounded rectangle] (eps) {\( \{b,f_1,f_4,g_3,h_2,h_5\} \)};
	  \node[state, above of=eps,rounded rectangle] (h12) {\( \{b,f_1,h_2\} \)};
      \node[state, below of=eps,rounded rectangle] (g13) {\( \{b,f_4,g_3,h_5\} \)};
      \node[state, right of=g13,rounded rectangle] (g23) {\( \{a\} \)};
	  \node[state, left of=g13,rounded rectangle] (h15) {\( \{b,f_4,h_5\} \)};
	  \draw (eps) ++(-2cm,0cm) node {\( b \)}  edge[->] (eps);
	  \draw (h12) ++(2cm,0cm) node {\( b \)}  edge[->] (h12);
	  \draw (h15) ++(0cm,-1cm) node {\( b \)}  edge[->] (h15);
	  \draw (g23) ++(1cm,0cm) node {\( a \)}  edge[->] (g23);
	  \draw (g13) ++(0cm,-1cm) node {\( b \)}  edge[->] (g13);
      \path[->]
		%
		(h12) edge[->,loop,above] node {\( f,h \)} ()
		(h12) edge[->,below right] node {\( f,h \)} (eps)
	    %
		(h15) edge[->, in=135,out=-135,loop,left] node {\( f,h \)} ()
		(h15) edge[->,above left] node {\( f,h \)} (eps)
		(h15) edge[->,above left] node[pos=0.98] {\( f,h \)} (g13)
		%
	  ;
      \draw (eps) ++(1.75cm,-0.75cm)  edge[->,in=0,out=90] node[above right,pos=0] {\( g \)} (eps) edge (g13) edge (g23);
      \draw (g13) ++(1.5cm,1cm)  edge[->,in=90,out=145] node[above right,pos=0] {\( g \)} (g13) edge (g13) edge (g23);
    \end{tikzpicture}
  }
  \caption{L'automate des follows de \( E \).}%
  \label{fig r t e3}
\end{figure}
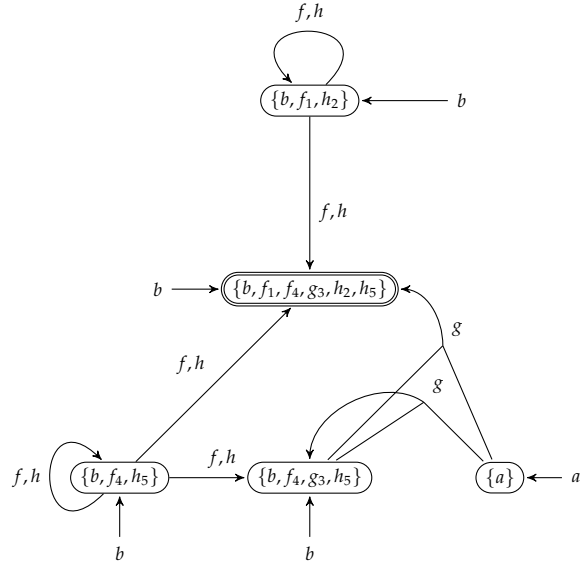

\begin{example}\label{exp follow automaton}
  L'automate des follows \( \mathcal{F}_E \) associée à l'expression
  \begin{equation*}
    {E}={({f(a)}^{*_a}\cdot_a b+ h(b))}^{*_b}+{g(c,a)}^{*_c}\cdot_c {({f(a)}^{*_a}\cdot_a b+ h(b))}^{*_b}
  \end{equation*}
  de l'Exemple~\ref{Pos Automat} est représenté Figure~\ref{fig r t e3}.
  Rappelons que les ensembles \( \mathrm{Follow} \) vérifient les égalités suivantes~:
  \begin{align*}
    \mathrm{Follow}(\overline{E},f_1,1)&=\mathrm{Follow}(\overline{E},h_2,1),\\
    \mathrm{Follow}(\overline{E},f_4,1)&=\mathrm{Follow}(\overline{E},h_5,1).
  \end{align*}
  Ainsi, l'automate des \( \mathrm{Follow} \) de \( E \) est isomorphe à l'automate obtenu en fusionnant les états \( f_1^1 \) avec \( h_2^1 \) et \( f_4^1 \) avec \( h_5^1 \) de l'automate des positions de \( E \).
  Les ensembles d'états sont
  \begin{align*}
    Q &=\{ \{a\},\{b,f_1,h_2\},\{b,f_1,h_2,g_3,f_4,h_5\},\{b,g_3,f_4,h_5\}, \{b,f_4,h_5\} \},\\
    F & =\{ \{b,f_1,h_2,g_3,f_4,h_5\} \}.
  \end{align*}
  L'ensemble des transitions est constitué des transitions

  \centerline{\( \begin{array}{cc}
     (\{b,f_1,h_2\},f,\{b,f_1,h_2,g_3,f_4,h_5\}) &
     (\{b,f_1,h_2\}, f,\{b,f_1,h_2\})\\
     (\{b,f_1,h_2\}, h, \{b,f_1,h_2,g_3,f_4,h_5\}) &
     (\{b,f_1,h_2\}, h, \{b,f_1,h_2\})\\
     ((\{b,g_3,f_4,h_5\},g_3^2), g,\{b,g_3,f_4,h_5\}) &
     ((\{b,g_3,f_4,h_5\},g_3^2), g, \{b,f_1,h_2,g_3,f_4,h_5\})\\
     (\{b,f_4,h_5\}, f ,\{b,f_1,h_2,g_3,f_4,h_5\}) &
     (\{b,f_4,h_5\},f,\{b,g_3,f_4,h_5\})\\
     (\{b,f_4,h_5\},f,\{b,f_4,h_5\}) &
     (\{b,f_4,h_5\},h,\{b,f_1,h_2,g_3,f_4,h_5\})\\
     (\{b,f_4,h_5\},h,\{b,g_3,f_4,h_5\}) &
     (\{b,f_4,h_5\},h,\{b,f_4,h_5\})\\
     (a,\{a\})&
     (b,\{b,f_1,h_2,g_3,f_4,h_5\})\\
     (b,\{b,f_1,h_2\})&
     (b,\{b,g_3,f_4,h_5\})\\
     (b,\{b,f_4,h_5\})
  \end{array} \)
}

  Le nombre d'états est \( |Q|=5 \) et le nombre de transitions est \( |\delta|=17 \).
\end{example}

\section{Les automates des c-continuations et des termes dérivés}

Les résultats de cette section sont tirés de~\cite{MOZ14b, MOZ17b, MOZ14, MOZ17}, écrits en collaboration avec Nadia Ouali-Sebti et Djelloul Ziadi.

L'automate des c-continuations (pour continuations canoniques) est un automate isomorphe à l'automate des positions dû à Champarnaud et Ziadi~\cite{CZ02}.
Cet automate permet également de faire le lien entre les méthodes de positions et les méthodes de dérivation.
En effet, il ne s'agit plus là de déterminer les successeurs directs de positions,
mais de calculer, par des formules relativement proches de la dérivation,
une expression dénotant le langage des mots suivants une position donnée dans un mot du langage.

Pour cela, Champarnaud et Ziadi s'inspirent du calcul des continuations de Berry et Sethi~\cite{BS86},
et montrent comment calculer un représentant unique (canonique) de ces continuations; en effet,
les continuations d'une position donnée ne sont pas toutes égales, mais équivalentes, modulo l'associativité,
la commutativité et l'idempotence de la somme.

Ainsi, la c-continuation \( c_a(E) \) d'une expression linéaire \( E \) par rapport à une position \( a \) de \( E \) se calcule inductivement comme suit~:
\begin{align*}
  c_a(a) &= \varepsilon, &
  c_a(F+G) &=
    \begin{cases}
      c_a(F) & \text{ si } a \text{ est une position de } F,\\
      c_a(G) & \text{sinon,}
    \end{cases}\\
  c_a(F^*) &= c_a(F)\cdot F^*, &
  c_a(F \cdot G) &=
    \begin{cases}
      c_a(F) \cdot G & \text{ si } a \text{ est une position de } F,\\
      c_a(G) & \text{sinon.}
    \end{cases}
\end{align*}

Une fois ces continuations calculées, il suffit alors de construire l'automate des c-continuations de la façon suivante~:
les états sont des couples (positions, c-continuations) dont la finalité dépend de la présence du mot vide dans le langage dénoté par la c-continuations, et les transitions sont de la forme \( ((\_,E_1), a, (a, c_a(E))) \) si \( a \) appartient à \( \mathrm{First}(E_1) \).

Le point fort de cette méthode est qu'elle permet de construire un autre automate, l'automate des termes dérivés, par un quotient très simple et ainsi de faire le lien entre les méthodes de position et celles de dérivation.

L'automate des termes dérivés est un automate construit à partir de l'opération de dérivation partielle, due à Antimirov~\cite{Ant96}.
La dérivée partielle d'une expression \( E \) par rapport à un symbole \( a \) est un ensemble \( \frac{\delta}{\delta_a}(E) \) d'expressions dont la somme \( S \) vérifie la propriété suivante~:
\begin{equation*}
  L(S) = \{ w \mid aw \in L(E)\}.
\end{equation*}
Cette opération se définit inductivement comme suit~:
\begin{align*}
  \frac{\delta}{\delta_a}(0) &= \emptyset, &
  \frac{\delta}{\delta_a}(\varepsilon) &= \emptyset,\\
  \frac{\delta}{\delta_a}(b) &=
    \begin{cases}
      \{\varepsilon \} & \text{ si } a = b,\\
      \emptyset & \text{sinon,}
    \end{cases}, &
  \frac{\delta}{\delta_a}(F+G) &= \frac{\delta}{\delta_a}(F) \cup \frac{\delta}{\delta_a}(G),\\
  \frac{\delta}{\delta_a}(F\cdot G) &=
    \begin{cases}
      \frac{\delta}{\delta_a}(F)\cdot G \cup \frac{\delta}{\delta_a}(G) & \text{ si } \mathrm{Null}(F),\\
      \frac{\delta}{\delta_a}(F)\cdot G & \text{sinon,}
    \end{cases} &
  \frac{\delta}{\delta_a}(F^*) &= \frac{\delta}{\delta_a}(F)\cdot F^*,
\end{align*}
où
\begin{equation*}
  \{F_1,\ldots,F_n\} \cdot G = \{F_1\cdot G, \ldots, F_n \cdot G\}.
\end{equation*}
Cette opération s'étend par linéarité sur les ensembles d'expressions~:
\begin{equation*}
  \frac{\delta}{\delta_{a}}(\{E_1,\ldots,E_n\}) = \frac{\delta}{\delta_{a}}(E_1) \cup\cdots\cup \frac{\delta}{\delta_{a}}(E_n),
\end{equation*}
et inductivement sur les mots~:
\begin{align*}
  \frac{\delta}{\delta_{\varepsilon}}(E) &= \{E\}, &
  \frac{\delta}{\delta_{aw}} &= \frac{\delta}{\delta_{w}} \circ \frac{\delta}{\delta_{a}}.
\end{align*}
Chaque expression contenue dans une dérivée partielle est appelée \firstocc{terme dérivé}.
De plus, le cardinal de l'ensemble des termes dérivés est nécessairement inférieur à \( (n+1) \), où \( n \) est le nombre des occurrences des symboles de l'expression.

Une fois ces termes dérivés calculés, l'automate des termes dérivés est construit ainsi~: les états sont les termes dérivés,
dont la finalité dépend de la présence du mot vide dans les langages que ces termes dénotent, et les transitions sont de la forme \( (E_1, a, E_2) \) où \( E_2 \) est un terme dérivé de \( E_1 \) par rapport à \( a \).

L'inconvénient de cette construction est qu'elle est coûteuse en temps~: en effet, en considérant \( n \) le nombre de symboles d'une expression et \( m \) la taille de son arbre syntaxique, la construction de l'automate des termes dérivés admet une complexité de l'ordre de \( n^3\times m^2 \).
L'utilisation des c-continuations permet de réduire cette complexité.

Pour cela, une fois l'automate des c-continuations calculé, il suffit de considérer l'équivalence d'états \( \sim \) définie par
\begin{equation*}
  (\_, E_1) \sim (\_, E_2) \Leftrightarrow \mathrm{h}(E_1) = \mathrm{h}(E_2)
\end{equation*}
qui est ainsi une congruence et de calculer le quotient.
Cela permet d'atteindre alors une complexité de construction quadratique par rapport à la taille de l'arbre syntaxique de l'expression.
Cette méthode a même été améliorée pour atteindre une complexité dépendant de la taille de l'automate de sortie (méthode \emph{output sensitive})~\cite{KOZ08}.

Afin d'étendre ces résultats aux expressions d'arbres, nous procéderons de la façon suivante.
Tout d'abord, nous expliciterons les formules des c-continuations ainsi que l'automate des c-continuations.
Ensuite, nous présenterons l'automate des termes dérivés de Kuske et Meinecke.
Enfin, nous illustrerons le lien morphique entre les deux.

\subsection{Les c-continuations d'arbres}

\textbf{\emph{N.~B.}~:} Dans cette section, nous considérons le quotient \( 0\cdot_c{E}=0 \).
Une expression sans occurrence de \( 0 \) ou réduite à  \( 0 \) est dite \firstocc{réduite}.

Les formules inductives de calculs des c-continuations d'une expression linéaire d'arbres sont relativement proches de celles de mots.
La seule différence est que, comme dans le cas du calcul des \( \mathrm{Follow} \), l'arité de la position doit être prise en compte en ajoutant un indice indiquant le sous-arbre choisi.
Ainsi, pour \( E \) une expression linéaire non-nulle, \( f \) un symbole \( m \)-aire (\( m\neq 0 \)) de \( E \)
et \( k \) un indice entre \( 1 \) et \( m \), la c-continuation de \( E \) par rapport à \( f^k \) se calcule inductivement comme suit~:

  \begin{table}[H]
    \begin{align*}
       C_{f^k}(g({E}_1, \ldots,{E}_m))&=
       \begin{cases}
       {E}_k & \text { si } f=g,\\
       C_{f^k}({E}_j)& \text{ si } f\in \Sigma_{{E}_j},
       \end{cases}\\
       C_{f^k}({E}_1+{E}_2)&=
       \begin{cases}
          C_{f^k}({E}_1)& \text{ si } f\in \Sigma_{{E}_1},\\
          C_{f^k}({E}_2)& \text{ si } f\in \Sigma_{{E}_2},
       \end{cases}\\
       C_{f^k}({E}_1\cdot_c {E}_2)&=
       \begin{cases}
          C_{f^k}({E}_1)\cdot_{c} {E}_2 &\text{ si } f\in \Sigma_{{E}_1},\\
          C_{f^k}({E}_2)             &\text{ si } f\in \Sigma_{{E}_2} \text{ et } c\in \mathrm{Last}(E_1),\\
          0 					   &\text{ sinon,}
       \end{cases}\\
       C_{f^k}({{E}_1}^{*_c})&=C_{f^k}({{E}_1})\cdot_{c} {{E}_1}^{*_c}.
    \end{align*}
    \caption{Définition inductive des c-continuations.}
  \end{table}
\textbf{\emph{N.~B.}~:}Pour la suite de cette section, dans un souci d'harmonisation nous poserons \( C_{\varepsilon}({E})={E} \).

Une fois les c-continuations calculées, il suffit de les combiner en suivant le même principe que dans le cas des mots afin de construire un automate.
\begin{definition}\label{def aut c cont lin}
  Soit \( {E}\neq 0 \) une expression rationnelle linéaire sur un alphabet \( \Sigma \).
  L'\firstocc{automate des c-continuations}{automate!des c-continuations} de \( E \) est l'automate  \( \mathcal{C}_E=(Q,\Sigma,\{\mathrm{cont}_{\varepsilon}\},\delta) \) défini par
 \begin{align*}
   Q &= \{\mathrm{cont}_{f^k} \mid f\in \Sigma_m, 1\leq k\leq m\}\cup \{\mathrm{cont}_{\varepsilon}\}, \\
   \delta &= \{(
     \mathfrak{C}_g,
     g,
     \mathrm{cont}_x
     )\mid  g\in{\Sigma}_m, m\geq 1, {g}\in\mathrm{First}(C_{x}({E})) \} \\
      & \quad \cup \{(c,\mathrm{cont}_x)\mid c\in L(C_{x}({E})) \cap\Sigma_0\}
  \end{align*}
  où pour toute position \( g \) d'arité \( m \), pour tout indice \( j \) entre \( 1 \) et \( m \),
  \begin{align*}
    \mathfrak{C}_g &= (\mathrm{cont}_{g^1},\dots,\mathrm{cont}_{g^m}), &
    \mathrm{cont}_{g^j} &= (g^j,C_{g^j}({E})).
  \end{align*}
\end{definition}
Enfin, l'\firstocc{automate des c-continuations}\ d'une expression \( E \) non linéaire est obtenue en appliquant le morphisme alphabétique \( \mathrm{h} \) sur \( \mathcal{C}_{\overline{E}} \).

\begin{example}\label{ex c-cont}
  Soient
  \begin{align*}
    {E} &= {({f(a)}^{*_a}\cdot_a b+ h(b))}^{*_b}+{g(c,a)}^{*_c}\cdot_c {({f(a)}^{*_a}\cdot_a b+ h(b))}^{*_b},\\
    \overline{E} &= {({f_1(a)}^{*_a}\cdot_a b+ h_2(b))}^{*_b} + {g_3(c,a)}^{*_c} \cdot_c {({f_4(a)}^{*_a}\cdot_a b+ h_5(b))}^{*_b}
  \end{align*}
  les expressions définies dans l'Exemple~\ref{Pos Automat}.
  Posons alors~:
  \begin{align*}
     F _1 &= {({f_1(a)}^{*_a}\cdot_a b+ h_2(b))}^{*_b},\\
     F _2 &= {g_3(c,a)}^{*_c},\\
     F _3 &= {({f_4(a)}^{*_a}\cdot_a b+ h_5(b))}^{*_b}.
  \end{align*}
  Ainsi,
  \begin{equation*}
     \overline{E} = F_1 + F_2 \cdot_c F_3.
  \end{equation*}
  Les c-continuations de \( \overline{E} \) sont les suivantes~:
    \begin{align*}
      C_{f^1_1}(\overline{E}) &=((a\cdot_a {f_1(a)}^{*_a})\cdot_a b)\cdot_b  F _1,\\
      C_{h^1_2}(\overline{E})&= b\cdot_b  F _1,\\
      C_{g^1_3}(\overline{E})&= (c\cdot_c  F _2)\cdot_c  F _3,\\
      C_{g^2_3}(\overline{E})&= (a\cdot_c  F _2)\cdot_c  F _3,\\
      C_{f^1_4}(\overline{E})&=((a\cdot_a {f_4(a)}^{*_a})\cdot_a b)\cdot_b  F _3,\\
      C_{h^1_5}(\overline{E})&= b\cdot_b  F _3.
    \end{align*}
  L'ensemble des états de l'automate \( \mathcal{C}_E \) est
  \begin{equation*}
    Q = \{\mathrm{cont}_\varepsilon, \mathrm{cont}_{f^1_1}, \mathrm{cont}_{h^1_2}, \mathrm{cont}_{g^1_3},
    \mathrm{cont}_{g^2_3},  \mathrm{cont}_{f^1_4}, \mathrm{cont}_{h^1_5}\}.
  \end{equation*}
  Les transitions de \( \delta \) sont

  \centerline{
    \( \begin{array}{cccc}
      (a,\mathrm{cont}_{g^2_3}), &
      (b, \mathrm{cont}_{f^1_1}), &
      (b, \mathrm{cont}_{g^1_3}), &
      (b, \mathrm{cont}_{\varepsilon}),
    \end{array} \)
  }

  \centerline{\( \begin{array}{ccc}
    (b, \mathrm{cont}_{h^1_2}), &
    (b, \mathrm{cont}_{h^1_5}), &
    (b, \mathrm{cont}_{f^1_4}),
    \end{array} \)
  }

  \centerline{\( \begin{array}{cc}
    (\mathrm{cont}_{f^1_4}, f, \mathrm{cont}_{g^1_3}), &
    (\mathrm{cont}_{f^1_4}, f, \mathrm{cont}_{f^1_4}),\\
    (\mathrm{cont}_{f^1_4}, f, \mathrm{cont}_{h^1_5}), &
    (\mathrm{cont}_{f^1_1}, f, \mathrm{cont}_{h^1_2}),\\
    (\mathrm{cont}_{f^1_1}, f, \mathrm{cont}_{\varepsilon}), &
    (\mathrm{cont}_{f^1_1}, f, \mathrm{cont}_{f^1_1}),\\
    (\mathrm{cont}_{f^1_4}, f , \mathrm{cont}_{\varepsilon}), &
    ((\mathrm{cont}_{g^1_3}, \mathrm{cont}_{g^2_3}), g, \mathrm{cont}_{\varepsilon}),\\
    (\mathrm{cont}_{h^1_5}, h, \mathrm{cont}_{\varepsilon}), &
    ((\mathrm{cont}_{g^1_3}, \mathrm{cont}_{g^2_3}), g, \mathrm{cont}_{g^1_3}),\\
    (\mathrm{cont}_{h^1_2}, h, \mathrm{cont}_{h^1_2}), &
    (\mathrm{cont}_{h^1_2}, h, \mathrm{cont}_{\varepsilon}),\\
    (\mathrm{cont}_{h^1_2}, h, \mathrm{cont}_{f^1_1}), &
    (\mathrm{cont}_{h^1_5}, h, \mathrm{cont}_{h^1_5}),\\
    (\mathrm{cont}_{h^1_5}, h, \mathrm{cont}_{f^1_4}), &
    (\mathrm{cont}_{h^1_5}, h, \mathrm{cont}_{g^1_3}).
  \end{array} \)}

  Le nombre d'états est \( |Q|=7 \) et le nombre de transitions est \( |\delta|=23 \).
  L'automate des c-continuations de \( E \) est représenté Figure~\ref{fig Cfk e}.
\end{example}


\begin{figure}[H]
  \centerline{
	\begin{tikzpicture}[node distance=2.5cm,bend angle=30,transform shape,scale=1]
	  \node[accepting,state,rounded rectangle] (eps) {\( \mathrm{cont}_{\varepsilon} \)};
	  \node[state, above left of=eps,rounded rectangle] (f11) {\( \mathrm{cont}_{f^1_1} \)};
	  \node[state, above right of=eps,rounded rectangle] (h12) {\( \mathrm{cont}_{h^1_2} \)};
      \node[state, below of=eps,rounded rectangle] (g13) {\( \mathrm{cont}_{g^1_3} \)};
      \node[state, right of=g13,node distance=3.5cm,rounded rectangle] (g23) {\( \mathrm{cont}_{g^2_3} \)};
	  \node[state, below left of=g13,node distance=3.5cm,rounded rectangle] (h15) {\( \mathrm{cont}_{h^1_5} \)};
      \node[state, below right of=g13,node distance=3.5cm,rounded rectangle] (f14) {\( \mathrm{cont}_{f^1_4} \)};
	  \draw (eps) ++(-2cm,0cm) node {\( b \)}  edge[->] (eps);
	  \draw (f11) ++(-2cm,0cm) node {\( b \)}  edge[->] (f11);
	  \draw (h12) ++(2cm,0cm) node {\( b \)}  edge[->] (h12);
	  \draw (h15) ++(0cm,-1cm) node {\( b \)}  edge[->] (h15);
	  \draw (g23) ++(2cm,0cm) node {\( a \)}  edge[->] (g23);
	  \draw (g13) ++(-2cm,0cm) node {\( b \)}  edge[->] (g13);
	  \draw (f14) ++(0cm,-1cm) node {\( b \)}  edge[->] (f14);
      \path[->]
        (f11) edge[->,below left] node {\( f_1 \)} (eps)
		(f11) edge[->,loop,above] node {\( f_1 \)} ()
		(h12) edge[->,bend right,above] node {\( h_2 \)} (f11)
		%
		(h12) edge[->,loop,above] node {\( h_2 \)} ()
		(h12) edge[->,below right] node {\( h_2 \)} (eps)
		(f11) edge[->,bend right,above] node {\( f_1 \)} (h12)
	    %
		(h15) edge[->, in=135,out=-135,loop,left] node {\( h_5 \)} ()
		(h15) edge[->,above left] node {\( h_5 \)} (eps)
		(h15) edge[->,above left] node {\( h_5 \)} (g13)
		(h15) edge[->,bend right,above] node {\( h_5 \)} (f14)
		(f14) edge[->,in=45,out=-45,loop,right] node {\( f_4 \)} ()
		(f14) edge[->,bend right,above] node {\( f_4 \)} (h15)
		(f14) edge[->,above right] node {\( f_4 \)} (eps)
		(f14) edge[->,above right] node {\( f_4 \)} (g13)
	  ;
      \draw (eps) ++(1.75cm,-0.75cm)  edge[->,in=0,out=90] node[above right,pos=0] {\( g_3 \)} (eps) edge[] (g13) edge[] (g23);
      \draw (g13) ++(1.5cm,1cm)  edge[->,in=90,out=145] node[above right,pos=0] {\( g_3 \)} (g13) edge[] (g13) edge[] (g23);
    \end{tikzpicture}
  }
  \caption{L'automate des c-continuations de \( \overline{E} \).}%
  \label{fig Cfk e}
\end{figure}

Cet automate est isomorphe à l'automate des positions de \( {E} \).
Pour le montrer, il suffit de considérer le lien suivant entre les fonctions de positions et les c-continuations~:
	\begin{equation*}
		\mathrm{Follow}(E,f,k)=\mathrm{First}(C_{f^k}( {{E}})).
	\end{equation*}
Il s'ensuit alors que
\begin{theorem}
  L'automate \( \mathcal{C}_E \) reconnaît \( L(E) \).
\end{theorem}
Montrons alors comment quotienter cet automate afin de calculer un équivalent à l'automate des termes dérivés.
Pour cela, commençons par rappeler la définition de cet automate, proposée par Kuske et Meinecke.

\subsection{L'automate des termes dérivés de Kuske et Meinecke}\label{subsec:derivPart}

Dans~\cite{KM11}, Kuske et Meinecke étendent la notion de dérivation partielle des expressions de mots (\cite{Ant96})
à la dérivation partielle d'expressions d'arbres dans le but de calculer depuis une expression \( E \) un automate d'arbre reconnaissant \( L(E) \).
Cependant, de par la présence de symboles non unaires, les dérivées partielles ne sont plus des ensembles d'expressions, mais des ensembles de \( n \)-uplets d'expressions.
Ainsi, la dérivée partielle d'une expression \( E \) par un symbole non nullaire \( f \) d'arité \( k \) est l'ensemble de \( k \)-uplets \( f^{-1}(E) \) défini inductivement comme suit~:
  \begin{align*}
    f^{-1}(0) & =\emptyset \\
    f^{-1}(F+G)&= f^{-1}( F ) \cup  f^{-1}( G )\\
    f^{-1}({ F }^{*_c})&= f^{-1}({ F })\cdot_{c} { F }^{*_c}\\
    f^{-1}(g({E}_1, \ldots,{E}_n))&=
      \begin{cases}
      \{({E}_1, \ldots,{E}_n)\} &\text { si } f=g,\\
      \emptyset &\text { sinon,}
    \end{cases}\\
    f^{-1}( F \cdot_c  G )&=
      \begin{cases}
        f^{-1}( F )\cdot_c  G &\text { si } c \notin L(F),\\
        f^{-1}( F )\cdot_c  G \cup f^{-1}( G ) &\text{ sinon,}
      \end{cases}
  \end{align*}
où pour tout \( n \)-uplet d'expressions \( \mathcal{N}=(E_1,\ldots,E_n) \), pour tout ensemble \( \mathcal{S} \) de \( n \)-uplets d'expressions~:
\begin{align*}
  \mathcal{N}\cdot_c F &= (E_1\cdot_c F ,\ldots,E_n\cdot_c F),\\
  \mathcal{S}\cdot_c F &= \{\mathcal{N}\cdot_c F \mid \mathcal{N}\in \mathcal{S}\}.
\end{align*}
La fonction \( f^{-1} \) s'étend pour tout ensemble \( S \) d'expressions rationnelles par
\begin{equation*}
  f^{-1}(S)=\bigcup_{{E}\in S}f^{-1}({E}).
\end{equation*}
Afin de déterminer l'ensemble des termes dérivés, il n'est pas nécessaire de prendre en compte la structure des arbres pour déterminer les dérivations successives.
En effet, après avoir calculé un \( k \)-uplet d'expressions, il n'est pas nécessaire de dériver le \( k \)-uplet, mais chaque expression.
L'opération de dérivation est par essence indépendante du contexte.
Ainsi, il suffit de répéter la dérivation par des symboles, ce qui revient à \emph{oublier} la structure d'opérade de l'ensemble des arbres en oubliant la graduation de l'alphabet.
On se ramène ainsi à une dérivation d'expressions d'arbres par des mots.

Par conséquent, la \firstocc{dérivée partielle}{dérivée partielle} d'une expression \( E \) par rapport à un mot \( w \) de \( \Sigma_{\geq 1}^* \) ici considéré comme le monoïde libre, est l'ensemble d'expressions dénoté par \( \partial_w(E) \) inductivement défini par~:
  \begin{align*}
   \partial_w(E)=
     \begin{cases}
        \{E\} &\text{ si } w=\varepsilon,\\
        \mathrm{Set}(f^{-1}(\partial_{u}(E)))&\text{ si } w=uf, f\in \Sigma_{\geq 1},u\in\Sigma_{\geq 1}^*, f^{-1}(\partial_{u}(E))\neq \emptyset,\\
        \{0\}&\text{ si } w=uf, f\in \Sigma_{\geq 1},u\in\Sigma_{\geq 1}^*,f^{-1}(\partial_{u}(E))=\emptyset,
    \end{cases}
  \end{align*}
où pour tout \( k \)-uplets d'expressions \( (E_1,\ldots,E_k) \), pour tout ensemble \( S \) de \( k \)-uplets d'expressions,
\begin{align*}
  \mathrm{Set}(S) &= \bigcup_{c \in S} \mathrm{Set}(c),\\
  \mathrm{Set}((E_1,\ldots,E_n)) &= \{E_1,\ldots,E_n\}.
\end{align*}
Une fois les termes dérivés calculés, il ne reste plus qu'à les utiliser pour construire l'automate recherché.
Ainsi, l'\firstocc{automate des termes dérivés}{automate!des termes dérivés} d'une expression \( E \) est l'automate \( \mathcal {A}_E=(Q,\Sigma,Q_T,\delta) \) défini par~:
\begin{align*}
  Q  &= \bigcup_{w\in \Sigma^*_{\geq 1}}\partial_{w}(E),\\
  Q_T &= \{E\},\\
  \delta &= \{((G_1,\dots, G_m),f,F)\mid  F \in Q,f\in\Sigma_m, m\geq 1,( G_1,\dots, G_m)\in f^{-1}(F)\} \\
  & \quad \cup \{(c,F)\mid F \in Q, c\in(L(F)\cap\Sigma_0)\}.
\end{align*}

\begin{figure}[H]
  \centerline{
	\begin{tikzpicture}[node distance=2.5cm,bend angle=30,transform shape,scale=1]
	  \node[accepting,state] (eps) {\( q_0 \)};
      \node[state, below of=eps] (g13) {\( q_3 \)};
      \node[state, right of=g13,node distance=3.5cm] (g23) {\( q_4 \)};
	  \node[state, below left of=g13,node distance=3.5cm] (h15) {\( q_2 \)};
      \node[state, below right of=g13,node distance=3.5cm] (f14) {\( q_1 \)};
	  \draw (eps) ++(-1cm,0cm) node {\( b \)}  edge[->] (eps);
	  \draw (h15) ++(0cm,-1cm) node {\( b \)}  edge[->] (h15);
	  \draw (g23) ++(1cm,0cm) node {\( a \)}  edge[->] (g23);
	  \draw (g13) ++(-1cm,0cm) node {\( b \)}  edge[->] (g13);
	  \draw (f14) ++(0cm,-1cm) node {\( b \)}  edge[->] (f14);
      \path[->]
		(h15) edge[->, in=135,out=-135,loop,left] node {\( h \)} ()
		(h15) edge[->,above left] node {\( h \)} (eps)
		(h15) edge[->,above left] node {\( h \)} (g13)
		(h15) edge[->,bend right,above] node {\( h \)} (f14)
		(f14) edge[->,in=45,out=-45,loop,right] node {\( f \)} ()
		(f14) edge[->,bend right,above] node {\( f \)} (h15)
		(f14) edge[->,above right] node {\( f \)} (eps)
		(f14) edge[->,above right] node {\( f \)} (g13)
	  ;
      \draw (eps) ++(1.75cm,-0.75cm)  edge[->,in=0,out=90] node[above right,pos=0] {\( g \)} (eps) edge[] (g13) edge[] (g23);
      \draw (g13) ++(1.5cm,1cm)  edge[->,in=90,out=145] node[above right,pos=0] {\( g \)} (g13) edge[] (g13) edge[] (g23);
    \end{tikzpicture}
  }
  \caption{L'automate des termes dérivés \( \mathcal{A}_E \).}%
  \label{fig r t e3 eq}
\end{figure}
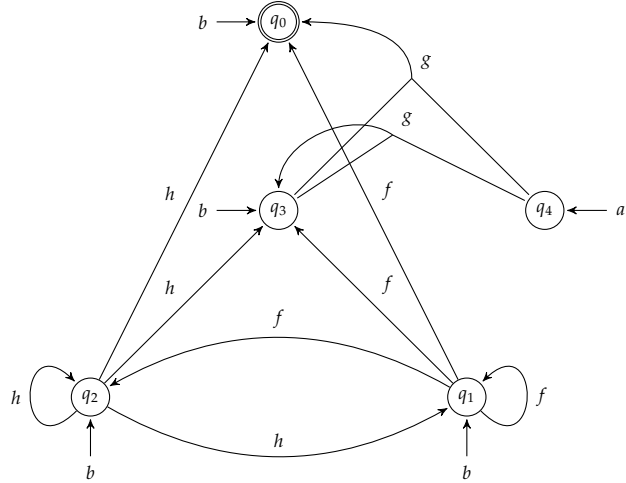

\begin{example}\label{exp equation automaton}
  Soit \( E= F + G\cdot_c  F  \) l'expression rationnelle de l'Exemple~{\ref{Pos Automat}} avec
    \begin{align*}
       F &={({f(a)}^{*_a}\cdot_a b+ h(b))}^{*_b}, &
       G&={g(c,a)}^{*_c}.
    \end{align*}
    Les dérivées partielles de \( E \) sont les ensembles
    \begin{align*}
      \partial_h(E) &=\{b\cdot_b  F \}, &
      \partial_f(E)&=\{((a\cdot_a{f(a)}^{*_a})\cdot_a b)\cdot_b  F \},\\
      \partial_{ff}(E)&=\{((a\cdot_a{f(a)}^{*_a})\cdot_a b)\cdot_b  F \}, &
      \partial_{fh}(E)&=\{b\cdot_b  F \},\\
      \partial_g(E)&=\{(a\cdot_c  G)\cdot_c  F , (c\cdot_c  G)\cdot_c  F \}, &
      \partial_{hf}(E)&=\{((a\cdot_a{f(a)}^{*_a})\cdot_a b)\cdot_b  F \},\\
      \partial_{gh}(E)&=\{b\cdot_b  F \}, &
      \partial_{hh}(E)&=\{((a\cdot_a{f(a)}^{*_a})\cdot_a b)\cdot_b  F \},\\
      \partial_{gf}(E)&=\{((a\cdot_a{f(a)}^{*_a})\cdot_a b)\cdot_b  F \}, &
      \partial_{gg}(E)&=\{(a\cdot_c  G)\cdot_c  F , (c\cdot_c  G)\cdot_c  F \}.
    \end{align*}
    Les éléments de l'ensemble des états \( Q \) sont alors les états suivants~:
    \begin{align*}
      q_0 &= E, & q_1 &= ((a\cdot_a{f(a)}^{*_a})\cdot_a b)\cdot_b  F, \\
      q_2 &= b\cdot_b  F, & q_3 &= (c\cdot_c  G)\cdot_c  F, \\
      q_4 &= (a\cdot_c  G)\cdot_c  F.
    \end{align*}
    Le seul état final est \( q_0 \).
    Les transitions sont~:
    \begin{align*}
      (b, q_0) &&
      (b, q_1) &&
      (b, q_3) \\
      (b, q_2) &&
      (a, q_4) &&
      (q_1, f, q_0)\\
      (q_1, f, q_1) &&
      (q_1, f, q_2) &&
      (q_4, f, q_1) \\
      (q_0, h, q_2) &&
      (q_2, h, q_1) &&
      (q_2, h, q_2)\\
      (q_4, h, q_2) &&
      (q_3,q_4,g,q_0) &&
      (q_3,q_4,g,q_4).
    \end{align*}
    Le nombre d'états est \( |Q|=5 \) est le nombre de transitions est \( 15 \).
    L'automate des termes dérivés de \( E \) est représenté Figure~\ref{fig r t e3 eq}.
\end{example}

\subsection{Lien entre termes dérivés et c-continuations}

Afin d'étendre la méthode de conversion de l'automate des c-continuations en automate des termes dérivés, on peut tout simplement étendre la congruence classique, c'est-à-dire la congruence Top-Down \( \sim \) suivante définie pour tout deux états \( (f^k,C_{f^k}(\overline{E})) \) et \( (g^p,C_{g^p}(\overline{E})) \) par
\begin{align*}
  (f^k,C_{f^k}(\overline{E})) \sim (g^p,C_{g^p}(\overline{E})) & \Leftrightarrow \mathrm{h}(C_{f^k}(\overline{E}))=\mathrm{h}(C_{g^p}(\overline{E})).
\end{align*}
Avec cette congruence, l'automate quotient \( {(\mathcal{C}_E)}_\sim \) est isomorphe à l'automate des termes dérivés \( \mathcal{A}_E \).

\begin{figure}[H]
	\centerline{
	\begin{tikzpicture}[node distance=2.5cm,bend angle=30,transform shape,scale=1]
		\node[accepting,state,rounded rectangle] (eps) {\( \{\mathrm{h}(C_{\varepsilon}(E))\} \)};
			\node[state, below of=eps,rounded rectangle] (g13) {\( \{\mathrm{h}(C_{g_3^1}(E))\} \)};
			\node[state, right of=g13,node distance=3.5cm,rounded rectangle] (g23) {\( \{\mathrm{h}(C_{g_3^2}(E))\} \)};
		\node[state, below left of=g13,node distance=3.5cm,rounded rectangle] (h15) {\( \{\mathrm{h}(C_{h_2^1}(E))\} \)};
			\node[state, below right of=g13,node distance=3.5cm,rounded rectangle] (f14) {\( \{\mathrm{h}(C_{f_1^1}(E))\} \)};
		\draw (eps) ++(-2cm,0cm) node {\( b \)}  edge[->] (eps);
		\draw (h15) ++(0cm,-1cm) node {\( b \)}  edge[->] (h15);
		\draw (g23) ++(2cm,0cm) node {\( a \)}  edge[->] (g23);
		\draw (g13) ++(-2cm,0cm) node {\( b \)}  edge[->] (g13);
		\draw (f14) ++(0cm,-1cm) node {\( b \)}  edge[->] (f14);
			\path[->]
		(h15) edge[->, in=135,out=-135,loop,left] node {\( h \)} ()
		(h15) edge[->,above left] node {\( h \)} (eps)
		(h15) edge[->,above left] node {\( h \)} (g13)
		(h15) edge[->,bend right,above] node {\( h \)} (f14)
		(f14) edge[->,in=45,out=-45,loop,right] node {\( f \)} ()
		(f14) edge[->,bend right,above] node {\( f \)} (h15)
		(f14) edge[->,above right] node {\( f \)} (eps)
		(f14) edge[->,above right] node {\( f \)} (g13)
		;
			\draw (eps) ++(1.75cm,-0.75cm)  edge[->,in=0,out=90] node[above right,pos=0] {\( g \)} (eps) edge[] (g13) edge[] (g23);
			\draw (g13) ++(1.5cm,1cm)  edge[->,in=90,out=145] node[above right,pos=0] {\( g \)} (g13) edge[] (g13) edge[] (g23);
		\end{tikzpicture}
	}
	\caption{L'automate \({ (\mathcal{C}_E)}_\sim \).}%
	\label{fig r t e31}
\end{figure}
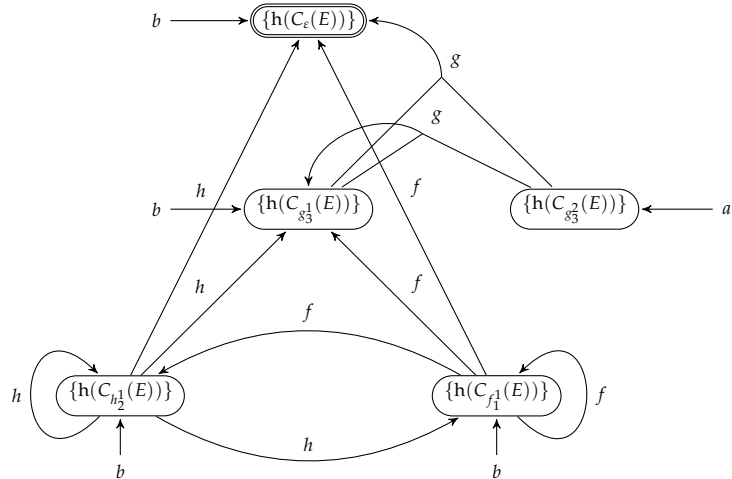

\begin{example}
  Considérons les c-continuations de l'Exemple~\ref{ex c-cont}.
  Par définition de la relation \( \sim \) sur les états de l'automate \( \mathcal{E} \) (Figure~\ref{fig Cfk e}),
	\begin{align*}
	  \mathrm{h}(C_{f^1_1}(\overline{E})) &= \mathrm{h}(C_{f^1_4}(\overline{E})), &
	  \mathrm{h}(C_{h^1_2}(\overline{E})) &= \mathrm{h}(C_{h^1_5}(\overline{E})).
	\end{align*}
	L'automate \( {(\mathcal{C}_E)}_\sim \), isomorphe à \( \mathcal{A}_E \), est représenté Figure~\ref{fig r t e31}.
	Le nombre d'états est \( |Q|=5 \) et le nombre de transitions est \( |\delta|=15 \).
\end{example}

Enfin, en utilisant cette congruence, mais également en utilisant des algorithmes classiques (tels que l'algorithme de minimisation d'automates acycliques de mots de Revuz~\cite{Rev92}),
des structures avancées (telles que la ZPC de Ziadi \emph{et.\ al}~\cite{ZPC97}) et des encodages d'expressions (compression de pseudo-continuations de~\cite{KOZ08}), nous avons montré,
avec Djelloul Ziadi et Nadia Ouali-Sebti~\cite{MOZ17b}, que l'automate des termes dérivés pouvaient être calculé en temps et en espace de l'ordre de \( n\times m \), où \( n \) est le nombre de termes dérivés de \( E \) et \( m \) la taille de l'arbre syntaxique de \( E \).


\section{Un automate plus petit}

Les résultats de cette section sont tirés de~\cite{MOZ17}, écrit avec la collaboration de Nadia Ouali-Sebti et de Djelloul Ziadi.

Dans les sections précédentes, nous avons vu que l'automate des \( \mathrm{Follow} \) et l'automate des termes dérivés étaient des quotients distincts de l'automate des positions par les relations \( \sim_{\mathcal{F}} \) et \( \sim \).

Ces deux relations étant distinctes, on peut les combiner pour avoir une relation plus grossière afin d'obtenir un automate encore plus réduit.

C'est ce que proposent dans~\cite{GLRA11} Garc\'{\i}a \emph{et al.}, où ils  donnent un nouvel algorithme pour obtenir un automate depuis une expression.
Leur méthode est basée sur le calcul d'une relation \( \equiv_V \), jonction des relations \( \sim_{\mathcal{F}} \) et \( \sim \).

Appliquons alors la méthode aux automates d'arbres directement.
L'idée est d'étendre tout d'abord la relation \( \sim_{\mathcal{F}} \) aux c-continuations comme suit~:
\begin{align*}
  C_{f^k}(\overline{E}) \sim_{\mathcal{F}} C_{g^p}(\overline{E})\Leftrightarrow \mathrm{Follow}(C_{f^k}(\overline{E}),f,k)= \mathrm{Follow}(C_{g^p}(\overline{E}),p).
\end{align*}
Une fois cela fait, il ne reste plus qu'à combiner les relations \( \sim_{\mathcal{F}} \) et \( \sim \) en une relation \( \equiv_V \)~:
\begin{align*}
  C_{f^k}(\overline{E}) \equiv_V C_{g^p}(\overline{E}) \Leftrightarrow
  \begin{cases}
    (\exists C_{h^l}(\overline{E}) \sim_{\mathcal{F}} C_{f^k}(\overline{E}) \mid~ C_{h^l}(\overline{E})\sim C_{g^p}(\overline{E}))\\
    \quad \vee (\exists C_{h^l}(\overline{E}) \sim_{\mathcal{F}} C_{g^p}(\overline{E})\mid~ C_{h^l}(\overline{E})\sim  C_{f^k}(\overline{E}))
  \end{cases}
\end{align*}
et à calculer le quotient de l'automate des c-continuations par cette nouvelle relation.

Plus formellement, l'\firstocc{automate réduit}{automate!réduit} \( \mathcal{V} \), associé à une expression \( E \), est l'automate \( (Q,\Sigma,\{ {[\overline{E}]}_{\equiv_V}\},\delta) \) défini par
 \begin{align*}
   Q &=  \{\mathrm{contEq}_{f^k} \mid f\in {\Sigma_{\overline{E}}}_m, 1\leq k\leq m\}\cup \{ {[\overline{E}]}_{\equiv_V}\} \\
   \delta  &= \{(\mathfrak{C}_g,\mathrm{h}(g), \mathrm{contEq}_{x}) \mid  g\in{\Sigma_{\overline{E}}}_m, m\geq 1, g \in \mathrm{First}(C_{x}(\overline{E}))\} \\
      & \qquad \cup \{(\mathrm{contEq}_{x},c)\mid c\in L(C_{x}(\overline{E}))\cap\Sigma_0\}
\end{align*}
où
\begin{align*}
  \mathrm{contEq}_{g^j} &= {[C_{g^j}(\overline{E})]}_{\equiv_V},\\
  \mathfrak{C}_g &= (\mathrm{contEq}_{g^1}, \ldots, \mathrm{contEq}_{g^m}).
\end{align*}
En tant que conséquence directe, on peut montrer que l'automate \( \mathcal{V} \) est plus petit que l'automate des follows ou que l'automate des termes dérivés.
En effet, si deux continuations admettent les mêmes ensembles de Follows, ils sont fusionnés dans l'automate des follows et ainsi dans \( \mathcal{V} \).
Par conséquent, deux états distincts de \( \mathcal{V} \) admettent des ensemble de Follows différents.

D'une façon similaire, si deux états distincts de l'automate des continuations sont égaux une fois délinéarisés, ces états sont fusionnés dans \( \mathcal{V} \).
Ainsi, deux états \( \sim \)-équivalents dans l'automate des c-continuations sont fusionnés dans l'automate des termes dérivés tout comme dans \( \mathcal{V} \).
Par conséquent, deux états \( \sim \)-équivalents ne peuvent apparaître dans des états distincts de \( \mathcal{V} \).

Finalement, remarquons que les expressions fusionnées admettent les mêmes prédécesseurs une fois délinéarisées, puisque \( \sim_{\mathcal{F}} \) est une relation de similarité, et que \( \sim \) préserve la structure syntaxique des expressions et ainsi les résultats de dérivation.

Par conséquent, on peut en déduire les liens morphiques entre l'automate \( \mathcal{V} \), l'automate des termes dérivés et l'automate des follows, le premier étant un quotient des suivants.
\begin{example}
  Considérons l'expression de l'Exemple~\ref{Pos Automat}:
  \begin{align*}
    {E}&={({f(a)}^{*_a}\cdot_a b+ h(b))}^{*_b}+{g(c,a)}^{*_c}\cdot_c {({f(a)}^{*_a}\cdot_a b+ h(b))}^{*_b},\\
    \overline{E}&={({f_1(a)}^{*_a}\cdot_a b+ h_2(b))}^{*_b}+{g_3(c,a)}^{*_c}\cdot_c {({f_4(a)}^{*_a}\cdot_a b+ h_5(b))}^{*_b}.
  \end{align*}
  Pour rappel, les c-continuations de \( {E} \) sont les suivantes~:
  \begin{align*}
    C_{f^1_1}(\overline{E})&=((a\cdot_a {f_1(a)}^{*_a})\cdot_a b)\cdot_b {({f_1(a)}^{*_a}\cdot_a b+ h_2(b))}^{*_b},\\
    C_{h^1_2}(\overline{E})&= b\cdot_b  {({f_1(a)}^{*_a}\cdot_a b+ h_2(b))}^{*_b},\\
    C_{g^1_3}(\overline{E})&= (c\cdot_c {g_3(c,a)}^{*_c})\cdot_c  {({f_4(a)}^{*_a}\cdot_a b+ h_5(b))}^{*_b},\\
    C_{g^2_3}(\overline{E})&= ({(a\cdot_c g_3(c,a))}^{*_c})\cdot_c {({f_4(a)}^{*_a}\cdot_a b+ h_5(b))}^{*_b},\\
    C_{f^1_4}(\overline{E})&= ((a\cdot_a {f_4(a)}^{*_a})\cdot_a b)\cdot_b {({f_4(a)}^{*_a}\cdot_a b+ h_5(b))}^{*_b},\\
    C_{h^1_5}(\overline{E})&= b\cdot_b {({f_4(a)}^{*_a}\cdot_a b+ h_5(b))}^{*_b}.
  \end{align*}
  Par application de \( \sim_{\mathcal{F}} \), on obtient~:
  \begin{align*}
    C_{f^1_1}(\overline{E})&\sim_{\cal F} C_{h^1_2}(\overline{E}), & C_{f^1_4}(\overline{E})&\sim_{\cal F} C_{h^1_5}(\overline{E}).
  \end{align*}
  Les états \( C_{g^1_3}(\overline{E}) \) et \( C_{g^2_3}(\overline{E}) \) sont indépendants.

  \noindent Le nombre des états est \( |Q|=5 \) et le nombre de transitions est \( |\delta|=15 \).
  L'automate quotient par \( \sim_{\mathcal{F}} \), isomorphe à l'automate des \( \mathrm{Follow} \), est représenté Figure~\ref{fig Cfk ef}.

  \noindent De plus,
	\begin{align*}
	  \mathrm{h}(C_{f^1_1}(\overline{E})) &= \mathrm{h}(C_{f^1_4}(\overline{E})), &
	  \mathrm{h}(C_{h^1_2}(\overline{E})) &= \mathrm{h}(C_{h^1_5}(\overline{E})).
	\end{align*}
  Ainsi,
  \begin{equation*}
    C_{f^1_1}(\overline{E}) \equiv C_{h^1_2}(\overline{E}) \equiv C_{h^1_4}(\overline{E}) \equiv C_{h^1_5}(\overline{E}).
  \end{equation*}

  L'automate réduit de \( E \) est représenté Figure~\ref{fig Cfk efe}.
  Le nombre d'états est  \( |Q|=4 \) et le nombre de transitions est \( |\delta|=14 \).
\end{example}
\begin{figure}[H]
  \centerline{
  \begin{tikzpicture}[node distance=2.5cm,bend angle=30,transform shape,scale=1]
    \node[accepting,state,rounded rectangle] (eps) {\( \{C_{\varepsilon}(\overline{E})\} \)};
    \node[state, above of=eps,rounded rectangle] (h12) {\( \{C_{f_1^1}(\overline{E}),C_{h_2^1}(\overline{E})\} \)};
      \node[state, below of=eps,rounded rectangle,node distance=3.5cm] (g13) {\( \{C_{g_3^1}(\overline{E})\} \)};
      \node[state, right of=g13,rounded rectangle,node distance=3.5cm] (g23) {\( \{C_{g_3^2}(\overline{E})\} \)};
    \node[state, left of=g13,rounded rectangle,node distance=3.5cm] (h15) {\( \{C_{f_4^1}(\overline{E}),C_{h_5^1}(\overline{E})\} \)};
    \draw (eps) ++(-2cm,0cm) node {\( b \)}  edge[->] (eps);
    \draw (h12) ++(2cm,0cm) node {\( b \)}  edge[->] (h12);
    \draw (h15) ++(0cm,-1cm) node {\( b \)}  edge[->] (h15);
    \draw (g23) ++(2cm,0cm) node {\( a \)}  edge[->] (g23);
    \draw (g13) ++(0cm,-1cm) node {\( b \)}  edge[->] (g13);
      \path[->]
    %
    (h12) edge[->,loop,above] node {\( f,h \)} ()
    (h12) edge[->,below right] node {\( f,h \)} (eps)
      %
    (h15) edge[->, in=147,out=-147,loop,left] node {\( f,h \)} ()
    (h15) edge[->,above left] node {\( f,h \)} (eps)
    (h15) edge[->,above left] node {\( f,h \)} (g13)
    %
    ;
      \draw (eps) ++(1.75cm,-0.75cm)  edge[->,in=0,out=90] node[above right,pos=0] {\( g \)} (eps) edge[] (g13) edge[] (g23);
      \draw (g13) ++(1.5cm,1cm)  edge[->,in=90,out=145] node[above right,pos=0] {\( g \)} (g13) edge[] (g13) edge[] (g23);
    \end{tikzpicture}
  }
  \caption{L'automate \( {(\mathcal{C}_E)}_{\sim_{\mathcal{F}}} \).}%
  \label{fig Cfk ef}
\end{figure}
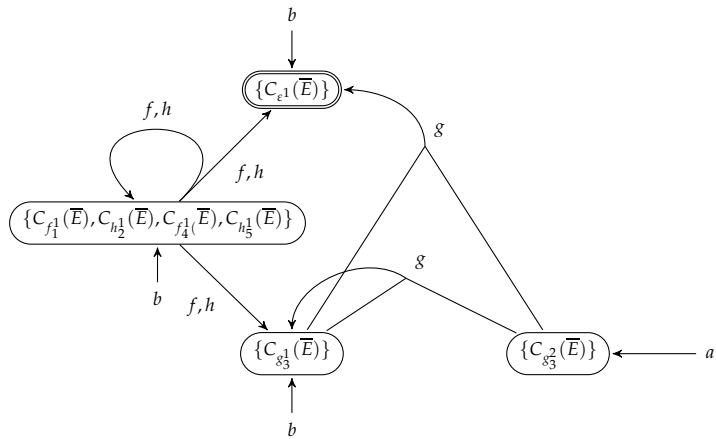
\begin{figure}[H]
  \centerline{
  \begin{tikzpicture}[node distance=2.5cm,bend angle=30,transform shape,scale=1]
    \node[accepting,state,rounded rectangle] (eps) {\( \{C_{\varepsilon^1}(\overline{E})\} \)};
      \node[state, below of=eps,rounded rectangle,node distance=3.5cm] (g13) {\( \{C_{g_3^1}(\overline{E})\} \)};
      \node[state, right of=g13,rounded rectangle,node distance=3.5cm] (g23) {\( \{C_{g_3^2}(\overline{E})\} \)};
    \node[state, below left of=eps,rounded rectangle] (h15) {\( \{C_{f_1^1}(\overline{E}),C_{h_2^1}(\overline{E}),C_{f_4^1(}\overline{E}),C_{h_5^1}(\overline{E})\} \)};
    \draw (eps) ++(0cm,1cm) node {\( b \)}  edge[->] (eps);
    \draw (h15) ++(0cm,-1cm) node {\( b \)}  edge[->] (h15);
    \draw (g23) ++(2cm,0cm) node {\( a \)}  edge[->] (g23);
    \draw (g13) ++(0cm,-1cm) node {\( b \)}  edge[->] (g13);
      \path[->]
    %
      %
    (h15) edge[->,loop,above] node {\( f,h \)} ()
    (h15) edge[->,below right] node {\( f,h \)} (eps)
    (h15) edge[->,below left] node {\( f,h \)} (g13)
    %
    ;
      \draw (eps) ++(1.75cm,-0.75cm)  edge[->,in=0,out=90] node[above right,pos=0] {\( g \)} (eps) edge[] (g13) edge[] (g23);
      \draw (g13) ++(1.5cm,1cm)  edge[->,in=90,out=145] node[above right,pos=0] {\( g \)} (g13) edge[] (g13) edge[] (g23);
    \end{tikzpicture}
  }
  \caption{L'automate \( \mathcal{V} \).}%
  \label{fig Cfk efe}
\end{figure}

\chapter{Les constructions Bottom-Up}\label{chapBotUp}

Les méthodes de construction Bottom-Up permettent de réaliser une distinction plus fine
en termes de propriétés que celles relatives aux automates descendants; l'approche Bottom-Up permet en effet de considérer les automates
déterministes comme reconnaisseurs de tout langage rationnel d'arbres, ce qui permet d'étendre les études possibles sur les langages et expressions d'arbres.

Par exemple, en étendant la caractérisation des automates de Glushkov de Caron et Ziadi~\cite{CZ00},
il serait ensuite possible de s'intéresser à des propriétés de langages (telle que la 1-non-ambiguïté
d'un langage~\cite{BW98}) ou à des optimisations de mises à jour d'expressions et d'automates~\cite{BDHLM04}.

Cela permet également de considérer d'autres méthodes de construction d'automates, tel que la méthode de dérivation de Brzozowski~\cite{Brz64}.
Avant d'aborder cette méthode par dérivation, commençons par présenter les méthodes ascendantes de construction d'automates d'arbres.


\section{L'automate des positions Bottom-Up}

Les résultats de cette section sont tirés de~\cite{AMZ18a,AMZ18b}, écrits avec la collaboration de Samira Attou et de Djelloul Ziadi.

Comme dans le cas de l'approche Top-Down, il est possible de définir une méthode de
construction Bottom-Up d'automates basée sur la construction de l'automate des positions.
Il suffit de considérer alors les parents des différents symboles pouvant apparaître dans les arbres
du langage au lieu des fils pour le calcul des transitions.

\textbf{\emph{N.B.}}: Dans ce qui suit, nous ne considérons que des expressions
dont les sous-expressions de la forme \(E_1\cdot_c E_2\) satisfont toutes le fait que \( c \) apparaisse dans \( E_1 \).

Comme rappelé dans la Section~\ref{sec aut pos}, la construction de l'automate des positions
est basée sur le calcul de fonctions de positions.
La seule différence ici avec la version Top-Down est l'utilisation non plus de la fonction \(\mathrm{Follow}\),
mais d'une fonction déterminant l'ensemble des parents possibles d'un symbole.
De plus, afin de conserver une sémantique Bottom-Up, nous renommerons \(\Racine \) la fonction
\(\mathrm{First}\).

  Soient \( E \) une expression linéaire sur un alphabet gradué \( \Sigma \) et \( f \) un symbole de \( \Sigma_k \).
  L'ensemble \firstocc{\( \Racine(E) \)}{ensemble!des racines d'un arbre}, sous-ensemble de \( \Sigma \), contient les racines des arbres de \( L(E) \), c'est-à-dire
  \begin{equation}\label{eq def root}
    \Racine(E) = \{\racine(t) \mid t\in L(E)\}.
  \end{equation}
  L'ensemble \firstocc{\( \Parent(E,f) \)}{ensemble!des parents d'un arbre}, sous-ensemble de \( \Sigma\times\mathbb{N} \), contient le couple \( (g,i) \) s'il existe un arbre dans \( L(E) \) avec un n{\oe}ud étiqueté par \( g \) dont le \( i \)-ième fils est un n{\oe}ud étiqueté par \( f \)~:
 \begin{equation}\label{eq def father}
    \Parent(E,f)=\bigcup_{t\in L(E)} \parent(t,f).
  \end{equation}
  \begin{example}\label{ex:calcul fonctions glushkov}
    Considérons l'alphabet gradué défini par \( \Sigma_2=\{f\} \), \( \Sigma_1=\{g\} \), et \( \Sigma_0=\{a,b\} \).
    Soient \( E \) et \( \overline{E} \) les expressions définies par
    \begin{equation*}
      E = {(f(a,a)+g(b))}^{*_a}\cdot_b f(g(a),b), \quad
      \overline{E} = {(f_1(a,a)+g_2(b))}^{*_a}\cdot_b f_3(g_4(a),b).
    \end{equation*}
     Ainsi,
     \begin{gather*}
       \Racine(\overline{E}) = \{a,f_1,g_2\},\\
       \begin{aligned}
         \Parent(\overline{E},f_1) &= \{(f_1,1),(f_1,2)\}, &
         \Parent(\overline{E},a) &= \{(f_1,1),(f_1,2),(g_4,1)\},\\
         \Parent(\overline{E},g_2) &= \{(f_1,1),(f_1,2)\}, &
         \Parent(\overline{E},b) &= \{(f_3,2)\},\\
         \Parent(\overline{E},f_3) &= \{(g_2,1)\}, &
         \Parent(\overline{E},g_4) &= \{(f_3,1)\}.
       \end{aligned}
     \end{gather*}
 \end{example}
La fonction \(\Racine \) se calcule d'une façon similaire à la fonction \(\mathrm{First}\)
(voir Table~\ref{table:first}).
La fonction \(\Parent \) se calcule comme suit:
\begin{table}[H]
  \begin{align*}
    \Parent(g(E_1,\ldots,E_n),f) &= \bigcup_{i\leq n} \Parent(E_i,f) \cup \{(g,i)\mid f\in\Racine(E_i)\},\\
    \Parent(E_1+E_2,f) &= \Parent(E_1,f)\cup \Parent(E_2,f),\\
    \Parent(E_1\cdot_c E_2,f) &=
        (\Parent(E_1,f) \mid f \neq c) \cup \Parent(E_2,f)\\
      & \qquad \cup (\Parent(E_1,c) \mid f\in\Racine(E_2))\\
    \Parent(E_1^{*_c},f) &= \Parent(E_1,f) \cup (\Parent(E_1,c) \mid f\in\Racine(E_1)),\\
  \end{align*}
  \caption{Définition inductive de \(\Parent \).}
\end{table}

Une fois ces fonctions calculées, il est alors possible de définir un automate
reconnaissant le langage dénoté par l'expression de départ.
Pour cela, on construit un automate dont les états sont les positions de l'expression (y compris les positions d'arité \(0\)),
les états correspondant aux racines étant finaux, puis les transitions sont
déterminées à l'aide de la fonction \(\Parent \).

  \begin{definition}\label{def glush bu}
    L'\firstocc{automate des positions Bottom-Up}{Construction d'automates!automate des positions Bottom-Up} \( \mathcal{P}_{E} \) d'une expression linéaire \( E \) sur un alphabet \( \Sigma \) est l'automate \( (\Sigma,\mathrm{Pos}(E),\Racine(E),\delta) \) défini par
    \begin{equation*}
      ((f_1,\ldots,f_n),g,h) \in \delta  \Leftrightarrow h = g \wedge \forall i \leq n,      (g,i)\in\Parent(E,f_i).
    \end{equation*}
  \end{definition}
  Remarquons qu'en conséquence de la linéarité de \( E \), \( \mathcal{P}_{E} \) est nécessairement déterministe.
  \begin{example}
    L'automate des positions Bottom-Up \( (\mathrm{Pos}(\overline{E}),\mathrm{Pos}(\overline{E}),\Racine(\overline{E}),\delta) \) de l'expression \( \overline{E} \) définie dans l' Exemple~\ref{ex:calcul fonctions glushkov} est définie comme suit~:
    \begin{gather*}
      \begin{aligned}
        \mathrm{Pos}(E) & = \{a,b,f_1,g_2,f_3,g_4\}, &
      \Racine(\overline{E}) &= \{a,f_1,g_2\},
      \end{aligned}\\
      \begin{aligned}
        \delta &= \{(a,a), (b,b),  ((a,a),f_1,f_1), ((a,f_1),f_1,f_1), ((a,g_2),f_1,f_1), ((f_1,a),f_1,f_1),\\
        & \qquad  ((f_1,f_1),f_1,f_1), ((f_1,g_2),f_1,f_1), ((g_2,a),f_1,f_1), ((g_2,f_1),f_1,f_1),\\
        & \qquad  ((g_2,g_2),f_1,f_1), (f_3, g_2,g_2),((b,g_4),f_3,f_3), (a,g_4,g_4)\}.
      \end{aligned}
    \end{gather*}
  \end{example}
  L'automate des positions Bottom-Up d'une expression (non nécessairement linéaire) \( E \) est obtenu en calculant tout d'abord l'automate des positions Bottom-Up de son expression linéarisée \( \overline{E} \) et ensuite en appliquant le morphisme de délinéarisation \( \mathrm{h} \).

  Un des avantages de cette méthode est qu'elle permet d'utiliser la structure
  des expressions rationnelles afin de calculer un automate compressé de façon directe.
  En effet, les expressions rationnelles factorisent naturellement la structure de transitions d'un automate des positions.
  \begin{definition}\label{def glu comp}
    L'\firstocc{automate compressé des positions Bottom-Up}{Construction d'automates!automate des positions Bottom-Up!compressé} \( \mathcal{C}(E) \) d'une expression linéaire \( E \) est l'automate \( (\Sigma,\mathrm{Pos}(E),\Racine(E),\delta) \) défini par
    \begin{equation*}
      \delta = \{(Q_1,\ldots,Q_k,f,\{f\}) \mid Q_i = \{g \mid (f,i) \in \Parent(E,g)\} \}.
    \end{equation*}
  \end{definition}
  \begin{example}
    Considérons l'expression \( \overline{E} \) définie dans l'Exemple~\ref{ex:calcul fonctions glushkov}.
    L'automate compressé de \( \overline{E} \) est représenté Figure~\ref{linear_compressed_automata}.
  \end{example}
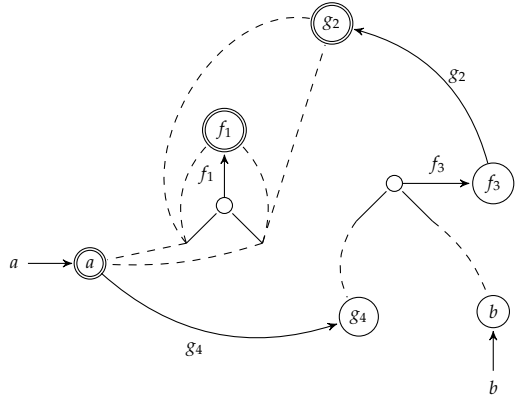
\begin{figure}[H]
  \centerline{
  \begin{tikzpicture}[node distance=2.5cm,bend angle=30,transform shape,scale=1]
    \node[accepting ,state] (f1)  {\( f_1 \)} ;
    \node[state,double,  above right of=f1, node distance =2cm] (g_2) {\( g_2 \)};
     \node[state, below right of=g_2, node distance = 3cm] (f_3) {\( f_3 \)};
    \node[state, below of= f_3,node distance=1.7cm] (b) {\( b \)};
    \node[state,  below left of=f_3] (g_4) {\( g_4 \)};
    \node[state,  below of=f1,node distance = 1cm] (cerc) {};
    \node[state,  left of=f_3,node distance = 1.3cm] (cerc1) {};
    \node[state, double,  below left of=f1,node distance = 2.5cm] (a) {\( a \)};
    \draw (a) ++(-1cm,0cm) node {\( a \)}  edge[->] (a);
    \draw (b) ++(0cm,-1cm) node {\( b \)}  edge[->] (b);
    \path[->]
     (f_3) edge[->,  bend right,right] node {\( g_2 \)} (g_2)
     (a) edge[->,below left,bend right] node {\( g_4 \)} (g_4)
     (cerc) edge[->, left] node {\( f_1 \)} (f1)
     (cerc1) edge[->, above] node {\( f_3 \)} (f_3);
    \draw (cerc) ++(0.5 cm,-0.5cm)edge node [above,pos=0.5] {} (cerc)  edge[dashed]node[left,above,pos=0.6]{} (g_2)  edge[dashed,bend left=10] node[left,above,pos=0.5]{}(a)edge[dashed,bend right=30] node[left,above,pos=0.5]{}(f1);
    \draw (cerc) ++(-0.5 cm,-0.5cm)edge node [above,pos=0.5] {} (cerc)  edge[dashed,bend left=70]node[left,above,pos=0.6]{} (g_2)  edge[dashed] node[left,above,pos=0.5]{}(a)
    edge[dashed,bend left=30] node[left,above,pos=0.5]{}(f1);
    \draw (cerc1) ++(-0.5 cm,-0.5cm)edge node [above,pos=0.5] {} (cerc1)  edge[dashed,bend right]node[left,above,pos=0.6]{} (g_4);
    \draw (cerc1) ++(0.5 cm,-0.5cm)edge node [above,pos=0.5] {} (cerc1)  edge[dashed,bend left=15]node[left,above,pos=0.6]{} (b);
    \end{tikzpicture}
    }
    \caption{L'automate compressé de l'expression \( {(f_1(a,a)+g_2(b))}^{*_a}\cdot_b f_3(g_4(a),b) \).}%
    \label{linear_compressed_automata}
\end{figure}
  L'automate des positions compressé d'une expression non nécessairement linéaire \( E \) est obtenu en calculant tout d'abord l'automate des positions compressé de sa linéarisée \( \overline{E} \) puis en appliquant le morphisme de délinéarisation \( \mathrm{h} \).


\section{L'automate des parents}\label{sec:father}

Les résultats de cette section sont tirés de~\cite{AMZ18a,AMZ18b}, écrits avec la collaboration de Samira Attou et de Djelloul Ziadi.

  Dans cette section, nous allons définir l'automate des parents associé à une expression.
  Cet automate est l'extension du cas classique de l'automate (de mots) des Follows~\cite{IY03} déjà présenté dans le cas des automates Top-Down dans la Section~\ref{sec:follows}.

  Afin de faciliter l'écriture des définitions, nous allons inclure le calcul de la fonction \( \Racine \) dans celui de la fonction \( \Parent \) en ajoutant un symbole unaire \( \$ \) qui n'est pas dans \( \Sigma \) tout en haut de l'arbre syntaxique d'une expression, c'est-à-dire considérer une expression \( \$(E) \) à la place de l'expression \(E\).
  En effet,
  \begin{equation}
    f \in \Racine(E) \Leftrightarrow (\$, 1) \in \Parent(\$(E),f).
  \end{equation}
  Autrement dit, les deux conditions suivantes sont équivalentes~:
    \begin{enumerate}
      \item \( f \) est un état final de l'automate des positions Bottom-Up,
      \item \( (\$, 1) \) est dans \( \Parent(\$(E),f) \).
    \end{enumerate}
  À l'aide de cette formule étendue, on peut alors calculer très simplement un automate plus petit que l'automate des positions, comme suit.
  \begin{definition}\label{def father aut}
    L'\firstocc{automate des parents}{Construction d'automates!automate des parents} d'une expression linéaire \( E \) sur un alphabet gradué \( \Sigma \) est l'automate \( \mathcal{F}_E=(\Sigma,Q,F,\delta) \) défini par
    \begin{gather*}
      \begin{aligned}
        Q &= \{\Parent(\$(E),f) \mid f\in\Sigma \}, &
        F &= \{q\in Q \mid \$ \in q\},
      \end{aligned}\\
      ((\Parent(\$(E),f_1),\ldots,\Parent(\$(E),f_n)),h,\Parent(\$(E),g)) \in \delta \\
      \Leftrightarrow h = g \wedge \forall i \leq n,      (g,i)\in\Parent(E,f_i).
    \end{gather*}
  \end{definition}
  En conséquence de la linéarité de \( E \), \( \mathcal{P}_{E} \) est nécessairement déterministe.

  Comme dans le cas de l'automate des \(\mathrm{Follows}\), nous allons montrer que l'automate des parents est un quotient de l'automate (Bottom-Up) des positions.
  Pour cela, nous commençons par définir une congruence basée sur la fonction \(\Parent \).

  \begin{definition}
    La \firstocc{congruence des parents}{congruence!Bottom-Up!des parents} associée à une expression linéaire \( E \) sur un alphabet \( \Sigma \) est la congruence \( \sim \) définie pour toutes deux positions \(p\) et \(p'\) par
    \begin{equation*}
      p \sim p' \Leftrightarrow \Parent(\$(E),p) = \Parent(\$(E),p').
    \end{equation*}
  \end{definition}
  D'une façon équivalente, la congruence des parents est le \firstocc{noyau}{fonction!noyau d'une} de la fonction envoyant un symbole \( p \) sur \( \Parent(\$(E),p) \).
  On pourra vérifier simplement que la congruence des parents d'une expression linéaire \( E \) est une congruence Bottom-Up pour la fonction de transition de l'automate des positions Bottom-Up de \( E \).
  De plus, l'automate des parents associé à une expression linéaire \( E \) est isomorphe au quotient de l'automate des positions Bottom-Up de \( E \) par la congruence des parents.
  Ainsi, l'automate des parents d'une expression linéaire \( E \) reconnaît \( L(E) \).

  En appliquant le morphisme de délinéarisation \( \mathrm{h} \) sur \( \mathcal{F}_{\overline{E}} \), on obtient l'automate des parents de l'expression \( E \), automate qui reconnaît \( L(E) \).
  \begin{example}
    L'automate des parents \( (\mathrm{Pos}(\overline{E}),{\mathrm{Pos}(\overline{E})}_\sim,{\Racine(\overline{E})}_\sim,\delta) \) de l'expression \( \overline{E} \) définie dans l'Exemple~\ref{ex:calcul fonctions glushkov}
    est obtenu en fusionnant les états \( f_1 \) et \( g_2 \) de \( \mathcal{P}_E \), c'est à dire~:
    \begin{gather*}
      \begin{aligned}
        \mathrm{Pos}(E) & = \{[a],[b],\{f_1,g_2\},[f_3],[g_4]\}, &
      \Racine(\overline{E}) &= \{[a],[f_1]\},
      \end{aligned}\\
      \begin{aligned}
        \delta &= \{(a,[a]), (b,[b]),  (([a],[a]),f_1,[f_1]), (([a],[f_1]),f_1,[f_1]),  (([f_1],a),f_1,[f_1]),\\
        & \qquad  (([f_1],[f_1]),f_1,[f_1]), ([f_3],g_2,[g_2]),(([b],[g_4]),f_3,[f_3]), ([a],g_4,[g_4])\}.
      \end{aligned}
    \end{gather*}
  \end{example}

  Comme dans le cas de l'automate des positions, l'approche Bottom-Up permet de
  bénéficier facilement de la structure des expressions pour \emph{factoriser}
  les transitions dans un automate compressé.
  Il est ainsi possible de calculer un automate compressé plus petit que l'automate compressé des positions, comme suit.
  \begin{definition}\label{def fat comp}
    L'\firstocc{automate compressé des parents}{Construction d'automates!automate des parents!compressé} \( \mathcal{CF}(E) \) d'une expression linéaire \( E \) est l'automate défini par le quintuplet \( (\Sigma,\mathrm{Pos}(E),\Racine(E),\delta) \) où
    \begin{equation*}
      \delta = \{(Q_1,\ldots,Q_k,f,\{\Parent(\$(E),f)\})
        \mid Q_i = \{\Parent(\$(E),g) \mid (f,i) \in \Parent(\$(E),g)\} \}.
    \end{equation*}
  \end{definition}
  En conséquence de l'Équation~\eqref{eq:extdeltaEnsComp}, la définition d'une congruence Bottom-Up pour \( A \) est exactement la même que celle de l'Équation~\eqref{eq def bot up cong}, en étendant la définition de la congruence aux ensembles de cardinalité inférieures à 0 comme suit:
  \begin{align*}
    \emptyset &\sim \emptyset, & \emptyset &\not \sim \{ \_ \}, & \{p\} \sim \{q\} &\Leftrightarrow p\sim q.
  \end{align*}
  Ainsi, le \firstocc{quotient}{quotient!d'un automate!compressé} d'un automate compressé déterministe \( A=(\Sigma,Q,F,\delta) \) par une congruence Bottom-Up \( \sim \) est l'automate compressé \( A_\sim=(\Sigma,Q_\sim, F_\sim, \delta') \) où
  \begin{equation*}
    \delta'((Q_1,\ldots,Q_m),f) = \{\phi(q)\mid q\in \delta((q_1,\ldots,q_m),f) \wedge  \forall i\leq m, [q_i] \in Q_i\}.
  \end{equation*}
  On pourra alors vérifier que l'automate compressé des parents est un quotient de l'automate des positions compressé par la congruence des parents.
  Ainsi, l'automate compressé des parents et l'automate des parents d'une expression linéaire \( E \) reconnaissent \( L(E) \).

\begin{figure}[H]
  \centerline{
  \begin{tikzpicture}[node distance=2.5cm,bend angle=30,transform shape,scale=1]
    \node[accepting ,state, rounded rectangle] (f1)  {\( \{f_1,g_2\} \)} ;
     \node[state, below right of=g_2, node distance = 3cm, rounded rectangle] (f_3) {\( [f_3] \)};
    \node[state, below of= f_3,node distance=1.7cm, rounded rectangle] (b) {\( [b] \)};
    \node[state,  below left of=f_3, rounded rectangle] (g_4) {\( [g_4] \)};
    \node[state,  below of=f1,node distance = 1cm] (cerc) {};
    \node[state,  left of=f_3,node distance = 1.3cm] (cerc1) {};
    \node[state, double,  below left of=f1,node distance = 2.5cm, rounded rectangle] (a) {\( [a] \)};
    \draw (a) ++(-1cm,0cm) node {\( a \)}  edge[->] (a);
    \draw (b) ++(0cm,-1cm) node {\( b \)}  edge[->] (b);
    \path[->]
     (f_3) edge[->,  bend right,above right] node {\( g_2 \)} (f1)
     (a) edge[->,below left,bend right] node {\( g_4 \)} (g_4)
     (cerc) edge[->, left] node {\( f_1 \)} (f1)
     (cerc1) edge[->, above] node {\( f_3 \)} (f_3);
    \draw (cerc) ++(0.5 cm,-0.5cm)edge node [above,pos=0.5] {} (cerc)  edge[dashed,bend left=10] node[left,above,pos=0.5]{}(a)edge[dashed,bend right=30] node[left,above,pos=0.5]{}(f1);
    \draw (cerc) ++(-0.5 cm,-0.5cm)edge node [above,pos=0.5] {} (cerc)  edge[dashed] node[left,above,pos=0.5]{}(a)
    edge[dashed,bend left=30] node[left,above,pos=0.5]{}(f1);
    \draw (cerc1) ++(-0.5 cm,-0.5cm)edge node [above,pos=0.5] {} (cerc1)  edge[dashed,bend right]node[left,above,pos=0.6]{} (g_4);
    \draw (cerc1) ++(0.5 cm,-0.5cm)edge node [above,pos=0.5] {} (cerc1)  edge[dashed,bend left=15]node[left,above,pos=0.6]{} (b);
    \end{tikzpicture}
    }
    \caption{L'automate compressé des parents de \( {(f_1(a,a)+g_2(b))}^{*_a}\cdot_b f_3(g_4(a),b) \).}%
    \label{linear_compressed_automata_father}
  \end{figure}
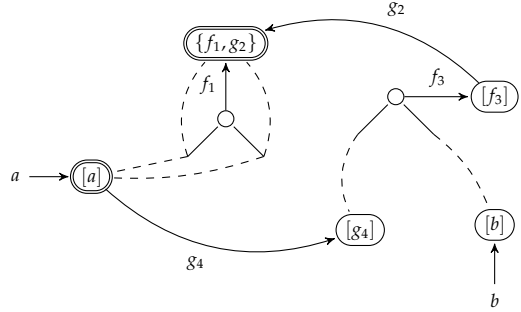

  En appliquant le morphisme de délinéarisation \( \mathrm{h} \) sur \( \mathcal{CF}_{\overline{E}} \), on obtient l'automate compressé des parents d'une expression \( E \) quelconque, automate reconnaissant \( L(E) \).
  \begin{example}
    Considérons l'expression \( \overline{E} \) définie dans l'Exemple~\ref{ex:calcul fonctions glushkov}.
    L'automate compressé des parents de \( \overline{E} \) est représenté Figure~\ref{linear_compressed_automata_father}.
  \end{example}


\section{Le quotient et la dérivation \emph{Bottom-Up}}

Les résultats de cette section sont tirés de~\cite{CMOZ17} et de~\cite{AMZ19}, écrits avec la collaboration de Samira Attou, de Jean-Marc Champarnaud, de Nadia Ouali-Sebti et de Djelloul Ziadi.

Dans cette section, nous allons nous intéresser à une nouvelle forme de quotient de langages, le quotient \emph{Bottom-Up}.
Remarquons qu'ici, nous n'allons pas produire un automate fini reconnaissant le langage dénoté par l'expression de départ.
En effet, rien n'assure que le nombre de dérivées produites est fini.
Cependant, nous allons pouvoir simuler la construction d'une sous-partie finie de cet automate,
ce qui permettra d'en déduire un algorithme purement syntaxique de test d'appartenance.
Nous verrons également une construction inductive permettant, lorsqu'un point fixe existe, de construire un automate équivalent à celui des dérivées dans le cas des mots.
Enfin, une des particularités de cette méthode est qu'elle est compatible avec tous les opérateurs booléens, y compris le complémentaire.

La méthode de dérivation est la suivante.
Au lieu de retirer les symboles depuis la racine, comme dans le cas de la dérivation partielle présentée
dans la Section~\ref{subsec:derivPart}, nous allons retirer les symboles depuis les feuilles.
Un n{\oe}ud pourra alors être retiré une fois tous ses fils éliminés.
Cela implique alors de considérer les arbres de différentes arités, et non plus seulement les arbres d'arité nulle.

Ainsi, dans la suite de cette section, nous considérerons des arbres \( k \)-aires, c'est-à-dire des arbres auxquels il y aura \( k \) feuilles manquantes.
Par exemple, alors que \( f(a,g(a,b)) \) est d'arité nulle, l'arbre \( f(\varepsilon,g(a,b)) \) est unaire et \( f(\varepsilon,g(\varepsilon,\varepsilon)) \) ternaire.
Pour rappel, pour un entier \( k \) donné, nous noterons \( \Sigma^*_k \) l'ensemble des arbres \( k \)-aires sur un alphabet gradué \( \Sigma \).

La composition des arbres peut être facilement étendue à la composition d'ensembles d'arbres de même arité, c'est-à-dire à la composition d'un langage d'arbres \( k \)-aires \( S \) avec une liste de \( k \) arbres \( l=(t_1,\ldots,t_k) \) en posant
\begin{equation*}
  S\circ l = \bigcup_{t\in S} \{t\circ l\}.
\end{equation*}
Cependant, cette extension est assez contraignante en termes de construction: le même ordre est utilisé en termes de composition d'arbres.
Par exemple, lorsque \( S=\{f(\varepsilon,\varepsilon),g(\varepsilon,\varepsilon)\} \), même si \( f(\varepsilon,\varepsilon)\circ (a,b)=f(a,b) \) et \( g(\varepsilon,\varepsilon)\circ (b,a)=g(b,a) \), il n'existe pas de liste \( l \) telle que \( S\circ l = \{f(a,b),g(b,a)\} \).
De même, il n'existe pas de liste depuis laquelle obtenir l'ensemble \( \{h(a),h(b)\} \) depuis l'ensemble \( \{h(\varepsilon)\} \) par composition.
Cette restriction réduit l'expressivité des formules dans la suite de cette étude.
Une façon de résoudre cela est d'étendre la notation en indexant les \( \varepsilon \), et en considérant alors l'ordre des indices lors de la composition.

Plus concrètement, le symbole \( \varepsilon \) d'un arbre \( k \)-aire peut être remplacé
par \( k \) occurrences de symboles distincts \( \varepsilon_{x_1} \), \( \ldots \), \( \varepsilon_{x_k} \),
où \( x_1 \), \( \ldots \), \( x_k \) sont \( k \) entiers de \( \mathbb{N}\setminus \{0 \} \).
Ainsi, pour un arbre \( k \)-aire \( t \), nous noterons \( \mathrm{Ind}_\varepsilon(t) \) l'ensemble \( \{x_1,\ldots,x_k\} \) de ses \( \varepsilon \)-indices.
Ce sous-ensemble fini et naturellement ordonné de \( \mathbb{N} \) contient les indices \( x_l \) des symboles \( \varepsilon_{x_l} \) apparaissant dans \( t \).
\begin{definition}\label{def tree with eps index}
Un arbre \( k \)-aire \( t \) avec l'ensemble \( R \) de ses \( \varepsilon \)-indices est défini inductivement comme suit
\begin{itemize}
  \item \( t=\varepsilon_j \) avec \( j \) un entier; dans ce cas, \( k=1 \) et \( R=\{j\} \);
  \item \( t=f(t_1,\ldots,t_l) \) avec \( f \) un symbole de \( \Sigma_l \) et pour \( 1\leq j\leq l \), \( t_j \) est un \( n_j \)-aire arbre d'\( \varepsilon \)-indices \( R_j \), tel que \( 1\leq j\leq j'\leq l \), \( R_j\cap R_{j'}=\emptyset \);
  dans ce cas, \(k = \sum_{1\leq j\leq l} n_j \) et \( R=\bigcup_{1\leq j\leq l}R_j \).
\end{itemize}
\end{definition}

Ainsi, la composition  \( t\circ(t'_1,\ldots,t'_k) \) substitue \( t'_l \) à \( \varepsilon_{x_l} \) dans \( t \), où \( \mathrm{Ind}_\varepsilon(t)=\{x_1,\ldots,x_k\} \).
Remarquons que tout arbre \( k \)-aire satisfait \( \mathrm{Card}(\mathrm{Ind}_\varepsilon(t))=k \), et que les occurrences ne sont pas nécessairement indexées par leur ordre d'apparence dans \( t \) ni de \( 1 \) à \( l \).

Par exemple, soit
\begin{equation*}
  t=f(\varepsilon_5,g(\varepsilon_3,\varepsilon_8)).
\end{equation*}
Alors
\begin{equation*}
  t\circ(a,b,a)=f(b,g(a,a)).
\end{equation*}
Dans cet exemple,
\begin{equation*}
  \mathrm{Ind}_\varepsilon(t)=\{x_1=3,x_2=5,x_3=8\}
\end{equation*}
et ainsi \( \varepsilon_3 \) (resp. \( \varepsilon_5 \), \( \varepsilon_8 \)) est substitué par \( a \) (resp. \( b \), \( a \)).
Remarquons alors que l'arbre vide n'est plus un élément identité, puisque son action peut modifier l'indexation.

Par conséquent, la composition \( \circ \) est redéfinie inductivement comme suit:
pour tout arbre \( m \)-aire \( t=f(t_1,\ldots,t_n) \) avec \( \mathrm{Ind}_\varepsilon(t)=\{x_1,\ldots,x_m\} \),
pour tout \( m \) arbres \( t'_1 \), \( \ldots \), \( t'_m \) avec des ensembles d'\( \varepsilon \)-indices distincts deux à deux,
on pose:
  \begin{align}
    \varepsilon_1\circ(t'_1) &= t'_1, &
    t \circ(t'_1,\ldots,t'_m)&=f({(t_j\circ{(t'_k)}_{x_k\in\mathrm{Ind}_\varepsilon(t_j)})}_{1\leq j\leq n}).
   \label{eq calc ind compo}
  \end{align}

Un langage d'arbre \( L \) est dit \firstocc{homogène}{langage d'arbres!homogène} si tous les arbres qu'il contient sont de même arité et possèdent le même ensemble d'\( \varepsilon \)-indices, et \( k \)-\firstocc{homogène}\ s'il est homogène et que les arbres qu'il contient sont d'arité \( k \).
Dans ce cas, on note \( \mathrm{Ind}_\varepsilon(L) \) cet ensemble.

L'ensemble des langages d'arbres sur un alphabet gradué \( \Sigma \) est noté \( \mathcal{L}(\Sigma) \), et l'ensemble des langages \( k \)-homogènes par \( {\mathcal{L}(\Sigma)}_k \) pour tout entier \( k\geq 0 \).
Remarquons que l'union de deux langages \( k \)-homogènes de mêmes \( \varepsilon \)-indices \( R \) est un langage \( k \)-homogène de mêmes \( \varepsilon \)-indices.

La composition \( \circ \) est étendue en une opération de  \( {\mathcal{L}(\Sigma)}_k\times {(\mathcal{L}(\Sigma))}^k \) vers \( \mathcal{L}(\Sigma) \): pour tout langage \( L \) de \( {\mathcal{L}(\Sigma)}_k \),
pour tout \( k \) langages \( L_1 \), \( \ldots \), \( L_k \) de \( \mathcal{L}(\Sigma) \) tels que
\( \mathrm{Ind}_\varepsilon(L_i)\cap\mathrm{Ind}_\varepsilon(L_j)=\emptyset \) pour tout \( 1\leq i<j\leq k \),
\begin{equation*}
  L\circ (L_1,\ldots,L_k)=\{t\circ(t_1,\ldots,t_k)\mid t\in L, t_i\in L_i, i\leq k\}.
\end{equation*}
Remarquons que si \( L_j \) est \( l_j \)-homogène pour tout entier \( 1\leq j\leq k \), alors \( L\circ (L_1,\ldots,L_k) \) est \( \sum_{1\leq j\leq k} l_j \)-homogène.
Cette composition peut être itérée pour produire de nouveaux ensembles.
Considérons un langage \( 1 \)-homogène \( L \) d' \( \varepsilon \)-index \( \{x\} \) et un entier \( n \).
La \firstocc{composition itérée}\  \( n_\circ \) est récursivement définie par
\begin{align*}
  L^{0_\circ}&=\{\varepsilon_x\}, & L^{n+1_\circ}&=L^{n_\circ}\cup L^{n_\circ}\circ L.
\end{align*}
D'une façon similaire à l'étoile de Kleene obtenue depuis la concaténation dans le cas des mots, la composition peut être répétée un nombre arbitraire de fois sur des langages unaires, définissant alors une nouvelle opération.
La \firstocc{fermeture de composition}\ de \( L \) est le langage \( L^\circledast=\bigcup_{k\geq 0} L^{k_\circ} \).
Remarquons que \( L^\circledast \) est \( 1 \)-homogène d'\( \varepsilon \)-index \( \{x\} \).

\begin{example}
  Considérons les ensembles \( S_1=\{f(a,b)\} \) et \( S_2=\{f(\varepsilon_1,b)\} \).
  Alors:
  \begin{align*}
    S_1^{*_a} &=\{a,f(a,b),f(f(a,b),b),\ldots \}, \\
    S_2^\circledast &= \{\varepsilon_1,f(\varepsilon_1,b),f(f(\varepsilon_1,b),b),\ldots \}.
  \end{align*}
\end{example}

Le quotient Bottom-Up d'un langage d'arbres par rapport à un arbre est une opération qui supprime des n{\oe}uds internes des arbres.
La partie restante est habituellement appelée \firstocc{contexte}{contexte} dans la littérature~\cite{CDGLJLTT07}; ici nous parlerons d'arbres \( k \)-aires, puisque nous considérons le paramètre \( k \).
Par conséquent, nous réinterpréterons des résultats classiques en utilisant cette notion de quotient.
Ainsi, le quotient est l'opération duale de la composition: le quotient d'un arbre \( t \) par un arbre \( t' \) est l'opération produisant l'ensemble \( t'^{-1}(t) \) des arbres \( t'' \) contenant une occurrence d'\( \varepsilon_1 \) et tel que la substitution de \( \varepsilon_1 \) par \( t' \) dans \( t'' \) produit \( t \).
En conséquence directe, puisque \( \varepsilon_1 \) peut apparaître dans \( t \), la production de \( t'' \) nécessite une réindexation des \( \varepsilon \)-indices.
Dans la suite, nous les incrémenterons.
\begin{example}
  Soit \( t=
    f(g(b,b),\varepsilon_1,h(g(b,b)))
   \) un arbre sur \( \Sigma \), où \( b\in\Sigma_0 \), \( h\in\Sigma_1 \), \( g\in\Sigma_2 \) et \( f\in\Sigma_3 \). Soit
  \( t'=g(b,b)
   \). Alors
  \begin{equation*}
    t'^{-1}(t)=
    \{f(\varepsilon_1,\varepsilon_2,h(g(b,b)))
    ,
    f(g(b,b),\varepsilon_2,h(\varepsilon_1))
     \}.
  \end{equation*}
  En effet, pour tout \( t'' \) de \( t'^{-1}(t) \),
  \begin{equation*}
    t''\circ(t',\varepsilon_1
    )=t.
  \end{equation*}
\end{example}

Formalisons alors cette notion de quotient:
Soient \( t \) un arbre \( k \)-aire de \( \Sigma^* \) et \( t' \) un arbre \( k' \)-aire de \( \Sigma^* \) tels que \( \mathrm{Ind}_\varepsilon(t')\subset \mathrm{Ind}_\varepsilon(t) \).
Soient
\begin{align*}
  R&=\mathrm{Ind}_\varepsilon(t),\\
  R'&=\mathrm{Ind}_\varepsilon(t'),\\
  R''&=\{{(x_z)}_{1\leq z\leq k'-k}\}=R\setminus R'.
\end{align*}
Le \firstocc{quotient}{Bottom-Up quotient} de \( t \) par rapport à \( t' \) est le langage d'arbre \( k-k'+1 \)-homogène \( t'^{-1}(t) \) contenant tous les arbres \( t'' \) satisfaisant les deux conditions suivantes:
\begin{align}
   t= t''\circ(t',{(\varepsilon_{x_z})}_{1\leq z\leq k-k'}),
   \quad
    \mathrm{Ind}_\varepsilon(t'')=\{1,{(x_z+1)}_{1\leq z\leq k'-k}\}\label{eq def quot tree}
\end{align}
Par conséquent,

\begin{minipage}{0.45\linewidth}
\begin{align}
  \varepsilon_j^{-1}(\varepsilon_l)&=
  \begin{cases}
    \varepsilon_1 & \text{ si }j=l,\\
    \emptyset & \text{sinon,}
  \end{cases}
  \label{eq def quot eps}
\end{align}
\end{minipage}
\hfill
\begin{minipage}{0.45\linewidth}
\begin{align}
  t^{-1}(t')=\{\varepsilon_1\} & \Leftrightarrow t=t'.\label{eq def quot t par t}
\end{align}
\end{minipage}

\begin{definition}\label{def quot lang}
  Soit \( \Sigma \) un alphabet gradué.
  Soit \( L \) un langage d'arbre de \( \mathcal{L}(\Sigma) \) et \( t \) un arbre de \( \Sigma^* \).
  Le \firstocc{quotient Bottom-Up}\ de \( L \) par \( t \) est le langage d'arbre
  \begin{equation*}
    t^{-1}(L)=\bigcup_{t'\in L} t^{-1}(t').
  \end{equation*}
\end{definition}

En conséquence directe de l'Équation~\eqref{eq def quot t par t}, l'appartenance d'un arbre à un langage d'arbre peut être redéfini en termes de quotient.
\begin{proposition}\label{prop eq membership t eps}
  Soit \( L \) un langage sur un alphabet gradué \( \Sigma \) et \( t \) un arbre de \( \Sigma^* \).
  Alors
  \begin{equation*}
    t\in L \Leftrightarrow \varepsilon_1\in t^{-1}(L).
  \end{equation*}
\end{proposition}

Ces notions permettent de définir un automate d'arbre isomorphe à l'automate déterministe minimal associé à un langage.
Soit \( L \) un langage d'arbre de \( {\mathcal{L}(\Sigma)}_0 \).
L'\firstocc{automate Bottom-Up quotient de}\  \( L \) est l'automate \( A_L=(\Sigma,Q,F,\delta) \) défini par
\begin{align*}
  Q&=\{t^{-1}(L)\mid t\in T_{\Sigma}\},\\
  F&=\{L'\in Q \mid\varepsilon_1\in L'\},\\
  \delta(t_1^{-1}(L),\ldots,t_k^{-1}(L), f)&=\{{f(t_1,\ldots,t_k)}^{-1}(L)\}.
\end{align*}
Il peut être montré qu'il n'existe pas d'automate déterministe plus petit
reconnaissant \( L \)(voir l'automate des contextes de~\cite{CGLTT03}).


  \subsection{Formules inductives du quotient pour les arbres}

  Pour définir inductivement le quotient d'un arbre par un autre arbre, il faut bien évidemment prendre en compte la notion d'\( \varepsilon \)-indices:
  en effet, quotienter un arbre revient à remplacer un sous-arbre par une occurrence d'\( \varepsilon_1 \).
  Cependant, si ce symbole apparaît dans l'arbre de départ, l'objet obtenu ne correspond plus à la définition d'un arbre, puisque tous les indices d'\( \varepsilon \) doivent être distincts, selon la Définition~\ref{def tree with eps index}.
  Une des solutions les plus simples est alors d'incrémenter tous les indices des \( \varepsilon \) de l'arbre de départ.

  Ainsi, pour un arbre \( k \)-aire \( t \) et un entier \( z \) donnés, nous noterons \( \mathrm{Inc}_\varepsilon(z,t) \) la substitution de tout symbole \( \varepsilon_x \) par \( \varepsilon_{x+z} \) dans \( t \), étendu linéairement sur un langage d'arbre \( L \) comme
  \begin{align*}
    \mathrm{Inc}_\varepsilon(z,L) &=\{\mathrm{Inc}_\varepsilon(z,t)\mid t\in L\}.
  \end{align*}
  Trivialement, il s'ensuit que
  \begin{align}
    (\mathrm{Inc}_\varepsilon(1,t))\circ({(\varepsilon_{j})}_{j\in\mathrm{Ind}_\varepsilon(t)}) &=t.
    \label{eq inc var}
  \end{align}

  La définition du quotient d'un arbre par un autre arbre (Définition~\ref{def quot lang}), n'est pas une formule inductive.
  Par définition, il faut donc considérer les deux parties de l'induction:
  le quotient par un arbre vide, puis le quotient par un symbole composé avec d'autres arbres.
  Le quotient Bottom-Up d'un arbre par rapport à un symbole composé peut être inductivement défini comme suit: puisque le quotient est l'opération duale de la composition, calculer le quotient d'un arbre \( t \) par rapport à un arbre \( t' \) correspond à la substitution d'une occurrence de \( t' \) dans \( t \) par \( \varepsilon_1 \) et à l'incrémentation des \( \varepsilon \)-indices.
  Dans ce cas, on remarquera que le quotient par un symbole \( f \) d'arité \( k \) correspond au quotient par l'arbre \( f(\varepsilon_1,\ldots,\varepsilon_k) \).

  \begin{example}
    Considérons l'arbre \( t=f(h(a),\varepsilon_1,a) \) avec \( f \) un symbole ternaire, \( h \) unaire et \( a \) \( 0 \)-aire.
    Ainsi
    \begin{align*}
      a^{-1}(t) &= \{f(h(\varepsilon_1),\varepsilon_2,a),f(h(a),\varepsilon_2,\varepsilon_1)\},\\
      a^{-1}(a^{-1}(t)) &= \{f(h(\varepsilon_2),\varepsilon_3,\varepsilon_1),f(h(\varepsilon_1),\varepsilon_3,\varepsilon_2)\}.
    \end{align*}
  \end{example}

  \begin{proposition}\label{prop calc quot symb}
    Soient \( \Sigma \) un alphabet gradué, \( k \) un entier et \( \alpha \) un symbole de \( \Sigma_k \).
    Alors:
	  \begin{align*}
	    \alpha^{-1}(\varepsilon_x)&=\emptyset,\\
	    \alpha^{-1}(f(t_1,\ldots,t_n)) &= \bigcup_{1\leq j\leq n} f(t'_1,\ldots,t'_{j-1},\alpha^{-1}(t_j),t'_{j+1},\ldots,t'_n)\\
      & \qquad \cup \{\varepsilon_1\mid f=\alpha \wedge  (t_1,\ldots,t_n)=(\varepsilon_1,\ldots,\varepsilon_n)\},
	  \end{align*}
	  avec \( x \) un entier dans \( \mathbb{N} \), \( f \) un symbole de \( \Sigma_n \), \( t_1,\ldots,t_n \) \( n \) arbres de \( \Sigma^* \) et \( \forall 1\leq z\leq n \), \( t'_z=\mathrm{Inc}_\varepsilon(1,t_{z}) \).
	\end{proposition}
  Selon la définition du quotient (Équation~\eqref{eq def quot tree} et Définition~\ref{def quot lang}), quotienter par un \( \varepsilon \) indicé revient à modifier les \( \varepsilon \)-indices, que ce soit pour un arbre comme pour un langage.
  \begin{example}
    Considérons l'arbre \( t=f(h(a),\varepsilon_2,\varepsilon_5) \) avec \( f \) un symbole ternaire, \( h \) unaire et \( a \) \( 0 \)-aire.
    Alors \( \varepsilon_5^{-1}(t)=\{f(h(a),\varepsilon_3,\varepsilon_1)\} \).
  \end{example}
  \begin{proposition}\label{prop bot up quot eps}
    Soit \( \Sigma \) un alphabet gradué.
    Soit \( L \) un langage \( k \)-homogène avec \( \mathrm{Ind}_\varepsilon(L)=\{j_1,\ldots,j_k\} \).
    Soit \( j \) un entier.
    Alors:
    \begin{equation*}
      \varepsilon_j^{-1}(L)=
        \begin{cases}
          L\circ (
            \varepsilon_{j_1+1},\ldots,\varepsilon_{j_{z-1}+1},
            \varepsilon_1,
            \varepsilon_{j_{z+1}+1},\ldots,\varepsilon_{j_{k}+1})
          & \text{ si } j=j_z\in\mathrm{Ind}_\varepsilon(L),\\
          \emptyset & \text{sinon.}
        \end{cases}
    \end{equation*}
  \end{proposition}
  Pour terminer, quotienter un arbre \( t \) par un arbre \( f(t_1,\ldots,t_k) \) d'arité \( 0 \) peut se décrire inductivement comme suit: tout d'abord, le quotient \( U_k \) de \( t \) par \( t_k \) produit un ensemble d'arbres tels que la substitution de \( \varepsilon_1 \) par \( t_k \) produit \( t \).
  Puis le quotient \( U_{k-1} \) de \( U_k \) par \( t_{k-1} \) produit un ensemble d'arbres tels que la substitution d'\( \varepsilon_2 \) par \( t_{k-1} \) puis d'\( \varepsilon_1 \) par \( t_k \) produit \( t \).
  Par application successive, le quotient \( U_1 \) de \( U_2 \) par \( t_1 \) produit un ensemble d'arbres tels que la substitution d'\( \varepsilon_k \) par \( t_k \), \( \ldots \), puis de \( \varepsilon_1 \) par \( t_1 \) produit \( t \).
  Finalement, le quotient \( V \) de \( U_1 \) par \( f \) (équivalent au quotient par \( f(\varepsilon_1,\ldots,\varepsilon_k) \)) est calculé, produisant un ensemble d'arbres tels que la substitution d'\( \varepsilon_1 \) par \( f(\varepsilon_1,\ldots,\varepsilon_k) \) produit un arbre tel que la substitution de \( \varepsilon_k \) par \( t_k \), \( \ldots \), puis de \( \varepsilon_1 \) par \( t_{1} \) produit \( t \); par définition, \( V={f(t_1,\ldots,t_k)}^{-1}(t) \).
  Remarquons que la présence d'\( \varepsilon \)-indices implique leur réindexation:
	\textbf{(I)} si \( t \) contient une occurrence d'un arbre vide (c'est-à-dire d'un sous-arbre \( \varepsilon_j \) pour un entier \( j \)), alors ces indices doivent être incrémentés \( k+1 \) fois par \( 1 \), soit par ses \( k+1 \) opérations de quotients;
  par conséquent, pour quotienter par \( f(t_1,\ldots,t_k) \), si une occurrence de \( \varepsilon_j \) est présente dans \( t \), alors l'ensemble \( V \) obtenu par quotient de \( U_1 \) par \( f \) contient des arbres	avec une occurrence de \( \varepsilon_{j+k+1} \), à ré-indexer en \( \varepsilon_{j+1} \);
	\textbf{(II)} si \( f(t_1,\ldots,t_k) \) contient un arbre vide, \( \varepsilon_j \) dans \( t_l \) par exemple, alors l'ensemble \( U_{l+1} \), contenant les arbres vides \( (\varepsilon_1,\ldots,\varepsilon_{k-l}) \) (si \( t \) contient des occurrences de \( (t_{l+1},\ldots,t_k) \)) et l'arbre vide \( \varepsilon_{j+k-l} \), ne peut pas être quotienté par \( t_l \):
  si \( t_l \) est présent dans \( t \), alors ses \( \varepsilon \)-indices ont été incrémentés, et alors \( \mathrm{Inc}_\varepsilon(k-l,t_l) \) doit être considéré pour quotienter \( U_{l+1} \).

  \begin{example}
    Considérons l'arbre \( t=f(h(a),\varepsilon_1,g(a,b)) \) avec \( f \) un symbole ternaire, \( g \) binaire, \( h \) unaire, \( a \) et \( b \) \( 0 \)-aires.
    Alors:
    \begin{align*}
      b^{-1}(t) &= \{f(h(a),\varepsilon_2,g(a,\varepsilon_1))\},\\
      a^{-1}(b^{-1}(t)) &= \{f(h(\varepsilon_1),\varepsilon_3,g(a,\varepsilon_2)), f(h(a),\varepsilon_3,g(\varepsilon_1,\varepsilon_2))\},\\
      {g(a,b)}^{-1}(t) &= g^{-1}(a^{-1}(b^{-1}(t))) \\
        &= \{f(h(a),\varepsilon_2,\varepsilon_1)\}.
    \end{align*}
  \end{example}

  Plus formellement, le quotient est défini inductivement comme suit.
	\begin{proposition}\label{prop quot arbre wrt arbre}
	  Soit \( \Sigma \) un alphabet gradué.
	  Soit \( t=f(t_1,\ldots,t_k) \) un arbre \( l \)-aire de \( \Sigma^* \) avec \( f \) un symbole de \( \Sigma_k \) et \( (t_1,\ldots,t_k) \) \( k \) arbres de \( \Sigma^* \) différents de \( (\varepsilon_1,\ldots,\varepsilon_k) \).
	  Soit \( u \) un arbre de \( \Sigma^* \) avec \( \mathrm{Ind}_\varepsilon(u)=\{x_1,\ldots,x_n\} \).
	  Soient
    \begin{equation*}
      \{y_1,\ldots,y_{n-l}\}=\mathrm{Ind}_\varepsilon(u)\setminus \mathrm{Ind}_\varepsilon(t)
    \end{equation*}
	  et \( \forall 1\leq j\leq  k \), \( t'_j=\mathrm{Inc}_\varepsilon(k-j,t_j) \).
    Alors:
	  \begin{equation*}
	      t^{-1}(u) =(f^{-1}({t'_1}^{-1}(\cdots ({t'_k}^{-1}(u))\cdots))\circ(\varepsilon_1,{(\varepsilon_{y_z+1})}_{1\leq z \leq n-l }).
	  \end{equation*}
	\end{proposition}

	L'indexation des \( \varepsilon \) joue un rôle fondamental pendant les calculs: elle est nécessaire pour considérer la non-commutativité de l'opérade libre des arbres (c'est-à-dire que \( f(a,b)\neq f(b,a) \)).
	\begin{example}
	  Considérons l'arbre \( t=g(h(a),b) \).
	  Alors:
	  \begin{align*}
	    b^{-1}(t)&= \{g(h(a),\varepsilon_1)\},\\
	    a^{-1}(t)&=\{g(h(\varepsilon_1),b)\},\\
	    a^{-1}( b^{-1}(t))&= \{g(h(\varepsilon_1),\varepsilon_2)\},\\
	    b^{-1}(a^{-1}(t))&= \{g(h(\varepsilon_2),\varepsilon_1)\},\\
      {h(a)}^{-1}(b^{-1}(t))&=  h^{-1}(g(h(\varepsilon_1),\varepsilon_2))\circ(\varepsilon_1,\varepsilon_2)\\
	    &= \{g(\varepsilon_1,\varepsilon_3) \circ (\varepsilon_1,\varepsilon_2)\} \\
	    &= \{g(\varepsilon_1,\varepsilon_2)\},\\
      {h(b)}^{-1}(a^{-1}(t))&=g(h^{-1}(h(\varepsilon_2)),\varepsilon_2) \\
	    &= \emptyset,\\
	    {g(h(a),b)}^{-1}(t) &=g^{-1}({h(a)}^{-1}( b^{-1}(t))))\\
      & = g^{-1}(g(\varepsilon_1,\varepsilon_2))\\
      & = \{\varepsilon_1\}.
	  \end{align*}
	  Ainsi,
	  \begin{align*}
      \{\varepsilon_1\} &= {g(h(a),b)}^{-1}((g(h(a),b))\\
      & \neq {g(h(b),a)}^{-1}((g(h(a),b)))=\emptyset.
    \end{align*}
	\end{example}

  \subsection{Formules inductives du quotient pour les opérations sur les langages}

  \subsubsection{L'union}

  En conséquence directe de la Définition~\ref{def quot lang}, la formule inductive de l'union est la suivante.
  \begin{lemma}\label{lem quot union}
    Soit \( \Sigma \) un alphabet gradué.
    Soient \( t \) un arbre de \( \Sigma^* \), et \( L_1 \) et \( L_2 \) deux langages sur \( \Sigma \).
    Alors:
      \begin{equation*}
        t^{-1}(L_1\cup L_2)=t^{-1}(L_1)\cup t^{-1}(L_2).
      \end{equation*}
  \end{lemma}
  Ainsi, puisque l'union est distributive sur la composition, en conséquence directe du Lemme~\ref{lem quot union} de la Proposition~\ref{prop quot arbre wrt arbre}, le résultat suivant est vérifié.
  \begin{corollary}\label{cor deriv langage arbre}
	  Soit \( \Sigma \) un alphabet gradué.
	  Soit \( t=f(t_1,\ldots,t_k) \) un arbre \( l \)-aire tel que \( f \) est dans \( \Sigma_k \) et \( (t_1,\ldots,t_k) \) est un \( k \)-uplet d'arbres dans \( \Sigma^* \) distinct de \( (\varepsilon_1,\ldots,\varepsilon_k) \).
	  Soit \( L \) un langage \( k \)-homogène sur \( \Sigma \) avec \( \mathrm{Ind}_\varepsilon(L)=\{x_1,\ldots,x_k\} \).
	  Soient \( \{y_1,\ldots,y_{n-l}\}=\mathrm{Ind}_\varepsilon(L)\setminus \mathrm{Ind}_\varepsilon(t) \)
	  et \( \forall 1\leq j\leq  k \), \( t'_j=\mathrm{Inc}_\varepsilon(k-j,t_j) \).
    Alors:
	  \begin{equation*}
	    t^{-1}(L)=(f^{-1}({t'_1}^{-1}(\cdots ({t'_k}^{-1}(L))\cdots))\circ(\varepsilon_1,{(\varepsilon_{y_z+1})}_{1\leq z\leq n-l}).
	  \end{equation*}
  \end{corollary}
  Selon le Corollaire~\ref{cor deriv langage arbre}, il suffit alors de montrer comment calculer inductivement le quotient d'un langage par rapport à un symbole de \( \Sigma \).

  \subsubsection{Le produit de substitution}

  \textbf{\emph{N.B.:}} Dans la suite de cette partie, afin de préserver l'homogénéité, nous considérons que l'opération de substitution \(L_1 \cdot_a L_2\) pour un symbole nullaire \(a\) n'est définie que si \(L_1\) est homogène et si \(L_2\) est \(0\)-homogène.

  Calculer le \( b \)-produit revient à remplacer toutes les occurrences d'un symbole \( 0 \)-aire \( b \) dans un arbre \( t \) par un langage \( L \).
  Ainsi, quotienter \( t \) par un symbole \( \alpha \) est réalisé en suivant les conditions suivantes:
  \textbf{(1)} les occurrences d'\( \alpha \) devant être effacées par le quotient dans \( t\cdot_b L \) peuvent apparaître dans \( L \).
    Cependant, calculer directement \( t\cdot_b \alpha^{-1}(L) \) peut produire un langage contenant des arbres avec plusieurs occurrences d'\( \varepsilon_1 \).
    Ainsi, il faut tout d'abord effacer une occurrence de \( b \) dans \( t \) en calculant \( b^{-1}(t) \), puis considérer la substitution des autres occurrences de \( b \) par \( L \) dans \( t \), puis composer cette nouvelle occurrence d'\( \varepsilon_1 \) dans \( b^{-1}(t) \) avec le quotient de  \( L \), soit
    \begin{equation*}
      (b^{-1}(t)\cdot_b L)\circ_1 \alpha^{-1}(L),
    \end{equation*}
    où \(\circ_1\) est la composition partielle définie par
    \begin{equation*}
      L \circ_1 L' = L\circ (L', {(\varepsilon_l)}_{j_2 \leq l \leq j_k})
    \end{equation*}
    avec \(\mathrm{Ind}_\varepsilon(L)= \{j_1,\ldots,j_k\} \);
    \textbf{(2)} lorsque \( \alpha\neq b \), les occurrences d'\( \alpha \) retirées lors du quotient de \( t\cdot_b L \) peuvent apparaître également dans \( t \).
    Dans ce cas, une occurrence d'\( \alpha \) doit être substituée par \( \varepsilon_1 \), et les occurrences de \( b \) dans \( t \) toujours remplacées par \( L \), soit \( \alpha^{-1}(t)\cdot_b L \).

  \begin{example}
    Considérons l'arbre \( t=f(h(a),\varepsilon_1,g(a,b))\cdot_a b = f(h(b),\varepsilon_1,g(b,b)) \) avec \( f \) un symbole ternaire, \( g \) binaire, \( h \) unaire, \( a \) et \( b \) \( 0 \)-aires.
    Alors:
    \begin{align*}
      b^{-1}(t) &= b^{-1}(f(h(a),\varepsilon_1,g(a,b)))\cdot_a \{b\} \\
      & \qquad \cup (a^{-1}(f(h(a),\varepsilon_1,g(a,b)))\cdot_a \{b\})\circ \varepsilon_1 \\
      &= \{f(h(a),\varepsilon_2,g(a,\varepsilon_1))\} \cdot_a \{b\} \\
      & \qquad \cup (\{f(h(\varepsilon_1),\varepsilon_2,g(a,b)), f(h(a),\varepsilon_2,g(\varepsilon_1,b))\} \cdot_a \{b\})\circ \varepsilon_1 \\
      &= \{f(h(b),\varepsilon_2,g(b,\varepsilon_1)),f(h(\varepsilon_1),\varepsilon_2,g(b,b)), f(h(b),\varepsilon_2,g(\varepsilon_1,b))\}.
    \end{align*}
  \end{example}


  Récapitulons ainsi dans le lemme suivant.
  \begin{lemma}\label{lem quot tree cdotb}
    Soit \( \Sigma \) un alphabet gradué.
    Soient \( t \) un arbre \( k \)-aire de \( \Sigma^* \) et \( L \) un langage \( 0 \)-homogène.
    Soient \( \alpha \) un symbole de \( \Sigma \) et \( b \)un symbole de \( \Sigma_0 \).
    Alors:
    \begin{equation*}
      \alpha^{-1}(t\cdot_b L)=
        \begin{cases}
          (b^{-1}(t)\cdot_b L) \circ_1 b^{-1}(L) & \text{ si }\alpha=b,\\
          \alpha^{-1}(t)\cdot_b L \cup (b^{-1}(t)\cdot_b L) \circ_1 \alpha^{-1}(L) & \text{ si }\alpha\in\Sigma_0\setminus \{b\},\\
          \alpha^{-1}(t)\cdot_b L & \text{sinon.}
        \end{cases}
    \end{equation*}
  \end{lemma}
  Ainsi, en conséquence direct du lemme précédent, puisque
  \begin{equation*}
    L\cdot_b L'=\bigcup_{t\in L} t\cdot_b L',
  \end{equation*}
  \begin{proposition}\label{prop quot cdotb lang}
    Soit \( \Sigma \) un alphabet gradué.
    Soient \( L_1 \) un langage \( k \)-homogène, \( L_2 \) un langage \( 0 \)-homogène,
    \( \alpha \) un symbole de \( \Sigma \) et \( b \) un symbole de \( \Sigma_0 \).
    Alors:
    \begin{align*}
      \alpha^{-1}(L_1\cdot_b L_2)&=
        \begin{cases}
          (b^{-1}(L_1)\cdot_b L_2) \circ_1 b^{-1}(L_2) & \text{ si }\alpha=b,\\
          \alpha^{-1}(L_1)\cdot_b L_2
            \cup (b^{-1}(L_1)\cdot_b L_2) \circ_1 \alpha^{-1}(L_2) & \text{ si }\alpha\in\Sigma_0\setminus \{b\},\\
          \alpha^{-1}(L_1)\cdot_b L_2 & \text{sinon.}
        \end{cases}
    \end{align*}
  \end{proposition}

  \subsubsection{Le produit de composition}

  Composer un arbre \( k \)-aire \( t \), qui satisfait  \( \mathrm{Ind}_\varepsilon(t)=\{x_1,\ldots,x_k\} \), avec \( k \) arbres \( t_1,\ldots,t_k \) est l'action d'accrocher ces arbres à \( t \) aux positions où les symboles \( \varepsilon_{x_1},\ldots,\varepsilon_{x_k} \) apparaissent.
  Ainsi, l'arbre obtenu \( t' \) peut être vu comme un arbre avec une partie supérieure contenant \( t \) et des parties inférieures contenant exactement les arbres \( t_1,\ldots,t_k \).
  Par conséquent, si \( \alpha \) apparaît dans un arbre inférieur \( t_j \), cet arbre doit être quotienté par \( \alpha \) alors que les autres parties se voient toutes \( \varepsilon \)-incrémentées.
  De plus, si \( n \) arbres de \( t_1,\ldots,t_k \) sont égaux à \( \varepsilon_1,\ldots,\varepsilon_n \), par exemple \( t_{p_1},\ldots,t_{p_n} \), et si \( t'=\alpha(\varepsilon_{x_{p_1}},\ldots,\varepsilon_{x_{p_n}}) \) apparaît dans \( t \),
  alors \( t' \) doit être substitué par \( \varepsilon_1 \) et les autres parties inférieures \( t_j \) avec \( j\neq p_m \), \( m\in \{1,\ldots,n\} \) \( \varepsilon \)-incrémentées, puisque l'opération inverse produit \( t \).

  \begin{example}
    Considérons l'arbre \( t=f(h(a),h(\varepsilon_4),\varepsilon_2)\circ (a,\varepsilon_1) = f(h(a),h(\varepsilon_1),a) \) avec \( f \) un symbole ternaire, \( h \) unaire et \( a \) \( 0 \)-aire.
    Alors:
    \begin{align*}
      h^{-1}(t) &=  {h(\varepsilon_4)}^{-1}(t) \circ (\varepsilon_1,a) \\
      &=  \{f(h(a),\varepsilon_1,\varepsilon_3)\} \circ (\varepsilon_1,a) \\
      &=  \{f(h(a),\varepsilon_1,a)\}
    \end{align*}
  \end{example}

  \begin{lemma}\label{lem quot tree compos}
    Soit \( \Sigma \) un alphabet gradué.
    Soient \( t \) un arbre \( k \)-aire avec \( \mathrm{Ind}_\varepsilon(t)=\{j_1,\ldots,j_k\} \) et \( t_1,\ldots,t_k \) \( k \) arbres.
    Soit \( \alpha \)un symbole de \( \Sigma_n \).
    Alors:
    \begin{align*}
      \alpha^{-1}(t\circ(t_1,\ldots,t_k))&=
          \bigcup_{1\leq j\leq k} t\circ ( {( \mathrm{Inc}_\varepsilon(1,t_l) )}_{1\leq l\leq j-1},\alpha^{-1}(t_j), {( \mathrm{Inc}_\varepsilon(1,t_l) )}_{j+1\leq l\leq k})\\
          &\quad \cup
            \begin{cases}
               {\alpha( {(\varepsilon_{j_{p_l}})}_{1\leq l\leq n})}^{-1}(t) \circ(\varepsilon_1,{(\mathrm{Inc}_\varepsilon(1,t_l))}_{1\leq l\leq k\mid \forall j, l\neq p_j}) \\
               \quad \text{ si }\forall 1\leq l\leq n,  \exists 1\leq p_l\leq k, t_{p_l}= \varepsilon_l, \\
               \emptyset  \quad \text{ sinon.}
            \end{cases}
    \end{align*}
  \end{lemma}
  Ainsi, puisque \( L\circ(L_1,\ldots,L_k)=\bigcup_{t\in L,(t_1,\ldots,t_k)\in L_1\times\cdots\times L_k} t\circ (t_1,\ldots,t_k) \),
  \begin{proposition}\label{prop quot lang circ}
    Soit \( \Sigma \) un alphabet gradué.
    Soient \( L \) un langage \( k \)-homogène avec \( \mathrm{Ind}_\varepsilon(L)=\{j_1,\ldots,j_k\} \), \( \alpha \) un symbole de \( \Sigma_n \) et \( L_1,\ldots,L_k \) \( k \) langages.
    Alors:
    \begin{align*}
      \alpha^{-1}(L\circ(L_1,\ldots,L_k))&=  \bigcup_{1\leq j\leq k} L\circ({(\mathrm{Inc}_\varepsilon(1,L_l))}_{1\leq l\leq j},\alpha^{-1}(L_j),{(\mathrm{Inc}_\varepsilon(1,L_l))}_{j+1\leq l\leq k}) \\
      &\quad \cup
        \begin{cases}
          {\alpha({(\varepsilon_{j_{p_l}})}_{1\leq l\leq n})}^{-1}(L) \circ(\varepsilon_1,{(\mathrm{Inc}_\varepsilon(1,L_l))}_{1\leq l\leq k\mid \forall z, l\neq p_z}))\\
          \quad \text{ si } \forall 1\leq l\leq n, \exists 1\leq p_l\leq k, \varepsilon_l \in L_{p_l}\\
          \emptyset \quad \text{ sinon.}
        \end{cases}
    \end{align*}
  \end{proposition}

  \subsubsection{La composition itérée}
  La composition itérée ne peut être quotientée que par un symbole \( \alpha \) d'arité \( n=0 \) ou \( n=1 \) uniquement, puisque tout langage
  obtenu par composition itérée est \(1\)-homogène.

  Si \( n=1 \), puisque \( L^{\circledast} \) est obtenue par l'application de la composition un nombre arbitraire de fois, alors quotienter par \( \alpha \) revient à quotienter un arbre \( t \) de \( L \) par
  \( \alpha \) puis le raccrocher au langage obtenu par une application d'un nombre arbitraire de composition, c'est-à-dire \( L^{\circledast} \).
  D'une façon équivalente, l'occurrence d'\( \alpha \) à effacer apparaît dans une partie inférieure d'un arbre.
  Cependant, lorsque \( n=0 \), l'occurrence d'\( \alpha \) à effacer peut apparaître à n'importe quel niveau: elle peut être localisée dans une partie supérieure de \( L^{\circledast} \) tout comme dans une partie inférieure de \( L^{\circledast} \), par exemple lorsque l'arbre \( t \) à quotienter appartient à \( L^{\circledast}\circ \{t'\}\circ L^{\circledast} \) avec \( t' \) un arbre de \( L \) dans lequel apparaît une occurrence d'\( \alpha \).
  Dans ce cas, \( t' \) doit être quotienté par \( \alpha \), créant alors une occurrence de \( \varepsilon_1 \), ce qui force l'ancien unique indice \( \varepsilon \) de \( L^{\circledast} \) à être incrémenté, en conformité avec la définition du quotient.
  \begin{example}
    Considérons le langage
    \begin{align*}
      L &= {(f(h(a),\varepsilon_1,a))}^\circledast  \\
       &= \{\varepsilon_1,f(h(a),\varepsilon_1,a),f(h(a),f(h(a),\varepsilon_1,a),a),\ldots \}
    \end{align*}
    avec \( f \) un symbole ternaire, \( h \)  unaire et \( a \) \( 0 \)-aire.
    Alors:
    \begin{align*}
      a^{-1}(L) &= ({f(h(a),\varepsilon_1,a)}^\circledast \circ a^{-1}(f(h(a),\varepsilon_1,a))) \circ (\varepsilon_1,{f(h(a),\varepsilon_2,a)}^{\circledast})\\
      &= ({f(h(a),\varepsilon_1,a)}^\circledast \circ \{f(h(\varepsilon_1),\varepsilon_2,a),f(h(a),\varepsilon_2,\varepsilon_1))\})\\
      & \qquad \circ (\varepsilon_1,{f(h(a),\varepsilon_2,a)}^{\circledast})\\
      &= ({f(h(a),\varepsilon_1,a)}^\circledast \\
      & \qquad \circ \{f(h(\varepsilon_1),{f(h(a),\varepsilon_2,a)}^{\circledast},a),f(h(a),{f(h(a),\varepsilon_2,a)}^{\circledast},\varepsilon_1)\}) \\
      &= \{f(h(\varepsilon_1),\varepsilon_2,a), f(h(a),\varepsilon_2,\varepsilon_1), f(h(\varepsilon_1),f(h(a),\varepsilon_2,a),a),\\
      & \qquad   f(h(a),f(h(a),\varepsilon_2,a),\varepsilon_1), f(h(a),f(h(\varepsilon_1),\varepsilon_2,a),a),\\
      & \qquad  f(h(a),f(h(a),\varepsilon_2,\varepsilon_1),a),\ldots \}
    \end{align*}
  \end{example}
  Ainsi:
  \begin{proposition}\label{prop bot up quot star rond}
    Soit \( \Sigma \) un alphabet gradué.
    Soit \( L \) un langage \( 1 \)-homogène.
    Soit \( \alpha \) un symbole de \( \Sigma \).
    Alors:
    \begin{equation*}
      \alpha^{-1}(L^{\circledast})=
        \begin{cases}
          (L^\circledast\circ (\alpha^{-1}(L)))\circ(\varepsilon_1, \mathrm{Inc}_{\varepsilon}(1,L^\circledast)) & \text{ si }\alpha\in\Sigma_0,\\
          (L^\circledast\circ (\alpha^{-1}(L))) & \text{ sinon.}
        \end{cases}
    \end{equation*}
  \end{proposition}

  \subsubsection{La substitution itérée}

  \textbf{\emph{N.B.:}} Dans la suite de cette partie, afin de préserver l'homogénéité, nous considérons que l'opération de substitution itérée \(L^{*_a}\) pour un symbole nullaire \(a\) n'est définie que si \(L\) est \(0\)-homogène.

  Dans le cas de la substitution itérée, deux cas sont à considérer lors du quotient par \( \alpha \):
  lorsque \( b=\alpha \), alors une occurrence de \( b \) dans un arbre de \( L \) doit être transformée en \( \varepsilon_1 \),
  alors que les autres peuvent encore être substituées par \( L \).
  Mais lorsque \( \alpha\neq b \), alors la situation est plus complexe.
  Comme dans le second cas de la composition itérée, l'occurrence d'\( \alpha \) à retirer peut apparaître n'importe où:
  elle peut être localisée sous une partie supérieure de \( L^{*_b} \) une fois substituée depuis une occurrence de \( b \), mais également sous une partie inférieure de \( L^{*_b} \), si elle contient aussi une occurrence de \( b \).
  Cela peut se produire lorsque l'arbre \( t \) à quotienter appartient à \( L^{*_b}\cdot_b \{t'\}\cdot_b L^{*_b} \) où \( t' \) est un arbre de \( L \) contenant \( \alpha \).
  Dans ce cas, \( L \) doit tout d'abord être quotienté par \( b \) pour créer une nouvelle occurrence de \( \varepsilon_1 \), où le quotient \( \alpha^{-1}(L) \) sera alors raccroché.
  Puis un \( b \)-produit est ajouté, puisque toute occurrence de \( b \) peut toujours être substituée par \( L^{*_b} \).

  \begin{example}
    Considérons le langage
    \begin{align*}
      L &={(f(h(a),b,a))}^{*_a}\\
        &=\{a,f(h(a),b,a),f(h(f(h(a),b,a)),b,a),f(h(a),b,f(h(a),b,a)),\ldots \}
    \end{align*}
     avec \( f \) un symbole ternaire, \( h \) unaire, \( a \) et \( b \) \( 0 \)-aires.
     Alors:
    \begin{align*}
      b^{-1}(L) &= {(a^{-1}(f(h(a),b,a)))}^\circledast \circ b^{-1}(f(h(a),b,a)) \cdot_a {(f(h(a),b,a))}^{*_a}\\
      &= {\{f(h(\varepsilon_1),b,a),f(h(a),b,\varepsilon_1)\}}^\circledast \circ \{f(h(a),\varepsilon_1,a)\} \cdot_a {(f(h(a),b,a))}^{*_a}\\
      &= \{f(h(a),\varepsilon_1,a), f(h(f(h(a),\varepsilon_1,a)),b,a), f(h(f(h(a),b,a)),\varepsilon_1,a),\ldots \}
    \end{align*}
  \end{example}
  Par conséquent:
  \begin{proposition}\label{prop bot up quot star symb}
    Soit \( \Sigma \) un alphabet gradué.
    Soit \( L \) un langage \( 0 \)-homogène.
    Soient \( \alpha \) un symbole de \( \Sigma \) et \( b \) un symbole de \( \Sigma_0 \).
    Alors:
    \begin{equation*}
      \alpha^{-1}(L^{*_b})=
        \begin{cases}
          {(b^{-1}(L))}^\circledast\cdot_b L^{*_b} & \text{ si }\alpha=b,\\
          ({(b^{-1}(L))}^\circledast \circ (\alpha^{-1}(L))) \cdot_b L^{*_b} & \text{sinon.}
        \end{cases}
    \end{equation*}
  \end{proposition}


\subsection{Un test d'appartenance purement syntaxique}
  Un langage rationnel peut être associé à au moins une combinaison d'opérations rationnelles, comme dans le cas des expressions rationnelles.
  Nous appellerons cette combinaison \firstocc{arbre syntaxique}\ d'un langage.
  Les formules de la section précédente permettent de réaliser le test d'appartenance sur un arbre syntaxique d'un langage puisque, selon la Proposition~\ref{prop eq membership t eps}, il suffit de tester si \( \varepsilon_1 \) est dans \( t^{-1}(L) \) pour décider de l'appartenance de \( t \) à \( L \).
  Montrons alors comment calculer le booléen \( \mathrm{Null}(L)=(\varepsilon_1\in L) \) récursivement.
  \begin{table}[H]
    \begin{align*}
      \mathrm{Null}(\emptyset) &=\mathrm{Faux},\\
      \mathrm{Null}(\{t\}) &=
        \begin{cases}
          \mathrm{Vrai} & \text{si } t=\varepsilon_1,\\
          \mathrm{Faux} & \text{sinon,}
        \end{cases}\\
      \mathrm{Null}(L\cup L') &= \mathrm{Null}(L)\vee \mathrm{Null}(L'),\\
      \mathrm{Null}(L\cdot_a L') &= \mathrm{Null}(L),\\
      \mathrm{Null}(L^{*_a}) &= \mathrm{Faux},\\
      \mathrm{Null}(L^\circledast) &=\mathrm{Vrai},\\
      \mathrm{Null}(L\circ(L_1,\ldots,L_k) &=
        \begin{cases}
          \mathrm{Null}(L)\wedge \mathrm{Null}(L_1) & \text{si } k=1\\
          \mathrm{Faux} & \text{sinon.}
        \end{cases}
    \end{align*}
    \caption{Définition inductive de \(\mathrm{Null}\).}
  \end{table}

  En conséquence directe,
  \begin{theorem}
    Pour tout arbre syntaxique d'un langage \( L \), l'appartenance d'un arbre \( t \) à \( L \) peut être décidé en utilisant les formules de quotient.
  \end{theorem}

\begin{figure}[H]
    \centerline{
	    \begin{tikzpicture}[node distance=2.5cm,bend angle=30,transform shape,scale=0.8]
	  \node[state] (X1) {\( X_1 \)};
	  \node[state,accepting, right of=X1] (X2) {\( X_2 \)};
	  \node[state,accepting, right of=X2] (X3) {\( X_3 \)};
	  \draw (X1) ++(-1cm,0cm) node {\( a \)}  edge[->] (X1);
	  \draw (X2) ++(-0cm,-1cm) node {\( b \)}  edge[->] (X2);
      \path[->]
        (X1) edge[->,above] node {\( h \)} (X2)
        (X2) edge[->,above] node {\( h \)} (X3)
        (X3) edge[->,in=45,out=-45,loop,right] node {\( h \)} ()
	  ;
      \draw (X2) ++(1cm,0.75cm)  edge[->,in=90,out=45,looseness=2] node[right,pos=0] {\( f \)} (X2) edge[out=225,in=60] (X2) edge[out=225,in=0] (X2);
    \end{tikzpicture}
    }
  \caption{L'automate minimal de \( L_1 \).}%
  \label{fig buquotDFA ex}
  \end{figure}
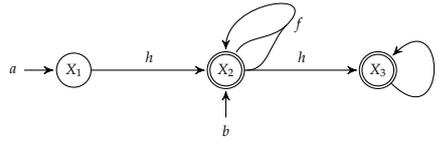

 \begin{example}
  Considérons l'alphabet \( \Sigma \) avec \( \Sigma_0=\{a,b\} \), \( \Sigma_1=\{h\} \) et \( \Sigma_2=\{f\} \).
  Calculons alors les quotients du langage
  \begin{equation*}
    L_1=h^{\circledast}\circ L_2
  \end{equation*}
  avec
  \begin{equation*}
    L_2={(h(a)+f(b,b))}^{*_b}.
  \end{equation*}
  Tout d'abord, considérons les langages
  \begin{align*}
    L_3&={(f(\varepsilon_1,b) + f(b,\varepsilon_1))}^{\circledast}, &
    L'_3&={(f(\varepsilon_2,b) + f(b,\varepsilon_2))}^{\circledast}.
  \end{align*}
  Alors:
  \begin{align*}
    b^{-1}(L_3) &= (L_3 \circ (f(\varepsilon_1,\varepsilon_2)+f(\varepsilon_2,\varepsilon_1)))\circ(\varepsilon_1,L'_3)\\
    & = L_3 \circ (f(\varepsilon_1,L'_3)+f(L'_3,\varepsilon_1))\\
    & = L_4,\\
    f^{-1}(L_4)&=L_3 \circ (f^{-1}f(\varepsilon_1,L'_3)+f^{-1}f(L'_3,\varepsilon_1))\\
    & =L_3.
  \end{align*}
  Considérons maintenant \( L_2={(h(a)+f(b,b))}^{*_b} \).
  Alors:
  \begin{align*}
    a^{-1}(L_2) &=(L_3\circ h(\varepsilon_1)) \cdot_b  L_2,\\
    {h(a)}^{-1}(L_2)&=L_3 \cdot_b  L_2,\\
    b^{-1}(L_2)&=L_3  \cdot_b  L_2,\\
    b^{-1}(b^{-1}(L_2))&=(b^{-1}(L_3)  \cdot_b  L_2)\circ_1 (b^{-1}(L_2))\\
    &=(L_4  \cdot_b  L_2)\circ_1 (L_3\cdot_b L_2)\\
    &=(L_4 \circ_1 L_3) \cdot_b L_2\\
    &=(L_3 \circ (f(L_3,L'_3)+f(L'_3,L_3)) ) \cdot_b L_2,\\
    {f(b,b)}^{-1}(L_2)&=L_3\cdot_b L_2,\\
    a^{-1}(a^{-1}(L_2)) &= a^{-1}( (L_3\circ h(\varepsilon_1)) ) \cdot_b L_2\ \cup ( b^{-1}( L_3\circ h(\varepsilon_1))  \cdot_b L_2) \circ_1 a^{-1}(L_2)\\
    &= (L_4 \circ h(\varepsilon_2))  \cdot_b L_2) \circ_1 a^{-1}(L_2)\\
    &= (L_4 \circ h(a^{-1}(L_2))) \cdot_b L_2\\
    &= L_5,\\
    {f(a,a)}^{-1}(L_2) &= f^{-1}(L_5)\\
    & = \emptyset.
  \end{align*}
  Finalement, considérons \( L_1=h^{\circledast}\circ L_2 \).
  \begin{align*}
    a^{-1}(L_1) &=h^{\circledast}\circ a^{-1}(L_2)\\
    &=h^{\circledast}\circ ((L_3\circ h(\varepsilon_1)) \cdot_b  L_2)\\
    &=X_1,\\
    {h(a)}^{-1}(L_1) &=h^{\circledast}\circ {h(a)}^{-1}(L_2)\\
    &=h^{\circledast}\circ(L_3 \cdot_b  L_2)\\
    & =X_2,\\
    b^{-1}(L_1)&= h^{\circledast}\circ b^{-1}(L_2)\\
    &=   h^{\circledast}\circ(L_3  \cdot_b  L_2)\\
    & =X_2,\\
    {h(b)}^{-1}(L_1) &= h^{-1}(X_2)\\
     &=h^{-1}h^{\circledast}\\
     & =h^{\circledast}\\
     &=X_3,\\
    {h(h(b))}^{-1}(L_1) &=h^{-1}h^{\circledast}\\
    &=h^{\circledast}\\
    &=X_3,\\
    {f(b,b)}^{-1}(L_1)&=h^{\circledast}\circ {f(b,b)}^{-1}(L_2)\\
    &=h^{\circledast}\circ( L_3\cdot_b L_2)\\
    &=X_2.
  \end{align*}
  Remarquons également que
  \begin{equation*}
    {f(a,a)}^{-1}(L_1)= h^{\circledast}\circ( {f(a,a)}^{-1}(L_2))=\emptyset.
  \end{equation*}
  Ainsi, l'automate quotient de \( L_1 \) est représenté Figure~\ref{fig buquotDFA ex}, avec
  \begin{align*}
    \mathrm{Null}(X_2) &=\mathrm{Null}(X_3)\\
    & =\mathrm{Vrai},\\
    \mathrm{Null}(X_1) &=\mathrm{Faux}.
  \end{align*}
  \end{example}

  \subsection{Extension aux opérations booléennes}

  Il est possible d'étendre la formule de quotient à toutes les opérations booléennes.
  En effet, l'union de langage n'est pas la seule opération compatible.
  Le complémentaire l'est également.
  Cependant, l'opération que nous allons considérer n'est pas le complémentaire \emph{universel}.
  Nous allons considérer une version restreinte, le \emph{complémentaire gradué}, restreint aux langages homogènes.

  Par exemple, pour un langage \(k\)-homogène \(L\) donné, on note \(\neg L\) l'ensemble
  \begin{equation}\label{eqdef homogene comp}
    \{t \in {\Sigma^*}_k \mid t\notin L, \mathrm{Ind}_{\varepsilon}(t) = \mathrm{Ind}_{\varepsilon}(L) \}.
  \end{equation}
  En restreignant d'une façon similaire l'union à des paires de langages ayant les mêmes \(\varepsilon \)-indices, on peut redéfinir les opérations booléennes classiques comme combinaisons d'unions et de complémentaires (par exemple la différence symétrique).

  Le quotient d'un langage complémenté se calcule selon la formule classique suivante.
  \begin{proposition}\label{prop lang complem}
    Soient \(L\) un langage homogène sur un alphabet \(\Sigma \) et \(t \in  \Sigma^*\).
    Ainsi
    \begin{equation*}
      t^{-1}(\neg L) = \neg(t^{-1}(L)).
    \end{equation*}
  \end{proposition}
  Ainsi, en conséquence du quotient d'une union,
  \begin{corollary}\label{cor:quot op bool}
    Soient \((L_1,\ldots,L_k)\) des langages \(k'\)-homogènes de mêmes \(\varepsilon \)-indices et \(\mathrm{op}\) une opération booléenne.
    Alors pour tout arbre \(t\) de \(\Sigma^*\)
    \begin{equation*}
      t^{-1}(\mathrm{op}(L_1,\ldots,L_k)) = \mathrm{op}(t^{-1}(L_1),\ldots,t^{-1}(L_k)).
    \end{equation*}
  \end{corollary}

  \subsection{Expressions étendues homogènes}

  L'intégration des opérateurs (homogènes) booléens dans les expressions d'arbres nécessite une légère modification syntaxique.
  Supposons par exemple le test d'appartenance de l'arbre vide \(\varepsilon_j\) dans le complémentaire d'un langage vide.
  Sa présence dépend en fait des \(\varepsilon \)-indices \emph{choisis} dans la définition de l'ensemble vide.

  Afin de lever toute ambiguïté, nous indexerons les occurrences de l'ensemble vide par l'ensemble des \(\varepsilon \)-indices nécessaire à la définition de son homogénéité.

  Ainsi, une \firstocc{expression étendue homogène}{expression!d'arbres!homogène étendue} \(E\) sur un alphabet \(\Sigma \) est définie inductivement comme suit
  \begin{gather}
    \begin{aligned}
      E &= f(E_1, \ldots, E_n), & E &= \varepsilon_j, &
      E &= \emptyset_{\mathcal{I}}, \\
      E &= \mathrm{op}(E_1,\ldots,E_n), &
      E &= E' \circ (E_1,\ldots, E_n), & E &= E_1^{\circledast},\\
      E &= E_1 \cdot_a E_2, &  E &= E_1^{*_a},
    \end{aligned}
  \end{gather}
  où \(f\) est un symbole de \(\Sigma_n\), \((E', E_1, \ldots, E_n)\) sont \((n+1)\) expressions sur \(\Sigma \), \(j\) est un entier positif, \(\mathcal{I}\) est un ensemble d'entiers, \(\mathrm{op}\) est un opérateur booléen \(n\)-aire et \(a\) est un symbole de \(\Sigma_0\).
  L'expression \(\emptyset_\emptyset \) sera aussi notée \(\emptyset \).

  La définition syntaxique précédente n'assure pas que l'expression permette de dénoter un langage homogène.
  Il faut s'assurer en effet qu'elle soit valide, c'est-à-dire que les \(\varepsilon \)-indices des expressions la composant soient inter-compatibles.
  Par exemple, dans une composition \(E\circ(E_1,\ldots,E_n)\), il faut que:
  \begin{itemize}
    \item \(E\) soit valide,
    \item il doit y avoir exactement \(n\) \(\varepsilon\)-indices différents apparaissant dans \(E\),
    \item les ensembles d'\(\varepsilon \)-indices des expressions \(E_1,\ldots,E_n\) doivent être deux à deux disjoints.
  \end{itemize}
  Dans le cas d'un opérateur booléen, ses opérandes doivent être valides et avoir les mêmes \(\varepsilon \)-indices.
  Enfin, les opérateurs classiques (d'expressions d'arbres) doivent satisfaire les propriétés d'homogénéité décrites dans les sections précédentes.

  Ainsi, le langage \(L(E)\) dénoté par une expression valide \(E\) d' \(\varepsilon \)-indices \(\mathcal{I}\) est défini inductivement comme suit
  \begin{gather*}
    \begin{aligned}
      L(f(E_1, \ldots, E_n)) &= f(L(E_1), \ldots, L(E_n)), & L(\varepsilon_j) &= \{\varepsilon_j\},\\
      L(\mathrm{op}(E_1,\ldots,E_n)) &= \mathrm{op}'(L(E_1), \ldots,L(E_n)), &
      L(\emptyset_{\mathcal{I}}) &= \emptyset,\\
      L(E' \circ (E_1,\ldots, E_n)) &= L(E') \circ (L(E_1), \ldots L(E_n)), & L(E_1^{\circledast}) &= {(L(E_1))}^{\circledast},\\
      L(E_1 \cdot_a E_2) &= L(E_1) \cdot_a L(E_2), &  L(E_1^{*_a}) &= {(L(E_1))}^{*_a},
    \end{aligned}
  \end{gather*}
  où \(f\) est un symbole de \(\Sigma_n\), \((E', E_1, \ldots, E_n)\) sont \((n+1)\) expressions d'arbres sur \(\Sigma \), \(j\) est un entier positif, \(\mathrm{op}\) est un opérateur booléen \(n\)-aire, \(\mathrm{op}'\) est l'opération booléenne \(n\)-aire sur des langages homogènes d'\(\varepsilon \)-indices \(\mathcal{I}\) associée à \(\mathrm{op} \) et \(a\) est un symbole de \(\Sigma_0\).

  Il suffit alors de transcrire au niveau des expressions les formules de quotient afin d'obtenir les formules de dérivation d'expression.

  La dérivation d'une expression \(E\) valide par \(\varepsilon_j\), notée \(d_{\varepsilon_j}(E)\) est obtenue en incrémentant les \( \varepsilon \)-indices de \(E\) par \(1\) excepté \(\varepsilon_j\), remplacé par \(\varepsilon_1\).


  La dérivation d'une expression \(E\) valide, dont les \(\varepsilon \)-indices sont \( \{1,\ldots,n\} \),  par un symbole \(\alpha \) de \(\Sigma_n \) est l'expression inductivement calculée comme suit:
    \begin{align*}
      d_\alpha(\emptyset_{\mathcal{I}}) &= \emptyset_{\{1\} \cup \{i+1 \mid i > n, i\in\mathcal{I}\}},\\
      d_\alpha(\varepsilon_1) &= \emptyset_{\{1\}}, \\
      d_\alpha(\alpha(\varepsilon_1,\ldots,\varepsilon_n)) &= \varepsilon_1,\\
      d_\alpha(f(E_1,\ldots,E_m)) &= \sum_{1\leq j\leq n} f(E'_1, \ldots, E'_{j-1},
      d_\alpha(t_j),
      E'_{j+1},
      \ldots,
      E'_m),\\
      d_\alpha(\mathrm{op}(E_1,\ldots,E_k)) &= \mathrm{op}(d_\alpha(E_1), \ldots, d_\alpha(E_k)),\\
      d_\alpha(E_1\cdot_b E_2)&=
        \begin{cases}
          (d_b(E_1)\cdot_b E_2) \circ_1 d_b(E_2) & \text{ if }\alpha=b,\\
          d_\alpha(E_1)\cdot_b E_2 + (d_b(E_1)\cdot_b E_2) \circ_1 d_\alpha(E_2) & \text{ if }\alpha\in\Sigma_0\setminus \{b\},\\
          d_\alpha(E_1)\cdot_b E_2 & \text{otherwise,}\\
        \end{cases}\\
      d_\alpha(E\circ(E_1,\ldots,E_k))&=  \sum_{1\leq j\leq k} E\circ({(\mathrm{Inc}_\varepsilon(1,E_l))}_{1\leq l\leq j},d_\alpha(E_j),{(\mathrm{Inc}_\varepsilon(1,E_l))}_{j+1\leq l\leq k}) \\
      &\quad +
        \begin{cases}
          d_{
            \alpha({(\varepsilon_{j_{p_l}})}_{1\leq l\leq n})
          }
          (E) \circ(\varepsilon_1,{(\mathrm{Inc}_\varepsilon(1,E_l))}_{1\leq l\leq k\mid \forall z, l\neq p_z})\\
          \quad \text{ if } \forall 1\leq l\leq n, \exists 1\leq p_l\leq k, \varepsilon_l \in L(E_{p_l})\\
          \emptyset_{\mathrm{Ind}_\varepsilon(E\circ(E_1,\ldots,E_k))\setminus \{1,\ldots,n\}} \quad \text{ otherwise,}
        \end{cases}\\
      d_\alpha(E^{\circledast})&=
        \begin{cases}
          (E^\circledast\circ (d_\alpha(E)))\circ(\varepsilon_1, \mathrm{Inc}_{\varepsilon}(1,E^\circledast)) & \text{ if }\alpha\in\Sigma_0,\\
          (E^\circledast\circ (d_\alpha(E))) & \text{ otherwise,}
        \end{cases}\\
      d_\alpha(E^{*_b})&=
        \begin{cases}
          {(d_b(E))}^\circledast\cdot_b E^{*_b} & \text{ if }\alpha=b,\\
          ({(d_b(E))}^\circledast \circ (d_\alpha(E))) \cdot_b E^{*_b} & \text{otherwise,}\\
        \end{cases}
    \end{align*}
    où
    \(d_{\alpha(\varepsilon_{j_1},\ldots,\varepsilon_{j_n})(E)}
    =
      d_{\alpha}(
        d_{\varepsilon_{j_{1} + n - 1}}(
          \cdots
            d_{\varepsilon_{j_{n-1} + 1}}(
              d_{\varepsilon_{j_n}}(E)
            )
          \cdots
        )
      )
    \),
    où pour tout entier \(i\) l'expression \(E'_i\) vaut  \(\mathrm{Inc}_\varepsilon(1,E_i)\),
    et où
    \(\circ_1\) est la composition partielle
    \(
      E \circ_1 E' = E\circ (E', {(\varepsilon_l)}_{j_2 \leq l \leq j_k})
    \)
    avec \(\mathrm{Ind}_\varepsilon(E)= \{j_1,\ldots,j_k\} \).

  La dérivation par un arbre \(t=f(t_1,\ldots,t_n)\) de \(\Sigma^*\) tel que
    \(
      \mathrm{Ind}_\varepsilon(t)\subseteq \mathrm{Ind}_\varepsilon(E).
    \) est l'expression définie par
    \begin{equation*}
      d_t(E)=(d_f(d_{t'_1}(\cdots (d_{t'_k}(E))\cdots))\circ(\varepsilon_1,{(\varepsilon_{y_z+1})}_{1\leq z\leq n-l})),
    \end{equation*}
    où  \( \{y_1,\ldots,y_{n-l}\} = \mathrm{Ind}_\varepsilon(E)\setminus \mathrm{Ind}_\varepsilon(t) \) et \(\forall 1\leq j\leq  k \), \( t'_j=\mathrm{Inc}_\varepsilon(k-j,t_j) \).

  La dérivation par un arbre est une implantation du calcul du quotient:
  en effet, la dérivée d'une expression valide \(E\) par rapport à un arbre \(t\) dénote \(t^{-1}(L(E))\).

  \begin{example}\label{ex:derivative}
    Considérons l'alphabet gradué définit par \( \Sigma_2=\{f\} \),  \( \Sigma_1=\{g\} \) et \( \Sigma_0=\{a,b,c\} \) et soit \( E \) l'expression définie par
    \begin{equation*}
      E = E_1 \cdot_a E_2,
    \end{equation*}
    avec \(E_1 = \neg({g(a)}^{*_a})\) et \(E_2 = f(f(a,a),a) \).
    Montrons comment calculer la dérivée de \( E \) par \(t = f(f(a,a),a)\).
    Tout d'abord, calculons la dérivée de \(E_2\) par \(t\):
    \begin{align*}
       d_a(E_2) &=  f(f(\varepsilon_1,a)+f(a,\varepsilon_1),a)+f(f(a,a),\varepsilon_1),\\
       d_a(d_a(E_2)) &=  f(f(\varepsilon_2,\varepsilon_1)+f(\varepsilon_1,\varepsilon_2),a)
       + f(f(\varepsilon_2,a)+ f(a,\varepsilon_2),\varepsilon_1)\\
       & \qquad + f(f(\varepsilon_1,a)+f(a,\varepsilon_1),\varepsilon_2),\\
       d_a(d_a(d_a(E_2))) &=  f(f(\varepsilon_3,\varepsilon_2)+f(\varepsilon_2,\varepsilon_3),\varepsilon_1)
       + f(f(\varepsilon_3,\varepsilon_1)+f(\varepsilon_1,\varepsilon_3),\varepsilon_2)\\
       & \qquad + f(f(\varepsilon_2,\varepsilon_1)+f(\varepsilon_1,\varepsilon_2),\varepsilon_3),\\
       d_{f(a,a)}(d_a(E_2)) &= d_{f(\varepsilon_1, \varepsilon_2)}(d_a(d_a(d_a(E_2)))) \circ (\varepsilon_1, \varepsilon_2)\\
       &=
       (\emptyset_{1,4} + \emptyset_{1,4} + f(\emptyset_1 + \varepsilon_1 ,\varepsilon_4))
       \circ (\varepsilon_1, \varepsilon_2) = f(\varepsilon_1, \varepsilon_2),\\
       d_{t}(E_2) &= d_{f(\varepsilon_1, \varepsilon_2)} (d_{f(a,a)}(d_a(E_2))) = \varepsilon_1.
    \end{align*}
    Par soucis de concision, posons
    \begin{align*}
      E' &= \neg({g(\varepsilon_1)}^{\circledast})\cdot_a E_2, &
      E'' &= \neg(\emptyset_{\{1,2\}})\cdot_a E_2.
    \end{align*}
    Ainsi:
    \begin{align*}
      d_a(E) &= E' \circ d_a(E_2),\\
      d_{f(a,a)}(d_{a}(E)) &= E'' \circ (f(\varepsilon_1, a),\varepsilon_4 ) \circ  (\varepsilon_1, \mathrm{Inc}_{\varepsilon}(1,d_a(E_2)))
      + E' \circ d_{f(a,a)}(d_a(E_2)) ,\\
      d_t(E) &= E'.
    \end{align*}
  \end{example}

  \subsection{Construction d'automate associée}
  Montrons alors comment construire à partir d'une expression \(E\) avec \(\mathrm{Ind}_\varepsilon(E) = \emptyset \) un automate à partir d'un processus itératif.

  Pour une expression \(E\) valide sur un alphabet \(\Sigma \), calculons tout d'abord l'ensemble
  \begin{equation*}
    D_0(E) = \{d_a(E)\mid a\in\Sigma_0\}.
  \end{equation*}
  Depuis cet ensemble, considérons l'automate \(A_0=(\Sigma,D_0(E),F_0,\delta_0)\) où
  \begin{align*}
    F_0 &= \{E'\in D_0(E) \mid \varepsilon_1 \in L(E')\}, &
    \delta_0 &= \{(a, d_a(E)) \mid a \in \Sigma_0\}.
  \end{align*}
  À partir de cette étape, on peut choisir une fonction \(\mathrm{tree}_0\) associant toute expression \(E'\) de \(D_0(E)\) avec un arbre \(t\) tel que
  \begin{equation*}
    \mathrm{tree}_0(E') = t \Rightarrow d_t(E) = E',
  \end{equation*}
  en choisissant pour toute expression \(E'\) de \(D_0(E)\) un symbole \(a\in\Sigma_0\) tel que \((a,E') \in \delta_0\).
  Depuis cette base d'induction, considérons l'ensemble de transitions \(\delta_n\) défini inductivement comme suit
  \begin{align}
    \begin{split}
      \delta_n = \{((E'_1, \ldots, E'_m), f, d_t(E)) \mid
          & t = f(\mathrm{tree}_{n-1}(E'_1),\ldots, \mathrm{tree}_{n-1}(E'_m)),\\
          & f \in \Sigma_m,\\
          & E'_1,\ldots,E'_m \in D_{n-1}(E)\}.
      \label{eq:delta}
    \end{split}
  \end{align}
  Considérons également l'ensemble
  \(
    D_n(E) = D_{n-1}(E) \cup \pi_3(\delta_n)
  \),
  où \(\pi_3\) est la projection classique
  \(
    \pi_3(X) = \{z \mid (\_,\_,z) \in X\}
    \).
  Clairement, il est encore possible de choisir une fonction \(\mathrm{tree}_n\) associant à toute expression \(E'\) de \(D_n(E)\) un arbre \(t\) rel que
  \begin{equation*}
    \mathrm{tree}_n(E') = t \Rightarrow d_t(E) = E',
  \end{equation*}
  en choisissant une transition \(((E'_1, \ldots, E'_m), f, E')\) in \(\delta_n\) pour toute expression \(E'\) de \(D_{n}(E)\setminus (D_{n-1}(E))\)
  et en définissant \(t\) comme \(f(\mathrm{tree}_{n-1}(E'_1), \ldots, \mathrm{tree}_{n-1}(E'_k))\).
  Puis, en considérant l'ensemble
  \begin{equation}
    F_n = \{E' \in D_n(E) \mid \varepsilon_1 \in L(E')\},
    \label{eq:final states}
  \end{equation}
  on peut définir l'automate \(A_n=(\Sigma,D_n(E),F_n,\delta_n)\).
  Finalement, soit \(A(E)\) le point fixe, s'il existe, de cette construction (en fonction du choix des fonctions \(\mathrm{tree}_*\)), appelé \firstocc{automate des dérivées Bottom-Up}{automate!d'arbres!des dérivées Bottom-Up} de \(E\).

  Cette construction produit par définition un automate déterministe.
  De plus, la validité de la construction ne dépend pas des choix des fonctions \(\mathrm{tree}_*\).
  Enfin, l'automate des dérivées Bottom-Up d'une expression \(E\) reconnaît \(L(E)\).

  On peut remarquer que lorsque \(\Sigma_n=\emptyset \) pour tout \(n\geq 2\), les règles ACI de la somme (associativité, commutativité et idempotence) suffisent à la démonstration de l'existence du point fixe, d'une façon similaire à celle de la construction de l'automate (de mots) de Brzozowski~\cite{Brz64}.
Cependant, ce n'est pas garanti dans le cas général.

  \begin{example}\label{ex:cons}
    Considérons les quatre expressions définies dans l'Exemple~\ref{ex:derivative}, c'est-à-dire
    \begin{align*}
      E   &= E_1 \cdot_a E_2, & E_1 &= \neg({g(a)}^{*_a}), &
      E_2 &= f(f(a,a),a), & E'  &= \neg({g(\varepsilon_1)}^{\circledast})\cdot_a E_2.
    \end{align*}
    Montrons comment calculer l'automate associé à \(E\).
    Tout d'abord, calculons \( A_0 = (\Sigma, D_0(E), F_0, \delta_0) \):
    par définition,
    \(
       D_0(E) = \{d_a(E), d_b(E), d_c(E)\}.
       \)
    Ainsi,
    \begin{gather*}
      \begin{aligned}
        d_a(E) &= E' \circ d_a(E_2), &
        d_b(E) &= d_c(E) = \neg(\emptyset_{\{1\}}) \cdot_a E_2,
      \end{aligned}\\
      \begin{aligned}
      D_0(E) &= \{ d_a(E), d_b(E)\}, &
      F_0 &=\{d_b(E)\}, &
      \delta_0 &= \{(a, d_a(E)), (b, d_b(E)), (c, d_b(E))\}.
      \end{aligned}
    \end{gather*}
    L'automate \(A_0\) est représenté Figure~\ref{fig a0}.
    \begin{figure}[H]
      \centering
      \begin{tikzpicture}[node distance=2.5cm,bend angle=30,transform shape,scale=1]
        \node[state, rounded rectangle] (1)  {\( d_a(E) \)} ;
        \node[state, rounded rectangle,double, right of = 1,node distance=3cm] (2) {\( d_b(E) \)};
        \draw (1) ++(-1cm,0cm) node {\( a \)}  edge[->] (1);
        \draw (2) ++(1.5cm,0cm) node {\( b,c \)}  edge[->] (2);
      \end{tikzpicture}
      \caption{L'automate \(A_0\).}%
      \label{fig a0}
    \end{figure}
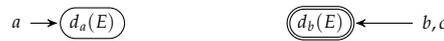
    Choisissons alors une fonction \(\mathrm{tree}_0\):
    \(
      \mathrm{tree}_0(d_a(E)) = a,  \mathrm{tree}_0(d_b(E)) = b.
    \)
    Montrons alors comment calculer \( A_1=(\Sigma, D_1(E), F_1, \delta_1) \).
    Selon l'Équation~\eqref{eq:delta}, il suffit de calculer les dérivées de \(E\) par les arbres de l'ensemble \( \{f(a,a), f(a,b), f(b,a), f(b,b), g(a), g(b)\} \):
    \begin{gather*}
      \begin{aligned}
        d_{f(a,a)}(E) &= E' \circ f(\varepsilon_1, a), &
        d_{f(b,b)}(E) &= d_{g(b)}(E) = d_{b}(E),
      \end{aligned}\\
      d_{f(a,b)}(E) = d_{f(b,a)}(E) = d_{g(a)}(E) = \emptyset_{\{1\}}.
    \end{gather*}
    Cela produit un nouvel état, \(E' \circ f(\varepsilon_1, a)\), associé à \(f(a,a)\) par la fonction \(\mathrm{tree}_1\), et trois nouvelles transitions:
    \begin{equation*}
      \delta_1 = \delta_0 \cup \{(d_a(E), d_a(E), f, d_{f(a,a)}(E)), (d_b(E), d_b(E), f, d_{b}(E)), (d_b(E), g , d_b(E))\}.
    \end{equation*}
    L'automate \(A_1\) est représenté Figure~\ref{fig a1}, où l'état-puits \(\emptyset_1\) et ses transitions sont absentes.
    \begin{figure}[H]
      \centering
      \begin{tikzpicture}[node distance=2.5cm,bend angle=30,transform shape,scale=1]
        \node[state, rounded rectangle] (1)  {\( d_a(E) \)} ;
        \node[state, rounded rectangle, right of=1, node distance=4cm] (3) {\( d_{f(a,a)}(E) \)};
        \node[state, rounded rectangle,double, right of=3,node distance=4cm] (2) {\( d_b(E) \)};

        \draw (1) ++(-1cm,0cm) node {\( a \)}  edge[->] (1);
        \draw (2) ++(1.5cm,0cm) node {\( b,c \)}  edge[->] (2);
        \path[->]
          (2) edge[->, loop above] node {\( g \)} ();
        \draw (3) ++(-2 cm,0cm)  edge[->]   node[above,pos=0.5] {\( f \)} (3) edge[dashed, bend left]node[pos=0.5,below]{\tiny{2}} (1) edge[dashed, bend right] node[pos=0.5,above]{\tiny{1}}(1);
        \draw (2) ++(-2cm,0cm) edge[->,bend left=-40]   node[below,pos=0.5] {\( f \)} (2) edge[dashed, bend left=-10]node[pos=0.5,below]{\tiny{1}}(2) edge[dashed, bend left=10] node[pos=0.5,above]{\tiny{2}}(2);
      \end{tikzpicture}
      \caption{L'automate \(A_1\).}%
      \label{fig a1}
    \end{figure}
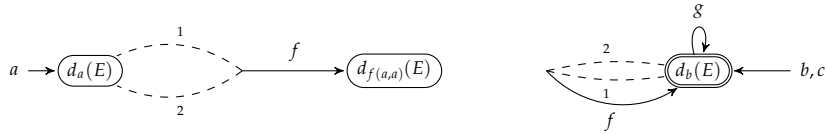
    L'automate \(A_2\) est calculé depuis les dérivées par les arbres de l'ensemble
    \begin{equation*}
    \{f(f(a,a),f(a,a)), f(f(a,a),a),  f(f(a,a),b), f(a,f(a,a)), f(b,f(a,a)), g(f(a,a))\}:
    \end{equation*}
    \(
      d_{f(f(a,a),a)}(E) = d_{f(f(a,a),f(a,a))}(E) = E'\) et les quatre autres sont égales à
      \(\emptyset_{\{1\}} \).
    Cela produit ainsi un nouvel état, \(E'\), associé à \(t = f(f(a,a),a)\) par la fonction \(\mathrm{tree}_2\), et une nouvelle transition:
      \(\delta_2 = \delta_1 \cup \{(d_{f(a,a)}(E), d_a(E), f, d_{t}(E))\} \).
    L'automate \(A_2\) est représenté Figure~\ref{fig a2}.
    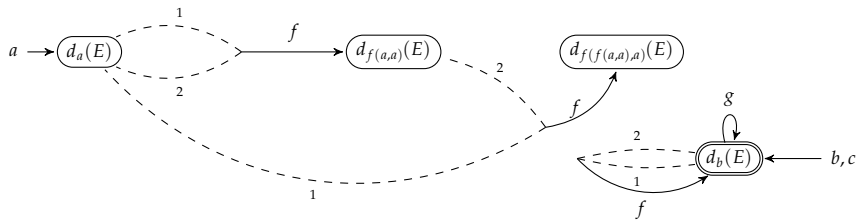
\begin{figure}[H]
      \centering
      \begin{tikzpicture}[node distance=2.5cm,bend angle=30,transform shape,scale=1]
        \node[state, rounded rectangle] (1)  {\( d_a(E) \)} ;
        \node[state, rounded rectangle, right of=1, node distance=4cm] (3) {\( d_{f(a,a)}(E) \)};
        \node[state, rounded rectangle, right of=3,node distance = 3cm] (4) {\( d_{f(f(a,a),a)}(E) \)};
        \node[state, rounded rectangle,double, below right of=4,node distance=2cm] (2) {\( d_b(E) \)};

        \draw (1) ++(-1cm,0cm) node {\( a \)}  edge[->] (1);
        \draw (2) ++(1.5cm,0cm) node {\( b,c \)}  edge[->] (2);
        \path[->]
          (2) edge[->, loop above] node {\( g \)} ();
        \draw (3) ++(-2 cm,0cm)  edge[->]   node[above,pos=0.5] {\( f \)} (3) edge[dashed, bend left]node[pos=0.5,below]{\tiny{2}} (1) edge[dashed, bend right] node[pos=0.5,above]{\tiny{1}}(1);
        \draw (2) ++(-2cm,0cm) edge[->,bend left=-40]   node[below,pos=0.5] {\( f \)} (2) edge[dashed, bend left=-10]node[pos=0.5,below]{\tiny{1}}(2) edge[dashed, bend left=10] node[pos=0.5,above]{\tiny{2}}(2);
        \draw (4) ++(-1 cm, -1cm)  edge[->,bend right]   node[left,pos=0.5] {\( f \)} (4) edge[dashed, bend right=20]node[pos=0.5,above]{\tiny{2}} (3) edge[dashed, bend left=40] node[pos=0.5,below]{\tiny{1}}(1);
      \end{tikzpicture}
      \caption{L'automate \(A_2\).}%
      \label{fig a2}
    \end{figure}
    L'automate \(A_3\) est obtenu en calculant les dérivées par les arbres de l'ensemble \( \{f(t, t), f(t, a), f(t, b),\ldots \} \).
    Il y a seulement quatre calculs retournant des expressions différentes de \( \emptyset_{\{1\}} \):
      \(d_{f(t,t)}(E) = d_{f(t,b)}(E) = d_{f(b, t)}(E) = d_b(E)\) et
      \(d_{g(t)}(E) = d_t(E)\), produisant les transitions suivantes:
    \begin{equation*}
      \delta_3 = \delta_2 \cup \{(d_{t}(E), d_t(E), f, d_b(E)), (d_{t}(E), d_b(E), f, d_b(E)), (d_{b}(E), d_t(E), f, d_b(E)), (d_{t}(E), g, d_t(E))\},
    \end{equation*}
    calculs ne produisant pas de nouveaux états.
    Ainsi la construction s'arrête et l'automate des dérivées de \( E \) est \(A_3\), représenté Figure~\ref{fig1}.

  \begin{figure}[H]
    \centering
    \begin{tikzpicture}[node distance=2.5cm,bend angle=30,transform shape,scale=1]
      \node[state, rounded rectangle] (1)  {\( d_a(E) \)} ;
      \node[state, rounded rectangle, above of=1,node distance=2cm] (3) {\( d_{f(a,a)}(E) \)};
      \node[state, rounded rectangle, right of=1,node distance = 3cm] (4) {\( d_{f(f(a,a),a)}(E) \)};
      \node[state, rounded rectangle,double, above right of=4,node distance=3cm] (2) {\( d_b(E) \)};

      \draw (1) ++(0cm,-1cm) node {\( a \)}  edge[->] (1);
      \draw (2) ++(1cm,0cm) node {\( b,c \)}  edge[->] (2);
      \path[->]
        (2) edge[->, loop above] node {\( g \)} ()
        (4) edge[->, loop below] node {\( g \)} ();
      \draw (3) ++(-1 cm,-1cm)  edge[->,bend left=40]   node[above,pos=0.5] {\( f \)} (3) edge[dashed, bend left=-20]node[pos=0.5,below]{\tiny{1}} (1) edge[dashed, bend left=0] node[pos=0.5,above]{\tiny{2}}(1);
      \draw (4) ++(-1 cm, -1cm)  edge[->,bend left=0]   node[left,pos=0.5] {\( f \)} (4) edge[dashed, bend right=20]node[pos=0.5,below]{\tiny{2}} (3) edge[dashed, bend left=20] node[pos=0.5,left]{\tiny{1}}(1);
      \draw (2) ++(-2 cm,-0.5cm)  edge[->,bend left=40]   node[above,pos=0.5] {\( f \)} (2) edge[dashed, bend left=10] node[pos=0.5,right]{\tiny{2}}(4) edge[dashed, bend left=-10] node[pos=0.5,left]{\tiny{1}}(4);
      \draw (2) ++(-1cm,-1cm)  edge[->,bend left=30]   node[above,pos=0.5] {\( f \)} (2) edge[dashed, bend left=-20] node[pos=0.5,right]{\tiny{2}}(4) edge[dashed, bend left=15] node[pos=0.5,right]{\tiny{1}}(2);
      \draw (2) ++(-0.5cm,-2cm) edge[->,bend left=-20]   node[right,pos=0.5] {\( f \)} (2) edge[dashed, bend left=20] node[pos=0.5,above]{\tiny{1}}(4) edge[dashed, bend left=15] node[pos=0.5,right]{\tiny{2}}(2);
      \draw (2) ++(1cm,-1.5cm) edge[->,bend left=-40]   node[right,pos=0.5] {\( f \)} (2) edge[dashed, bend left=-10]node[pos=0.5,right]{\tiny{1}}(2) edge[dashed, bend left=10] node[pos=0.5,left]{\tiny{2}}(2);
    \end{tikzpicture}
    \caption{L'automate des dérivées Bottom-Up de \(E\).}%
    \label{fig1}
  \end{figure}
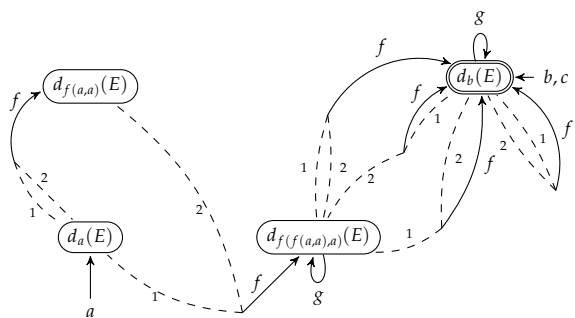
\end{example}

\chapter{Conclusion}

Dans cette section, nous avons rappelé diverses constructions d'automates classiques de la théorie des automates en les appliquant sur les expressions d'arbres afin de construire des automates ascendants comme descendants.

Ces constructions semblent n'être que des adaptations (quelques fois plus techniques) des algorithmes connus sur des éléments plus structurés que les mots (les arbres).
Cette remarque peut être également faite au niveau même des automates de mots, en considérant les différentes techniques déployées pour construire des automates déterministes, non-déterministes ou à multiplicités depuis une expression par dérivation par exemple.
Ainsi, une question naturelle se pose: quel pourrait être le degré d'abstraction nécessaire pour unifier les constructions sur les arbres et les mots, mais aussi en termes de propriétés ou de type d'automates?

Nous essaierons ainsi de proposer un projet de recherche dans ce sens dans la partie suivante.
Tout d'abord, nous montrerons comment décrire des automates de mots à l'aide de théorie des catégories afin d'unifier les différents types d'automates de mots (déterministes, non-déterministes, à multiplicités, alternants, \emph{etc.}).
Puis dans un second temps, nous nous intéresserons à la théorie des catégories enrichies afin d'essayer de proposer une structure permettant d'unifier les automates de mots et d'arbres.
Enfin, nous proposerons une structure adaptée pour les expressions rationnelles, et étendrons les méthodes de construction précédentes sur celles-ci.
Toutes ces structures seront implantées en Haskell afin de fournir des outils pratiques de manipulation de ces notions.


\part{Les automates catégoriques}\label{partAutCat}

\chapter{Présentation}

\begin{flushright}
  \emph{Aucune généralisation n'est totalement vraie, même pas celle-ci.}\\
   Oliver Wendell Holmes
\end{flushright}


Dans les sections précédentes, on a pu voir les similitudes entre les constructions classiques d'automates depuis une expression rationnelle de mots avec celles concernant les arbres.
Il a ainsi été possible de réutiliser la notion de position et de successeur, soit ascendant (parents) soit descendant (fils), pour construire un automate d'arbres.
La notion de dérivation est également réutilisable (dans le cas de la dérivation partielle descendante) sur les expressions pour décrire d'une façon presque directe la fonction de transition d'un automate descendant.
La similitude entre la dérivation et la fonction de transition d'un automate se renforce d'autant plus lors de leurs extensions respectives de l'action d'un symbole à celle d'une structure \emph{composite} (arbres ou mots).
Comme nous l'avons suggéré, ces extensions sont directement représentables par des notions de théorie des catégories (plus précisément des adjonctions).

Ces observations laissent présager une factorisation possible de ces méthodes afin de construire indistinctement des automates d'arbres, de mots, voire d'autres structures par des méthodes unifiées, en utilisant des outils de théorie des catégories, permettant de manipuler et de composer les morphismes entre différents objets.
C'est ainsi le but du projet de recherche proposé dans la suite de ce document.

Nous allons essayer de mettre en place, \emph{via} des outils de théorie des catégories, une description unifiée, dans un premier temps, des automates de mots (déterministes, non-déterministes, à multiplicités, \emph{etc.}).
L'approche suggérée dans la suite, à savoir manipuler des outils de théorie des catégories, n'est pas nouvelle.
En effet, de nombreux auteurs ont déjà proposé des descriptions catégoriques des automates (voir par exemple l'utilisation des catégories pour réaliser de l'apprentissage par automates dans~\cite{UrSc19}, basée sur une description algébrique des automates de~\cite{ArMa75}; voir également~\cite{HKRSS20} pour une définition des automates d'arbres selon la même description; voir finalement~\cite{BBHKKPS20} pour une approche unificatrice des méthodes algébriques et coalgébriques).
Dans la suite, nous proposerons une définition d'automates similaire à celle de Colcombet et Petrisan~\cite{CP17}, construisant des automates à partir de morphismes de catégories.

Nous montrerons ensuite comment généraliser cette structure afin d'englober également les automates d'arbres ascendants ou descendants en utilisant des éléments de théorie des catégories enrichies.
Cette deuxième extension pourra mettre en lumière la compatibilité avec l'approche fonctorielle de Colcombet et Petrisan: si un automate de mot peut être décrit comme un foncteur, les automates \emph{enrichis} seront descriptibles en termes de foncteurs enrichis.

Enfin, nous proposerons également une structure commune pour les expressions, basées sur des produits tensoriels, permettant de factoriser expressions d'arbres et de mots.

Toutes ces notions seront implantées en Haskell, langage de programmation fonctionnelle à évaluation paresseuse (voir par exemple~\cite{Kur18} pour un bon survol des notions de base de la programmation Haskell).
Des notions de théorie des catégories ont déjà été implantées dans des modules de ce langage (voir par exemple~\cite{HSS17} pour une implantation des automates utilisant la théorie des catégories pour l'apprentissage par automates).
L'utilisation de la théorie des catégories est à la base même de ce langage, où les notions de foncteurs ou de monades sont fondamentales.
Ce lien entre théorie des catégories et programmation est une thématique de recherche très active, que ce soit en Haskell (voir par exemple le lien entre catégories enrichies et accesseurs d'enregistrement décrit dans~\cite{CEGLMPR20}) mais également dans d'autres langages de programmation, comme en Python (voir par exemple l'utilisation des catégories monoïdales en traitement automatique du langage naturel avec l'implantation décrite dans~\cite{FTC20}).
Des implantations d'automates utilisant des outils de théorie des catégories ont egalement été proposées~\cite{}

 Cependant, nous n'utiliserons généralement pas de modules pré-existants pour l'implantation des outils de théorie des catégories.
 À la place, dans un souci de compréhension, nous construirons tout au long de ce document les modules que nous utiliserons afin d'illustrer les notions théoriques par leurs implantations, mais également pour décrire les différents mécanismes de programmation fonctionnelle avancée mis en {\oe}uvre dans ce document au fur et à mesure de leur utilisation.
Les modules ainsi implantés ne seront pas nécessairement optimisés;
nous soulignerons alors lors des différentes réalisations les améliorations possibles de l'utilisation de modules plus standards et avancés.
Enfin, nous transcrirons l'implantation dans ce document par l'intermédiaire de \emph{codes Pseudo-Haskell}.

\section*{Qu'est-ce qu'un code Pseudo-Haskell?}

On entend dans la suite de ce document par code \emph{Pseudo-Haskell} un morceau de code écrit en utilisant la syntaxe du langage Haskell, mais simplifié dans sa notation pour le rendre accessible à des non-initiés\footnote{Par exemple, moins de contraintes explicitées, pas de \emph{pragmas} déclarés, pas de \emph{forall} explicites, moins de déclarations de modules, \emph{etc.}}.
Ainsi, les codes présentés dans ce document ne sont pas directement compilables\footnote{La version non-simplifiée, c'est-à-dire tout le code source, dont ces codes sont tirés l'est.}.

La compilation et l'utilisation des sources peut se faire en utilisant \href{https://docs.haskellstack.org/en/stable/README/}{stack}.
Une fois installé et mis à jour, cet outils permet la compilation automatique des différents modules, et le chargement des modules suffisants à l'interprétation du code.
Il permet également de générer et de consulter la documentation de ce projet.

Les conventions de notations utilisées par la suite sont les suivantes:
\begin{itemize}
  \item une fonction anonyme se note \texttt{\textbackslash x -> val} où \texttt{x} est un paramètre et \texttt{val} le résultat;
  \item une fonction de la forme \texttt{\textbackslash x1 -> \ldots \textbackslash xn -> val} peut être notée \texttt{\textbackslash x1 \ldots xn -> val};
  \item une fonction nommée \texttt{f = \textbackslash x1 \ldots xn -> val} peut être écrite \texttt{f x1 xn = val};
  \item une expression de la forme \texttt{let x = y in z} est du sucre syntaxique pour la fonction définie comme \texttt{(\textbackslash x -> z) y};
  \item un commentaire est précédé par \texttt{{-}{-}};
  \item l'opérateur \texttt{\$} est un opérateur d'évaluation (c'est-à-dire que \texttt{f \$ x = f x}) associatif à droite et de priorité la plus faible possible;
  \item dans un filtrage de motif, le symbole \texttt{\_} est un joker;
  \item le type d'une expression est indiqué par les symboles \texttt{::};
  \item le type noté \(()\) est le type contenant l'unique élément \(()\);
  \item un type algébrique de données est introduit par le mot clé \texttt{data} suivi par le nom du constructeur de types et les types dont il dépend;
    ses valeurs sont définies par une énumération (symbole \texttt{|}) de constructions produits toutes introduites par un constructeur de valeur, suivi des types des valeurs contenues;
    entre accolades peut alors apparaître le nom d'une fonction de projection;
  \item une fonction, un constructeur de valeur ou un constructeur de type binaire dont le nom est un mot (suite de caractères alphabétiques) est préfixe, et infixe si écrit entre \emph{backticks} (caractère \`{});
  \item une fonction, un constructeur de valeur ou un constructeur de type binaire autre (par exemple la fonction \texttt{+}) est infixe, et préfixe si écrit entre parenthèses;
  \item le constructeur de types de liste, unaire et noté \texttt{[]}, peut être utilisé en position misfixe (le type \texttt{[] a} pourra être noté \texttt{[a]});
  \item le constructeur \texttt{(->)} est le constructeur binaire du type des fonctions;
  \item une \emph{classe de types} est une relation entre plusieurs types permettant de décrire l'existence de fonctions (\emph{ad-hoc}) polymorphes, \emph{concrétisée} par la déclaration d'instances;
     une \emph{super-classe} peut-être définie en précédant la déclaration par le symbole \texttt{=>};
     une méthode (c'est-à-dire une fonction d'une classe) peut avoir une définition par défaut et dans ce cas son corps apparaît dans la définition de la classe;
  \item une classe peut apparaître comme une \firstocc{contrainte} (condition nécessaire sur un type) dans une signature de type, précédant cette signature est étant suivie par \texttt{=>};
  \item les propriétés mathématiques classiques (par exemple l'associativité de la loi de composition d'un semigroupe) non testées au niveau des types ne seront pas nécessairement explicitement rappelées\footnote{voir conclusion Chapitre~\ref{chapterConclu}\label{footnoteTypeLevelPerspectives}}; ainsi, il est possible de construire des valeurs ne vérifiant pas les propriétés mathématiques nécessaires pour assurer la validité des constructions;
  \item chaque partie de pseudo-code est tirée d'un fichier dont le nom est précédé par le mot-clé \texttt{module}; les fonctions de ce module sont utilisables soit par la compilation d'un programme les utilisant, mais aussi \emph{via} l'interpréteur \texttt{ghci}. Pour cela, il suffit d'utiliser la commande \texttt{stack ghci src/<chemin du fichier>};
  \item toute extension à ces conventions sera décrite en commentaires.
\end{itemize}

Différents exemples seront donnés tout au long du document.
Pour les exécuter, il est possible:
\begin{itemize}
  \item soit d'utiliser l'interpréteur \texttt{ghci} en tapant dans un terminal ouvert depuis le dossier source la commande précédente \texttt{stack ghci src/<chemin du fichier>};
  \item soit d'exécuter le programme principal listant les exemples, après compilation pour bénéficier d'optimisations, en tapant \texttt{stack build} (pour la compilation) puis \texttt{stack exec Main} pour l'exécution du programme principal.
\end{itemize}

Toutes les fonctions Haskell (de base ou de modules non développés pour ce document) utilisées dans ce document ne sont pas nécessairement explicitées; cependant, toutes sont accessibles sur le site \href{https://hoogle.haskell.org/}{Hoogle}.
Les fonctions développées dans le cadre de ce document sont toutes commentées; la documentation, que l'on peut générer et consulter \emph{via} la commande \texttt{stack haddock {-}{-}open} depuis la racine des sources est un complément essentiel de la lecture de ce document.

\begin{implementationBox}[]{Exemple jouet}
  \begin{minted}[xleftmargin=1em,   autogobble, fontsize=\footnotesize]{haskell}
    -- Type des Booléens
    data Bool = True | False

    -- Classe des types dont l'égalité est décidable
    class Eq a where
      -- l'égalité
      (==) :: a -> a -> Bool
      -- la différence
      (/=) :: a -> a -> Bool
      a /= b = not $ a == b

    -- L'égalité des Booléens est décidable
    instance Eq Bool where
      True  == True   = True
      False == False  = True
      _     == _      = False

    -- Description des ordres
    data Ordering = LT | EQ | GT

    -- Classe des types ordonnés
    class Eq a => Ord a where
      compare :: a -> a -> Ordering

    -- Les Booléens sont ordonnés
    instance Ord Bool where
      False `compare` True  = LT
      True  `compare` False = GT
      _     `compare` _     = EQ

    -- Le constructeur de type neutre, et la définition
    -- de la fonction runIdentity de signature Identity a -> a.
    -- Identity a est isomorphe à a.
    data Identity a = Identity {runIdentity :: a}

    -- S'il y a une égalité décidable pour un type a,
    -- l'égalité est décidable pour le type Identity a
    instance Eq a => Eq (Identity a) where
      x == y = runIdentity x == runIdentity y

    -- Les classes des semigroupes et des monoïdes
    class Semigroup a where
      -- l'opération du semigroupe (doit être associative)
      (<>) :: a -> a -> a
    class Semigroup a => Monoid a where
      -- élément neutre du monoïde
      mempty :: a

    -- Une liste d'éléments de type a est
    -- * soit la liste vide []
    -- * soit un élément suivi d'une liste (constructeur binaire (:)).
    data [a] = [] | a : [a]

    -- Une liste est un monoïde; plus précisément,
    -- il s'agit du monoïde libre.
    instance Semigroup [a] where
      []       <> ys = ys
      (x : xs) <> ys = x : (xs <> ys)
    instance Monoid [a] where
      mempty = []
  \end{minted}
\end{implementationBox}

\chapter{Les automates de mots}\label{chapAutCatMots}

Dans cette section, nous allons nous intéresser à la notion d'automates de mots.
Pour cela, nous allons étudier les différentes façons de combiner les transitions d'automates (déterministes, non déterministes, à multiplicités, \emph{etc.}) en les interprétant à la lueur de la théorie des catégories, théorie de la composition par excellence.
Nous allons donc nous intéresser aux points communs de chacune des compositions possibles, afin de factoriser des opérations communes à ces machines dont les fonctionnements semblent pourtant différents.

Nous commencerons donc par rappeler la définition de catégorie, puis nous montrerons comment définir la notion d'automates sur une catégorie.
L'utilisation de cette structure nous permettra de factoriser les structures d'automates de mots (déterministes, non déterministes, à multiplicités, \emph{etc.}).
Nous verrons également que plusieurs algorithmes classiques (déterminisation, complétion, et autres conversions) sont des applications d'une même transformation.

Tout au long de cette partie, nous manipulerons des concepts avancés de programmation fonctionnelle mettant en {\oe}uvre des concepts de théorie des catégories;
nous nous efforcerons alors de donner des exemples concrets de construction, afin d'illustrer le plus possible ces notions.


\section{Automates sur une catégorie}\label{sec:automatesMotsCat}

Une \firstocc{catégorie}{catégorie} \(\mathcal{C}\) est constituée
\begin{itemize}
  \item d'une classe \(\mathrm{Obj}_\mathcal{C}\), les \firstocc{objets}{catégorie!objets} de \(\mathcal{C}\);
  \item d'un ensemble \(\mathrm{Hom}_\mathcal{C}(A, B)\) de \firstocc{morphismes}{catégorie!morphismes} pour tous deux objets \(A\) et \(B\) de \(\mathcal{C}\);
  \item d'une fonction \(\circ_\mathcal{C}\) de \firstocc{composition}{catégorie!composition} de \(\mathrm{Hom}_\mathcal{C}(B, C) \times \mathrm{Hom}_\mathcal{C}(A, B) \rightarrow \mathrm{Hom}_\mathcal{C}(A, C) \) pour tous trois objets \(A\), \(B\) et \(C\) de \(\mathcal{C}\)
\end{itemize}
satisfaisant les conditions suivantes:
\begin{itemize}
  \item la composition est associative;
  \item il existe pour tout objet \(A\) un élément \(\mathrm{id}_A\), neutre pour la composition: pour tous deux morphismes \(f\) de \(\mathrm{Hom}_\mathcal{C}(A, B)\) et \(g\) de \(\mathrm{Hom}_\mathcal{C}(B, A)\),
  \begin{align*}
    f \circ_{\mathcal{C}} \mathrm{id}_A &= f, &
    \mathrm{id}_A \circ_{\mathcal{C}} g &= g.
  \end{align*}
\end{itemize}
En absence d'ambiguïté, on notera \(\mathrm{id}_A\) par \(\mathrm{id}\), \(\mathrm{Hom}_\mathcal{C}\) par \(\mathrm{Hom}\) et \(\circ_{\mathcal{C}}\) par \( \circ \).

\begin{implementationBox}[label={codeCat}]{Catégorie}
  \begin{minted}[xleftmargin=1em, autogobble, fontsize=\footnotesize]{haskell}
    -- module Category.CategoryGen

    -- Une catégorie est un constructeur de types binaire :
    -- * dont on sait composer les valeurs
    -- * qui contient un élément pointé pour chaque type
    class CategoryGen cat where
      -- le morphisme identité
      id  :: cat a a
      -- la composition de morphisme
      (.) :: cat b c -> cat a b -> cat a c
  \end{minted}
\end{implementationBox}

\begin{implementationBox}[label={codeTypeCat}]{Catégorie des types}
  \begin{minted}[xleftmargin=1em, autogobble, fontsize=\footnotesize]{haskell}
    -- module Category.CategoryGen

    -- Le constructeur du type des fonctions est une catégorie
    instance CategoryGen (->) where
      -- le morphisme identité est la fonction identité
      id :: a -> a
      id x = x

      -- la composition est la composition usuelle des fonctions
      (.) :: (a -> b) -> (c -> a) -> c -> b
      (.) f g x = f $ g x
  \end{minted}
\end{implementationBox}
\begin{remarqueBox}[]{Monoïde comme catégorie}
  On peut interpréter la notion de catégorie comme une extension de la notion de monoïde.
  En effet, une catégorie avec un unique objet est équivalente à un monoïde, les éléments du monoïde étant les morphismes de la catégorie, se composant associativement et ayant nécessairement un élément neutre.
\end{remarqueBox}
\begin{implementationBox}[label={codeCat1objMonoid}]{Catégorie à un objet}
  \begin{minted}[xleftmargin=1em, autogobble, fontsize=\footnotesize]{haskell}
    -- module Category.CategoryGen

    -- Type des catégories avec un unique objet.
    data MonoCat cat a = MonoCat {runCat :: cat a a}

    -- Une catégorie avec un seul objet est un semigroupe
    instance CategoryGen cat => Semigroup (MonoCat cat a) where
      (<>) :: MonoCat cat a -> MonoCat cat a -> MonoCat cat a
      MonoCat m1 <> MonoCat m2 = MonoCat $ m2 . m1

    -- Une catégorie avec un seul objet est un monoïde
    instance CategoryGen cat => Monoid (MonoCat cat a) where
      mempty :: MonoCat cat a
      mempty = MonoCat id
  \end{minted}
\end{implementationBox}
\noindent La structure d'automate peut-être alors réinterprétée, d'une façon équivalente à~\cite{CP17}, à l'aide de ces notions.
Pour cela, il suffit de décrire:
\begin{itemize}
  \item une configuration initiale;
  \item une fonction de transition associant à un symbole une modification de configuration;
  \item une transformation finale des configurations.
\end{itemize}
Une fois une catégorie choisie, il suffit alors de choisir un objet \(Q\) de celle-ci pour représenter les configurations, un objet \(V\) pour représenter les \emph{poids} et un morphisme de \(\mathrm{Hom}(Q, V)\) pour \emph{interpréter} une configuration.
Cependant, il faut remarquer que la catégorie choisie n'est pas nécessairement celle des ensembles.
Ainsi, \(Q\) n'est pas nécessairement un ensemble et la configuration initiale ne peut alors pas être défini comme un élément de \(Q\).
Il suffit alors de définir la configuration initiale comme un morphisme de \(\mathrm{Hom}(I,Q)\) avec \(I\) un objet de la catégorie ambiante.
Le cas d'une configuration initiale choisie dans un ensemble pourra être retrouvée par l'isomorphisme classique entre \(() \rightarrow Q\) et \(Q\).
\begin{definition}
  Un \firstocc{automate}{catégorie!automate sur une} sur une catégorie \(\mathcal{C}\) est défini par:
  \begin{itemize}
    \item un \firstocc{alphabet} \(\Sigma \);
    \item par trois objets: \(I\) l' \firstocc{objet initial}, \(Q\) l'\firstocc{objet état} et \(V\) l'\firstocc{objet final};
    \item un \firstocc{morphisme initial} de \(\mathrm{Hom}(I, Q)\);
    \item une \firstocc{fonction de transition} \( \delta \) de \(\Sigma \rightarrow \mathrm{Hom}(Q, Q)\);
    \item un \firstocc{morphisme final} de \(\mathrm{Hom}(Q, V)\).
  \end{itemize}
\end{definition}
\begin{implementationBox}[label={codeAut}]{Automate}
  \begin{minted}[xleftmargin=1em, autogobble, fontsize=\footnotesize]{haskell}
    -- module Automata.Automaton

    -- Le type de données des automates avec
    -- * cat une catégorie
    -- * init l'objet initial
    -- * symbols l'alphabet
    -- * state l'objet état
    -- * value l'objet final.
    data Automaton cat init symbols state value = Auto {
      -- le morphisme initial
      initial :: cat init state,
      -- la fonction de transition
      delta   :: symbols -> cat state state,
      -- le morphisme d'évaluation
      final   :: cat state value
    }
  \end{minted}
\end{implementationBox}
Remarquons alors que l'extension de la fonction de transition des symboles aux mots relève alors de la composition des morphismes de la catégorie ambiante.
Par exemple, le morphisme correspondant à un mot de deux symboles \(a\cdot b\) serait la composition \(\delta(b) \circ \delta(a)\).
Cette extension d'une fonction depuis symboles en une fonction depuis des mots repose sur une construction fondamentale de la théorie des catégories: l'\firstocc{adjonction fonctorielle}{foncteur!adjonction}.

Un \firstocc{foncteur}{foncteur} \(F\) entre deux catégories \(\mathcal{C}\) et \(\mathcal{C}'\) envoie
\begin{itemize}
  \item tout objet \(A\) de \(\mathcal{C}\) sur un objet \(F(A)\) de \(\mathcal{C}'\),
  \item tout morphisme \(f\) de \(\mathrm{Hom}_\mathcal{C}(A, B)\) vers un morphisme \(F(f)\) contenu dans \(\mathrm{Hom}_{\mathcal{C}'}(F(A), F(B))\)
\end{itemize}
en compatibilité avec
\begin{itemize}
  \item la composition: pour tous deux morphismes \(f\) et \(g\) de \(\mathcal{C}\),
    \begin{equation*}
      F(f \circ_{\mathcal{C}} g) = F(f) \circ_{\mathcal{C}'} F(g);
    \end{equation*}
  \item le morphisme identité:
  \begin{equation*}
    F(\mathrm{id}_{\mathcal{C}}) = \mathrm{id}_{\mathcal{C}'}.
  \end{equation*}
\end{itemize}


\begin{remarqueBox}[]{Foncteur identité}
  Le plus simple des foncteurs est le foncteur \(1\), \firstocc{foncteur identité} envoyant tout objet sur lui-même et tout morphisme sur lui-même.
\end{remarqueBox}

\begin{remarqueBox}[]{Composition de foncteurs}
  La composition de deux foncteurs \(F\) d'une catégorie \(\mathcal{C}'\) vers une catégorie \(\mathcal{C}''\) et \(G\) d'une catégorie \(\mathcal{C}\) vers \(\mathcal{C}'\), notée \(FG\), envoyant un objet \(A\) vers \(F(G(A))\) et un morphisme \(f\) vers \(F(G(f))\) est elle aussi un foncteur.
\end{remarqueBox}
\begin{implementationBox}[label={codeFoncteur}]{Foncteur}
  \begin{minted}[xleftmargin=1em, autogobble, fontsize=\footnotesize]{haskell}
    -- module Category.FunctorGen

    -- Soient f un constructeur unaire de types, et cat et cat'
    -- deux catégories. Ainsi, f est un foncteur de cat vers cat'
    -- s'il existe une fonction envoyant les morphismes de
    -- cat vers ceux de cat'.
    class (CategoryGen cat, CategoryGen cat') => FunctorGen f cat cat' where
      fmap :: cat a b -> cat' (f a) (f b)
  \end{minted}
\end{implementationBox}
\begin{remarqueBox}[]{Classe de types des foncteurs}
  En Haskell, la classe de types des foncteurs, \texttt{Functor}, ne permet que de manipuler la catégorie des types et est définie par la fonction
  \begin{center}
    \texttt{fmap:: (a -> b) -> f a -> f b}
  \end{center}
  La classe \texttt{FunctorGen} définie précédemment est plus flexible en permettant de définir des foncteurs pour d'autres catégories (par exemple des sous-catégories de la catégorie des types).
\end{remarqueBox}
Deux des foncteurs les plus simples à décrire sont les (endo) foncteurs identité et des listes sur la catégorie des types.
\begin{implementationBox}[label={codeFoncteurListe}]{Foncteurs liste et identité}
  \begin{minted}[xleftmargin=1em, autogobble, fontsize=\footnotesize]{haskell}
    -- module Category.OfStructures.CategoryOfTypes

    -- Endofoncteur des listes
    instance FunctorGen [] (->) (->) where
      fmap :: (a -> b) -> [a] -> [b]
      fmap _ []       = []
      fmap f (x : xs) = f x : fmap f xs

    -- Endofoncteur identité
    instance FunctorGen Identity (->) (->) where
      fmap :: (a -> b) -> Identity a -> Identity b
      fmap f (Identity a) = Identity $ f a
  \end{minted}
\end{implementationBox}
\begin{remarqueBox}[]{Types Algébriques de Données Généralisés}
  Les types algébriques de données définis précédemment peuvent être généralisés en \emph{Generalized Algebraic Data Types} (GADTs), permettant d'ajouter des informations supplémentaires sur les types pour renforcer le moteur d'inférence de typage.
  Par exemple, des contraintes peuvent être ajoutées dans la définition du type de données.

  Dans ce cas, la syntaxe utilisée sera la suivante: comme précédemment, le mot clé \texttt{data} suivi du nom \texttt{T} du type et de ses paramètres \texttt{a\textsubscript{1} \ldots a\textsubscript{n}}.
  Pour différencier la syntaxe avec les ADT classiques, le symbole \texttt{=} est remplacé par le mot clé \texttt{where}.
  Suivent ensuite les constructeurs de valeurs, sous la syntaxe suivante: nom du constructeur, puis son type, type d'une fonction dont les entrées sont les champs du constructeur de valeur, pouvant être nommés (pour définir simultanément des fonctions de projection), et dont la sortie est le type de données défini.

  Par exemple, la définition suivante
  \begin{minted}[xleftmargin=1em, autogobble, fontsize=\footnotesize]{haskell}
    data Expr a where
      IfThenElse :: Expr Bool -> Expr a -> Expr a -> Expr a
      Sum :: Expr Int -> Expr Int -> Expr Int
      Neg :: Expr Bool -> Expr Bool
      And :: Expr Bool -> Expr Bool -> Expr Bool
      Val :: a -> Expr a
  \end{minted}
  permet de définir facilement des expressions typées, dont les opérateurs ne sont compatibles qu'avec une sous-partie des valeurs inductivement définies, compatibilité testée à la compilation.
  Ainsi, les termes suivants sont bien formées:
  \begin{minted}[xleftmargin=1em, autogobble, fontsize=\footnotesize]{haskell}
    true = Val True
    one = Val 1
    two = Sum one one
    false = Neg true
    one' = IfThenElse true one two
  \end{minted}
  mais pas celui-ci:
  \begin{minted}[xleftmargin=1em, autogobble, fontsize=\footnotesize]{haskell}
    bad = IfThenElse one one one
  \end{minted}
\end{remarqueBox}
Les deux constructeurs de types \texttt{[]} et \texttt{Identity} permettent également définir des foncteurs entre la catégorie des types et celle des monoïdes.
\begin{implementationBox}[]{Foncteurs liste et identité}
  \begin{minted}[xleftmargin=1em, autogobble, fontsize=\footnotesize]{haskell}
    -- module Category.OfStructures.CategoryOfMonoids

    -- Définition du type des morphismes de monoïdes, dont les valeurs
    -- sont définies par constructeur de valeurs MonoidMorph
    -- contenant une fonction entre deux monoïdes a et b,
    -- et définition de la fonction de projection runMorph
    data MonoidMorph a b where
      MonoidMorph ::
        (Monoid a, Monoid b) => {runMorph :: a -> b} -> MonoidMorph a b

    -- Définition de la catégorie des monoïdes
    instance CategoryGen MonoidMorph where
      id = MonoidMorph id
      f . g = MonoidMorph $ runMorph f . runMorph g

    -- Le constructeur [] est un foncteur
    -- de la catégorie des types vers celle
    -- des monoïdes
    instance FunctorGen [] (->) MonoidMorph where
      fmap = MonoidMorph . fmap

    -- Tout endofoncteur de la catégorie des types est un foncteur de
    -- la catégorie des monoïdes vers la catégorie des types.
    -- Est inclus le constructeur Identity
    instance FunctorGen f (->) (->) => FunctorGen f MonoidMorph (->) where
      fmap = fmap . runMorph
  \end{minted}
\end{implementationBox}


%
Il existe un lien particulier entre le foncteur des listes et le foncteur identité.
Toute fonction envoyant un type sur un monoïde peut être étendue en un morphisme de monoïde entre le monoïde libre (dont l'ensemble sous-jacent est celui des listes) et ce monoïde; et cette extension peut être inversée.
\begin{implementationBox}[label={}]{Relation liste et identité}
  \begin{minted}[xleftmargin=1em, autogobble, fontsize=\footnotesize]{haskell}
    -- Promotion d'une fonction en morphisme de monoïde
    promoteFunToMorph :: Monoid b => (a -> b) -> MonoidMorph [a] b
    promoteFunToMorph f = MonoidMorph $ foldMap f
      where
      -- Fonction utilitaire, déjà définie en Haskell
      -- [voir Data.Foldable]
      foldMap :: Monoid b => (a -> b) -> [a] -> b
      foldMap _ []       = mempty
      foldMap f (x : xs) = f x <> foldMap f xs

    -- Destitution d'un morphisme en une fonction
    lowerMorphToFun :: MonoidMorph [a] b -> a -> b
    lowerMorphToFun (MonoidMorph f) = f . return
      where
      -- Fonction utilitaire, déjà définie en Haskell
      -- dans le module de base Prelude
      return :: a -> [a]
      return a = [a]
  \end{minted}
\end{implementationBox}
C'est cette relation particulière entre foncteurs qui est appelée adjonction.
Plus formellement, soient \(F\) un foncteur entre deux catégories \(\mathcal{C}\) et \(\mathcal{C}'\) et \(G\) un foncteur entre \(\mathcal{C}'\) et \(\mathcal{C}\).
Ces deux foncteurs sont dits \firstocc{adjoints}{foncteur!adjonction!par isomorphisme} si pour tous deux objets \(A\) de \(\mathcal{C}\) et \(B\) de \(\mathcal{C}'\) les ensembles \(\mathrm{Hom}_\mathcal{C}(A, G(B))\) et \(\mathrm{Hom}_{\mathcal{C}'}(F(A), B)\) sont isomorphes.
L'isomorphisme d'ensemble n'est pas la seule façon de décrire une adjonction entre foncteurs.
On peut également étudier le lien entre les compositions de ces foncteurs (c'est-à-dire les (endo) foncteurs \(FG\) et \(GF\)) et les foncteurs identités.
Pour cela, il suffit d'étudier les morphismes entre ces foncteurs, appelés \firstocc{transformations naturelles}.

Soient \(F\) et \(G\) deux foncteurs entre deux catégories \(\mathcal{C}\) et \(\mathcal{C}'\).
Une \firstocc{transformation naturelle}{transformation naturelle} de \(F\) en \(G\) est une famille \( \alpha \) de morphisme de \(\mathcal{C}'\) indexée par les objets de \(\mathcal{C}\) faisant commuter le diagramme suivant pour tout morphisme \(h\) de \(\mathcal{C}\):

\begin{center}
   \begin{tikzpicture}
     \matrix (m) [matrix of math nodes,row sep=3em,column sep=4em,minimum width=2em]{
       F(A) & G(A) \\
       F(B) & G(B)\rlap{.} \\
     };
     \path[-stealth]
       (m-1-1)
         edge node [above] {\( \alpha_A \)} (m-1-2)
         edge node [left] {\( F(h) \)} (m-2-1)
       (m-1-2)
         edge node [right] {\( G(h) \)} (m-2-2)
       (m-2-1)
         edge node [below] {\( \alpha_B \)} (m-2-2)
         ;
   \end{tikzpicture}
\end{center}

\begin{remarqueBox}[]{Transformation naturelle}
  En Pseudo-Haskell, une transformation naturelle de \texttt{f} en \texttt{g} est une fonction polymorphe \texttt{t} de type \texttt{(f a -> g a)} satisfaisant l'équation
  \begin{equation*}
    \texttt{t . fmap h = fmap h . t}
  \end{equation*}
  pour tout morphisme \texttt{h}.
\end{remarqueBox}
Une adjonction peut alors être définie à l'aide des transformations naturelles.
Soient \(F\) un foncteur entre deux catégories \(\mathcal{C}\) et \(\mathcal{C}'\) et \(G\) un foncteur entre \(\mathcal{C}'\) et \(\mathcal{C}\).
Ces deux foncteurs sont dits \firstocc{adjoints}{foncteur!adjonction!par transformation naturelle} s'il existe deux transformations naturelles
\begin{itemize}
  \item \( \epsilon \) entre \(FG\) et \(1\)
  \item \(\eta \) entre \(1\) et \(GF\)
\end{itemize}
satisfaisant pour tous deux objets \(X\) de \(\mathcal{C}'\) et \(Y\) de \(\mathcal{C}\)
\begin{align*}
  \mathrm{id}_{F(Y)} &= \epsilon_{F(Y)} \circ F(\eta_Y), &
  \mathrm{id}_{G(X)} &= G(\epsilon_X) \circ \eta_{G(X)}.
\end{align*}

\begin{implementationBox}[label={codeAdjonction}]{Adjonction}
  \begin{minted}[xleftmargin=1em,  mathescape=true, autogobble, fontsize=\footnotesize]{haskell}
    -- module Category.AdjunctionGen

    -- Adjonction entre les foncteurs f et g, respectivement
    -- foncteurs de cat vers cat' et de cat' vers cat
    class (FunctorGen f cat cat', FunctorGen g cat' cat)
      => AdjunctionGen f g cat cat' where

      -- Transformation $\eta$ de l'adjonction
      unit ::  cat a (g (f a))

      -- Transformation $\epsilon$ de l'adjonction
      counit ::  cat' (f (g a)) a

      -- Isomorphisme de l'adjonction
      promote :: cat a (g b) -> cat' (f a) b
      promote f = counit . fmap f

      -- Inverse de la fonction promote
      lower :: cat' (f a) b -> cat a (g b)
      lower h = fmap h . unit
  \end{minted}
\end{implementationBox}
\begin{remarqueBox}[]{Classe de types et code par défaut}
  Dans le Pseudo-Code~\ref{codeAdjonction}, les promotions et destitutions de morphismes se déduisent directement de l'unité et de la co-unité de l'adjonction.
  Ainsi, nous ne donnerons dans la suite, pour la définition des adjonctions que nous allons utiliser, que la définition des morphismes unité et co-unité de l'adjonction.
\end{remarqueBox}
\begin{implementationBox}[label={codeAdjonctionListIdentity}]{Adjonction Liste Identité}
  \begin{minted}[xleftmargin=1em, autogobble, fontsize=\footnotesize]{haskell}
    -- module Category.OfStructures.CategoryOfMonoids

    -- Définition de l'adjonction Liste - Identité
    instance AdjunctionGen [] Identity (->) MonoidMorph where
      unit :: a -> Identity [a]
      unit = Identity . return

      counit ::  MonoidMorph [Identity a] a
      counit = MonoidMorph $ foldMap runIdentity

    -- Fonction auxiliaire pour éviter les appels aux constructeurs de
    -- types et aux fonctions de projection
    promoteFun
      :: CategoryGen cat => (b -> cat a a) -> [b] -> cat a a
    promoteFun f = runCat . (runMorph $ promote $ Identity . MonoCat . f)
  \end{minted}
\end{implementationBox}

Cette dernière fonction, \texttt{promoteFun}, découlant directement d'une adjonction, est la fonction idéale pour promouvoir la fonction de transition \(\delta \) d'un automate considérant des symboles en entrée en une fonction considérant des mots.

\begin{implementationBox}[label={codeDeltaExtension}]{Promotion de \(\delta \)}
  \begin{minted}[xleftmargin=1em, autogobble, fontsize=\footnotesize]{haskell}
    -- module Automata.Automaton

    -- Fonction de transition depuis les mots associée
    -- à un automate
    delta'
      :: CategoryGen cat
      => Automaton cat init symbols state value
      -> [symbols]
      -> cat state state
    delta' = promoteFun . delta
  \end{minted}
\end{implementationBox}
Avec le formalisme choisi dans cette section, la configuration initiale de l'automate est un morphisme.
La fonction \texttt{delta'} permet alors de calculer la configuration obtenue après la lecture d'un mot, par sa simple composition avec le morphisme initial.
\begin{implementationBox}[label={codeConfigAtteinte}]{Configuration associée à un mot}
  \begin{minted}[xleftmargin=1em, autogobble, fontsize=\footnotesize]{haskell}
    -- module Automata.Automaton

    -- Calcul de la configuration associée à un mot pour un automate donné
    getConfig
      :: CategoryGen cat
      => Automaton cat init symbols state value
      -> [symbols]
      -> cat init state
    getConfig auto w = delta' auto w . initial auto
  \end{minted}
\end{implementationBox}


Le poids d'un mot est alors la composition de la configuration atteinte et du morphisme final.
\begin{implementationBox}[label={CodePoidsMot}]{Poids d'un mot}
  \begin{minted}[xleftmargin=1em, autogobble, fontsize=\footnotesize]{haskell}
    -- module Automata.Automaton

    -- Poids d'un mot
    weight
      :: CategoryGen cat
      => Automaton cat init symbols state value
      -> [symbols]
      -> cat init value
    weight auto w = final auto . getConfig auto w
  \end{minted}
\end{implementationBox}
Pour améliorer la lisibilité, définissons une classe de conversion de types.
\begin{implementationBox}[]{Équivalence de types}
  \begin{minted}[xleftmargin=1em, autogobble, fontsize=\footnotesize]{haskell}
    -- module Tools.TypeEq

    -- Classes de conversion de types
    class Castable a b where
      -- Première conversion
      cast :: a -> b
      -- Seconde conversion
      castInv :: b -> a

    -- module Automata.Automaton

    -- Poids d'un mot comme élément d'un type équivalent au
    -- type cat init value
    weightValue
      :: (CategoryGen cat, Castable (cat init value) value')
      => Automaton cat init symbols state value
      -> [symbols]
      -> value'
    weightValue auto = cast . weight auto

    -- Synonyme dans le cas d'un poids booléen.
    recognizes
      :: (Castable (cat init value) Bool, CategoryGen cat)
      => Automaton cat init symbols state value
      -> [symbols]
      -> Bool
    recognizes = weightValue
  \end{minted}
\end{implementationBox}

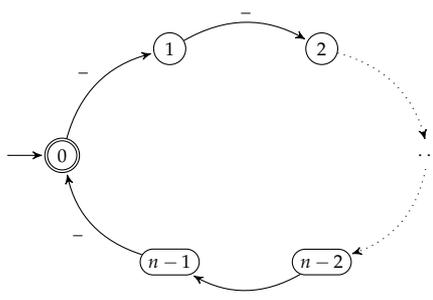
\begin{figure}[H]
  \centerline{
    \begin{tikzpicture}[node distance=2cm, bend angle=30]
    \node[initial left, state, accepting] (S0) {\(0\)};
    \node[state, above right of=S0] (S1) {\(1\)};
    \node[state, right of=S1] (S2) {\(2\)};
    \node[below right of=S2] (Sdots) {\(\cdots\)};
    \node[state, rounded rectangle, below left of=Sdots] (Sn-2) {\(n - 2\)};
    \node[state, rounded rectangle, left of=Sn-2] (Sn-1) {\(n - 1\)};
    \path[->]
      (S0)   edge [above left, bend left] node {\(\_\)} (S1)
      (S1)   edge [above, bend left] node {\(\_\)} (S2)
      (S2)   edge [above right, bend left, dotted] node {} (Sdots)
      (Sdots)   edge [below right, bend left, dotted] node {} (Sn-2)
      (Sn-2)   edge [below, bend left] node {\(\_\)} (Sn-1)
      (Sn-1)   edge [below left, bend left] node {\(\_\)} (S0)
      ;
    \end{tikzpicture}
  }
  \caption{Un reconnaisseur de mots de longueurs modulaires.}%
  \label{fig ex aut long mod}
\end{figure}

Le type des automates les plus simples à implanter est alors le type des automates déterministes complets.
Ci-dessous sa définition, ainsi que des fonctions utilitaires de construction, mais également les algorithmes classiques de complémentation et de combinaison booléenne binaire.
\begin{implementationBox}[]{Automates déterministes complets}
  \begin{minted}[xleftmargin=1em, autogobble, fontsize=\footnotesize]{haskell}
    -- module Automata.DFAComp

    -- Synonyme de type pour la lisibilité
    type DFAComp symbol state = Automaton (->) () symbol state Bool

    -- Fonction de construction usuelle d'un automate déterministe complet
    -- depuis :
    -- * un état initial
    -- * une fonction associant à un symbole et à un état un unique état
    -- * une fonction associant à un état sa finalité
    pack
      :: state
      -> (symbol -> state -> state)
      -> (state -> Bool)
      -> DFAComp symbol state
    pack = Auto . const
      where
      -- fonction usuelle déjà définie en Haskell
      -- dans le module de base Prelude
      const x _ = x

    -- Équivalence classique entre les types (() -> a) et a,
    -- permettant l'utilisation de la fonction recognizes
    -- avec les automates déterministes complets,
    -- en castant les morphismes de type () -> Bool
    -- en éléments de type Bool
    instance Castable (() -> a) a where
      cast f = f ()
      castInv a () = a

    -- Complémenter un DFA complet, c'est inverser la finalité des états
    -- de celui-ci.
    complement :: DFAComp symbol state -> DFAComp symbol state
    complement (Auto i d f) = Auto i d (not . f)
      where
      -- fonction usuelle déjà définie en Haskell
      -- dans le module de base Prelude
      not True = False
      not False = True

    -- Combinaison binaire de DFAs complets en fonction d'une
    -- fonction booléenne donnée.
    booleanBinaryCombination
      :: (Bool -> Bool -> Bool)
      -> DFAComp symbols state
      -> DFAComp symbols state'
      -> DFAComp symbols (state, state')
    booleanBinaryCombination g (Auto i1 d1 f1) (Auto i2 d2 f2) = Auto i d f
     where
      -- la configuration initiale est le couple des configurations initiales
      i () = (i1 (), i2 ())
      -- la fonction de transition est obtenue par un parcours
      -- des automates en parallèle
      d a (p, q) = (d1 a p, d2 a q)
      -- la finalité est obtenue en appliquant la fonction
      -- booléenne donnée en paramètre
      f (p, q) = g (f1 p) (f2 q)

    -- Union de deux DFAs complets,
    -- en utilisant la disjonction booléenne (||).
    (<+>)
      :: DFAComp symbols state
      -> DFAComp symbols state'
      -> DFAComp symbols (state, state')
    (<+>) = booleanBinaryCombination (||)

    -- Intersection de deux DFAs complets,
    -- en utilisant la conjonction booléenne (&&).
    (<&>)
      :: DFAComp symbols state
      -> DFAComp symbols state'
      -> DFAComp symbols (state, state')
    (<&>) = booleanBinaryCombination (&&)
  \end{minted}
\end{implementationBox}
Afin de concrétiser l'utilisation de ces fonctions, considérons la construction d'automate suivante basée sur les automates de la famille définie Figure~\ref{fig ex aut long mod}, où les transitions étiquetées par le symbole \( \_ \) représentent des transitions définies quel que soit le symbole de l'alphabet considéré:
un automate déterministe de cette famille reconnaît les mots qui ont une longueur congrue à un entier fixé.
Le Pseudo-Code~\ref{pseudoCodeDetComp}\footnote{Le module exemple peut être chargé dans l'interpréteur \texttt{ghci} par la commande \texttt{stack ghci src/HDRExample/DFAComp.hs}.} illustre comment combiner (d'une façon Booléenne) différents automates de cette famille.


\begin{implementationBox}[label={pseudoCodeDetComp}]{Automates déterministes complets}
  \begin{minted}[xleftmargin=1em, autogobble, fontsize=\footnotesize]{haskell}
    -- module HDRExample.DFAComp

    -- Fonction de transition définissant les transitions de la forme
    -- (p, _, (p + 1) % n)
    deltaMod :: Integral state => state -> symbol -> state -> state
    deltaMod n _ p = (p + 1) `mod` n

    -- a1 reconnait les mots de longueurs paires
    -- a2 reconnait les mots de longueurs multiples de 4
    -- a3 reconnait les mots de longueurs multiples de 8
    a1, a2, a3 :: DFAComp symbol Int
    a1 = pack 0 (deltaMod 2) (== 0)
    a2 = pack 0 (deltaMod 4) (== 0)
    a3 = pack 0 (deltaMod 8) (== 0)

    -- a4 reconnait les mots de longueurs paires non multiples
    -- de 4 ou multiples de 8
    a4 :: DFAComp symbols ((Int, Int), Int)
    a4 = a1 <&> complement a2 <+> a3

    -- Fonction d'affichage de test pour un mot donné
    -- Par exemple,
    -- test "aa"
    -- > a4 recognizes "aa": True
    -- test "aaaa"
    -- > a4 recognizes "aaaa": False
    -- test "aaaaaaaa"
    -- > a4 recognizes "aaaaaaaa": True
    test :: Show symbols => [symbols] -> IO ()
    test w =
      putStrLn $
        mconcat ["a4 recognizes ", show w, ": ", show (a4 `recognizes` w)]
  \end{minted}
\end{implementationBox}
On peut ainsi représenter les automates complets déterministes en utilisant la catégorie des types, dont les objets sont les types et les morphismes les fonctions.
Dans la section suivante, on va s'intéresser aux autre types d'automates classiques, que l'on va pouvoir représenter \emph{via} d'autres catégories connues.

\section{Autres types d'automates de mots}\label{sec:automatesMotsKleisli}
En choisissant d'autres catégories, il est possible de représenter une très grande variété d'automates très différents les uns des autres: déterministes (non nécessairement complets), non déterministes, à multiplicités, alternants, \emph{etc}.
Ces différents types d'automates ont cependant des similarités dans leurs fonctionnements: tout comme les automates déterministes complets, le rôle d'un symbole est de transformer une configuration en une autre: un ensemble d'états en un ensemble d'états pour un automate non déterministe, une combinaison linéaire d'états en une combinaison linéaire d'états pour un automate à multiplicités, \emph{etc.}; cependant, cette transformation s'exprime par l'action d'un symbole sur un état.
Dans de tels automates, la fonction de transition est alors exprimée comme une fonction du type \texttt{symbol -> state -> m state}, où \texttt{m} est un constructeur de types unaire.
Une succession d'action par les symboles (cas de l'action par les mots) nécessite alors le fait que les fonctions du type \texttt{state -> m state} puissent alors se composer.
Cette composition est possible dans le cas de foncteurs particuliers: les \firstocc{monades}.

\subsection{Monades et catégories de Kleisli}

Formellement, une \firstocc{monade}{monade} sur une catégorie \(\mathcal{C}\) est un endofoncteur \(T\) de \(\mathcal{C}\) tel qu'il existe deux transformations naturelles
  \begin{itemize}
    \item \(\eta \) (\texttt{return}\footnote{déjà mentionnée dans le cas des listes} en Haskell) entre \(1\) et \(T\),
    \item \(\mu \) (\texttt{join} en Haskell) entre \(TT\) et \(T\),
  \end{itemize}
  faisant commuter les deux diagrammes suivants:
  \begin{center}
    \begin{tikzpicture}
      \matrix (m) [matrix of math nodes,row sep=3em,column sep=4em,minimum width=3em]{
        TTT(X) & TT(X)\\
        TT(X) & T(X)\rlap{,} \\
      };
      \path[-stealth]
        (m-1-1)
          edge node [above] {\( T(\mu_X) \)} (m-1-2)
          edge node [left] {\( \mu_{T(X)} \)} (m-2-1)
        (m-1-2)
          edge node [right] {\( \mu_X \)} (m-2-2)
        (m-2-1)
          edge node [below] {\( \mu_X \)} (m-2-2)
          ;
    \end{tikzpicture}
    \begin{tikzpicture}
      \matrix (m) [matrix of math nodes,row sep=3em,column sep=4em,minimum width=3em]{
        T(X) & TT(X)\\
        TT(X) & T(X)\rlap{.}\\
      };
      \path[-stealth]
        (m-1-1)
          edge node [above] {\( \eta_{T(X)} \)} (m-1-2)
          edge node [left] {\( T(\eta_X) \)} (m-2-1)
          edge node [above right] {\( \mathrm{Id}_{T(X)} \)} (m-2-2)
        (m-1-2)
          edge node [right] {\( \mu_X \)} (m-2-2)
        (m-2-1)
          edge node [below] {\( \mu_X \)} (m-2-2)
          ;
    \end{tikzpicture}
  \end{center}

\begin{implementationBox}[]{Monade}
  \begin{minted}[xleftmargin=1em, autogobble, fontsize=\footnotesize]{haskell}
    -- module Category.MonadGen

    -- Soit m un endofoncteur sur une catégorie cat.
    -- m est une monade s'il existe deux fonctions permettant:
    -- * de promouvoir une valeur en son type monadique (return)
    -- * de supprimer un niveau monadique dans les types de valeur (join)
    class FunctorGen m cat cat => MonadGen m cat where
      -- Promotion monadique.
      return :: cat a (m a)

      -- Suppression d'un niveau monadique.
      join :: cat (m (m a)) (m a)
  \end{minted}
\end{implementationBox}
\begin{remarqueBox}[label={DefClassiqueMonadeHaskell}]{Monades en Haskell}
  La classe de types \texttt{Monad} existe dans les modules de base Haskell.
  Classiquement, cette classe est définie par la fonction \texttt{return} comme ci-avant, et par la fonction \texttt{bind} (noté également par l'opérateur \texttt{(>>=)}) de signature \texttt{(m a -> (a -> m b) -> m b)}, sémantiquement définie comme
    \begin{minted}[xleftmargin=1em, autogobble, fontsize=\footnotesize]{haskell}
      m_a >>= f = join $ fmap f m_a
    \end{minted}
  L'opérateur \texttt{(=<<)}, de signature \texttt{((a -> m b) -> m a -> m b)}, lui est équivalent, ne changeant que par la position des paramètres.
  Dans l'implantation fournie, il est défini pour toute monade de la façon suivante:
  \begin{minted}[xleftmargin=1em, autogobble, fontsize=\footnotesize]{haskell}
    (=<<) :: cat a (m b) -> cat (m a) (m b)
    (=<<) f =  join . fmap f
  \end{minted}
\end{remarqueBox}
Comme pour les foncteurs, deux des monades les plus simples à définir sont la monade des Listes et la monade Identité.
\begin{implementationBox}[]{Exemples de monades: Listes et Identité}
  \begin{minted}[xleftmargin=1em, autogobble, fontsize=\footnotesize]{haskell}
    -- module Category.OfStructures.CategoryOfTypes

    -- L'endofoncteur des listes est une monade.
    instance MonadGen [] (->) where
      return x = [x]

      join [] = []
      join (xs : xss) = xs <> join xss

    -- L'endofoncteur Identité est une monade.
    instance MonadGen Identity (->) where
      return = Identity
      join (Identity x) = x
  \end{minted}
\end{implementationBox}

\begin{remarqueBox}[]{La monade IO}
  Une monade très utile en Haskell est la monade \texttt{IO}, la monade des effets de bords.
  Elle permet en effet de \emph{marquer} sans ambiguïté les valeurs associées à des expressions \emph{impures}, pouvant mettre en péril la transparence référentielle.
  Cette information permet alors au compilateur d'effectuer des optimisations en déterminant d'une façon plus simple les expressions dont l'interprétation peut déclencher des actions.

  Ainsi, une expression permettant de lire une chaîne de caractères depuis l'entrée standard serait de type \texttt{IO String}, une valeur entière générée aléatoirement de type \texttt{IO Int}, et un programme de type \texttt{IO ()}.
\end{remarqueBox}

\begin{remarqueBox}[label={RemMonadeCompo}]{Composition de monades}
  Contrairement aux foncteurs, la composition de deux monades (en tant que foncteur) n'est pas nécessairement une monade (voir~\cite{KS18} pour le cas surprenant des ensembles d'ensembles).
\end{remarqueBox}

D'autres foncteurs remarquables sont également des monades; c'est le cas pour le constructeur \texttt{Maybe} (constructeur de type présent dans les modules de base d'Haskell, correspondant à l'ajout d'un élément au type) ou pour les constructeurs d'ensembles finis.

\begin{remarqueBox}[]{Les ensembles en Haskell}
  Le langage Haskell possède de nombreuses implantations des ensembles mathématiques, finis ou non.
  Dans la suite, nous considérerons le constructeur de types \texttt{HashSet}, permettant de construire des ensembles finis efficaces en termes de recherche en utilisant des fonctions de hachage.
  Ce constructeur est disponible dans le module \href{https://hackage.haskell.org/package/unordered-containers-0.2.10.0/docs/Data-HashSet.html}{\texttt{Data.HashSet}} du package \href{http://hackage.haskell.org/package/unordered-containers}{\texttt{unordered-containers}}.
  Ce module, appelé \texttt{Set} par la suite, contient de nombreuses fonctions utilitaires, telles que \texttt{map} (l'application d'une fonction sur les éléments d'un ensemble, définition d'un foncteur), \texttt{singleton} (renvoyant un ensemble de cardinalité 1 à partir d'une valeur) ou \texttt{fold} (méthode de la classe \href{http://hackage.haskell.org/package/base-4.12.0.0/docs/Data-Foldable.html}{\texttt{Foldable}} combinant les valeurs d'un ensemble si elles sont d'un type monoïdal).
\end{remarqueBox}

\begin{implementationBox}[]{Exemples de monades}
  \begin{minted}[xleftmargin=1em, autogobble, fontsize=\footnotesize]{haskell}
    -- module Category.OfStructures.CategoryOfTypes

    -- Type de données ajoutant une valeur particulière à un type,
    -- défini dans les modules de base d'Haskell.
    data Maybe a = Nothing | Just a

    -- Le constructeur Maybe est un foncteur.
    instance FunctorGen Maybe (->) (->) where
      fmap _ Nothing  = Nothing
      fmap f (Just x) = Just $ f x

    -- Le foncteur Maybe est une monade.
    instance MonadGen Maybe (->) where
      return = Just

      join Nothing        = Nothing
      join (Just x)       = x

    -- Le constructeur HashSet est un foncteur.
    instance FunctorGen HashSet (->) (->) where
      fmap = Set.map

    -- Le foncteur HashSet est une monade.
    instance MonadGen HashSet (->) where
      return = Set.singleton
      join   = fold
  \end{minted}
\end{implementationBox}
Un des grands intérêts des monades (en programmation fonctionnelle notamment) est qu'elles permettent de composer très facilement les morphismes du type \texttt{a -> m b}.
En effet, cette combinaison est décrite par une catégorie associée à toute monade, la \firstocc{catégorie de Kleisli}.

Formellement, la \firstocc{catégorie de Kleisli}{catégorie!de Kleisli} associée à une monade \((T, \mu, \eta)\) sur une catégorie \(\mathcal{C}\) est la catégorie \(\mathcal{C}_T\) dont:
\begin{itemize}
  \item les objets sont les objets de \(\mathcal{C}\),
  \item les morphismes entre deux objets \(A\) et \(B\) sont les morphismes de \(\mathcal{C}\) entre \(A\) et \(T(B)\),
  \item la composition de deux morphismes \(f\) de \(\mathrm{Hom}(B, T(C))\) et \(g\) de \(\mathrm{Hom}(A, T(B))\) est définie par
  \begin{equation*}
    f \circ_{\mathcal{C}_T} g = \mu \circ_{\mathcal{C}} T(f) \circ_{\mathcal{C}} g;
  \end{equation*}
  \item l'identité est définie par la transformation naturelle \(\eta \):
  \begin{equation*}
    id_A = \eta_A.
  \end{equation*}
\end{itemize}

\begin{implementationBox}[label={CodeCatKleisli}]{Catégorie de Kleisli}
  \begin{minted}[xleftmargin=1em, autogobble, fontsize=\footnotesize]{haskell}
    -- module Category.Kleisli

    -- Type des morphismes de Kleisli, définissant:
    -- * le constructeur de types Kleisli;
    -- * la fonction de projection runKleisli.
    data KleisliCat m cat a b = Kleisli {runKleisli :: cat a (m b)}

    -- Le constructeur de types KleisliCat m cat, constructeur binaire,
    -- est une catégorie.
    instance MonadGen m cat => CategoryGen (KleisliCat m cat) where
      id = Kleisli return
      (Kleisli f) . (Kleisli g) = Kleisli ((=<<) f . g)
  \end{minted}
\end{implementationBox}

\begin{remarqueBox}[]{Composition de Kleisli}
  La fonction utilisée pour composer les morphismes de la catégorie de Kleisli, la fonction associant à deux morphismes \texttt{f} et \texttt{g} l'expression \texttt{((=<<) f.g)}, est habituellement notée \texttt{(<=<)} et appelée \emph{fish operator} ou composition de Kleisli.
  On pourra la retrouver dans la définition de la classe \texttt{MonadGen} du module \texttt{Category.MonadGen}.
\end{remarqueBox}


\subsection{Automates et catégories de Kleisli}

Les catégories de Kleisli permettent de définir les automates classiques comme des cas particuliers du type des automates.
Considérons par exemple un automate du type \texttt{Automaton (KleisliCat Maybe (->)) () symbol state ()}.
Un tel automate est défini (voir Code~\ref{codeAut}) par trois éléments:
\begin{itemize}
  \item le morphisme \texttt{initial} de la forme \texttt{Kleisli i} où \texttt{i} est du type \texttt{() -> Maybe state}, isomorphe au type \texttt{Maybe state}; la configuration initiale d'un tel automate est alors un élément de type \texttt{Maybe state}, c'est-à-dire soit un état \texttt{p} de type \texttt{state} (\texttt{Just p}), soit rien (\texttt{Nothing});
  \item la fonction \texttt{delta} associant à chaque symbole un morphisme \texttt{Kleisli d\_a} où \texttt{d\_a} est du type \texttt{state -> Maybe state}, c'est-à-dire une fonction partielle entre \texttt{state} et \texttt{state}; ainsi \texttt{delta} associe à un symbole et à un état soit un état \texttt{p} de type \texttt{state} (\texttt{Just p}), soit rien (\texttt{Nothing});
  \item le morphisme \texttt{final} de la forme \texttt{Kleisli f} où \texttt{f} est une fonction de type \texttt{state -> Maybe ()}, c'est-à-dire associant à chaque état une valeur d'un type, \texttt{Maybe ()}, à deux valeurs, et donc isomorphe aux Booléens: soit \texttt{Nothing} (\texttt{False}), soit \texttt{Just ()} (\texttt{True}).
\end{itemize}
Un tel automate est alors un automate déterministe non complet.
D'une façon similaire, la catégorie de Kleisli de la monade des ensembles permet de représenter les automates non déterministes.
\begin{implementationBox}[]{Automates classiques comme automates de Kleisli}
  \begin{minted}[xleftmargin=1em, autogobble, fontsize=\footnotesize]{haskell}
    -- module Automata.KleisliAutomata

    -- Un DFA non complet est un automate sur la catégorie de Kleisli
    -- associée à la monade Maybe.
    type DFA symbol state =
      Automaton (KleisliCat Maybe (->)) () symbol state ()

    -- Un NFA est un automate sur la catégorie de Kleisli
    -- associée à la monade des ensembles.
    type NFA symbols state =
      Automaton (KleisliCat HashSet (->)) () symbols state ()
  \end{minted}
\end{implementationBox}
Par ce formalisme, certaines fonctions peuvent être factorisées entre tous les automates sur des catégories de Kleisli, quelle que soit la monade associée.
C'est le cas des deux fonctions suivantes.
\textbf{NB:} La fonction \texttt{convert} utilisée dans la fonction \texttt{applyFunctor} est une méthode de la classe \texttt{KleisliFunAux} du module \texttt{Category.Kleisli} permettant de définir des foncteurs entre catégories de Kleisli issues de monades pouvant être distinctes; les instances de cette classe utilisées par la suite sont disponibles dans le module \texttt{Category.OfStructures.CategoryOfTypes}.
\begin{implementationBox}[]{Automates de Kleisli}
  \begin{minted}[xleftmargin=1em, autogobble, fontsize=\footnotesize]{haskell}
    -- module Automata.KleisliAutomata

    -- Fonction de construction d'un automate de Kleisli
    -- quelle que soit la monade associée pour des foncteurs
    -- de la catégorie des types.
    packKleisli
      :: (Castable (m valeur) valeur')
      => m state
      -> (symbols -> state -> m state)
      -> (state -> valeur')
      -> Automaton (KleisliCat m (->)) init symbols state valeur
    packKleisli i d f =
      Auto (Kleisli (const i)) (Kleisli . d) (Kleisli (castInv . f))

    -- Application d'un foncteur f entre les catégories de Kleisli associées
    -- à deux monades m et h; permet de convertir les configurations
    --  de la forme (m state) en (h (f state)) tout en conservant l'état
    -- initial inchangé.
    applyFunctor
      :: (KleisliFunAux cat f m h
         , FunctorGen f (KleisliCat m cat) (KleisliCat h cat)
         , MonadGen m cat
         , MonadGen h cat
         )
      => Automaton (KleisliCat m cat) init symbols state valeur
      -> Automaton (KleisliCat h cat) init symbols (f state) (f valeur)
    applyFunctor (Auto (Kleisli i) d f) =
      Auto (Kleisli $ convert . i) (fmap . d) (fmap f)
  \end{minted}
\end{implementationBox}
L'introduction des fonctions techniques (dont la définition est rappelée ici de façon incomplète) \texttt{applyFunctor} et \texttt{convert} se justifie par le fait que leurs définitions renferment, d'une façon surprenante, des algorithmes classiques de manipulation d'automates.
Plus précisément, leur utilisation permet de rapprocher syntaxiquement des algorithmes classiques de conversion d'automates.
\begin{implementationBox}[label={codeAlgosClas}]{Algorithmes classiques et foncteurs}
  \begin{minted}[xleftmargin=1em, autogobble, fontsize=\footnotesize]{haskell}
    -- module Automata.KleisliAutomata

    -- Conversion d'un NFA en un DFA complet,
    -- passant d'une façon intermédiaire par un automate
    -- dont les états sont de types (Identity (HashSet a)).
    determinise
      :: (Eq state, Hashable state)
      => NFA symbol state
      -> DFAComp symbol (HashSet state)
    determinise = removeIdDFA . castFinality . applyFunctor

    -- Conversion d'un DFA en un DFA complet,
    -- passant d'une façon intermédiaire par un automate
    -- dont les états sont de types (Identity (Maybe a)).
    complete :: DFA symbol state -> DFAComp symbol (Maybe state)
    complete = removeIdDFA . castFinality . applyFunctor

    -- Conversion d'un NFA en un DFA
    -- dont les états sont de types (Maybe (HashSet a)).
    nfaToDFA
      :: (Hashable state, Eq state)
      => NFA symbol state
      -> DFA symbol (HashSet state)
    nfaToDFA = convertFinalityWith (Kleisli castInv) . applyFunctor
  \end{minted}
\end{implementationBox}
Dans les trois cas précédents, l'algorithme appliqué est le même: il s'agit d'une transformation fonctorielle de configuration de la forme \texttt{m state} en \texttt{h (f state)}:
\begin{itemize}
  \item pour la déterminisation (complète), transformation de morphismes de la forme \texttt{state -> HashSet state} en \texttt{state -> Identity (HashSet state)};
  \item pour la complétion, transformation de morphismes de la forme \texttt{state -> Maybe state} en \texttt{state -> Identity (Maybe state)};
  \item pour la déterminisation, transformation de morphismes de la forme \texttt{state -> HashSet state} en \texttt{state -> Maybe (HashSet state)}.
\end{itemize}
Dans les deux premiers cas, la monade \texttt{Identity}, induisant un isomorphisme de types, peut être éliminée: c'est le rôle de la fonction \texttt{removeIdDFA}, du module \texttt{Automata.DFAComp}, transformant un automate sur la catégorie de Kleisli associée à la monade \texttt{Identity} en un automate sur la catégorie des types.

Les algorithmes implantés dans le code source ne sont pas tous explicitement rappelés dans cette section de façon exhaustive, tout comme les différents types d'automates.
Par exemple, les automates alternants\footnote{Nous reviendrons sur ces automates en utilisant des opérades.} (dont les configurations sont exprimées par des expressions booléennes, formées elles-mêmes par un type monadique)
peuvent être convertis en automate non-déterministe (les configurations sont des ensembles de clauses conjonctives) ou en automate déterministe complet (les configurations sont des expressions) par une application de cette même fonction \texttt{applyFunctor} suivies de conversion mineures de types.

\begin{figure}[H]
  \centerline{
    \begin{tikzpicture}[node distance=2cm, bend angle=30]
    \node[initial left, state, accepting] (S0) {\(0\)};
    \node[initial above, state, above right of=S0] (S1) {\(1\)};
    \node[initial above, state, right of=S1] (S2) {\(2\)};
    \node[below right of=S2] (Sdots) {\(\cdots\)};
    \node[initial below, state, rounded rectangle, below left of=Sdots] (Sn-2) {\(n - 2\)};
    \node[initial below, state, rounded rectangle, left of=Sn-2] (Sn-1) {\(n - 1\)};
    \path[->]
      (S0)   edge [above left, bend left] node {\(a\)} (S1)
      (S1)   edge [above, bend left] node {\(a\)} (S2)
      (S1)   edge [above left, in=105, out=165, loop] node {\(b\)} ()
      (S2)   edge [above right, bend left, dotted] node {\(a\)} (Sdots)
      (S2)   edge [above right, in=15, out=75, loop] node {\(b\)} ()
      (Sdots)   edge [below right, bend left, dotted] node {\(a\)} (Sn-2)
      (Sn-2)   edge [below, bend left] node {\(a\)} (Sn-1)
      (Sn-2)   edge [below right, in=-15, out=-75, loop] node {\(b\)} ()
      (Sn-1)   edge [below left, bend left] node {\(a\)} (S0)
      (Sn-1)   edge [below left, in=-105, out=-165, loop] node {\(b\)} ()
      ;
    \end{tikzpicture}
  }
  \caption{Un (e famille d') automate non-déterministe dont la déterminisation atteint la borne.}%
  \label{fig ex aut explo comb}
\end{figure}
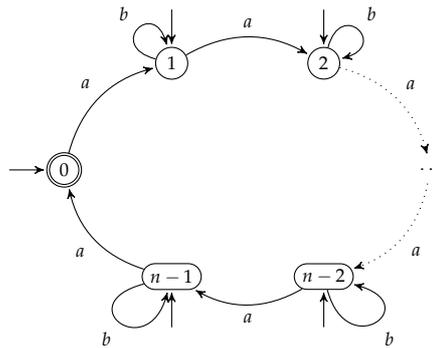

Concluons cette sous-section par l'implantation d'un cas concret classique.
Intéressons-nous à la famille d'automates de la Figure~\ref{fig ex aut explo comb}, connue pour être une famille d'automates dont la déterminisation atteint la borne exponentielle classique.
Les propriétés d'évaluation paresseuse d'Haskell, couplées à la mémoïsation (simple à mettre en {\oe}uvre en utilisant le module \texttt{Data.Function.Memoize} du package \href{https://hackage.haskell.org/package/memoize-0.8.1}{memoize}) permettent de construire virtuellement l'automate déterminisé pour de grandes valeurs entières.
En effet, seule la partie utile (car utilisée) de l'automate sera réellement interprétée.
Pour illustrer cela, nous allons considérer le Pseudo-Code suivant\footnote{L'exemple peut être lancé dans l'interpréteur ghci par la commande \texttt{stack ghci src/HDRExample/KleisliAuto.hs}.}.
Le code est reporté ici à titre d'information, le point important étant l'exécution de la fonction \texttt{test} sur des valeurs croissantes.
La fonction \texttt{memoizeAndTraceAuto} utilisée ici est implantée dans le module \texttt{Automata.KleisliAutomata} et permet de mémoïser les calculs de transitions d'un automate.
\begin{remarqueBox}[]{La classe de type \texttt{Applicative}}
  La classe de types \texttt{Applicative} du module de base Haskell \texttt{Control.Applicative}, utilisée dans le pseudo-code suivant, est une sous-classe de \texttt{Functor} et une super-classe de \texttt{Monad}, permettant de promouvoir des fonctions de toute arité au niveau d'un foncteur.
  Elle permet de définir notamment les transformations suivantes:
  \begin{itemize}
    \item \texttt{(<*>) :: f (a -> b) -> f a -> f b} permettant d'évaluer une fonction contenue dans un conteneur fonctoriel;
    \item \texttt{liftA2 :: (a -> b -> c) -> f a -> f b -> f c } permettant de promouvoir une fonction binaire au niveau fonctoriel.
  \end{itemize}
\end{remarqueBox}

\begin{remarqueBox}[]{La notation \texttt{do}}
  En Haskell, les monades permettent de définir du sucre syntaxique permettant, par exemple, de simuler l'exécution séquentielle d'actions ou d'expressions enchaînées.

  Ainsi, un terme \texttt{t} de type \texttt{m a}, où \texttt{m} est une monade, peut s'écrire sous la forme
  \begin{minted}[xleftmargin=1em, autogobble, fontsize=\footnotesize]{haskell}
    t = do
      val1
      ...
      valn
  \end{minted}
  où \texttt{valn} est de type \texttt{m a} et
  où tout autre terme \texttt{vali} est soit
  \begin{itemize}
    \item de type \texttt{m b},
    \item de la forme \texttt{let x = val},
    \item de la forme \texttt{x <- val} où \texttt{val} est de type \texttt{m b}.
  \end{itemize}

  La transformation en code ``normal'' s'effectue de la façon suivante:
  \begin{itemize}
    \item un bloc \texttt{do} à une seule valeur \texttt{val} est transformé en \texttt{val};
    \item un bloc \texttt{do} de la forme
      \begin{minted}[xleftmargin=1em, autogobble, fontsize=\footnotesize]{haskell}
        t = do
          let x = val
          val2
          ...
          valn
      \end{minted}
      est transformé en
      \begin{minted}[xleftmargin=1em, autogobble, fontsize=\footnotesize]{haskell}
       t = let x = val in
         do
           val2
           ...
           valn
      \end{minted}
    \item un bloc \texttt{do} de la forme
      \begin{minted}[xleftmargin=1em, autogobble, fontsize=\footnotesize]{haskell}
        t = do
          x <- val
          val2
          ...
          valn
      \end{minted}
      est transformé en
      \begin{minted}[xleftmargin=1em, autogobble, fontsize=\footnotesize]{haskell}
       t = val >>= \x ->
         do
           val2
           ...
           valn
      \end{minted}
    \item un bloc \texttt{do} de la forme
      \begin{minted}[xleftmargin=1em, autogobble, fontsize=\footnotesize]{haskell}
        t = do
          val1
          val2
          ...
          valn
      \end{minted}
      est transformé en
      \begin{minted}[xleftmargin=1em, autogobble, fontsize=\footnotesize]{haskell}
       t = val1 >>= \_ ->
         do
           val2
           ...
           valn
      \end{minted}
  \end{itemize}
\end{remarqueBox}


\begin{implementationBox}[]{Exemple: déterminisation et mémoïsation}
  \begin{minted}[xleftmargin=1em,   mathescape=true, autogobble, fontsize=\footnotesize]{haskell}
    {-# LANGUAGE FlexibleContexts #-}
    {-# LANGUAGE TemplateHaskell  #-}

    module HDRExample.KleisliAuto where

    import           Automata.Automaton
    import           Automata.AutomatonTransition
    import           Automata.KleisliAutomata
    import           Category.Kleisli
    import           Control.Monad
    import           Data.Function.Memoize
    import           Data.Hashable
    import           Data.HashSet                   ( HashSet )
    import qualified Data.HashSet                  as Set
    import           Tools.ToString

    -- Déclaration d'un alphabet à deux lettres
    data Symbol = A | B
      deriving (Show, Read, Enum, Bounded)

    -- Instanciation automatique de la mémoïsation
    -- par programmation générique
    deriveMemoizable ''Symbol

    -- Instanciation de la mémoïsation des
    -- ensembles via leurs représentations
    -- en String
    instance (Show a, Read a, Eq a, Hashable a)
      => Memoizable (HashSet a) where
        memoize f = memoize (f . read) . show

    -- Fonction utilitaire pour la définition
    -- de la fonction de transition :
    -- incrémentation modulaire
    deltaMod :: (Hashable a, Integral a) => a -> a -> HashSet a
    deltaMod n p = Set.singleton $ (p + 1) `mod` n

    -- Fonction utilitaire pour la définition
    -- de la fonction de transition :
    -- identité sur les entiers différents de 0
    deltaIdPos :: (Eq a, Num a, Hashable a) => a -> HashSet a
    deltaIdPos 0 = Set.empty
    deltaIdPos p = Set.singleton p

    -- Automate atteignant la borne exponentielle de
    -- la déterminisation : ici, auto n est
    -- un NFA à n états dont le déterminisé atteint la
    -- borne de $2^n$ états
    auto :: (Hashable a, Integral a) => a -> NFA Symbol a
    auto n = packKleisli (Set.fromList [0 .. n - 1]) transFun (== 0)
     where
      transFun A = deltaMod n
      transFun B = deltaIdPos

    -- Déterminisé de l'automate calculé par auto n,
    -- ayant $2^n$ états.
    autoDet
      :: (Hashable state, Integral state)
      => state -> DFA Symbol (HashSet state)
    autoDet = nfaToDFA . auto

    -- Déterminisé de l'automate calculé par auto n,
    -- avec une fonction de transition mémoïsée :
    -- en mémoire, partie accessible de l'automate des parties
    -- utilisée. Affiche un calcul des transitions lors de
    -- son (seul et unique) calcul, le mémoïse ensuite.
    autoDetMemo
      :: (Read state, Show state, Hashable state, Integral state)
      => state
      -> DFA Symbol (HashSet state)
    autoDetMemo n = memoizeAndTraceAuto $ autoDet n

    -- Affichage du calcul de certaines configurations,
    -- traçant les uniques calculs des transitions avant leur mémoïsation.
    -- Le calcul étant paresseux, de grandes valeurs peuvent être utilisées.
    -- Plus précisément, test n b construit les DFA autoDet n et autoDetMemo n
    -- ayant tous deux $2^n$ états, puis détermine les configurations
    -- associées aux mots $A^n$ et $A^nBA^n$;
    -- si b est True, calcule en plus l'ensemble
    -- des états accessibles de l'automates, forçant l'évaluation et
    -- interprétant les résultats pour un nombre exponentiel de cas.
    --
    -- Lors de la première utilisation d'une transition étiquetée par x
    -- depuis un état ys de l'automate déterminisé s'affiche la chaine
    -- (x, fromList states)
    -- où states est la liste d'états correspondant à l'ensemble ys.
    -- Lors d'une utilisation ultérieure, le calcul ayant été mémoïsé,
    -- le calcul intermédiaire n'a plus lieu et rien n'est alors
    -- inscrit.
    --
    -- Par exemple, test 40 False affichera les configurations atteintes
    -- par les mots $A^{40}$ et $A^{40}BA^{40}$ dans les automates ("virtuels")
    -- autoDet 40 et autoDetMemo 40 (ayant $2^{40}$ états),
    -- en ne construisant "réellement" que les états utiles pour le parcours de
    -- ces mots.
    -- De plus, test 40 True réalisera les mêmes calculs, puis affichera
    -- la liste (de longueur exponentielle) des états accessibles de l'automate.
    test :: Int -> Bool -> IO ()
    test n b = do
      -- w = $A^n$
      let w  = replicate n A
      -- w2 = $A^nBA^n$
      let w2 = w <> [B] <> w
      -- DFA avec $2^n$ états
      let a1 = autoDet n
      -- DFA avec $2^n$ états, mémoïsé
      let a2 = autoDetMemo n
      -- application de print_config aux couples de paramètres
      -- ((aut, n), word) pour (aut, n) dans [(a1, 1), (a2, 2)] et word dans [w, w2]
      sequence_ $ print_config <$> [(a1, 1), (a2, 2)] <*> [w, w2]
      -- affichage conditionnel des états accessibles
      -- ($2^n$ états à afficher)
      when b $ putStrLn $ toString $ accessibleStates a2
     where
      -- fonction d'affichage d'une configuration d'un automate
      -- associée à un mot
      print_config (aut, n) word = do
        putStrLn
          $  "Configuration associated with "
          <> show word
          <> " in a"
          <> show n
          <> ":"
        putStrLn $ toString $ (runKleisli $ getConfig aut word) ()
  \end{minted}
\end{implementationBox}

Les automates Booléens ne sont pas les seuls concernés par cette description monadique.
Nous allons voir dans la sous-section suivante comment modéliser d'autres types d'automates en conservant le même formalisme.

\subsection{Monade des semimodules libres et automates à multiplicités}\label{subsecAutMult}
Les automates à multiplicités, eux-aussi, peuvent être définis à partir d'une monade: celle des semimodules libres sur un semianneau.
\textbf{NB:} dans la suite, nous allons définir la notion de semimodule par son implantation Haskell; l'implantation des semianneaux provient du module \texttt{Data.Semiring} du package \href{http://hackage.haskell.org/package/semirings}{\texttt{semirings}}, dont les fonctions seront préfixées par \texttt{S.} lors de leurs utilisations.
\begin{implementationBox}[]{Automates à multiplicités}
  \begin{minted}[xleftmargin=1em, autogobble, fontsize=\footnotesize]{haskell}
    -- module Algebra.Structures.Semimodule

    -- Un semimodule sur un semianneau est un monoide muni
    -- d'une action du semianneau telle que
    -- * action k (m1 <> m2) = action k m1 <> action k m2
    -- * action (k S.* k') m = action k (action k' m)
    -- * action S.one m = m
    -- * action (k1 S.+ k2) m = action k1 m <> action k2 m
    class (Semiring k, Monoid m) => Semimodule k m where
      action :: k -> m -> m
      actionR :: m -> k -> m

    -- Tout seminanneau est un semimodule sur lui-même.
    instance (Monoid k, Semiring k) => Semimodule k k where
      action = (S.*)
      actionR = action
  \end{minted}
\end{implementationBox}
Les structures algébriques dites libres sont souvent associées à des adjonctions fonctorielles donnant elles-mêmes lieu à des monades.
Dans le cas des semimodules, le semimodule libre sur un ensemble \texttt{m} sur un semianneau \texttt{k} est l'ensemble des combinaisons linéaires finies d'éléments de \texttt{m} à coefficients dans \texttt{k}.
En d'autres termes, une combinaison linéaire est une fonction partielle de \texttt{m} vers \texttt{k} dont une partie finie de \texttt{m} est le domaine.
Dans la suite, nous utiliserons pour les représenter une \emph{table associative}.
\begin{remarqueBox}[]{Les tables associatives en Haskell}
  Le langage Haskell possède de nombreuses implantations des tables associatives.
  Dans la suite, nous utiliserons le constructeur binaire de types \texttt{HashMap} du module \href{http://hackage.haskell.org/package/unordered-containers-0.2.10.0/docs/Data-HashMap-Strict.html}{\texttt{Data.HashMap.Strict}} du package \href{http://hackage.haskell.org/package/unordered-containers}{\texttt{unordered-containers}}.
  Les fonctions seront précédées par \texttt{Map.} lors de leur utilisation, telles que:
  \begin{itemize}
    \item \texttt{unionWith} réalisant l'union de deux tables en combinant leurs valeurs avec la fonction passée en paramètre le cas échéant,
    \item \texttt{empty} renvoyant une table associative vide,
    \item \texttt{singleton} associant une unique clé avec une unique valeur,
    \item \texttt{map} appliquant la fonction passée en paramètre à toutes les valeurs de la table,
    \item \texttt{insertWith} insérant un couple (clé, valeur) dans une table en combinant avec l'ancienne valeur le cas échéant,
    \item \texttt{foldlWithKey'} combinant tous les couples (clé, valeur) de la table sur un accumulateur.
  \end{itemize}
\end{remarqueBox}

\begin{figure}[H]
  \centerline{
    \begin{tikzpicture}[node distance=5cm, bend angle=30]
    \node[state] (P) {\(P\)};
    \node[left of=P, node distance=1cm] (sP) {\(2\)};
    \node[below of=P, node distance=1cm] (fP) {\(5\)};
    \node[state, above right of=P] (Q) {\(Q\)};
    \node[above left of=Q, node distance=1cm] (sQ) {\(3\)};
    \node[above of=Q, node distance=1cm] (fQ) {\(2\)};
    \node[state, below right of=Q] (R) {\(R\)};
    \node[below of=R, node distance=1cm] (sR) {\(5\)};
    \path[->]
      (P)   edge [above left, bend left] node {\(3a\)} (Q)
      (P)   edge [below] node {\(5a - b\)} (R)
      (P)   edge [below] node {} (fP)
      (sP)   edge [below] node {} (P)
      (P)   edge [above left, in=105, out=165, loop] node {\(2a+4b\)} ()
      (Q)   edge [above right, bend left] node {\(-a\)} (R)
      (Q)   edge [below right, bend left] node {\(4a\)} (P)
      (Q)   edge [below] node {} (fQ)
      (sQ)   edge [below] node {} (Q)
      (R)   edge [below left, bend left] node {\(3b\)} (Q)
      (R)   edge [above right, in=15, out=75, loop] node {\(5b\)} ()
      (sR)   edge [below] node {} (R)
      ;
    \end{tikzpicture}
  }
  \caption{Un automate à multiplicités.}%
  \label{fig aut mult}
\end{figure}
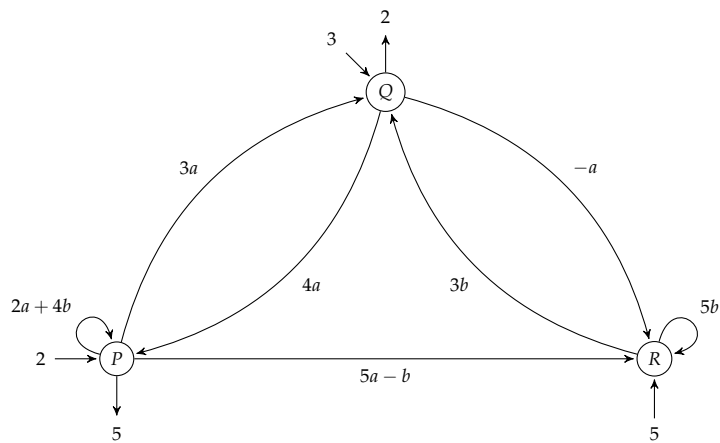

\begin{implementationBox}[label={semimodLibres}]{Les semimodules libres}
  \begin{minted}[xleftmargin=1em, autogobble, fontsize=\footnotesize]{haskell}
    -- module Algebra.Structures.Semimodule

    -- Type du semimodule libre sur le type m sur le semianneau k
    -- définissant:
    -- * le constructeur de valeur Free
    -- * la fonction de projection getMap pour obtenir la HashMap
    --    sous-jacente.
    data FreeSemimodule k m where
      Free :: Semiring k => {getMap :: HashMap m k} -> FreeSemimodule k m

    -- Le type FreeSemimodule k m est un semimodule.
    instance Semigroup (FreeSemimodule k m) where
      Free s1 <> Free s2 = Free $ Map.unionWith plus s1 s2

    instance Semiring k => Monoid (FreeSemimodule k m) where
      mempty = Free Map.empty

    instance Semiring k => Semimodule k (FreeSemimodule k m) where
      -- il suffit de multiplier tous les coefficients des éléments
      -- et de retirer ceux qui s'annulent
      action k m = Free $ Map.foldlWithKey' aux Map.empty $ getMap m
       where
        aux acc k_ v =
          let val = k S.* v in
          if val == zero then acc else Map.insert k_ val acc

      actionR m k = Free $ Map.foldlWithKey' aux Map.empty $ getMap m
       where
        aux acc k_ v =
          let val = v S.* k in
          if val == zero then acc else Map.insert k_ val acc

    -- Le constructeur FreeSemimodule k est un foncteur.
    instance FunctorGen (FreeSemimodule k) (->) (->) where
      fmap f (Free m) = Free $ Map.foldlWithKey' aux Map.empty m
        where aux acc k v = Map.insertWith plus (f k) v acc

    -- Le foncteur FreeSemimodule k est une monade.
    instance Semiring k => MonadGen (FreeSemimodule k) (->) where
      return a = Free $ Map.singleton a one

      join (Free m) =
        Free $ Map.foldlWithKey' aux Map.empty m
        where
          aux acc (Free _m) v =
            Map.unionWith (+) acc $ Map.map (S.* v) _m
  \end{minted}
\end{implementationBox}
L'utilisation de cette monade permet de manipuler des automates de Kleisli dont les configurations sont des combinaisons linéaires d'états, c'est-à-dire les automates à multiplicités, tout en conservant les fonctions déjà \emph{factorisées}, telles que la construction (\texttt{packKleisli}) ou l'application fonctorielle (\texttt{applyFunctor}).
Remarquons également l'égalité syntaxique des fonctions de conversion réciproques entre automates non-déterministes et à multiplicités Booléennes, ainsi que la grande proximité avec les algorithmes précédents de conversions d'automates Booléens (Code~\ref{codeAlgosClas}).
\begin{implementationBox}[]{Automates à multiplicités et conversions classiques}
  \begin{minted}[xleftmargin=1em, autogobble, fontsize=\footnotesize]{haskell}
    -- module Automata.KleisliAutomata

    -- Type des automates à multiplicités
    type WFA semiring symbol state
      = Automaton
        (KleisliCat (FreeSemimodule semiring) (->)) () symbol state ()

    -- Conversion classique d'un WFA à poids booléens en un NFA.
    booleanWFAToNFA :: WFA Bool symbol state -> NFA symbol state
    booleanWFAToNFA = castFinality . removeIdFromState . applyFunctor

    -- Conversion classique d'un NFA en un WFA à poids booléens.
    nfaToBooleanWFA :: NFA symbol state -> WFA Bool symbol state
    nfaToBooleanWFA = castFinality . removeIdFromState . applyFunctor
  \end{minted}
\end{implementationBox}
Les automates à multiplicités peuvent alors être utilisés comme suit\footnote{L'exemple peut être lancé dans l'interpréteur ghci par la commande \texttt{stack ghci src/HDRExample/KleisliWeightedAuto.hs}}.
Le Pseudo-Code~\ref{pseudoCodeAutMult} montre comment définir l'automate de la Figure~\ref{fig aut mult}.

\begin{implementationBox}[label={pseudoCodeAutMult}]{Exemple: Automates à multiplicités}
  \begin{minted}[xleftmargin=1em,   autogobble, mathescape=true, fontsize=\footnotesize]{haskell}
    {-# LANGUAGE DeriveGeneric #-}

    module HDRExample.KleisliWeightedAuto where

    import           Algebra.Structures.Semimodule
    import           Automata.Automaton             ( getConfig
                                                    , weightValue
                                                    )
    import           Automata.KleisliAutomata
    import           Category.Kleisli
    import           Data.Hashable                  ( Hashable )
    import qualified Data.HashMap.Strict           as Map
    import           Data.List                      ( intercalate )
    import           Data.Maybe                     ( fromMaybe )
    import           GHC.Generics                   ( Generic )
    import           Text.Read                      ( readMaybe )

    -- Déclaration d'un alphabet à trois lettres
    data Symbol = A | B | C
      -- instanciation automatique des classes
      -- * Show: conversion en String
      -- * Read: conversion depuis String
      -- * Eq: implantation de l'égalité
      -- * Generic: représentation canonique utilisée pour
      --    la génération automatique d'une fonction de
      --    hachage (voir plus loin)
      deriving (Show, Read, Eq, Generic)

    -- instanciation automatique d'une fonction de
    -- hachage
    instance Hashable Symbol

    -- Lecture d'une liste de symboles depuis une
    -- String de la forme "$X_1X_2\cdots$" avec $X_n$
    -- un caractère parmi 'A', 'B' ou 'C'
    stringToSymbol :: String -> Maybe [Symbol]
    stringToSymbol = mapM (readMaybe . return)

    -- Conversion d'une liste de Symbol en une String
    symbolListToString :: [Symbol] -> String
    symbolListToString [] = "$\varepsilon$"
    symbolListToString xs = concatMap show xs

    -- Types des états de l'automate
    data State = P | Q | R
      deriving (Show, Eq, Generic)

    instance Hashable State

    -- Multiplicité de l'automate
    type Weight = Int

    -- Exemples de combinaisons linéaires
    f1, f2, f3 :: FreeSemimodule Weight State
    -- f1 = 2 P + 3 Q + 5 R
    f1 = fromList [(P, 2), (Q, 3), (R, 5)]
    -- f2 = 4 P - R
    f2 = fromList [(P, 4), (R, -1)]
    -- f3 = 3 Q + 5 R
    f3 = fromList [(Q, 3), (R, 5)]

    -- Fonction de transition de l'automate
    delta :: Symbol -> State -> FreeSemimodule Weight State
    delta A P = f1
    delta B P = f2
    delta A Q = f2
    delta B R = f3
    delta _ _ = fromList []

    -- Poids de finalité
    fin :: State -> Weight
    fin P = 5
    fin Q = 2
    fin _ = 0

    -- Configuration initiale
    start :: FreeSemimodule Weight State
    start = f1

    -- Construction de l'automate
    auto :: WFA Weight Symbol State
    auto = packKleisli start delta fin

    -- Conversion d'une combinaison linéaire en String
    prettyPrint :: FreeSemimodule Weight State -> String
    prettyPrint (Free m) =
      intercalate " + "
        $ fmap (\(k, v) -> mconcat [show v, " ", show k])
        $ Map.toList m

    -- Affichage de la configuration et du poids
    -- associé à un mot entré sous la forme "$X_1X_2\cdots$"
    -- avec $X_n$ un caractère parmi 'A', 'B' ou 'C';
    -- si autre entrée, remplace l'entrée par le mot vide
    exWFA :: String -> IO ()
    exWFA xs = do
      let w = fromMaybe [] $ stringToSymbol xs
      putStrLn $ mconcat
        [ "configuration associated with "
        , symbolListToString w
        , ": "
        , prettyPrint (runKleisli (getConfig auto w) ())
        ]
      putStrLn $ mconcat
        [ "weight of "
        , symbolListToString w
        , ": "
        , show (weightValue auto w :: Weight)
        ]
  \end{minted}
\end{implementationBox}

D'autres monades bien connues car fréquemment utilisées en programmation fonctionnelle permettent également de modéliser des automates à multiplicités, telle que la monade \texttt{Writer} permettant de modéliser des automates séquentiels à multiplicités, ne nécessitant alors qu'une structure de monoïde et non de semianneau (voir la Section~\ref{sec::annexe:autMultWriter} en annexe).

\subsection{Monade State, transformers, et automates à pile}\label{subsecMonadState}
Les monades, les catégories de Kleisli associées et leur utilisation dans la définition d'automates sur des catégories permettent également de traverser la hiérarchie de Chomsky.
Par exemple, l'utilisation de la monade \texttt{State} et de ses \emph{transformers} permet de représenter les automates à pile par la définition d'une pile contextuelle.

\begin{remarqueBox}[]{Monades et transformers}
  Comme indiqué dans la Remarque~\ref{RemMonadeCompo}, les monades ne se composent généralement pas.
  Évidemment, certains couples particuliers de monades se composent.
  Les \emph{transformers} de monades permettent de construire de nouvelles monades en en combinant deux autres.
  Le cas des \emph{transformers} associés à la monade \texttt{State} du module \texttt{Control.Monad.State.Lazy} du package \texttt{mtl} est présenté dans la suite de cette sous-section.
\end{remarqueBox}

La monade \texttt{State} est une monade permettant de simuler des calculs dépendants de l'accès à une variable globale (au sens de la programmation impérative) ou de sa modification, tout en conservant un aspect fonctionnel pur.
Il s'agit, simplement et surprenamment, du type des fonctions polymorphes \texttt{s -> (a, s)}\footnote{En réalité, à un isomorphisme près.}.
\begin{implementationBox}[label={codeStateMonad}]{La monade State}
  \begin{minted}[xleftmargin=1em,  mathescape=true, autogobble, fontsize=\footnotesize]{haskell}

  -- module HDRExample.PushDownAutomata

  -- Le type State s a ne contient qu'un unique champ,
  -- une fonction du type s -> (a, s)
  newtype State s a = State {runState :: s -> (a, s)}

  -- Le type State s est un foncteur.
  instance FunctorGen (State s) (->) (->) where
    fmap f (State g) = State $ \s -> let (a, s') = g s in (f a, s')

  -- Le foncteur State s est une monade.
  instance MonadGen (State s) (->) where
    return a = State $ \s -> (a, s)
    join (State f) = State $ \s -> let (State g, s') = f s in g s'
  \end{minted}
\end{implementationBox}
Cette monade permet de modéliser simplement la dépendance à la pile du calcul de la fonction de transition d'un automate à pile déterministe; une telle fonction de transition est de la forme \(\Sigma \rightarrow Q \rightarrow \Gamma \rightarrow (\Gamma^*, Q)\), où \(\Sigma \) est l'alphabet des symboles, \(Q\) l'ensemble des états et \(\Gamma \) l'alphabet de pile.
Le calcul d'une transition dépend ainsi d'un symbole d'entrée, d'un état et du sommet de la pile, puis change l'état et remplace le sommet de la pile par un nouveau mot de pile, possiblement vide.
L'état de la pile peut être ainsi enregistré \emph{via} la monade \texttt{State}.
Cependant, la fonction de transition est partielle: elle n'est pas définie si la pile est vide.
Cela peut être résolu en combinant les monades \texttt{State} et \texttt{Maybe}.
\begin{implementationBox}[label={codeMaybeSTate}]{La monade MaybeState}
  \begin{minted}[xleftmargin=1em,  mathescape=true, autogobble, fontsize=\footnotesize]{haskell}

  -- module HDRExample.PushDownAutomata

  -- Il est possible de combiner la monade State et la monade Maybe
  -- pour former le constructeur MaybeState, qui ne contient qu'une
  -- fonction du type s -> Maybe (a, s)
  newtype MaybeState s a = MaybeState {runMaybeState :: s -> Maybe (a, s)}

  -- Le type MaybeState s est un foncteur.
  instance FunctorGen (MaybeState s) (->) (->) where
    fmap f (MaybeState g) = MaybeState $ (\(a, s') -> M.return (f a, s')) <=< g

  -- Le foncteur MaybeState s est une monade.
  instance MonadGen (MaybeState s) (->) where
    return a = MaybeState $ \s -> M.return (a, s)
    join (MaybeState f) = MaybeState $ (\(MaybeState g, s') -> g s') <=< f
  \end{minted}
\end{implementationBox}

\begin{figure}[H]
  \centerline{
    \begin{tikzpicture}[node distance=5cm, bend angle=30]
    \node[initial left, state] (S0) {\(0\)};
    \node[state, right of=S0] (S1) {\(1\)};
    \node[state, right of=S1] (S2) {\(2\)};
    \path[->]
      (S0)   edge [above, bend left] node {\((), b, \mathrm{pop}\)} (S1)
      (S0)   edge [above left, in=105, out=165, loop] node {\((), a, \mathrm{push} ()\)} ()
      (S1)   edge [above, bend left] node {\((), a, \mathrm{id}\)} (S2)
      (S1)   edge [above, in=120, out=60, loop] node {\((), b, \mathrm{pop}\)} ()
      (S2)   edge [above right, in=15, out=75, loop] node {\((), a + b, \mathrm{id}\)} ()
      ;
    \end{tikzpicture}
  }
  \caption{Un automate à pile déterministe.}%
  \label{fig det pda}
\end{figure}
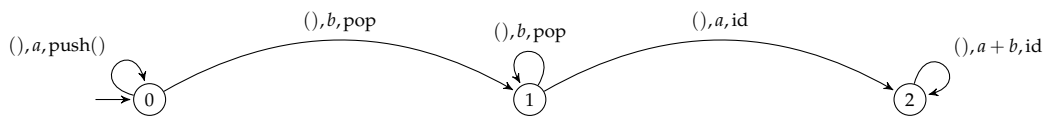

L'exemple suivant\footnote{qui peut être lancé dans l'interpréteur ghci par la commande \texttt{stack ghci src/HDRExample/PushDownAutomata.hs}} contient la définition des automates à pile déterministes comme automates sur la catégorie de Kleisli de la monade \texttt{MaybeState (Stack stackSymbol)}, où \texttt{Stack stackSymbol} est un type de pile polymorphe.
Le Pseudo-Code~\ref{pseudoCodeDetPDA} montre comment implanter l'automate à pile déterministe de la Figure~\ref{fig det pda}, reconnaissant les mots de la forme \(a^n{b}^{n + 1}\) par pile vide, avec pour alphabet de pile le type unitaire \texttt{()}.
Pour cela, les transitions sont étiquetées par des triplets de la forme \((s, x, f)\) avec \(s\) un symbole de sommet de pile, \(x\) un symbole de l'alphabet et \(f\) une action de remplacement parmi:
\begin{itemize}
  \item \(\mathrm{push}\ w \) remplaçant le sommet de pile par un mot \(w\) de symbole de pile;
  \item \(\mathrm{pop}\) retirant le symbole de sommet de pile;
  \item \(\mathrm{id}\) laissant le sommet de pile inchangé.
\end{itemize}

\begin{implementationBox}[label={pseudoCodeDetPDA}]{Exemple: automates à pile déterministe}
  \begin{minted}[xleftmargin=1em,  mathescape=true, autogobble, fontsize=\footnotesize]{haskell}
    {-# LANGUAGE FlexibleContexts      #-}
    {-# LANGUAGE LambdaCase            #-}
    {-# LANGUAGE MultiParamTypeClasses #-}
    {-# LANGUAGE PartialTypeSignatures #-}
    {-# LANGUAGE TupleSections         #-}
    {-# LANGUAGE TypeFamilies          #-}
    {-# LANGUAGE UndecidableInstances  #-}

    module HDRExample.PushDownAutomata where

    import           Automata.Automaton
    import           Category.FunctorGen           as F
    import           Category.Kleisli
    import           Category.MonadGen             as M
    import           Category.OfStructures.CategoryOfTypes
                                                  ( )
    import           Data.Maybe
    import           Text.Read

    ------------------- Fonctions utilitaires générales ------------------------

    -- Fonction utilitaire de composition d'une fonction unaire
    -- avec une fonction binaire, simplifiant la syntaxe
    (.:) :: (b -> c) -> (a1 -> a2 -> b) -> a1 -> a2 -> c
    (.:) = (.) . (.)

    ------------------- Définition d'un type de données pour les piles ---------

    -- Une pile est soit vide, soit un couple (élément, pile).
    data Stack a = Stack [a]
      deriving (Show)

    -- Teste si une pile est vide.
    isEmpty :: Stack a -> Bool
    isEmpty (Stack s) = null s

    -- Remplace le premier élément d'une pile par une liste (un mot) d'éléments.
    -- Ne change rien pour une pile vide.
    replaceHeadBy :: [a] -> Stack a -> Stack a
    replaceHeadBy _  (Stack []      ) = Stack []
    replaceHeadBy ys (Stack (_ : as)) = Stack (ys <> as)

    -- Crée une pile avec un unique élément.
    stackSingleton :: a -> Stack a
    stackSingleton a = Stack [a]

    ------- Fonctions utilitaires pour la construction d'automates à pile  -----

    -- Transforme un couple (état, symbole initial de pile) en un morphisme
    -- de la catégorie de Kleisli associée à la monade
    -- (MaybeState (Stack stackSymbol)).
    -- Permet de définir la configuration initiale d'un automate à pile
    -- déterministe.
    startK
      :: state
      -> stackSymbol
      -> KleisliCat (MaybeState (Stack stackSymbol)) (->) () state
    startK = Kleisli .: start
      where start i z () = MaybeState $ const $ Just (i, stackSingleton z)

    -- Transforme une fonction envoyant un triplet (symbol, state, stackSymbol)
    -- sur un couple (mot de stackSymbol, state) en une fonction envoyant un
    -- symbol sur un endomorphisme de la catégorie de Kleisli associée à
    -- la monade (MaybeState (Stack stackSymbol)).
    -- Permet de définir la fonction de transition d'un automate à pile
    -- déterministe.
    makeKTrans
      :: (symbol -> state -> stackSymbol -> ([stackSymbol], state))
      -> symbol
      -> KleisliCat (MaybeState (Stack stackSymbol)) (->) state state
    makeKTrans = Kleisli .: makeTrans
      where
        makeTrans trans x p = MaybeState $ \case
          Stack [] -> Nothing
          stack@(Stack (a : _)) ->
            let (bs, q) = trans x p a in Just (q, replaceHeadBy bs stack)

    -- Transforme un prédicat sur les états (finalité) en un morphisme
    -- de la catégorie de Kleisli associée à la monade
    -- (MaybeState (Stack stackSymbol)).
    -- Permet de définir la fonction de finalité d'un automate à pile
    -- déterministe.
    makeKFinality
      :: (state -> Bool)
      -> KleisliCat (MaybeState (Stack stackSymbol)) (->) state ()
    makeKFinality = Kleisli . makeFinality
      where
        makeFinality f p = MaybeState $ \s -> if f p then Just ((), s) else Nothing

    --------------------------- Automates à pile -----------------------------

    -- Synonyme de type pour les automates à pile déterministes
    type PDA stackSymbol symbol state
      = Automaton
          (KleisliCat (MaybeState (Stack stackSymbol)) (->))
          ()
          symbol
          state
          ()

    -- Construction d'un automate à pile
    makePDA
      :: state
      -> stackSymbol
      -> (symbol -> state -> stackSymbol -> ([stackSymbol], state))
      -> (state -> Bool)
      -> PDA stackSymbol symbol state
    makePDA i z trans finality =
      Auto (startK i z) (makeKTrans trans) (makeKFinality finality)

    -- Reconnaissance par pile vide
    emptyStackReco :: PDA stackSymbol symbol state -> [symbol] -> Bool
    emptyStackReco aut w =
      case (runMaybeState $ (runKleisli $ getConfig aut w) ()) (Stack []) of
        Nothing         -> False
        Just (_, stack) -> isEmpty stack

    ----------------- Utilisation des automates à pile déterministes------------

    -- alphabet binaire des symboles
    data Symbol = A | B
      deriving (Show, Read, Eq)

    -- Fonction utilitaire de conversion depuis une String
    stringToSymbol :: String -> Maybe [Symbol]
    stringToSymbol = mapM (readMaybe . return)

    -- Fonction utilitaire de conversion vers une String
    symbolListToString :: [Symbol] -> String
    symbolListToString [] = "$\varepsilon$"
    symbolListToString xs = concatMap show xs

    -- Exemple d'automate reconnaissant, par pile vide,
    -- les mots de la forme $A^nB^{n+1}$.
    auto :: PDA () Symbol Word
    auto = makePDA 0 () aux (const True)
     where
      -- par A, depuis 0, on arrive en 0 et on remplace le sommet de pile
      -- () par le mot ()() (push ())
      aux A 0 () = ([(), ()], 0)
      -- par B, depuis 0 ou 1, on arrive en 1 et on remplace le sommet de pile
      -- () par le mot vide (pop)
      aux B 0 () = ([], 1)
      aux B 1 () = ([], 1)
      -- pour les autres transitions, on va en 2 sans changer la pile
      aux _ _ _  = ([()], 2)

    -- Affiche la configuration associée et le résultat du
    -- test d'appartenance d'un mot w au langage reconnu par pile vide
    -- par l'automate auto, mot écrit depuis
    -- une String composée de caractères 'A' ou 'B'; remplace w par le mot
    -- vide en cas d'échec de lecture.
    test :: String -> IO ()
    test w = do
      let s = fromMaybe [] $ stringToSymbol w
      putStrLn $ mconcat
        [ "configuration associated with "
        , symbolListToString s
        , ":"
        , show ((runMaybeState $ (runKleisli $ getConfig auto s) ()) (Stack []))
        ]
      putStrLn $ mconcat
        [symbolListToString s, " is recognized ? ", show (emptyStackReco auto s)]
  \end{minted}
\end{implementationBox}


Remarquons alors que dans sa définition (Code~\ref{codeMaybeSTate}), la monade \texttt{MaybeState} n'utilise aucune spécificité de la monade \texttt{Maybe}: seules sont utilisées les fonctions \texttt{return} et \texttt{(<=<)}, définies pour toutes les monades.
On peut alors remplacer la monade Maybe par toute autre monade.
Ainsi, d'un point de vue théorique, il est possible de combiner toute monade avec la monade \texttt{State} en utilisant le \emph{transformer} \texttt{StateT} qui suit.
\begin{implementationBox}[]{Le combinateur StateT}
  \begin{minted}[xleftmargin=1em,  mathescape=true, autogobble, fontsize=\footnotesize]{haskell}

  -- module HDRExample.PushDownAutomata where

  -- Transformer associé à la monade State et au constructeur de type m,
  -- contenant une unique fonction du type s -> m (a, s)
  newtype StateT m s a = StateT {runStateT :: s -> m (a, s)}

  -- Si m est une monade, le constructeur de type StateT m s est un foncteur
  instance MonadGen m (->) => FunctorGen (StateT m s) (->) (->) where
    type FunctorConstraint (StateT m s) (->) (->) a
      = ( FunctorConstraint m (->) (->) (a, s)
        , FunctorConstraint m (->) (->) (m (a, s))
        , FunctorConstraint m (->) (->) (StateT m s a, s)
        )
    fmap f (StateT g) = StateT $ (\(a, s') -> M.return (f a, s')) <=< g

  -- Si m est une monade, le foncteur StateT m s est une monade
  instance MonadGen m (->) => MonadGen (StateT m s) (->) where
    return a = StateT $ \s -> M.return (a, s)
    join (StateT f) = StateT $ (\(StateT g, s') -> g s') <=< f
  \end{minted}
\end{implementationBox}
L'utilisation de ce \emph{transformer} permet alors d'utiliser des monades différentes de \texttt{Maybe} pour simuler des automates à piles d'autres sortes, telles que les non-déterministes.
\textbf{NB:} le problème de la partialité de la fonction de transition d'un automate à pile en cas de pile vide peut être résolu par l'utilisation de la classe de types \texttt{MonadPlus}, classe de la bibliothèque standard Haskell, rappelée dans l'exemple suivant avant son utilisation.

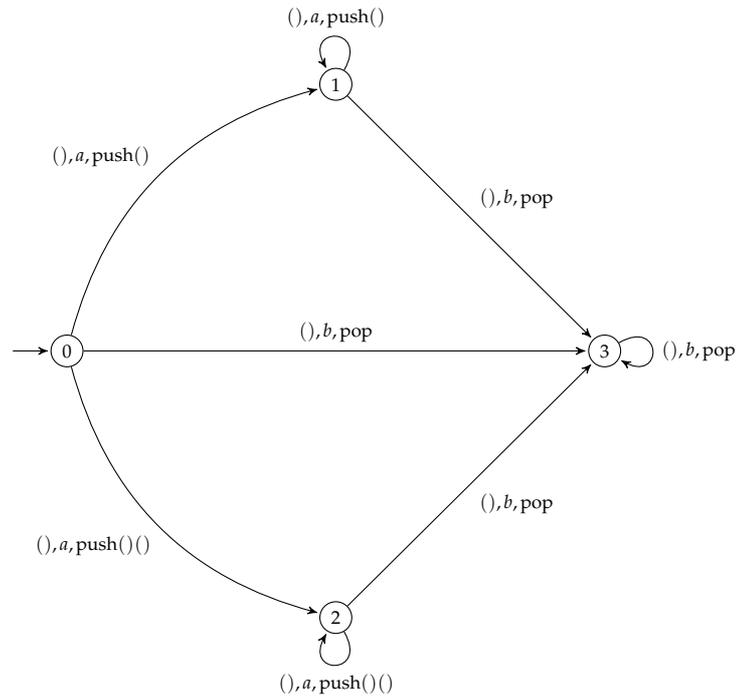
\begin{figure}[H]
  \centerline{
    \begin{tikzpicture}[node distance=5cm, bend angle=30]
    \node[initial left, state] (S0) {\(0\)};
    \node[state, above right of=S0] (S1) {\(1\)};
    \node[state, below right of=S0] (S2) {\(2\)};
    \node[state, above right of=S2] (S3) {\(3\)};
    \path[->]
      (S0)   edge [above left, bend left] node {\((), a, \mathrm{push} ()\)} (S1)
      (S0)   edge [below left, bend right] node {\((), a, \mathrm{push} () ()\)} (S2)
      (S0)   edge [above] node {\((), b, \mathrm{pop}\)} (S3)
      (S1)   edge [above, in=120, out=60, loop] node {\((), a, \mathrm{push} ()\)} ()
      (S1)   edge [above right] node {\((), b, \mathrm{pop}\)} (S3)
      (S2)   edge [below, in=-120, out=-60, loop] node {\((), a, \mathrm{push} () ()\)} ()
      (S2)   edge [below right] node {\((), b, \mathrm{pop}\)} (S3)
      (S3)   edge [right, in=-30, out=30, loop] node {\((), b, \mathrm{pop}\)} ()
      ;
    \end{tikzpicture}
  }
  \caption{Un automate à pile non-déterministe.}%
  \label{fig pda non det}
\end{figure}

Le Pseudo-Code~\ref{pseudoCodePdaNonDet} suivant montre comment implanter l'automate à pile non-déterministe de la Figure~\ref{fig pda non det} reconnaissant par pile vide le langage constitué des mots de la forme \(a^n{b}^{n+1}\) ou \(a^n{b}^{2n + 1}\).

\begin{implementationBox}[label={pseudoCodePdaNonDet}]{Exemple: automates à pile non déterministes}
  \begin{minted}[xleftmargin=1em,  mathescape=true, autogobble, fontsize=\footnotesize]{haskell}

  -- module HDRExample.PushDownAutomata [suite]

  ------- Utilitaires pour la construction d'automates de Kleisli à pile  -----

  startKT
    :: (_)
    => m state
    -> stackSymbol
    -> KleisliCat (StateT m (Stack stackSymbol)) (->) () state
  startKT = Kleisli .: start
    where start is z () = StateT $ const $ fmap (, stackSingleton z) is

  -- Classe des monades "combinables" et ayant une configuration vide.
  -- Utile pour généraliser l'absence de transition en cas de pile vide.
  -- Classe présente dans la bibliothèque standard Haskell.
  class MonadGen m (->) => MonadPlus m where
    -- configuration vide
    mzero :: m a
    -- combinaison de configurations
    mplus :: m a -> m a -> m a

  instance MonadPlus Maybe where
    mzero = Nothing
    mplus Nothing x = x
    mplus x       _ = x

  instance MonadPlus [] where
    mzero = []
    mplus = (<>)

  makeKTransT
    :: (MonadPlus m, _)
    => (symbol -> state -> stackSymbol -> m ([stackSymbol], state))
    -> symbol
    -> KleisliCat (StateT m (Stack stackSymbol)) (->) state state
  makeKTransT = Kleisli .: makeTrans
    where
      makeTrans trans x p = StateT $ \case
        Stack [] -> mzero
        stack@(Stack (a :  _)) ->
          fmap (\(bs, q) -> (q, replaceHeadBy bs stack)) $ trans x p a

  makeKTFinality
    :: (MonadPlus m, _)
    => (state -> Bool)
    -> KleisliCat (StateT m (Stack stackSymbol)) (->) state ()
  makeKTFinality = Kleisli . makeFinality
    where
      makeFinality f p = StateT $ \s -> if f p then return ((), s) else mzero

  ------------------- Automates de Kleisli à pile -----------------------------

  -- Exemple d'automate de Kleisli à pile: les automates
  -- à pile non déterministes à l'aide de la monade des listes.
  type NonDetPDA stackSymbol symbol state
    = Automaton
        (KleisliCat (StateT [] (Stack stackSymbol)) (->))
        ()
        symbol
        state
        ()

  -- Construction d'un automate à pile non déterministe
  makeNDetPDA
    :: [state]
    -> stackSymbol
    -> (symbol -> state -> stackSymbol -> [([stackSymbol], state)])
    -> (state -> Bool)
    -> NonDetPDA stackSymbol symbol state
  makeNDetPDA is z trans finality =
    Auto (startKT is z) (makeKTransT trans) (makeKTFinality finality)

  -- Reconnaissance par pile vide
  emptyStackNDetReco :: NonDetPDA stackSymbol symbol state -> [symbol] -> Bool
  emptyStackNDetReco aut w = any (\(_, stack) -> isEmpty stack)
    $ (runStateT $ (runKleisli $ getConfig aut w) ()) (Stack [])

  -------------- Utilisation des automates à pile non déterministes-------------

  -- Automate reconnaissant, par pile vide, les mots
  -- de la forme $A^nB^{n + 1}$ ou de la forme
  -- $A^{n}B^{2n + 1}$.
  auto2 :: NonDetPDA () Symbol Word
  auto2 = makeNDetPDA [0] () aux (== 0)
   where
    aux A 0 () = [([(), ()], 1), ([(), (), ()], 2)]
    aux B 0 () = [([], 3)]
    aux A 1 () = [([(), ()], 1)]
    aux B 1 () = [([], 3)]
    aux A 2 () = [([(), (), ()], 2)]
    aux B 2 () = [([], 3)]
    aux B 3 () = [([], 3)]
    aux _ _ _  = []

  -- Affiche la configuration associée et le résultat du
  -- test d'appartenance d'un mot w au langage reconnu par pile vide
  -- par l'automate auto2, mot écrit depuis
  -- une String composée de caractères 'A' ou 'B'; remplace w par le mot
  -- vide en cas d'échec de lecture.
  test2 :: String -> IO ()
  test2 w = do
    let s = fromMaybe [] $ stringToSymbol w
    putStrLn $ mconcat
      [ "configurations associated with "
      , symbolListToString s
      , ":"
      , show ((runStateT $ (runKleisli $ getConfig auto2 s) ()) (Stack []))
      ]
    putStrLn $ mconcat
      [ symbolListToString s
      , " is recognized ? "
      , show (emptyStackNDetReco auto2 s)
      ]
  \end{minted}
\end{implementationBox}

\section{Conclusion}

Dans cette section, nous avons vu comment construire des automates de mots les plus classiques de la littérature (déterministes, non déterministes, à poids, à pile) en utilisant des notions de théorie des catégories, et plus particulièrement les monades.

L'ajout d'un contexte, tel qu'une pile, a ainsi pu être modélisé \emph{via} l'utilisation de la monade \texttt{State} et de ses \emph{transformers}.
Remarquons alors \textbf{(1)} on peut utiliser toute autre monade à la place de celle des listes dans les automates à pile de Kleisli; \textbf{(2)} qu'une pile n'est pas le seul type de données qui peut servir de contexte: il n'y a pas de limites d'utilisation; ainsi, par l'adjonction d'une liste bi-infinie, ou de deux piles, il est alors possible de simuler une machine de Turing.

Nous avons également montré comment définir de nouveaux types d'automates, tels que les automates alternants généralisés, en modifiant très légèrement les monades sous-jacentes d'automates classiques de la littérature, comme les automates alternants.

Dans les sections suivantes, nous allons voir comment une généralisation de la notion de catégorie (les catégories \emph{enrichies}) permet de représenter des automates sur d'autres structures que les mots, en prenant l'exemple des arbres,
tout en conservant le pouvoir de factorisation offert par la notion de monade.
Pour cela, nous redéfinirons la notion d'opérade, structure algébrique permettant de représenter la composition des arbres, et nous étudierons son point commun avec la notion de monoïde.
Nous reviendrons ensuite sur cette notion pour représenter les expressions rationnelles.

\chapter{Extensions des automates de mots: les automates monoïdaux}\label{chapAutCatMonoidaux}

Dans cette section, nous allons nous intéresser à l'extension des automates de mots sur d'autres structures, tels que les arbres.
Pour cela, nous allons étudier une généralisation de la notion de monoïde, les monoïdes-objets, en comparant la structure de monoïde à celle d'opérade; cette notion nous permettra de généraliser tout d'abord des combinaisons classiques de la théorie des automates de mots, comme l'union ou la somme d'automates.

Une fois cela fait, nous utiliserons également cette généralisation des monoïdes pour construire des catégories dites \emph{enrichies}, où les morphismes n'existent plus nécessairement en tant qu'éléments d'ensembles; ces catégories permettront de définir des automates où les structures d'entrées ne sont plus nécessairement des mots.
Nous étudierons également la notion de monade enrichie afin d'expliciter et de retrouver différents types d'automates comme cas particuliers de la construction proposée.

Pour illustrer ces constructions, nous nous intéresserons aux cas particuliers permettant de retrouver les automates de mots et d'arbres, en tant que synonymes (ou spécialisations) des notions que nous aurons définies.


\section{Opérade, types gradués et programmation au niveau des types}

Comme rappelé dans les préliminaires (page~\pageref{pageOperades}), une opérade est une structure algébrique permettant de représenter la composition des arbres.
Cette structure est composée, comme peut l'être d'une façon différente un monoïde, d'un ensemble de valeurs, muni d'une opération \(\circ \) associative (ici verticalement et horizontalement), et d'un élément neutre.
La seule différence avec le produit d'un monoïde est que la composition d'une opérade n'est pas une opération binaire: si le produit d'un monoïde combine deux valeurs, la composition d'une opérade combine une valeur d'arité \(k\) avec \(k\) valeurs d'arité \((n_1, \ldots,n_k)\) pour former un arbre d'arité \(n_1 + \cdots + n_k\).

Cependant, une autre interprétation possible des opérations de combinaison peut permettre de regrouper ces deux opérations distinctes.
Le produit d'un monoïde sur un ensemble \(E\), de signature \(E\times E \rightarrow E\) peut être vu non pas comme une combinaison de deux valeurs de \(E\), mais comme une fonction envoyant \emph{une construction basée sur} \(E\) sur \(E\) lui-même.
Dans le cas des monoïdes, la construction sous-jacente est le produit cartésien d'ensembles.

Dans le cas des opérades, quelle pourrait être la construction sur l'ensemble?
Remarquons alors que la composition \(\circ \) peut se décomposer en deux sous-opérations: tout d'abord, elle produit un couple de valeurs de la forme \((e, (e_1, \ldots, e_k))\) où \(e\) est un élément de \(E\) d'arité \(k\), et où \((e_1,\ldots,e_k)\) sont \(k\) éléments de \(E\) d'arité \((n_1, \ldots,n_k)\); ensuite elle combine ces valeurs en un élément de \(E\) d'arité \(n_1 + \cdots + n_k\).

La construction intermédiaire, la formation du couple, peut être vue comme une opération binaire, mais sur les types!
Dans le cas des opérades, il suffit de produire une construction intermédiaire associant à deux ensembles gradués \(F\) et \(G\) l'ensemble gradué des couples de la forme \(c=(f,(g_1,\ldots,g_k))\), où \(f\) est un élément d'arité \(k\) de \(F\), \((g_1,\ldots,g_k)\) sont \(k\) éléments de \(G\) d'arité \((n_1, \ldots,n_k)\), et où \(c\) a pour arité (par définition) \(n_1 + \cdots + n_k\).
Cette opération est appelée \firstocc{produit tensoriel}{produit tensoriel!d'ensembles gradués}, noté classiquement \(\otimes \).

Ci-après se trouve une implantation de la notion de type gradué et du produit tensoriel associé en Haskell.
Afin de représenter la graduation des éléments au niveau des types, il peut être utile d'utiliser la programmation au niveau des types et une représentation des entiers au niveau des types; Ainsi, cela permettra de vérifier les futures compositions d'éléments à la compilation même (vérification par le moteur d'inférence de types).

\begin{remarqueBox}[label={RemProgType}]{Programmation au niveau des types et représentation des entiers}
  La programmation au niveau des types est une technique de programmation permettant de promouvoir des informations sur des valeurs au niveau des types, basé sur la théorie des types~\cite{ML84, hottbook}.
  Le langage Haskell est un très bon support pour implanter ces notions~\cite{Mag19} (voir également une présentation de cette notion en Idris dans~\cite{Bra17}, autre langage de programmation fonctionnelle très proche d'Haskell).

  Par exemple, on peut imaginer créer deux types numériques \texttt{Pair} et \texttt{Impair}, et une fonction de multiplication de signature possible
  \begin{itemize}
    \item \texttt{Pair -> Pair -> Pair};
    \item \texttt{Impair -> Impair -> Impair};
    \item \texttt{Impair -> Pair -> Pair};
    \item \texttt{Pair -> Impair -> Pair};
  \end{itemize}
  Une telle fonction est dite type-dépendante: son type de sortie dépend du type d'entrée.
  En Haskell, il est possible d'implanter cela à l'aide des classes de types, mais aussi de \emph{familles de types} (permettant de réaliser des fonctions au niveau des types).
  Par exemple, la fonction de multiplication peut être implantée comme suit:

    \begin{minted}[xleftmargin=1em,   autogobble, fontsize=\footnotesize]{haskell}
      {-# LANGUAGE TypeFamilies           #-}
      {-# LANGUAGE MultiParamTypeClasses  #-}

      -- Définition des types pour les entiers
      newtype Pair = Pair Word
      newtype Impair = Impair Word

      -- Fonction de conversion vers les entiers positifs
      class ToWord a where
        toWord :: a -> Word

      instance ToWord Pair where
        toWord (Pair x) = 2 * x

      instance ToWord Impair where
        toWord (Impair x) = 2 * x + 1

      -- Conversion en String
      instance Show Pair where
        show = show . toWord

      instance Show Impair where
        show = show . toWord

      -- Famille de types permettant de calculer
      -- le type de retour de la fonction de multiplication
      -- en fonction des entrées
      type family Combi a b where
        Combi Impair Impair = Impair
        Combi _ _ = Pair

      -- Déclaration de la classe de types pour
      -- implanter la fonction de multiplication
      -- dont le type de retour dépend du type
      -- des entrées
      class Mult a b where
        mult :: a -> b -> Combi a b

      -- Instances permettant de distinguer les différents
      -- cas de calculs
      instance Mult Pair Pair where
        mult (Pair x) (Pair y) = Pair $ 2 * x * y

      instance Mult Pair Impair where
        mult (Pair x) (Impair y) = Pair $ x * (2 * y + 1)

      instance Mult Impair Pair where
        mult (Impair y) (Pair x) = Pair $ x * (2 * y + 1)

      instance Mult Impair Impair where
        mult (Impair x) (Impair y) = Impair $ 2 * x * y + x + y

      -- cas incorrect détecté par le moteur d'inférence
      -- wrong :: Impair
      -- wrong = mult (Pair 2) (Pair 3)

      -- Affichage de quelques résultats où
      -- * Pair 2   = 4
      -- * Impair 2 = 5
      -- * Pair 3   = 6
      -- * Impair 3 = 7
      main :: IO ()
      main = do
        print $ mult (Pair 2) (Pair 3)     -- = 24
        print $ mult (Pair 2) (Impair 3)   -- = 28
        print $ mult (Impair 2) (Pair 3)   -- = 30
        print $ mult (Impair 2) (Impair 3) -- = 35
    \end{minted}

  Une autre application simple est par exemple de typer des listes en fonction de leurs longueurs; ainsi, une liste de type \texttt{longueur n}\footnote{Ces listes, dont la longueur est connue à la compilation, sont implantées dans le module \texttt{Tools.Vector}.} concaténée à une liste de type \texttt{longueur m} produit une liste de type \texttt{longueur (n + m)}.
  Remarquons alors que la fonction de concaténation est type-dépendante: le type de sortie \texttt{longueur (n + m)} dépend des types en entrée.
  Que devrait être alors le résultat de la concaténation avec une liste vide?

  Cette remarque soulève le problème de la représentation des entiers au niveau des types; le plus simple est alors d'utiliser l'arithmétique de Peano: un entier au niveau des types et soit \(0\), soit un successeur de \(0\).
  Cette représentation, appelée par la suite \texttt{Nat}, est implantée dans le module \texttt{Type.Natural} des sources de ce projet.
  Y sont également implantées les preuves (au sens de l'isomorphisme Curry-Howard) de l'égalité entre les types \(n\), \(n+0\) et \(0+n\), de l'associativité de la somme,
  \emph{etc}\footnote{Ces fonctions sont utilisées dans le code source pour lever les ambiguïtés induites par des égalités de types que le moteur d'inférence ne peut déduire seul (voir par exemple le module \texttt{Algebra.Structures.Operad});
  l'utilisation de ces fonctions sera cependant effacé des pseudo-codes de ce document, pour ne pas alourdir le propos.}.
  Dans la suite, pour lever les ambiguïtés, un entier \(n\) sera noté \texttt{Nat\_n} pour sa représentation au niveau des types.
\end{remarqueBox}


\begin{remarqueBox}[]{Types et genres}
  Si \texttt{Nat\_1}, \texttt{String} ou \texttt{Maybe} sont des types, ils ne sont pas comparables: ils ne sont pas du même \texttt{genre} (appelé \emph{kind} en Haskell).
  Attention: il y a nécessairement une hiérarchie au niveau des types; pour les mêmes raisons que l'ensemble des ensembles n'existe pas, il n'y a pas de types des types.
  Ici:
  \begin{itemize}
    \item \texttt{String} est un ensemble de valeurs: en Haskell, il a pour genre \texttt{*}, le genre des types contenant des valeurs;
    \item \texttt{Maybe} a pour genre \texttt{* -> *}, le type des constructeurs de type unaire: il attend un type de genre \texttt{*} pour renvoyer un type de genre \texttt{*}; par exemple \texttt{Maybe String} a pour genre \texttt{*};
    \item \texttt{Nat\_1} est (ici) de genre \texttt{Nat}; \texttt{Nat\_1} est bien un type, mais inhabité.
      Il en va de même pour les types \emph{listes de \texttt{Nat}}, tels que \texttt{[Nat\_1, Nat\_2,Nat\_1,Nat\_4]}, qui est un type du genre \texttt{[Nat]}.
  \end{itemize}
  Malheureusement, il n'y a pas (pour le moment) de \emph{type de genre} en Haskell, contrairement aux univers d'\href{https://agda.readthedocs.io/en/v2.6.0/language/universe-levels.html}{Agda}.
\end{remarqueBox}
\begin{implementationBox}[]{Types gradués}
  \begin{minted}[xleftmargin=1em, autogobble, fontsize=\footnotesize]{haskell}
    -- module Graded.Graded

    -- Un type est gradué s'il est instance de la classe suivante.
    class Graded a where

      -- Pour tout entier n, il existe un type de données pour les
      -- valeurs d'arité n de a
      data Graduation a (n :: Nat)

      -- Fonction de conversion entre les données graduées
      -- et les données non graduées de a
      fromGrad :: Graduation a n -> a

      -- Fonction transformant un élément de a en sa version graduée
      toGrad :: a -> Graduation a n

      -- Fonction permettant de récupérer à l'exécution la graduation
      -- d'un élément de type gradué
      graduation' :: Graduation a (n :: Nat) -> Natural n
      graduation' _ = singleton
  \end{minted}
\end{implementationBox}
Deux des constructions les plus simples de types gradués sont le type \texttt{()}, qui ne contient qu'un élément d'arité (arbitrairement choisie pour le moment) \(1\), et le type \texttt{Identity a} où \texttt{a} est un type gradué.
\begin{implementationBox}[]{Exemples de types gradués simples}
  \begin{minted}[xleftmargin=1em, autogobble, fontsize=\footnotesize]{haskell}
    -- module Graded.Graded

    -- Graduation pour le type ()
    instance Graded () where
      -- Il n'y a qu'une seule valeur graduée associée au type ()
      -- la valeur GradUn  d'arité 1
      data Graduation () n where
        GradUn ::Graduation () Nat_1

      -- Conversion de () à Graduation () Nat_1
      toGrad () = GradUn

      -- Conversion de Graduation () Nat_1 à ()
      fromGrad GradUn = ()

    -- Si a est un type gradué, alors Identity a est gradué
    instance Graded a => Graded (Identity a) where

      -- Les éléments d'arité n de Identity a sont de la forme
      -- GradId x, où x est un élément d'arité x de a;
      -- définition simultanée de la fonction de projection
      -- runId permettant d'extraire l'élément gradué
      data Graduation (Identity a) n where
        GradId ::{runId :: Graduation a n} -> Graduation (Identity a) n

      -- Conversions
      fromGrad (GradId x) = Identity $ fromGrad x
      toGrad (Identity x) = GradId $ toGrad x
  \end{minted}
\end{implementationBox}
Un produit tensoriel de types gradués est alors la construction d'un couple \(c\) composé d'un élément gradué \(e\) et d'une liste dont les \emph{contraintes} sont les suivantes:
\begin{itemize}
  \item sa longueur est l'arité de \(e\);
  \item les éléments qu'elle contient déterminent l'arité du couple, qui est égale à leur somme.
\end{itemize}
Pour réaliser cela au niveau des types, il suffit de créer le type des listes contenant des éléments d'un type gradué, type paramétré par l'arité des éléments qu'elles contiennent!
\textbf{NB:} Pour réaliser cela, nous utiliserons les fonctions et preuves (au sens de l'isomorphisme Curry-Howard) contenues dans le module \texttt{Type.NaturalList} des sources de ce projet.
\begin{implementationBox}[label={pseudoCodeGradVect}]{Les listes paramétrées par la graduation des éléments qu'elles contiennent}
  \begin{minted}[xleftmargin=1em, autogobble, fontsize=\footnotesize]{haskell}
    -- module Graded.GradedVector

    -- Type des listes paramétrées par la graduation des éléments
    -- qu'elles contiennent; ici ms est une liste (au niveau des types)
    -- de Nat
    data GradVect (ms :: [Nat]) a where
      -- la liste vide est paramétrée par la liste vide
      GradVNil ::GradVect  []  a
      -- une liste non vide est constituée d'un élément d'arité
      -- n1, d'une liste dont les arités sont une liste ms,
      -- et le paramètre d'une telle liste est l'ajout
      -- de n1 en tête de ms
      GradVCons :: Graduation a n1 -> GradVect ms a -> GradVect (n1 : ms) a
  \end{minted}
\end{implementationBox}
Ainsi, une liste de type \texttt{GradVect [Nat\_1, Nat\_4, Nat\_2] a} sera une liste contenant trois valeurs, d'arité respective 1, 4 et 2.
De même, si une liste est de type \texttt{GradVect ms ()}, alors la liste \texttt{ms} est une liste ne contenant que des occurrences d'arité \texttt{Nat\_1}, soit la valeur \texttt{GradUn}.
\begin{remarqueBox}[label={remGradVect}]{Le type \texttt{GradVect}}
  Le type \texttt{GradVect} est implanté dans le module \texttt{Graded.GradedVector}.
  Ce module contient de nombreuses fonctions utilitaires utilisées par la suite, telles que la fonction \texttt{mapVect} appliquant une fonction graduée sur tous les éléments, ou \texttt{replicateGrad} permettant de créer un vecteur de longueur fixé en répétant le même élément.
\end{remarqueBox}
L'utilisation de ces listes permet alors de définir simplement le produit tensoriel de types gradués, qui est lui-même gradué.
\textbf{NB:} Dans la suite sont utilisées des \emph{familles de types}, assimilables à des fonctions sur les types; ainsi:
\begin{itemize}
  \item \texttt{Length ms}, du module \texttt{Type.List} renvoie le \texttt{Nat} correspondant à la longueur d'une liste (au niveau des types) \texttt{ms};
  \item \texttt{First a} et \texttt{Second a}, du module \texttt{Category.CategoryGen} sont des fonctions de projections pour les couples de types;
  \item \texttt{SumNat ms}, du module \texttt{Type.Natural}, renvoie l'entier qui est la somme des \texttt{Nat} contenue dans une liste \texttt{ms}.
\end{itemize}
\begin{implementationBox}[]{Produit tensoriel des types gradués}
  \begin{minted}[xleftmargin=1em, autogobble, fontsize=\footnotesize]{haskell}
    -- Category.OfGraded.TensorProductOfGraded

    -- Le produit tensoriel d'un type a = (b, c) est composé
    -- d'un élément gradué de type b (= First a) dont l'arité est
    -- la longueur de la liste paramétrant le second composant,
    -- qui est une liste d'éléments gradués de c (= Second a);
    -- les arités des éléments qu'elle contient sont contenues dans ms
    data TensorProd a where
      TensorProd ::
        Graduation (First a) (Length ms)
          -> GradVect ms (Second a)
            -> TensorProd a

    -- Un produit tensoriel de type gradué est gradué
    instance Graded (TensorProd a) where

      -- Un produit tensoriel a pour graduation
      -- la somme des arités de sa seconde composante
      data Graduation (TensorProd a) arite where
        GradTensor ::
          Graduation (First a) (Length ms)
             -> GradVect ms (Second a)
               -> Graduation (TensorProd a) (SumNat ms)

      -- Fonctions de conversion
      toGrad (TensorProd x xs) = GradTensor x xs
      fromGrad (GradTensor x xs) = TensorProd x xs
  \end{minted}
\end{implementationBox}
La notion de \firstocc{produit tensoriel} n'est pas spécifique aux types gradués.
En effet, cette notion a un sens beaucoup plus général en théorie des catégories, et est une opération fondamentale dans la définition des \firstocc{catégories monoïdales}.


\section{Catégorie monoïdale}
  Nous avons vu comment construire un nouveau type gradué comme un produit tensoriel de deux types gradués.
  Mais cette construction permet également de construire des fonctions entre produits tensoriels.
  Par exemple, si \texttt{A}, \texttt{B}, \texttt{C} et \texttt{D} sont des types gradués, et si \texttt{f} et \texttt{g} sont des fonctions graduées (c'est-à-dire des fonctions préservant l'arité des éléments) respectivement de \texttt{A} vers \texttt{B} et de \texttt{C} vers \texttt{D},
  alors il est possible de construire une fonction graduée \texttt{f}\(\otimes \)\texttt{g} de \texttt{A}\(\otimes \)\texttt{C} vers \texttt{B}\(\otimes \)\texttt{D}: si \texttt{(a, (c\_1, \ldots, c\_k))} est un élément de \texttt{A}\(\otimes \)\texttt{C}, son image par \texttt{f}\(\otimes \)\texttt{g} est \texttt{(f a, (g c\_1, \ldots, g c\_k))}.

  Cette construction, envoyant alors des objets sur des objets et des fonctions sur des fonctions, est bel et bien un foncteur, depuis une catégorie \firstocc{produit}, définie comme suit.

  Le \firstocc{produit}{catégorie!produit} de deux catégories \(\mathcal{C}\) et \(\mathcal{D}\) est la catégorie \(\mathcal{C}\times \mathcal{D}\) dont:
  \begin{itemize}
    \item les objets sont les couples \((C, D)\) d'objets \(C\) de \(\mathcal{C}\) et \(D\) de \(\mathcal{D}\);
    \item les morphismes depuis un objet \((C_1, D_1)\) vers un objet \((C_2, D_2)\) sont les couples \((f, g)\) de morphismes \(f\) entre \(C_1\) et \(C_2\) et \(g\) entre \(D_1\) et \(D_2\);
    \item la composition de deux morphismes \((f,g)\) et \((f',g')\) est le morphisme \((f'\circ_{\mathcal{C}}f, g'\circ_{\mathcal{D}}g)\);
    \item le morphisme identité pour un objet \((C, D)\) est le morphisme \((\mathrm{Id}_C, \mathrm{Id}_D)\).
  \end{itemize}
  \begin{implementationBox}[]{Catégorie produit}
    \begin{minted}[xleftmargin=1em,   autogobble, fontsize=\footnotesize]{haskell}
      -- module Category.CategoryGen

      -- Type de données pour une catégorie produit
      data ProdCat cat cat' a b = Prod (cat (First a) (First b)) (cat' (Second a) (Second b))

      -- La construction ProdCat est une catégorie.
      instance (CategoryGen cat, CategoryGen cat') => CategoryGen (ProdCat cat cat') where

        id = Prod id id

        Prod f1 g1 . Prod f2 g2 = Prod (f1 . f2) (g1 . g2)
    \end{minted}
  \end{implementationBox}
  Cette construction de produit binaire peut être étendue sans difficulté à un produit d'une arité arbitraire.

  Un \firstocc{produit tensoriel} sur une catégorie \(\mathcal{C}\) est un foncteur de \(\mathcal{C}\times\mathcal{C}\) vers \(\mathcal{C}\)\footnote{aussi appelé \firstocc{bifoncteur} sur \(\mathcal{C}\).}.
  Cette construction de produit tensoriel binaire peut être étendue, elle aussi, sans difficulté à un produit tensoriel d'une arité arbitraire.

  Des constructions classiques de la théorie des ensembles, tels que le produit cartésien ou la somme disjointe, sont des produits tensoriels pour la catégorie des types, comme l'illustre le code suivant.
  \begin{implementationBox}[label={CodeProdTensTypes}]{Produits tensoriels de la catégorie des types}
    \begin{minted}[xleftmargin=1em,   autogobble, fontsize=\footnotesize]{haskell}
      -- Category.OfStructures.CategoryOfTypes

      -- Type de données pour le produit cartésien de types
      data Prod a = Product {runProduct :: (First a, Second a)}

      -- Le produit cartésien est un produit tensoriel.
      instance FunctorGen Prod (ProdCat (->) (->)) (->) where
        fmap (Prod f g) (Product (x, y)) = Product (f x, g y)

      -- Type de données pour la somme disjointe de types
      data Sum a = Fst (First a) | Snd (Second a)

      -- Type isomorphe au type des catégories (pour éviter
      -- un conflit des définitions de foncteurs)
      newtype SumTypeCat a b = SumTypeCat {runSum :: a -> b}
        deriving CategoryGen

      -- La somme disjointe est un produit tensoriel.
      instance FunctorGen Sum (ProdCat SumTypeCat SumTypeCat) SumTypeCat where
        fmap (Prod (SumTypeCat f) (SumTypeCat g)) = SumTypeCat $ \case
          Fst a -> Fst $ f a
          Snd b -> Snd $ g b
    \end{minted}
  \end{implementationBox}

  \begin{remarqueBox}[label={RemUnionInterAuto}]{Produits tensoriels et constructions d'automates}
    Un foncteur depuis une catégorie produit \texttt{ProdCat cat cat'} vers une catégorie \texttt{cat''} permet de combiner deux automates respectivement sur les catégories \texttt{cat} et \texttt{cat'} en un automate sur la catégorie \texttt{cat''}.

    Cette transformation est implantée par la fonction \texttt{functorProduct} du module \texttt{Automata.Automaton}:
      \begin{minted}[xleftmargin=1em,   autogobble, fontsize=\footnotesize]{haskell}
        -- Automata.Automaton

        -- Transforme deux automates en un automate réalisant le
        -- parcours parallèle de ces deux automates, dans la catégorie
        -- produit.
        parallelProduct
          :: Automaton cat init symbols state value
          -> Automaton cat' init' symbols state' value'
          -> Automaton
               (ProdCat cat cat')
               (init, init')
               symbols
               (state, state')
               (value, value')
        parallelProduct (Auto i d f) (Auto i' d' f') =
          Auto (Prod i i') (\a -> Prod (d a) (d' a)) (Prod f f')

        -- Applique un foncteur sur un automate.
        functorAut
          :: (FunctorGen f cat1 cat2)
          => Automaton cat1 a2 symbols a1 b
          -> Automaton cat2 (f a2) symbols (f a1) (f b)
        functorAut (Auto i d f) = Auto (F.fmap i) (F.fmap . d) (F.fmap f)

        -- Applique un foncteur depuis une catégorie produit
        -- sur le produit de deux automates
        -- donnés en paramètres
        functorProduct
          :: (FunctorGen bifunc (ProdCat cat cat') cat'')
          => Automaton cat init symbols state value
          -> Automaton cat' init' symbols state' value'
          -> Automaton
               cat''
               (bifunc (init, init'))
               symbols
               (bifunc (state, state'))
               (bifunc (value, value'))
        functorProduct a1 a2 = functorAut $ parallelProduct a1 a2
      \end{minted}

      Si le produit parallèle d'automates peut être utilisé pour définir des automates exotiques (voir par exemple la Section~\ref{sec::annexeProdParAut} pour un produit entre un automate alternant et un automate à multiplicités),
      on peut également l'utiliser pour retrouver des algorithmes bien connus.
      Les foncteurs \texttt{Prod} et \texttt{Sum} étant également des bifoncteurs pour certaines catégories de Kleisli (voir les implantations dans les modules \texttt{Category.OfStructures.CategoryOfTypes} et \texttt{Algebra.Structures.Semimodule}),
      la fonction \texttt{functorProduct} rend possible,
      dans le cadre des automates de Kleisli associés aux monades de la catégorie des types,
      la factorisation de transformations associées au produit cartésien et à la somme disjointe en deux fonctions du module \texttt{Automata.KleisliAutomata},
      les fonctions \texttt{binaryTransformationByProductWith} (créant l'automate produit par \texttt{functorProduct} puis utilisant des transformations naturelles des catégories monoïdales pour réduire la configuration initiale et la fonction paramètre pour calculer la sortie) et \texttt{binaryTransformationBySum} (créant l'automate somme par \texttt{functorProduct} puis réduisant la sortie par la fonction paramètre).

      Sans rentrer dans les détails très techniques, les transformations suivent les algorithmes classiques.
      Ainsi,
      \begin{itemize}
        \item l'intersection de deux automates non-déterministes, l'intersection de deux automates déterministes et le produit de Hadamard de deux automates à multiplicités sont implantés identiquement dans le module \texttt{Automata.KleisliAutomata} comme
          \begin{minted}[xleftmargin=1em,   autogobble, fontsize=\footnotesize]{haskell}
            -- module Automata.KleisliAutomata

            intersectionNFA = binaryTransformationByProductWith (const $ const ())

            productWFA = binaryTransformationByProductWith (const $ const ())

            intersectionDFA = binaryTransformationByProductWith (const $ const ())
          \end{minted}
        \item l'union de deux automates non-déterministes et la somme de deux automates à multiplicités sont implantés identiquement dans le module \texttt{Automata.KleisliAutomata} comme
          \begin{minted}[xleftmargin=1em,   autogobble, fontsize=\footnotesize]{haskell}
            -- module Automata.KleisliAutomata

            unionNFA a1 a2 =
              reduceInitialSum (<>) $ reduceFinalSum id id $ binaryTransformationBySum a1 a2

            sumWFA a1 a2 =
              reduceInitialSum (<>) $ reduceFinalSum id id $ binaryTransformationBySum a1 a2
          \end{minted}
      \end{itemize}
      La fonction \texttt{binaryTransformationBySum} est également utilisée pour calculer la concaténation de deux automates non-déterministes\footnote{Nous nous y intéresserons plus tard lorsque nous aborderons la construction inductive des automates depuis une expression rationnelle.}.
      Pour la version générale de la notion de combinaison Booléenne d'automates, voir le Pseudo-Code~\ref{CodeCombiGenBoolDFAs} en annexe.
  \end{remarqueBox}

  Au même titre qu'une opération ensembliste, un produit tensoriel peut vérifier des propriétés remarquables.
  Si une opération associative et unifère fonde un monoïde, un produit tensoriel naturellement associatif et unifère fonde une \firstocc{catégorie monoïdale}.

  Une \firstocc{catégorie monoïdale}{catégorie!monoïdale} est un quintuplet \((\mathcal{C}, \otimes, I, \alpha, \lambda, \rho)\) où
  \begin{itemize}
    \item \(\mathcal{C} \) est une catégorie;
    \item \(\otimes \) est un produit tensoriel sur \(\mathcal{C}\);
    \item \(I\), \firstocc{l'unité}, est un objet de \(\mathcal{C}\);
    \item \(\alpha \), \firstocc{l'associateur}, est un isomorphisme naturel entre deux produits tensoriels ternaires, dont chaque composant \(\alpha_{A,B,C}\) est un morphisme de \((A\otimes B) \otimes C\) vers \(A\otimes (B\otimes C)\);
    \item \(\lambda \), \firstocc{l'identité gauche}, est un isomorphisme naturel dont chaque composant \(\lambda_A\) est un morphisme de \(I\otimes A\) vers \(A\);
    \item \(\rho \), \firstocc{l'identité droite}, est un isomorphisme naturel dont chaque composant \(\rho_A\) est un morphisme de \(A\otimes I\) vers \(A\).
  \end{itemize}
  \begin{implementationBox}[]{Catégories monoïdales}
    \begin{minted}[xleftmargin=1em,   autogobble, fontsize=\footnotesize]{haskell}
      -- module Category.MonoidalCategory

      -- Classe de type des catégories monoïdales.
      class Bifunctor tensor cat => Monoidal cat tensor where

        -- Unité de la catégorie monoïdale
        type Unit cat tensor :: *

        -- Associateurs
        alpha :: cat (tensor (tensor (a, b), c)) (tensor (a, tensor (b, c)))
        alphaInv :: cat (tensor (a, tensor (b, c))) (tensor (tensor (a, b), c))

        -- Identités gauches
        lambda :: cat (tensor (Unit cat tensor, a)) a
        lambdaInv :: cat a (tensor (Unit cat tensor, a))

        -- Identités droites
        rho :: cat (tensor (a, Unit cat tensor)) a
        rhoInv :: cat a (tensor (a, Unit cat tensor))
    \end{minted}
  \end{implementationBox}
  Le produit cartésien ou la somme disjointe permettent de définir une catégorie monoïdale sur la catégorie des types.
  Si l'unité pour le produit cartésien est le type \texttt{()} à un seul élément, l'unité pour la somme disjointe est le type vide sans aucun élément, appelé \texttt{Void} en Haskell.
  \begin{remarqueBox}{Le type \texttt{Void}}
    Le type \texttt{Void} est un type inhabité, correspondant  à la constante fausse dans l'isomorphisme Curry-Howard.
    L'application de l'\emph{ex falso sequitur quodlibet} est la fonction \texttt{absurd} de signature \texttt{Void -> a}, définie pour tout type \texttt{a}.
  \end{remarqueBox}
  \begin{implementationBox}[label={RemVoid}]{Catégories monoïdales des types}
    \begin{minted}[xleftmargin=1em,   autogobble, fontsize=\footnotesize]{haskell}
      -- module Category.OfStructures.CategoryOfTypes

      -- La catégorie monoïdale associée au produit cartésien
      instance Monoidal (->) Prod where
        type Unit (->) Prod = ()

        alpha (Product (Product (a, b), c)) = Product (a, Product (b, c))
        alphaInv (Product (a, Product (b, c))) = Product (Product (a, b), c)

        lambda (Product ((), a)) = a
        lambdaInv a = Product ((), a)

        rho (Product (a, ())) = a
        rhoInv a = Product (a, ())

      -- La catégorie monoïdale associée à la somme disjointe
      instance Monoidal SumTypeCat Sum where
        type Unit SumTypeCat Sum = Void

        alpha = SumTypeCat $ \case
          Fst (Snd b) -> Snd $ Fst b
          Fst (Fst a) -> Fst a
          Snd c       -> Snd $ Snd c
        alphaInv = SumTypeCat $ \case
          Snd (Fst b) -> Fst $ Snd b
          Fst a       -> Fst $ Fst a
          Snd (Snd c) -> Snd c

        lambda = SumTypeCat $ \case
          Snd a -> a
          Fst v -> absurd v
        lambdaInv = SumTypeCat Snd

        rho       = SumTypeCat $ \case
          Fst a -> a
          Snd v -> absurd v
        rhoInv = SumTypeCat Fst
    \end{minted}
  \end{implementationBox}
  Le produit tensoriel de types gradués s'exprime également comme un foncteur.
  Pour cela, il faut d'abord définir la catégorie de départ sur laquelle définir un bifoncteur.
  Cette catégorie est la catégorie des types gradués, dont les morphismes sont les fonctions préservant l'arité.
  \begin{implementationBox}[]{Types gradués et produit tensoriel}
    \begin{minted}[xleftmargin=1em,   autogobble, fontsize=\footnotesize]{haskell}
      -- module module Category.OfGraded.CategoryOfGraded

      -- Type de données pour les fonctions préservant la graduation,
      -- et définition de la fonction de projection
      -- applyMorph
      data GradedMorph a b where
        GradedMorph ::
          (Graded a, Graded b)
            => {applyMorph ::
              forall n . Singleton n => Graduation a n -> Graduation b n}
                -> GradedMorph a b

      -- Les fonctions préservant la graduation forment une
      -- catégorie
      instance CategoryGen GradedMorph where

        GradedMorph f . GradedMorph g = GradedMorph $ f . g

        id = GradedMorph id

      -- Le produit tensoriel de type gradué est un bifoncteur
      instance FunctorGen TensorProd (ProdCat GradedMorph GradedMorph) GradedMorph where

        fmap (Prod (GradedMorph f) (GradedMorph g)) =
          GradedMorph $ \(GradTensor x ys) -> GradTensor (f x) (mapVect g ys)
    \end{minted}
  \end{implementationBox}
  Ce produit tensoriel est:
  \begin{itemize}
    \item associatif: pour un élément \(a\) d'arité \(k\), pour un vecteur \(b\) de \(k\) éléments \(b_1, \ldots, b_k\) d'arités respectives \(n_1, \ldots, n_k\), pour un vecteur \(c\) de \(n_1+\cdots+n_k\) éléments \(c_{1,1},\ldots,c_{1,n_1},\ldots, c_{k,1},\ldots,c_{k,n_k}\), l'associateur envoie l'élément \(((a, b), c)\)
      sur l'élément
      \begin{equation}\label{eq carve}
        (a, (b_1, (c_{1,1},\ldots,c_{1,n_1}), \ldots, b_k, (c_{k,1},\ldots,c_{k,n_k})));
      \end{equation}
      la transformation des deux vecteurs \(b\) et \(c\) en un vecteur de \(k\) produits tensoriels est implantée par la méthode \texttt{carve} de la classe \texttt{Carve} du module \texttt{Graded.GradedCoupleVector} (voir la Remarque~\ref{remCarve});
    \item unifère, en notant \(\ast \) l'élément d'arité 1 d'un ensemble à un élément:
      \begin{itemize}
        \item l'associateur gauche envoie un élément de la forme \((\ast, (x))\) sur \(x\);
        \item l'associateur droit envoie un élément de la forme \((x, (\ast, \ldots, \ast))\) sur \(x\).
       \end{itemize}
  \end{itemize}

  \begin{remarqueBox}[label={remCarve}]{Fonctions type-dépendantes et méthodes}
    Les fonctions type-dépendantes peuvent être implantées en Haskell par le biais des classes de types et des familles de types.

    Il suffit pour cela de définir les fonctions sur l'ensemble des types possibles, par induction sur la forme des types.

    C'est le cas par exemple de la méthode \texttt{carve} du module \texttt{Graded.GradedCoupleVector} permettant de recombiner les éléments d'un produit tensoriel.
    Pour cela sont définies des familles de types permettant d'exprimer le type de la sortie en fonction des types d'entrée.

    Le moteur d'inférence n'étant pas capable d'induire les équivalences de type non triviales, il peut être nécessaire de coder des preuves d'équivalence et des implications de contraintes.
    C'est le cas de la fonction \texttt{carveProof} du module \texttt{Graded.GradedCoupleVector}, permettant de justifier que la fonction \texttt{carve} est applicable pour toutes listes de \texttt{Nat} vérifiant des conditions minimales, à savoir que la longueur du vecteur \(c\) doit être la somme des arités des éléments du vecteur \(b\) de l'équation~\eqref{eq carve}.
  \end{remarqueBox}


  Le produit tensoriel précédent permet ainsi de définir une catégorie monoïdale sur les types gradués.

  \begin{implementationBox}[label={PseudoCodeCatMonGraded}]{Catégorie monoïdale des types gradués}
    \begin{minted}[xleftmargin=1em,   autogobble, fontsize=\footnotesize]{haskell}
      -- module module Category.OfGraded.CategoryOfGraded

      -- La catégorie monoïdale des types gradués
      instance Monoidal GradedMorph TensorProd where
        type Unit GradedMorph TensorProd = ()

        alpha = GradedMorph $ \(GradTensor (GradTensor a bs) cs) ->
          GradTensor a
            (mapVect (\(GradCouple b c) -> GradTensor b c) $ carve bs cs)

        lambda = GradedMorph $ \(GradTensor GradUn (GradVCons y GradVNil)) -> y
        lambdaInv =
          GradedMorph $ \y ->  GradTensor GradUn (GradVCons y GradVNil)

        rho = GradedMorph $ \(GradTensor x _) -> x
        rhoInv = GradedMorph $ \x ->
          GradTensor x $ replicateGrad (graduation' x) GradUn
    \end{minted}
  \end{implementationBox}
  Une fois fixé, un produit tensoriel permet également de mettre en lumière certains objets particuliers des catégories monoïdales.
  En effet, certains objets admettent une stabilité par rapport à un produit tensoriel; c'est le cas par exemple des monoïdes dans la catégorie monoïdale des types en considérant le produit cartésien comme produit tensoriel:
  la loi de composition interne d'un monoïde sur un ensemble sous-jacent \(M\) permet d'envoyer \(M\times M\) sur \(M\); son élément neutre, équivalent à un morphisme de \( \{ \ast \} \) (l'ensemble à un élément) vers \(M\), permet d'établir une propriété d'identité à partir des morphismes et du produit tensoriel.
  Pour résumer, il est possible de généraliser les monoïdes au sein de toutes catégories monoïdales; ces objets particuliers sont les \firstocc{monoïdes objets}, à la base de descriptions surprenantes de structures algébriques.

  Un \firstocc{monoïde objet}{monoïde objet} d'une catégorie monoïdale \((\mathcal{C}, \otimes, I, \alpha, \lambda, \rho)\) est un triplet \((M, \mu, \eta)\) avec:
  \begin{itemize}
    \item \(M\) un objet de \(\mathcal{C}\),
    \item \(\mu \) un morphisme de \(M\otimes M\) vers \(M\),
    \item \(\eta \) un morphisme de \(I\) vers \(M\)
  \end{itemize}
  faisant commuter les diagrammes suivants:

 \begin{minipage}{0.45\linewidth}
   \begin{equation*}
     \begin{tikzpicture}[baseline=(current  bounding  box.center)]
       \matrix (m) [matrix of math nodes,row sep=3em,column sep=3.5em,minimum width=3em]{
         M \otimes M \otimes M & M \otimes (M \otimes M) & M \otimes M\\
         M \otimes M & & M\rlap{,} \\
       };
       \path[-stealth]
         (m-1-1)
           edge node [above] {\( \alpha_{M,M,M} \)} (m-1-2)
           edge node [left] {\( \mu \otimes \mathrm{Id}_M \)} (m-2-1)
         (m-1-2)
           edge node [above] {\( \mathrm{Id}_M \otimes \mu \)} (m-1-3)
         (m-1-3)
           edge node [right] {\( \mu \)} (m-2-3)
         (m-2-1)
           edge node [below] {\( \mu \)} (m-2-3)
           ;
     \end{tikzpicture}
   \end{equation*}
 \end{minipage}
 \hfill
 \begin{minipage}{0.45\linewidth}
   \begin{equation*}
     \begin{tikzpicture}[baseline=(current  bounding  box.center)]
       \matrix (m) [matrix of math nodes,row sep=3em,column sep=3.5em,minimum width=3em]{
         I \otimes M & M \otimes M & M \otimes I \\
         & M\rlap{.} & \\
       };
       \path[-stealth]
         (m-1-1)
           edge node [above] {\( \eta \otimes \mathrm{Id}_M \)} (m-1-2)
           edge node [below left] {\( \lambda_M \)} (m-2-2)
         (m-1-2)
           edge node [left] {\(  \mu \)} (m-2-2)
         (m-1-3)
           edge node [above] {\( \mathrm{Id}_M \otimes \eta \)} (m-1-2)
           edge node [below right] {\( \rho_M \)} (m-2-2)
           ;
     \end{tikzpicture}
   \end{equation*}
 \end{minipage}

 \begin{implementationBox}[label={CategorieDesMonObj}]{Monoïdes objets et leurs catégories}
   \begin{minted}[xleftmargin=1em,   autogobble, fontsize=\footnotesize]{haskell}
     -- module Category.MonoidalCategory

    -- Classe de types des monoïdes objets d'une catégorie monoïdale
    class (Monoidal cat tensor) => MonoidObject m cat tensor where
      mu :: cat (tensor (m, m)) m
      eta :: cat (Unit cat tensor) m

    -- Type de données des morphismes de monoïdes objets,
    -- et définition de la fonction de projection
    -- runMonoidObjectMorph
    data MonoidObjectMorph cat tensor m m' where
      MonoidObjectMorph ::(MonoidObject m cat tensor, MonoidObject m' cat tensor)
        => {runMonoidObjectMorph :: cat m m'} -> MonoidObjectMorph cat tensor m m'

    -- Les morphismes de monoïdes objets forment une catégorie
    instance Monoidal cat tensor => CategoryGen (MonoidObjectMorph cat tensor) where

      id = MonoidObjectMorph id

      MonoidObjectMorph f . MonoidObjectMorph g = MonoidObjectMorph $ f . g
   \end{minted}
  \end{implementationBox}
  Si les monoïdes objets de la catégorie monoïdale des types munis du produit cartésien sont les monoïdes classiques, d'autres structures peuvent être définies par le biais des monoïdes objets.

  Par exemple, le principe d'Eckmann-Hilton~\cite{EH62} permet de montrer que les monoïdes objets de la catégorie monoïdale des monoïdes, construite avec le produit direct de monoïde en produit tensoriel, sont les monoïdes commutatifs.
  En effet, pour \((M, \cdot, \varepsilon)\) un monoïde et \((M, \cdot', \varepsilon')\), c'est-à-dire  \(\cdot' \) un morphisme de monoïde de \((M\times M, \odot, (\varepsilon, \varepsilon))\) dans \((M, \cdot, \varepsilon)\) et \(\varepsilon'\) un élément neutre pour \(\cdot'\)\footnote{Formellement, \(\varepsilon'\) est un morphisme du monoïde \((\{()\}, \_, ())\) à un élément dans \(M\), assimilable à un élément de \(M\).}, on a:
  \begin{align*}
    (x, x') \odot (y, y') &= (x \cdot y, x' \cdot y') & \text{(Définition \(\odot \))}\\
    (\cdot') ((x, x') \odot (y, y')) &= (\cdot') (x, x') \cdot (\cdot') (y, y') & \text{\((\cdot'\) morphisme de monoïde)}\\
    &= (x \cdot' x') \cdot (y \cdot' y') &\text{(Réécriture infixe)}
  \end{align*}
  Ainsi
  \begin{equation*}
    (\cdot') ((x \cdot y, x' \cdot y')) = (x \cdot y) \cdot' (x' \cdot y'),
  \end{equation*}
  c'est-à-dire
  \begin{equation*}
    (x \cdot y) \cdot' (x' \cdot y') = (x \cdot' x') \cdot (y \cdot' y').
  \end{equation*}
  Avec ceci, on peut montrer que
  \begin{itemize}
    \item \(\varepsilon = \varepsilon'\):
      \begin{align*}
        \varepsilon = \varepsilon \cdot \varepsilon &= (\varepsilon \cdot' \varepsilon') \cdot (\varepsilon' \cdot' \varepsilon)\\
        &= (\varepsilon \cdot \varepsilon') \cdot' (\varepsilon'\cdot \varepsilon) =  \varepsilon' \cdot' \varepsilon' = \varepsilon';
      \end{align*}
    \item \(\cdot = \cdot' \) et \(\cdot \) commute:
      \begin{alignat*}{3}
        x \cdot y &= (\varepsilon' \cdot' x) \cdot (y \cdot' \varepsilon' ) &
        &= (\varepsilon' \cdot y) \cdot' (x \cdot \varepsilon')\\
        &= (\varepsilon \cdot y) \cdot' (x \cdot \varepsilon) &
        &= y \cdot' x\\
        &= (y \cdot \varepsilon) \cdot' (\varepsilon \cdot x) &
        &= (y \cdot' \varepsilon) \cdot (\varepsilon \cdot' x)\\
        &= (y \cdot' \varepsilon') \cdot (\varepsilon' \cdot' x) &
        &= y \cdot x.
      \end{alignat*}
    \item \(\cdot \) est associative:
      \begin{align*}
        (x \cdot y) \cdot z &= (x \cdot y) \cdot (\varepsilon \cdot z)\\
        &= (x\cdot \varepsilon) \cdot (y \cdot z)
        = x \cdot (y\cdot z).
      \end{align*}
  \end{itemize}
  D'où \((M, \cdot, \varepsilon)\) est un monoïde commutatif.
  D'autres structures peuvent être définies à l'aide de monoïdes objets:
  \begin{itemize}
    \item les semi-anneaux sont les monoïdes objets de la catégorie des monoïdes commutatifs;
    \item les semi-algèbres sur un semi-anneau sont les monoïdes objets de la catégorie des semi-modules (sur le semi-anneau ambiant);
    \item les catégories monoïdales\footnote{strictes pour être précis, c'est-à-dire où les isomorphismes naturels sont des identités.} sont les monoïdes objets de la catégorie des (petites\footnote{distinction permettant d'éviter le paradoxe de Russel.}) catégories.
  \end{itemize}
  Y compris les monades:
  \begin{aquote}{Saunders MacLane, Categories for the Working Mathematician~\cite{MacLane71}}
    \emph{All told, a monad in [a category] \(\mathcal{C}\) is just a monoid in the category of endofunctors of \(\mathcal{C}\), with product \(\otimes \) replaced by composition of endofunctors and unit set by the identity endofunctor.}
  \end{aquote}


  \section{Retour aux opérades}\label{secOperadesReturn}

  Quant aux opérades, ce sont les monoïdes objets de la catégorie des types gradués.
  On peut donc les implanter ainsi.
  \begin{implementationBox}[label={CodeOperades}]{Opérades}
    \begin{minted}[xleftmargin=1em,   autogobble, fontsize=\footnotesize]{haskell}
      -- module Algebra.Structures.Operad

      -- Synonyme de type pour définir une opérade
      type Operad a = MonoidObject a GradedMorph TensorProd

      -- Synonyme de type pour les morphismes d'opérades
      type OperadMorphism o o' = MonoidObjectMorph GradedMorph TensorProd o o'

      -- Définition de la composition d'une opérade en fonction
      -- du morphisme mu du monoïde objet
      operadComposition
        :: ( Operad a
           , Graded a
           )
        => Graduation a (Length ns)
        -> GradVect ns a
        -> Graduation a (SumNat ns)
      operadComposition a as = applyMorph mu (GradTensor a as)

      -- Définition de l'élément neutre de l'opérade à partir
      -- du morphisme eta du monoïde objet
      operadUnit :: Operad a => Graduation a Nat_1
      operadUnit = applyMorph eta GradUn
    \end{minted}
   \end{implementationBox}
   Un exemple d'opérade non-triviale est l'opérade des fonctions \(n\)-aires sur un type donné munie de la composition de fonctions, implantée dans le module \texttt{Algebra.Functions.GradedFunctions}\footnote{Comme tous les modules développés pour ce document, la documentation est accessible \emph{via} la commande \texttt{stack haddock -open}.} sous le type général \texttt{MultiFun}.
   On peut également s'intéresser la construction de l'opérade libre depuis un type gradué.
   \begin{implementationBox}[]{Opérade libre}
     \begin{minted}[xleftmargin=1em,   autogobble, fontsize=\footnotesize]{haskell}
       -- module Algebra.Structures.Operad

       -- L'opérade libre sur le type gradué a est constitué soit de l'élément
       -- neutre Eps, soit d'une composition Op d'un élément de a d'arité n et
       -- d'un vecteur de n éléments de l'opérade libre.
       data FreeOperad a where
         Eps :: FreeOperad a
         Op :: Graduation a (Length arites) -> GradVect arites (FreeOperad a) -> FreeOperad a
     \end{minted}
  \end{implementationBox}
  Ce type peut être gradué comme suit.
  \begin{implementationBox}[]{Graduation de l'opérade libre}
    \begin{minted}[xleftmargin=1em,   autogobble, fontsize=\footnotesize]{haskell}
      -- module Algebra.Structures.Operad

      -- L'opérade libre sur a  est un type gradué .
      instance Graded a => Graded (FreeOperad a) where

        -- Valeurs du type gradué :
        data Graduation (FreeOperad a) n where
          -- GradEps, d'arité 1
          GradEps ::Graduation (FreeOperad a) Nat_1
          -- GradOp, composition d'un élément de a d'arité n
          -- et d'un vecteur de n éléments gradués de l'opérade libre;
          -- la graduation est alors la somme de celles des éléments
          -- du vecteur
          GradOp ::
            Graduation a (Length as)
              -> GradVect as (FreeOperad a)
                -> Graduation (FreeOperad a) (SumNat as)

        -- Fonctions de conversion
        fromGrad GradEps       = Eps
        fromGrad (GradOp a as) = Op a as
        toGrad Eps       = GradEps
        toGrad (Op a as) =  GradOp a as
    \end{minted}
   \end{implementationBox}
  Montrons alors que l'opérade libre est une opérade.
  Pour cela, définissons la fonction de composition.
  \textbf{NB:} Cette fonction utilise la méthode type-dépendante \texttt{carve} (voir Remarque~\ref{remCarve}) et la fonction \texttt{mapVect} (voir Remarque~\ref{remGradVect}).
  \begin{implementationBox}[]{L'opérade libre est une opérade}
    \begin{minted}[xleftmargin=1em,   autogobble, fontsize=\footnotesize]{haskell}
      -- module Algebra.Structures.Operad

      -- Composition d'un élément de l'opérade libre
      -- d'arité n et d'un vecteur de n éléments gradués
      -- pour produire un élément de l'opérade libre
      -- ayant pour arité la somme de celles des éléments
      -- du vecteur
      compo
        :: Graded a
        => Graduation (FreeOperad a) (Length ns)
        -> GradVect ns (FreeOperad a)
        -> Graduation (FreeOperad a) (SumNat ns)
      compo GradEps (GradVCons f GradVNil) = f
      compo (GradOp f fs) gs =
        GradOp f
          $ mapVect (\(GradCouple b ps) -> compo b ps)
            $ carve fs gs

      -- L'opérade libre est une opérade, c'est-à-dire un monoïde
      -- objet de la catégorie des types gradués.
      instance Graded a =>
        MonoidObject (FreeOperad a) GradedMorph TensorProd where

        mu  = GradedMorph $ \(GradTensor f xs) -> compo f xs
        eta = GradedMorph $ \GradUn -> GradEps
    \end{minted}
  \end{implementationBox}


Au même titre que le foncteur \texttt{List} est une monade de la catégorie des types, on peut montrer que le constructeur d'opérade libre est une monade de la catégorie des types gradués.
\begin{implementationBox}[]{Opérade libre et monade}
  \begin{minted}[xleftmargin=1em,   autogobble, fontsize=\footnotesize]{haskell}
    -- module Algebra.Structures.Operad

    -- L'opérade libre est un endofoncteur de la catégorie des types gradués.
    instance FunctorGen FreeOperad GradedMorph GradedMorph where

      fmap g@(GradedMorph f) = GradedMorph $ \case
        GradEps       -> GradEps
        (GradOp x xs) -> GradOp (f x) $ mapVect (applyMorph $ fmap g) xs

    -- Le foncteur opérade libre est une monade de
    -- la catégorie des types gradués.
    instance MonadGen FreeOperad GradedMorph where
      return = GradedMorph runId . unit
      join   = GradedMorph $ \case
        GradEps     -> GradEps
        GradOp x xs -> operadComposition x $ mapVect (applyMorph join) xs
  \end{minted}
\end{implementationBox}
Au même titre qu'il existe une adjonction entre les foncteurs \texttt{List} et Identité dans la catégorie des types et celle des monoïdes, on peut définir une adjonction entre l'opérade libre et le foncteur Identité dans la catégorie des types gradués et celle des opérades.
\begin{implementationBox}[label = {PseudoCodeAdjOperad}]{Adjonction opérade libre et identité}
  \begin{minted}[xleftmargin=1em,   autogobble, fontsize=\footnotesize]{haskell}
    -- module Category.OfGraded.CategoryOfGraded

    -- Le constructeur Identité est un foncteur de la catégorie des Opérades
    -- vers celle des types gradués
    instance FunctorGen Identity (MonoidObjectMorph GradedMorph TensorProd) GradedMorph where

      fmap (MonoidObjectMorph (GradedMorph f)) =
        GradedMorph $ \(GradId x) -> GradId $ f x

    -- module Algebra.Structures.Operad

    -- Le constructeur d'opérade libre est un foncteur de la catégorie des
    -- types gradués vers celle des opérades
    instance FunctorGen FreeOperad GradedMorph (MonoidObjectMorph GradedMorph TensorProd) where

      fmap = MonoidObjectMorph . fmap

    -- Adjonction opérade libre / identité
    instance AdjunctionGen FreeOperad Identity
      GradedMorph (MonoidObjectMorph GradedMorph TensorProd) where

      -- Unité de l'adjonction, envoyant un élément a d'arité n
      -- sur la composition de a avec un vecteur de longueur n
      -- contenant n fois la valeur GradEps
      unit = GradedMorph $ \a ->
        GradId $ GradOp a $ replicateGrad (graduation' a) GradEps

      -- Co-unité de l'adjonction, envoyant un élément de l'opérade
      -- libre construite sur une opérade a vers a
      counit = MonoidObjectMorph $ GradedMorph aux'
       where
        aux'
          :: (Operad a, Graded a)
          => Graduation (FreeOperad (Identity a)) n
          -> Graduation a n
        aux' GradEps = applyMorph eta GradUn
        aux' (GradOp (GradId a) as) =
          applyMorph mu $ GradTensor a $ mapVect aux' as
      \end{minted}
    \end{implementationBox}

\begin{figure}[H]
    \centerline{
      \begin{tikzpicture}[node distance=2.5cm, bend angle=30]
      \node[state] (A) {\(A\)};
      \node[state, right of=A] (B) {\(B\)};
      \node[state, right of=B] (C) {\(C\)};
      \node[state, right of=C] (D) {\(D\)};
      \node[state, right of=D] (E) {\(E\)};
      \node[state, above of=C, accepting] (S0) {\(0\)};
      \node[below of=C] (Start) {\(\wedge\)};
      \path[->]
        (A)   edge [above left] node {\(A\)} (S0)
        (A)   edge [left, in=210, out=150, loop] node {\(B, C, D, E\)} ()
        (B)   edge [above left] node {\(B\)} (S0)
        (B)   edge [left, in=210, out=150, loop] node {\(A, C, D, E\)} ()
        (C)   edge [left] node {\(C\)} (S0)
        (C)   edge [left, in=210, out=150, loop] node {\(A, B, D, E\)} ()
        (D)   edge [above right] node {\(D\)} (S0)
        (D)   edge [right, in=-30, out=30, loop] node {\(A, B, C, E\)} ()
        (E)   edge [above right] node {\(E\)} (S0)
        (E)   edge [right, in=-30, out=30, loop] node {\(A, B, C, D\)} ()
        (S0)   edge [above, in=120, out=60, loop] node {\(A, B, C, D, E\)} ()
        (Start) edge [above] node {} (A)
        (Start) edge [above] node {} (B)
        (Start) edge [above] node {} (C)
        (Start) edge [above] node {} (D)
        (Start) edge [above] node {} (E)
        ;
      \end{tikzpicture}
    }
    \caption{Un automate alternant reconnaissant les mots contenant tous les symboles de \( \{A,B,C,D,E \} \).}%
    \label{fig alternant}
  \end{figure}
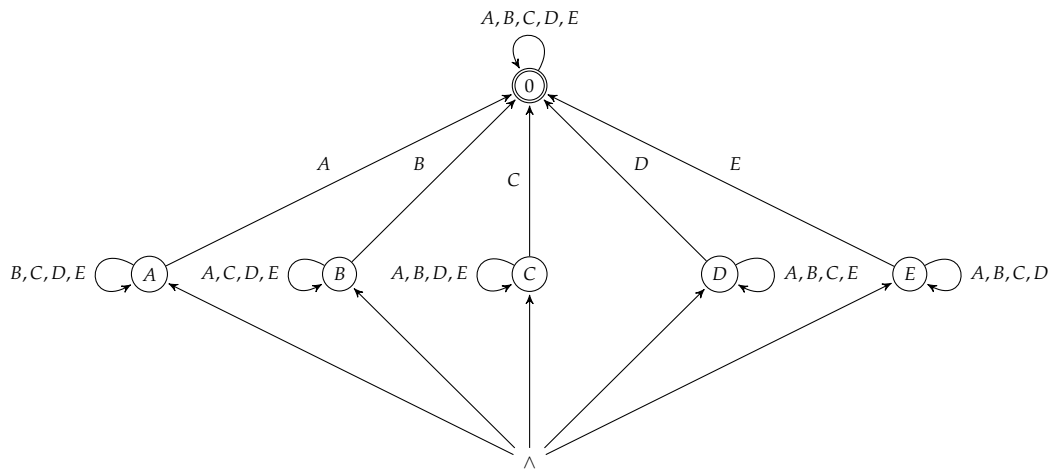

Cette définition monadique permet de promouvoir automatiquement toute fonction graduée entre un type gradué et une opérade en morphisme d'opérade.
\begin{implementationBox}[]{Promotion de morphisme}
  \begin{minted}[xleftmargin=1em,   autogobble, fontsize=\footnotesize]{haskell}
    -- module Category.OfGraded.CategoryOfGraded

    -- Promotion fonction graduée vers morphisme d'opérade
    promoteToOperadMorphism
      :: (Operad b, Graded b, Graded a)
      => (Graduation a n -> Graduation b n)
      -> OperadMorphism (FreeOperad a) b
    promoteToOperadMorphism f = promote $ GradedMorph $ GradId . f

    -- Promotion d'une fonction graduée entre un type gradué et
    -- une opérade en fonction graduée.
    promoteFunToOperad
      :: (Operad o, Graded o, Graded a)
      => (Graduation a n -> Graduation o n)
      -> Graduation (FreeOperad a) m
      -> Graduation o m
    promoteFunToOperad f =
      applyMorph $ runMonoidObjectMorph $ promoteToOperadMorphism f
\end{minted}
\end{implementationBox}
Comme indiqué dans le début de la section, nous utiliserons les opérades pour représenter les arbres, entrées des automates d'arbres.
Mais les arbres et les opérades peuvent également être utilisés pour modéliser des expressions comme éléments d'arité nulle, telles que les expressions Booléennes, formées par des opérateurs de négation, conjonction et disjonction, ainsi que des variables et des Booléens.
\begin{implementationBox}[label={CodeExprBool}]{Expressions Booléennes}
  \begin{minted}[xleftmargin=1em,   autogobble, fontsize=\footnotesize]{haskell}
  -- module Algebra.Structures.BoolExpr

  -- Définition du type des symboles pouvant apparaître dans
  -- une expression Booléenne.
  data Op a = BoolOp Bool | NotOp | AndOp | OrOp | VarOp a

  -- Définition de la graduation des symboles et fonctions
  -- de conversion
  instance Graded (Op a) where
    data Graduation (Op a) n where
      Var ::a -> Graduation (Op a) Nat_0
      Bool ::Bool -> Graduation (Op a) Nat_0
      Not ::Graduation (Op a) Nat_1
      And ::Graduation (Op a) Nat_2
      Or ::Graduation (Op a) Nat_2

    fromGrad (Var  v) = VarOp v
    fromGrad (Bool b) = BoolOp b
    fromGrad Not      = NotOp
    fromGrad And      = AndOp
    fromGrad Or       = OrOp
    toGrad (VarOp  b) = Var b
    toGrad (BoolOp b) = Bool b
    toGrad NotOp      = Not
    toGrad AndOp      = And
    toGrad OrOp       = Or

  -- Une expression booléenne est un élément de l'opérade libre
  -- construite depuis les symboles et d'arité 0.
  data BoolExpr a =
    BoolExpr {runBoolExpr :: Graduation (FreeOperad (Op a)) Nat_0}

  -- Si f est une fonction graduée envoyant les symboles gradués
  -- sur des fonctions n-aires sur un type b, alors il est possible
  -- d'envoyer une expression Booléenne (élément d'arité 0 de l'opérade
  -- libre) sur un élément de b (élément d'arité 0 de l'opérade
  -- des fonctions).
  -- NB: Graduation (MultiFun b b) n est le type des fonctions
  -- n-aires sur b implanté dans le module
  -- Algebra.Functions.GradedFunctions
  promoteOpFunToBoolExpr
    :: (Graduation (Op a) n -> Graduation (MultiFun b b) n)
    -> BoolExpr a
    -> b
  promoteOpFunToBoolExpr f = runGradFun . promoteFunToOperad f . runBoolExpr

  -- Conversion d'un élément d'arité 0 de l'opérade libre sur
  -- le type Op appliqué aux Booléens
  -- en un Booléen, par utilisation de la fonction promoteOpFunToBoolExpr :
  -- il suffit d'associer à chaque opérateur une fonction booléenne;
  -- le calcul est ensuite une conséquence de la composition de
  -- l'opérade des fonctions Booléennes n-aires.
  -- NB: Graduation (MultiFun Bool Bool) n est le type des fonctions
  -- Booléennes n-aires.
  instance Castable (Graduation (FreeOperad (Op Bool)) Nat_0) Bool where
    cast = runGradFun . promoteFunToOperad evalOpToBoolFun
     where
      evalOpToBoolFun
        :: Graduation (Op Bool) n -> Graduation (MultiFun Bool Bool) n
      evalOpToBoolFun (Var  b) = GradFun b
      evalOpToBoolFun (Bool b) = GradFun b
      evalOpToBoolFun Not      = GradFun not
      evalOpToBoolFun And      = GradFun (&&)
      evalOpToBoolFun Or       = GradFun (||)
    castInv = freeOpReturn . Bool

  -- Extension au type BoolExpr
  instance Castable (BoolExpr Bool) Bool where
    cast    = cast . runBoolExpr
    castInv = BoolExpr . castInv

  \end{minted}
\end{implementationBox}
Dans le module \texttt{Algebra.Structures.BoolExpr}, on pourra retrouver les définitions des types de classes montrant que le constructeur \texttt{BoolExpr} est un endofoncteur, et même une monade de la catégorie des types.
Cela implique donc que ce type définit une catégorie de Kleisli, mais aussi un automate de Kleisli qui lui est associé; dans ce cas, on obtient les automates alternants, dont les configurations sont des expressions Booléennes d'états.
On retrouve alors les algorithmes classiques (conversion en automate non-déterministe en utilisant des formes clausales (voir le module \texttt{Algebra.Structures.FND}), conversion en automate déterministe) dont la proximité syntaxique (voire l'identité) est à remarquer (voir Code~\ref{codeAlgosClas}).
\begin{implementationBox}[]{Automates alternants}
  \begin{minted}[xleftmargin=1em,   autogobble, fontsize=\footnotesize]{haskell}
  -- module Automata.KleisliAutomata

  type AFA symbol state
    = Automaton (KleisliCat BoolExpr (->)) () symbol state Bool

  -- Conversion d'un AFA en un DFA complet
  afaToDfaComp :: AFA symbol state -> DFAComp symbol (BoolExpr state)
  afaToDfaComp = removeIdDFA . castFinality . applyFunctor

  -- Conversion d'un AFA en un NFA.
  afaToNfa :: AFA symbol state -> NFA symbol (ClauseC state)
  afaToNfa =
    convertFinalityWith (Kleisli $ \x -> castInv (cast x :: Bool))
      . applyFunctor
  \end{minted}
\end{implementationBox}
Pour faciliter l'écriture des expressions booléennes, les fonctions suivantes sont définies dans le module \texttt{Algebra.Structures.BoolExpr}:
\begin{itemize}
\item \texttt{packVar}: promotion d'une variable en expression Booléenne atomique;
\item \texttt{packBool}: promotion d'un Booléen en expression Booléenne atomique;
\item \texttt{(<\textasciitilde>)}: négation d'une expression booléenne;
\item \texttt{(<\&{}\&{}>)}: conjonction de deux expressions booléennes;
\item \texttt{(<||>)}: disjonction de deux expressions booléennes.
\end{itemize}
L'exemple du Pseudo-Code~\ref{codeAFA} montre comment implanter l'automate alternant de la Figure~\ref{fig alternant} reconnaissant l'ensemble des mots contenant toutes les lettres d'un alphabet donné (\( \{A,B,C,D,E \} \)); cet exemple est exécutable \emph{via} la commande \texttt{stack ghci src/HDRExample/AFA.hs}.


  \begin{implementationBox}[label={codeAFA}]{Automates alternants}
    \begin{minted}[xleftmargin=1em,   autogobble, fontsize=\footnotesize]{haskell}
    {-# LANGUAGE DeriveGeneric #-}

    module HDRExample.AFA where

    import           Algebra.Structures.BoolExpr
    import           Algebra.Structures.FND
    import           Automata.Automaton             ( recognizes )
    import           Automata.AutomatonTransition
    import           Automata.KleisliAutomata
    import           Data.Hashable
    import           Data.List
    import           Data.Maybe
    import           GHC.Generics                   ( Generic )

    -- Description de l'alphabet via un type énuméré
    data Symbol = A | B | C | D | E
      deriving (Bounded, Enum, Show, Eq, Generic)
    instance Hashable Symbol

    -- Synonyme de type pour les états de l'automate
    type State = Maybe Symbol

    -- AFA reconnaissant les mots qui contiennent tous les
    -- symboles de l'alphabet:
    -- * les symboles sont de type Symbol
    -- * chaque état est soit un symbole, soit un état ajouté (Nothing)
    a :: AFA Symbol State
    a = packKleisli initial delta final
     where
      -- la combinaison initiale est la conjonction des états
      -- associés aux symboles
      initial = foldr1 (<&&>) $ fmap (packVar . Just) [A ..]
      -- Les chemins "reconnus" depuis chaque état associé à un symbole "X"
      -- sont les mots contenant au moins une occurrence de "X";
      -- une fois ce symbole lu, l'état associé à "X" est envoyé sur
      -- l'état Nothing
      delta x (Just y) | x == y    = packVar Nothing
                       | otherwise = packVar $ Just y
      delta _ Nothing = packVar Nothing
      -- le seul état final est l'état Nothing
      final = isNothing

    -- Transformation de l'AFA a en un NFA, dont les états sont les clauses
    -- conjonctives issues des configurations possibles de l'automate a.
    -- Cet automate est un NFA minimal (voir la Remarque suivante)
    b :: NFA Symbol (ClauseC State)
    b = afaToNfa a

    -- Affichage des nombres d'états et de transitions des automates a et b,
    -- et de résultats de tests d'appartenance pour des mots construits
    -- depuis l'alphabet fixé.
    testAFA :: IO ()
    testAFA = do
      -- affichage des nombres d'états et de transitions de a
      putStrLn $ "États de a : " ++ show (length $ accessibleStates a)
      putStrLn $ "\nTransitions de a : " ++ show (length $ getTransitions a)

      -- affichage des nombres d'états et de transitions de b
      putStrLn $ "\nÉtats de b : " ++ show (length $ accessibleStates b)
      putStrLn $ "\nTransitions de b : " ++ show (length $ getTransitions b)

      -- permuts est la liste des mots où chaque symbole apparait
      -- une et une seule fois
      let permuts = permutations [A ..]
      -- teste si tous les mots de permuts sont reconnus
      -- (doit afficher True deux fois)
      print $ all (recognizes a) permuts
      print $ all (recognizes b) permuts

      -- permuts' est la liste des mots de longueur au plus (n - 1)
      -- où chaque symbole apparait au plus une fois,
      -- avec n est le nombre de symboles
      let permuts' = subsequences [A ..] \\ permuts
      -- teste si un mot de permuts' est reconnu
      -- (doit afficher False deux fois)
      print $ any (recognizes a) permuts'
      print $ any (recognizes b) permuts'
    \end{minted}
  \end{implementationBox}
  \begin{remarqueBox}[]{Automates minimaux}
    Le problème du calcul d'un automate non-déterministe minimal est un problème difficile.
    Cependant, des méthodes existent pour définir une borne minimale du nombre d'états que tout automate non-déterministe reconnaissant un langage donné doit vérifier.
    Par exemple~\cite{Bir92, GS96}, il peut être montré que pour tout langage rationnel \(L\) tel qu'il existe un ensemble de \(n\) couples de mots \( \{(x_i, y_i) \mid 1 \leq i \leq n\} \) satisfaisant:
    \begin{itemize}
      \item \(xi w_i \in L\) pour \(1\leq i\leq n\),
      \item \(x_i w_j \notin L\) ou \(x_j w_i \notin L\)  pour \(1\leq i, j\leq n\),
    \end{itemize}
    tout automate non-déterministe reconnaissant \(L\) a au moins \(n\) états.
    Le langage reconnu par l'automate alternant de l'exemple précédent (voir Pseudo-Code~\ref{codeAFA}), sur un alphabet que nous noterons \(\Sigma \), est un bon support pour cette propriété.
    Ainsi, à chaque sous-ensemble \(P\) de l'alphabet \(\Sigma \) on peut associer le couple \((x_P, w_P)\) où \(x\) est le plus petit mot (au sens lexicographique) constitué des lettres de \(\Sigma \setminus P\) et où \(w\) est le plus petit mot constitué des lettres de \(P\).
    Par définition, \(x_P w_P\) est dans \(L\).
    Pour deux sous-ensembles distincts \(P\) et \(P'\) de \(\Sigma \), notons (sans perte de généralités) un élément \(a\) de \(P \setminus P'\).
    Le symbole \(a\) n'apparaît ni dans \(x_{P}\) ni dans \(w_{P'}\).
    Ainsi, \(x_P w_{P'}\) n'est pas dans \(L\).
    Tout automate non-déterministe reconnaissant le langage \(L\) a par conséquent au moins \(2^n\) états, où \(n\) est la taille de l'alphabet.
    L'automate alternant défini dans ce même exemple a, quant à lui, \( n + 1 \) états.
  \end{remarqueBox}
  Telles que définies dans le Pseudo-Code~\ref{CodeExprBool}, les expressions booléennes sont généralisables.
  Au lieu d'utiliser des opérateurs Booléens (c'est-à-dire des opérateurs associés à des fonctions Booléennes), rien ne nous empêche d'utiliser des opérateurs associés à des fonctions sur d'autres types.
  Ces expressions généralisées sont implantées dans le module \texttt{Algebra.Structures.GenExpr}, permettant de construire, similairement aux expressions booléennes, des expressions en tant qu'opérade libre dont les symboles sont soit des variables (\emph{via} un constructeur \texttt{packVar}), soit des opérateurs représentant des fonctions sur un type fixé (\emph{via} un constructeur \texttt{packFunction}).
  Par exemple, en considérant un type de symboles \texttt{State} et des termes \texttt{p, q, r, s} de type \texttt{State}, on peut définir l'expression
  \(\sqrt{\frac{p^2+q^2+r^2+s^2}{4}}\), qui une fois évaluée par une fonction d'interprétation associant aux termes \texttt{p, q, r, s} des nombres, calcule la moyenne quadratique de ces valeurs.
  Pour les mêmes raisons que celles énoncées précédemment pour les expressions booléennes, les expressions généralisées, une fois le type des valeurs fixées, forment une monade.

  L'exemple suivant (Pseudo-Code~\ref{codeGenAFA}) montre l'utilisation des automates de Kleisli leur étant associés, à savoir des automates alternants généralisés, où les configurations d'états sont des expressions généralisées sur un type \texttt{t}, et où le poids d'un mot est un élément de type \texttt{t}; plus particulièrement dans l'exemple, on associe à un mot la moyenne quadratique du nombre d'occurrences de ses voyelles.
  Pour cela, il suffit de construire un automate alternant avec:
  \begin{itemize}
    \item pour états les voyelles minuscules a, e, i, o, u, y;
    \item des transitions envoyant un état \(x\) par une lettre \(y\) sur l'expression généralisée \(1+x\) si \(x=y\), \(x\) sinon;
    \item comme poids final \(0\) pour tout état;
    \item comme configuration initiale l'expression
      \begin{equation*}
        \sqrt{\frac{a^2+e^2+i^2+o^2+u^2+y^2}{\mathrm{Ind}(a)+\cdots+\mathrm{Ind}(y)}}
      \end{equation*}
      où \(\mathrm{Ind}(n)\) renvoie \(0\) si \(n = 0\), \(1\) sinon.
  \end{itemize}

  \begin{implementationBox}[label={codeGenAFA}]{Automates alternants généralisés}
    \begin{minted}[xleftmargin=1em,   autogobble, mathescape=true, fontsize=\footnotesize]{haskell}
      module HDRExample.GenAFA where

      import           Algebra.Structures.GenExpr
      import           Automata.Automaton
      import           Automata.KleisliAutomata
      import           Data.Text.ICU.Char
      import qualified Data.Text                     as T
      import           Data.Text.ICU.Char
      import           Data.Text.ICU.Normalize
      import           Tools.Vector
      import           Type.Natural

      type Symbol = Char
      type State = Char

      -- fonction auxiliaire de composition d'une fonction
      -- unaire avec une fonction binaire
      (.:) :: (b -> c) -> (a1 -> a2 -> b) -> a1 -> a2 -> c
      (.:) = (.) . (.)

      -- automate calculant la moyenne quadratique des occurrences
      -- de caractères donnés dans une liste d'un mot
      -- en ne tenant compte que de ceux présents
      auto :: String -> GenAFA Double Symbol State
      auto symbols = packKleisli (initial_a symbols) delta_a final_a
       where
        -- si symbols' = [x1, ..., xn], alors
        -- initial_a symbols' = ((x1^2+...+xn^2)/(Ind(x1)+...+ Ind(xn))))^(1/2)
        -- où Ind(y1) vaut 0 si y vaut 0, 1 sinon.
        initial_a symbols' =
          packFunction one "^(1/2)" sqrt $ mkSingle $
            packFunction two "/" (/) $ mkTwo
              (foldr1 (packFunction two "+" (+) .: mkTwo) $
                fmap (packFunction one "^2" (** 2) . mkSingle . packVar) symbols'
              )
              (foldr1 (packFunction two "+" (+) .: mkTwo) $ fmap
                ( packFunction one "Ind" (\x -> if x == 0 then 0 else 1)
                . mkSingle
                . packVar
                )
                symbols'
              )
        -- chaque état agit comme un compteur pour chaque lettre
        -- s'incrémentant de 1 lors de la lecture de la "bonne" lettre
        delta_a c p | c == p    = packFunction one "1+" (1 +) $ mkSingle $ packVar c
                    | otherwise = packVar p
        -- la valeur "initiale" du compteur est 0
        final_a _ = 0 :: Double

      -- fonction utilitaire effaçant les accents des caractères
      removeAccents :: String -> String
      removeAccents =
        T.unpack . T.filter (not . property Diacritic) . normalize NFD . T.pack

      -- fonction associant à une chaine de caractères la moyenne quadratique
      -- de ses voyelles, où une voyelle est une lettre égale à un caractère
      -- de la chaine "aeiouy" après retrait d'un éventuel accent et unification
      -- de casse, en utilisant l'automate auto
      quadraticMeanOfVowels :: String -> Double
      quadraticMeanOfVowels =
        weightValue (auto "aeiouy") . fmap toLower . removeAccents
    \end{minted}
  \end{implementationBox}
  \begin{remarqueBox}[label={remTypeVector}]{Le type \texttt{Vector}}
    Le type \texttt{Vector n a}, où \texttt{n} est un \texttt{Natural} (Pseudo-Code~\ref{RemProgType}) et \texttt{a} un type, représente les vecteurs de longueur \texttt{n} d'éléments de \texttt{a}.
    Contrairement au type \texttt{GradVect ms a} du Pseudo-Code~\ref{pseudoCodeGradVect}, les éléments contenus ne sont pas gradués.
    Le type \texttt{Vector} est implanté dans le module \texttt{Tools.Vector}.
  \end{remarqueBox}
  Les opérades permettent donc de générer très facilement des types d'arbres par l'utilisation par exemple, des opérades libres et des adjonctions associées.
  Mais en tant que monoïdes objets, les opérades permettent également de combiner des morphismes (tels que des fonctions graduées) d'une autre façon que les compositions usuelles des catégories.

  L'introduction de ces compositions, permettant de combiner autre chose que des morphismes contenus dans un ensemble, est à la base d'une généralisation de la structure de catégories: les \firstocc{catégories enrichies}.


\section{Catégories enrichies}

  Une catégorie est définie, en partie, par sa composition, qui est une fonction permettant de combiner deux morphismes.
  Cette opération est par essence une opération binaire, c'est-à-dire une opération définissable à partir du produit cartésien de deux ensembles;
  en effet, pour rappel, la composition \(\circ \) d'une catégorie \(\mathcal{C}\) est une opération de signature \(\mathrm{Hom}_\mathcal{C}(B, C) \times \mathrm{Hom}_\mathcal{C}(A, B) \rightarrow \mathrm{Hom}_\mathcal{C}(A, C) \).

  De plus, le produit cartésien est une instance de produit tensoriel (voir Pseudo-Code~\ref{CodeProdTensTypes}).
  Une généralisation possible de la notion de catégorie peut alors être définie en considérant non plus le produit cartésien comme support de la composition, mais n'importe quel produit tensoriel d'une catégorie monoïdale.
  Ainsi, les objets à combiner ne sont plus des ensembles (structure algébrique des morphismes), mais des objets d'une catégorie monoïdale.
  De plus, le morphisme identité associé à un objet ne peut plus être un élément d'un ensemble de morphismes;
  cependant, comme un élément d'un ensemble \(X\) peut être vu comme un morphisme de l'ensemble à un élément (élément neutre du produit cartésien) vers \(X\), cette notion peut elle aussi être étendue.
  C'est ainsi qu'est définie une \firstocc{catégorie enrichie}.

  Une \firstocc{catégorie enrichie}{catégorie!enrichie} \(\mathcal{C}\) sur une catégorie monoïdale \(\mathfrak{M}=(\mathcal{M}, \otimes, I, \alpha, \lambda, \rho)\) est définie par:
  \begin{itemize}
    \item une classe \(\mathrm{Obj}_{\mathcal{C}}\), les \firstocc{objets} de \(\mathcal{C}\),
    \item pour tous deux objets \(A\) et \(B\) de \(\mathcal{C}\), un objet \(\mathrm{Hom}_{\mathcal{C}}(A, B)\) de \(\mathcal{M}\), le \firstocc{morphisme-objet entre} \(A\) et \(B\),
    \item  pour tout objet \(A\) de \(\mathcal{C}\), un morphisme \(\mathrm{id}_A\) de \(\mathrm{Hom}_{\mathcal{M}}(I, \mathrm{Hom}_{\mathcal{C}}(A, A))\), le \firstocc{morphisme identité} de \(A\),
    \item pour tous trois objets \(A\), \(B\) et \(C\) de \(\mathcal{C}\), un \firstocc{morphisme de composition} \(\circ_{A, B, C}\) défini comme un morphisme de \(\mathrm{Hom}_{\mathcal{M}}(\mathrm{Hom}_{\mathcal{C}}(B, C) \otimes \mathrm{Hom}_{\mathcal{C}}(A, B), \mathrm{Hom}_{\mathcal{C}}(A, C))\)
  \end{itemize}
  faisant commuter les diagrammes suivants, étendant respectivement l'associativité et l'unitarité de la composition\label{pagePropCatEnri}:
  \begin{equation*}
    \begin{tikzpicture}[baseline=(current  bounding  box.center)]
      \matrix (m) [matrix of math nodes,row sep=5em,column sep=6em,minimum width=3em]{
        \mathrm{Hom}_{\mathcal{C}}(C,D)\otimes\mathrm{Hom}_{\mathcal{C}}(B,C)\otimes
          \mathrm{Hom}_{\mathcal{C}}(A,B) & \mathrm{Hom}_{\mathcal{C}}(B,D)\otimes \mathrm{Hom}_{\mathcal{C}}(A,B)\\
         & \mathrm{Hom}_{\mathcal{C}}(A,D)\\
        \mathrm{Hom}_{\mathcal{C}}(C,D)\otimes(\mathrm{Hom}_{\mathcal{C}}(B,C)\otimes
          \mathrm{Hom}_{\mathcal{C}}(A,B)) & \mathrm{Hom}_{\mathcal{C}}(C,D)\otimes \mathrm{Hom}_{\mathcal{C}}(A,C)\rlap{,}\\
      };
      \path[-stealth]
        (m-1-1)
          edge node [above] {\( \circ_{B,C,D} \otimes \mathrm{id}_{\mathrm{Hom}_{\mathcal{C}}(A,B)} \)} (m-1-2)
          edge node [right] {\( \alpha_{\mathrm{Hom}_{\mathcal{C}}(C,D),\mathrm{Hom}_{\mathcal{C}}(B,C),\mathrm{Hom}_{\mathcal{C}}(A,B)} \)} (m-3-1)
        (m-1-2)
          edge node [right] {\( \circ_{A,B,D} \)} (m-2-2)
        (m-3-1)
          edge node [below] {\( \mathrm{id}_{\mathrm{Hom}_{\mathcal{C}}(C,D)} \otimes \circ_{A,B,C} \)} (m-3-2)
        (m-3-2)
          edge node [right] {\( \circ_{A,C,D} \)} (m-2-2)
          ;
    \end{tikzpicture}
  \end{equation*}
  \begin{equation*}
    \begin{tikzpicture}[baseline=(current  bounding  box.center)]
      \matrix (m) [matrix of math nodes,row sep=3em,column sep=4em,minimum width=3em]{
        I\otimes \mathrm{Hom}_{\mathcal{C}}(A,B) & & \mathrm{Hom}_{\mathcal{C}}(B,B) \otimes \mathrm{Hom}_{\mathcal{C}}(A,B)\\
        & \mathrm{Hom}_{\mathcal{C}}(A,B)\rlap{,} & \\
      };
      \path[-stealth]
        (m-1-1)
          edge node [below left] {\( \lambda_{\mathrm{Hom}_{\mathcal{C}}(A,B)} \)} (m-2-2)
          edge node [above] {\( \mathrm{id}_B \otimes \mathrm{id}_{\mathrm{Hom}_{\mathcal{C}}(A,B)}  \)} (m-1-3)
        (m-1-3)
          edge node [below right] {\( \circ_{A,B,B} \)} (m-2-2)
          ;
    \end{tikzpicture}
  \end{equation*}
  \begin{equation*}
    \begin{tikzpicture}[baseline=(current  bounding  box.center)]
      \matrix (m) [matrix of math nodes,row sep=3em,column sep=4em,minimum width=3em]{
        \mathrm{Hom}_{\mathcal{C}}(A,B) \otimes I & & \mathrm{Hom}_{\mathcal{C}}(A,B) \otimes \mathrm{Hom}_{\mathcal{C}}(A,A)\\
        & \mathrm{Hom}_{\mathcal{C}}(A,B)\rlap{.} & \\
      };
      \path[-stealth]
        (m-1-1)
          edge node [below left] {\( \rho_{\mathrm{Hom}_{\mathcal{C}}(A,B)} \)} (m-2-2)
          edge node [above] {\( \mathrm{id}_{\mathrm{Hom}_{\mathcal{C}}(A,B)}\otimes \mathrm{id}_A  \)} (m-1-3)
        (m-1-3)
          edge node [below right] {\( \circ_{A,A,B} \)} (m-2-2)
          ;
    \end{tikzpicture}
  \end{equation*}
  \begin{implementationBox}[]{Catégories enrichies}
    \begin{minted}[xleftmargin=1em, autogobble, fontsize=\footnotesize]{haskell}
      -- Enriched.Category.EnrichedCategory

      -- classe de type des catégories enrichies sur
      -- une catégorie monoïdale catMon
      class Monoidal catMon tensor
        => EnrichedCategory enrichedCat catMon tensor where

        -- l'identité pour un objet a est un morphisme de catMon
        -- entre l'unité de la catégorie monoïdale
        -- et le morphisme-objet enrichedCat a a
        id :: catMon (Unit catMon tensor) (enrichedCat a a)

        -- la composition associée aux objets a, b et c
        -- est un morphisme de catMon entre le produit tensoriel des
        -- morphismes-objets enrichedCat b c et enrichedCat a b vers
        -- le morphisme-objet enrichedCat a c
        (.) :: catMon (tensor (enrichedCat b c, enrichedCat a b)) (enrichedCat a c)
    \end{minted}
  \end{implementationBox}
  Remarquons que toute catégorie est équivalente à une catégorie enrichie sur la catégorie monoïdale des types.
  \begin{implementationBox}[label={enrichedCatTypeProd}]{Catégories enrichies}
    \begin{minted}[xleftmargin=1em, autogobble, fontsize=\footnotesize]{haskell}
      -- Category.OfStructures.CategoryOfTypes

      -- définition d'un nouveau type de données
      -- isomorphe à un type cat a b pour définir la
      -- notion de catégorie enrichie sur la catégorie
      -- des types (pour éviter une ambiguïté du
      -- moteur d'inférence)
      data EnrichedTypeCat cat a b = EnrichedTypeCat (cat a b)

      -- Fonction de projection pour récupérer le morphisme
      -- sous-jacent
      runEnrichedCat :: EnrichedTypeCat cat a b  -> cat a b
      runEnrichedCat (EnrichedTypeCat f) = f

      -- si cat a b est convertible en c (et inversement),
      -- c'est aussi le cas pour EnrichedTypeCat (cat a b)
      instance Castable (cat a b) c =>
        Castable (EnrichedTypeCat cat a b) c where
          cast (EnrichedTypeCat f) = cast f
          castInv x = EnrichedTypeCat $ castInv x

      -- si cat est une catégorie, c'est aussi
      -- le cas pour EnrichedTypeCat cat
      instance CategoryGen cat => CategoryGen (EnrichedTypeCat cat) where
        id = EnrichedTypeCat id
        EnrichedTypeCat f . EnrichedTypeCat g = EnrichedTypeCat $ f . g

      -- si cat est une catégorie, elle est équivalente à la
      -- catégorie enrichie sur la catégorie des types
      -- définie comme suit
      instance CategoryGen cat
        => EnrichedCategory (EnrichedTypeCat cat) (->) Prod where
          -- l'identité pour un objet a est la fonction
          -- envoyant () sur l'identité (de cat) sur a
          id () = id

          -- la composition associée à trois objets a, b et c
          -- est une fonction envoyant un produit de morphismes
          -- sur leur composition
          (.) (Product (EnrichedTypeCat f, EnrichedTypeCat g)) =
            EnrichedTypeCat $ f . g
    \end{minted}
  \end{implementationBox}
  On peut également s'intéresser à la définition d'une catégorie enrichie sur la catégorie monoïdale des types gradués (définie dans le Pseudo-Code~\ref{PseudoCodeCatMonGraded}).
  Une telle structure est ainsi définie, selon la définition classique de catégorie enrichie, par:
  \begin{itemize}
    \item une classe d'objets;
    \item pour deux objets \(A\) et \(B\), d'un type gradué paramétré par \(A\) et \(B\);
    \item pour tout objet \(A\), une fonction graduée (c'est-à-dire un morphisme de la catégorie des types graduées) de l'ensemble à 1 élément vers le type gradué associé à \(A\) et \(A\), qui équivaut à un choix d'un élément unaire du type gradué associé à \(A\) et \(A\);
    \item pour tous trois objets \(A\), \(B\) et \(C\), une composition envoyant le produit tensoriel du type gradué associé à \(B\) et \(C\) par celui associé à \(A\) et \(B\) sur le type gradué associé à \(A\) et \(C\),
        c'est-à-dire une fonction graduée envoyant un élément d'arité \(n\) du type gradué associé à \(B\) et \(C\) et \(n\) éléments d'arité \(k_1,\ldots,k_n\) du type gradué associé à \(A\) et \(B\) sur un élément d'arité \(k_1 + \cdots + k_n\) du type gradué associé à \(A\) et \(C\)
  \end{itemize}
  et respectant les diagrammes de cohérence des catégories enrichies, à savoir l'associativité et l'unitarité de la composition définies page~\pageref{pagePropCatEnri}.
  Par exemple, en considérant
  \begin{itemize}
    \item pour objets les types,
    \item pour morphismes-objets les types de fonctions (non nécessairement unaires) entre deux types \(A\) et \(B\), à savoir les fonctions de \(\bigcup_n A^n \rightarrow B\),
    \item pour identité la fonction identité (d'arité 1) (ou d'une façon équivalente la fonction graduée envoyant l'unique élément d'arité 1 de \(\textbb{1}\) vers la fonction identité),
    \item pour composition la composition de fonctions \(n\)-aires, composant une fonction d'arité \(n\) avec \(n\) fonctions (vue comme une fonction depuis le produit tensoriel des types gradués),
  \end{itemize}
  on obtient une catégorie enrichie sur la catégorie des types gradués.
  Notons qu'il est possible de définir une catégorie duale sur les fonctions de \(B \rightarrow A^n\); cette dualité sera utilisée pour distinguer les automates Bottom-Up des Top-Down.\\
  \textbf{NB:} Le Pseudo-Code suivant présente très partiellement l'implantation des fonctions \(n\)-aires réalisée en Haskell pour ce document.
  Notamment, il occulte la définition de la fonction \texttt{compo} de signature
  \texttt{Graduation (MultiFun b c) (Length arites)
    -> GradVect arites (MultiFun a b)}\\
\texttt{ -> Graduation (MultiFun a c) (SumNat arites)}, permettant de composer des fonctions \(n\)-aires, dont la technicité (fonction type-dépendante implantée par classe de types) n'a pas d'intérêt ici.
  \begin{implementationBox}[label={enrichedCatMultiFun}]{Catégorie enrichie des fonctions graduées}
    \begin{minted}[xleftmargin=1em, autogobble, fontsize=\footnotesize]{haskell}
      -- Algebra.Functions.GradedFunctions

      -- Familles de types (c'est-à-dire ensemble de types ici
      -- inductivement défini) représentant les fonctions
      -- graduées curryfiées de a^n vers b
      type family NFun n a b where
        -- le type des fonctions de a^0 vers b
        -- est b (case de base)
        NFun 'Z _ b = b
        -- étape d'induction
        NFun ('S n) a b = a -> NFun n a b

      -- Type de données des fonctions n-aires de a^n vers b
      data MultiFun (a :: Type) (b :: Type) where
        MultiFun :: NFun arite a b -> MultiFun a b

      -- Les fonctions n-aires sont graduées par leur arité
      instance Graded (MultiFun a b) where
        -- définition du type gradué, et de la fonction
        -- de projection runGradFun
        data Graduation (MultiFun a b) n where
          GradFun ::{runGradFun :: NFun n a b} -> Graduation (MultiFun a b) n

        -- Fonctions de conversion
        toGrad (MultiFun f) = GradFun f
        fromGrad (GradFun f) = MultiFun f

        -- Les fonctions graduées forment une catégorie enrichie
        -- sur les types gradués
        instance EnrichedCategory MultiFun GradedMorph TensorProd where
          -- l'identité est le morphisme envoyant GradUn,
          -- seul élément (d'arité 1) du type gradué associé
          -- à (), sur la fonction identité
          id  = GradedMorph $ \GradUn -> GradFun id
          -- la composition est la composition usuelle de fonction n-aires
          (.) = GradedMorph $ \(GradTensor f gs) -> compo f gs
    \end{minted}
  \end{implementationBox}
  \begin{remarqueBox}[]{Curryfication des fonctions n-aires}
    Afin de simplifier la curryfication des fonctions n-aires, processus type-dépendant par essence, nous utiliserons le type \texttt{Vector} du module \texttt{Tools.Vector} (Remarque~\ref{remTypeVector}), représentant des listes dont la longueur est connue.
    La curryfication et son inverse, la décurryfication, permettant de transformer une fonction n-aire en une fonction unaire sur un vecteur de longueur n (et inversement), seront utilisées sous le nom de \texttt{convertToVectMultiFun} et de \texttt{convertFromVectMultiFun} dans la suite de ce document.
    Ces fonctions sont implantées dans le module \texttt{Algebra.Functions.GradedFunctions}.
  \end{remarqueBox}
  On peut également s'intéresser à la catégorie des fonctions duales de type \(B\rightarrow A^n\).\\
  \textbf{NB:} Afin de conserver le même produit tensoriel et ainsi la même catégorie monoïdale, on encodera les fonctions entre \texttt{b} et \texttt{a\textsuperscript{n}} sous le nom \texttt{DualFunN a b}.
  \begin{implementationBox}[label={enrichedCatDualMultiFun}]{Catégorie enrichie des fonctions graduées (duales)}
    \begin{minted}[xleftmargin=1em, autogobble, fontsize=\footnotesize]{haskell}
      -- Algebra.Functions.GradedFunctionsTopDown

      -- Type de données des fonctions duales de B -> A^n.
      data DualFunN a b where
        DualFunN :: (b -> Vector n a) -> DualFunN a b

      -- Les fonctions duales sont graduées par la longueur
      -- du vecteur de sortie.
      instance Graded (DualFunN a b) where
        -- Type gradué associé
        data Graduation (DualFunN a b) n where
          DualGradFun ::(b -> Vector n a) -> Graduation (DualFunN a b) n

        -- Fonction de conversion
        toGrad (DualFunN f) = DualGradFun f
        fromGrad (DualGradFun f) = DualFunN f

      -- Les fonctions duales forment une catégorie enrichie
      -- sur la catégorie monoïdale des types gradués.
      instance EnrichedCategory DualFunN GradedMorph TensorProd where
        -- l'identité est la fonction envoyant GradUn (seul élément
        -- du type gradué associé à () et d'arité 1) à la fonction
        -- envoyant un élément sur le vecteur à un élément le
        -- contenant
        id  = GradedMorph $ \GradUn -> DualGradFun mkSingle

        -- La composition d'une fonction f de a vers b^n
        -- et de n fonctions duales de b vers c^{k_i} pour
        -- 0 < i < (n + 1) se décompose comme suit:
        -- * calcul du vecteur image obtenu par f
        -- * application du vecteur de n fonctions de type
        --     b -> c^{k_i}sur ce vecteur image
        -- * concaténation des vecteurs résultats
        --
        -- Ces deux dernières étapes sont réalisées dans la fonction
        -- applique définie ci-dessous.
        (.) = GradedMorph $ \(GradTensor f gs) -> compo f gs
         where
          compo (DualGradFun f) gs = DualGradFun $ applique gs . f
          applique
            :: GradVect ms (DualFunN b c)
            -> Vector (Length ms) c
            -> Vector (SumNat ms) b
          applique GradVNil VNil = VNil
          applique (GradVCons (DualGradFun g) gs) (VCons x xs) =
            vConcat (g x) $ applique gs xs
    \end{minted}
  \end{implementationBox}


\section{Automates généralisés}

La notion d'automate peut alors être généralisée en celle d'\firstocc{automate généralisé}, c'est-à-dire un automate sur une catégorie enrichie par une catégorie monoïdale.
D'une façon similaire à celle de l'extension de la notion de morphisme identité dans une catégorie enrichie,
les morphismes définissant la configuration initiale ou le morphisme vers l'objet final ne sont plus des éléments pris dans un ensemble;
ainsi, la même méthode de définition sera utilisée: nous n'utiliserons plus des morphismes, mais des morphismes de la catégorie monoïdale depuis l'unité vers un morphisme-objet.
De tels morphismes, de l'unité de la catégorie monoïdale vers un objet quelconque, sont appelés classiquement des \firstocc{éléments généralisés}{élément!généralisé}.
\begin{definition}
  Un \firstocc{automate généralisé}{automate!généralisé} sur une catégorie \(\mathcal{C}\) enrichie sur une catégorie monoïdale \(\mathfrak{M} = (\mathcal{M}, \otimes, I, \alpha, \lambda, \rho)\) est défini par
  \begin{itemize}
    \item un \firstocc{objet alphabet} \(\Sigma \) de \(\mathcal{M} \);
    \item par trois objets de \(\mathcal{C}\): \(S\) l' \firstocc{objet initial}, \(Q\) l'\firstocc{objet état} et \(V\) l'\firstocc{objet final};
    \item un \firstocc{morphisme initial} de \(\mathrm{Hom}_{\mathcal{M}}(I, \mathrm{Hom}_{\mathcal{C}}(S, Q))\);
    \item un \firstocc{morphisme de transition} \( \delta \) de \(\mathrm{Hom}_{\mathcal{M}}(\Sigma, \mathrm{Hom}_{\mathcal{C}}(Q, Q))\);
    \item un \firstocc{morphisme final} de \(\mathrm{Hom}_{\mathcal{M}}(I, \mathrm{Hom}_{\mathcal{C}}(Q, V))\).
  \end{itemize}
\end{definition}
\begin{implementationBox}[]{Automates généralisés}
  \begin{minted}[xleftmargin=1em, autogobble, fontsize=\footnotesize]{haskell}
    -- AutomataGen.AutomatonGen

    -- Type de données des automates généralisés sur une catégorie
    -- enrichedCat enrichie sur la catégorie monoïdale catMon:
    -- * init l'objet initial
    -- * symbols l'objet alphabet
    -- * state l'objet état
    -- * value l'objet final
    data AutomatonGen enrichedCat catMon tensor init symbols state value where
      Auto ::EnrichedCategory enrichedCat catMon tensor => {
        -- le morphisme initial
        initial :: catMon (Unit catMon tensor) (enrichedCat init state),
        -- le morphisme de transition
        delta   :: catMon symbols (enrichedCat state state),
        -- le morphisme final
        final   :: catMon (Unit catMon tensor) (enrichedCat state value)
      } -> AutomatonGen enrichedCat catMon tensor init symbols state value
  \end{minted}
\end{implementationBox}
On peut alors utiliser les catégories enrichies que nous avons définies, celles basées sur la catégorie des types munie du produit cartésien (Pseudo-code~\ref{enrichedCatTypeProd}) et
celles basées sur les fonctions graduées (Pseudo-code~\ref{enrichedCatMultiFun} et Pseudo-code~\ref{enrichedCatDualMultiFun}) pour définir des automates.
Tout d'abord, remarquons que les automates généralisés sur une catégorie enrichie \(\mathcal{C}\) sur la catégorie monoïdale des types munie du produit cartésien sont équivalents
aux automates sur une catégorie, puisque \(\mathcal{C}\) est équivalente à une catégorie (Pseudo-code~\ref{enrichedCatTypeProd}).
L'isomorphisme est présenté dans le code suivant.
\begin{implementationBox}[]{Automates sur une catégorie et automates généralisés}
  \begin{minted}[xleftmargin=1em, autogobble, fontsize=\footnotesize]{haskell}
  -- AutomataGen.WordAutomata.WordAut

  -- imports nommés de modules pour aider à la distinction
  import           Automata.Automaton            as A
  import           AutomataGen.AutomatonGen      as AG

  -- Synonyme de type pour les automates généralisés sur une
  -- catégorie enrichie sur la catégorie des types munie
  -- du produit cartésien
  type WordAut enrichedCat init symbols state value
    = AutomatonGen enrichedCat (->) Prod init symbols state value

  -- Transformation d'un automate sur une catégorie en un automate généralisé
  convertFromClassicalAutomaton
    :: CategoryGen cat
      => Automaton cat init symbols state value
      -> WordAut (EnrichedTypeCat cat) init symbols state value
  convertFromClassicalAutomaton (A.Auto i d f) = AG.Auto
    (const $ EnrichedTypeCat i)
    (EnrichedTypeCat . d)
    (const $ EnrichedTypeCat f)

  -- Transformation d'un automate généralisé en un automate sur une catégorie
  convertToClassicalAutomaton
    :: WordAut (EnrichedTypeCat cat) init symbols state value
      -> Automaton cat init symbols state value
  convertToClassicalAutomaton (AG.Auto i d f) = A.Auto
    (runEnrichedCat $ i ())
    (runEnrichedCat . d)
    (runEnrichedCat $ f ())

  -- Synonyme pour le calcul du poids depuis une liste de symboles (un mot).
  -- NB: utilise le constructeur List du module
  -- 'Category.OfStructures.CategoryOfTypes' existant pour définir
  -- l'adjonction entre les catégories (->) et
  -- 'MonoidObjectMorph (->) Prod' (équivalente à la catégorie des monoïdes)
  wordWeight
    :: (EnrichedCategory enrichedCat (->) Prod, _)
      => WordAut enrichedCat init symbols state value
        -> [symbols]
        -> enrichedCat init value
  wordWeight auto = weight auto . List
  \end{minted}
\end{implementationBox}
Dans le cas d'un automate généralisé sur la catégorie enrichie des fonctions \(n\)-aires, la structure d'automate est défini par:
\begin{itemize}
  \item un alphabet \(\Sigma \), qui est un type gradué;
  \item une fonction graduée initiale de l'ensemble gradué à un élément (d'arité \(1\)) vers les fonctions \(n\)-aires de \(S\) (l'objet initial) vers \(Q\) (l'objet état)
  (l'ensemble des fonctions de la forme \(S^n \rightarrow Q\)); d'une façon équivalente, il s'agit d'une fonction de \(S\) vers \(Q\);
  \item une fonction de transition graduée de \(\Sigma \) vers les fonctions \(n\)-aires de \(Q\) vers \(Q\) (l'ensemble des fonctions de la forme \(Q^n \rightarrow Q\));
  plus précisément, la fonction de transition envoie un symbole \(n\)-aire de \(\Sigma \) sur une fonction de \(Q^n \rightarrow Q\);
  \item une fonction graduée terminale de l'ensemble gradué à un élément (d'arité \(1\)) vers les fonctions \(n\)-aires de \(Q\) vers \(V\) (l'objet final) (l'ensemble des fonctions de la forme \(Q^n \rightarrow V\)); d'une façon équivalente, il s'agit d'une fonction de \(Q\) vers \(V\).
\end{itemize}


Il s'agit alors de \firstocc{Root-Weighted Tree Automata} (RWTA) séquentiels et complets~\cite{MOZ15}, réalisant des \firstocc{recognizable step functions}~\cite{DG07, DV06}.
Plus particulièrement, lorsque \(V\) est l'ensemble Booléen, il s'agit alors d'un automate d'arbre (Bottom-Up) déterministe complet, comme défini page~\pageref{defAutoArbreBUDet}.
La seule différence est la présence de la fonction initiale; son rôle est d'étendre la capacité de reconnaissance des automates d'arbres.
Au lieu de ne fixer un poids qu'à des automates nullaires, la fonction initiale permet de considérer un arbre \(n\)-aire,
de \emph{boucher} les arbres vides par des occurrences de \firstocc{variables}, éléments d'un ensemble \(S\), et de lui fixer ensuite un poids.
On retombe alors sur la définition initiale en fixant l'ensemble des variables à l'ensemble vide (\texttt{Void} en Haskell, voir Remarque~\ref{RemVoid}).
\begin{implementationBox}[]{Automates d'arbres Bottom-Up}
  \begin{minted}[xleftmargin=1em, autogobble, fontsize=\footnotesize]{haskell}
    -- AutomataGen.TreeAutomata.TreeAut

    -- Un automate d'arbres (général) est un automate généralisé sur une
    -- catégorie enrichie sur la catégorie monoïdale des fonctions graduées
    type TreeAut enrichedCat var symbols state value
      = AutomatonGen enrichedCat GradedMorph TensorProd var symbols state value

    -- AutomataGen.TreeAutomata.BottomUp.BottomUpCompDet

    -- Synonyme pour le type des RWTA séquentiels à variable
    type SeqRWTAWithVar weight var symbols state
      = TreeAut MultiFun var symbols state weight

    -- Synonyme pour le type des automates d'arbres Bottom-Up déterministe
    -- complet avec variables
    type BottomUpCompDFTAWithVar var symbols state
      = SeqRWTAWithVar Bool var symbols state

    -- Synonyme pour le type des automates d'arbres Bottom-Up déterministe
    -- complet sans variables
    type BottomUpCompDFTA symbols state = BottomUpCompDFTAWithVar Void symbols state

    -- Construit un RWTA séquentiel avec variables
    packRWTA
      :: (var -> state)
      -> GradedMorph symbols (MultiFun state state)
      -> (state -> weight)
      -> SeqRWTAWithVar weight var symbols state
    packRWTA i transMorph finalWeights =
      Auto (gradedMorphFromFun i) transMorph (gradedMorphFromFun finalWeights)

    -- Construit un automate d'arbres avec variables
    packVar
      :: (var -> state)
      -> GradedMorph symbols (MultiFun state state)
      -> (state -> Bool)
      -> BottomUpCompDFTAWithVar var symbols state
    packVar = packRWTA

    -- Construit un automate d'arbres sans variable
    pack
      :: GradedMorph symbols (MultiFun state state)
      -> (state -> Bool)
      -> BottomUpCompDFTA symbols state
    pack = packVar absurd
  \end{minted}
\end{implementationBox}
\begin{remarqueBox}[]{Cas des automates d'arbres Top-Down déterministes complets}
  D'une façon duale, les automates d'arbres Top-Down déterministes complets peuvent être définis comme des automates généralisés sur la catégorie enrichie des fonctions entre \(A\) et \(B^n\) (Pseudo-Code~\ref{enrichedCatDualMultiFun}).
  Ces automates sont implantés dans le module \texttt{AutomataGen.TreeAutomata.TopDown.TopDownCompDet}.
\end{remarqueBox}
Pour calculer le poids d'un mot ou d'un arbre (ou d'un élément d'une autre structure algébrique), il faut commencer par étendre le domaine du morphisme de transition.
Dans le cas des automates sur des catégories (non enrichies), la promotion de la fonction \(\delta \) (Pseudo-Code~\ref{codeDeltaExtension}) utilisait l'adjonction entre les foncteurs Liste (monoïde libre) et Identité (oubli).
Ce raisonnement est encore valable ici, en notant la remarque suivante.
\begin{remarqueBox}[]{Catégorie enrichie à un objet et monoïde objet}
  Au même titre qu'une catégorie à un seul objet est équivalente à un monoïde (Pseudo-Code~\ref{codeCat1objMonoid}), une catégorie enrichie avec un seul objet est un monoïde objet.
    \begin{minted}[xleftmargin=1em, autogobble, fontsize=\footnotesize]{haskell}
      -- Enriched.Category.EnrichedCategory

      instance EnrichedCategory enrichedCat catMon tensor
        => MonoidObject (enrichedCat a a) catMon tensor where
          mu  = (.)
          eta = id
    \end{minted}
\end{remarqueBox}


S'il est possible de transformer fonctoriellement l'objet alphabet \(\Sigma \) en monoïde objet libre \(\mathrm{Free}(\Sigma)\), c'est-à-dire si le constructeur de type Identité est un foncteur de la catégorie des monoïdes objets (Pseudo-Code~\ref{CategorieDesMonObj}) de \(\mathfrak{M}\) vers \(\mathcal{M}\)
et que ce foncteur admette un adjoint à gauche \(\mathrm{Free}\), il suffit d'utiliser cette adjonction pour promouvoir \(\delta \) en morphisme de monoïde objet de \(\mathrm{Free}(\Sigma)\), le monoïde objet libre, vers le monoïde objet défini par \(\mathrm{Hom}_{\mathcal{C}}(Q, Q)\).
Cela est réalisable en utilisant la même fonction de promotion, \texttt{promote}, du Pseudo-Code~\ref{codeAdjonction}.
\begin{implementationBox}[]{Promotion du morphisme de transitions}
  \begin{minted}[xleftmargin=1em, autogobble, fontsize=\footnotesize]{haskell}
    -- AutomataGen.AutomatonGen

    delta'
      :: ( AdjunctionGen f Identity catMon (MonoidObjectMorph catMon tensor)
         , EnrichedCategory enrichedCat catMon tensor
         )
      => AutomatonGen enrichedCat catMon tensor init symbols state value
      -> MonoidObjectMorph catMon tensor (f symbols) (enrichedCat state state)
    delta' auto = promote $ packId . delta auto
  \end{minted}
\end{implementationBox}
Ainsi, dans le cas des automates d'arbres, c'est l'adjonction du Pseudo-code~\ref{PseudoCodeAdjOperad}, entre fonctions graduées et morphismes d'opérades qui sera utilisée, permettant de déterminer l'action d'un arbre depuis la combinaison de chacun de ses symboles gradués.

On pourra remarquer que la seule différence avec la promotion de la fonction de transition des mots du Pseudo-Code~\ref{codeDeltaExtension} est le morphisme \texttt{packId} utile pour effacer la présence du constructeur \texttt{Identity}: mis à part ce détail technique mineur, les fonctions sont syntaxiquement les mêmes.

Cette proximité syntaxique n'est plus tout à fait la même pour les fonctions \texttt{getConfig} et \texttt{weight} des Pseudo-Codes~\ref{codeConfigAtteinte} et~\ref{CodePoidsMot}, bien que l'idée de base soit la même:
\begin{itemize}
  \item  pour \texttt{getConfig}, il suffit de combiner le morphisme initial de \(I\) vers \(\mathrm{Hom}_{\mathcal{C}}(S, Q)\) et le morphisme de transition de \(\Sigma \) vers \(\mathrm{Hom}_{\mathcal{C}}(Q, Q)\) en un morphisme de \(\Sigma \) vers \(\mathrm{Hom}_{\mathcal{C}}(S, Q)\);
  \item pour \texttt{weight}, il suffit de combiner le morphisme obtenu par \texttt{getConfig} de \(\Sigma \) vers \(\mathrm{Hom}_{\mathcal{C}}(S, Q)\) et le morphisme final de \(I\) vers \(\mathrm{Hom}_{\mathcal{C}}(Q, V)\) en un morphisme de \(\Sigma \) vers \(\mathrm{Hom}_{\mathcal{C}}(S, V)\).
\end{itemize}
Ces combinaisons sont définissables en toute généralité au niveau des catégories enrichies.
Dans le cas de \texttt{getConfig}, il suffit:
\begin{itemize}
  \item d'utiliser l'inverse de l'identité droite \(\rho^{-1}_{\Sigma}\), morphisme de \(\Sigma \) vers \(\Sigma\otimes I\);
  \item de calculer le produit tensoriel entre le morphisme initial et celui de transition pour former un morphisme de \(\Sigma\otimes I\) vers \(\mathrm{Hom}_{\mathcal{C}}(Q, Q) \otimes \mathrm{Hom}_{\mathcal{C}}(S, Q)\) et de le combiner à l'aide de la composition de la catégorie, elle-même morphisme de  \(\mathrm{Hom}_{\mathcal{C}}(Q, Q) \otimes \mathrm{Hom}_{\mathcal{C}}(S, Q)\) vers \(\mathrm{Hom}_{\mathcal{C}}(S, Q)\);
  \item de composer ces deux morphismes par la composition de la catégorie monoïdale.
\end{itemize}
Cette opération, combinaison d'un élément généralisé et d'un morphisme pour former un nouvel élément généralisé, est définie dans le module \texttt{Enriched.Category.EnrichedCategory} sous le nom de \texttt{prepend} (et \texttt{append} pour son équivalent de combinaison par un élément généralisé à droite).
\textbf{NB:} Pour la distinguer de celle de la catégorie enrichie, la composition de la catégorie monoïdale sera préfixée par \texttt{Cat.} dans le Pseudo-code suivant.
\begin{implementationBox}[label={codeCombiElemtGen}]{Combinaisons morphisme / élément généralisé }
  \begin{minted}[xleftmargin=1em, autogobble, fontsize=\footnotesize]{haskell}
    -- Enriched.Category.EnrichedCategory

    -- Combinaison d'un élément généralisé et d'un morphisme
    prepend
      :: EnrichedCategory enrichedCat catMon tensor
        => catMon (Unit catMon tensor) (enrichedCat x y)
        -> catMon t (enrichedCat y z)
        -> catMon t (enrichedCat x z)
    prepend i_m t_m = ((.) Cat.. fmap (Prod t_m i_m)) Cat.. rhoInv

    -- Combinaison d'un morphisme et d'un élément généralisé
    append
      :: EnrichedCategory enrichedCat catMon tensor
        => catMon t (enrichedCat x y)
        -> catMon (Unit catMon tensor) (enrichedCat y z)
        -> catMon t (enrichedCat x z)
    append t_m i_m = ((.) Cat.. fmap (Prod i_m t_m)) Cat.. lambdaInv
  \end{minted}
\end{implementationBox}
Ces deux opérations permettent alors de calculer le poids comme un morphisme entre l'objet libre (pouvant contenir les structures à pondérer) et l'objet final.
\begin{implementationBox}[]{Combinaisons morphisme / élément généralisé }
  \begin{minted}[xleftmargin=1em, autogobble, fontsize=\footnotesize]{haskell}
    -- AutomataGen.AutomatonGen

    getConfig
      :: ( EnrichedCategory enrichedCat catMon tensor
         , AdjunctionGen f Identity catMon (MonoidObjectMorph catMon tensor)
         , MonoidObject (f symbols) catMon tensor
         )
      => AutomatonGen enrichedCat catMon tensor init symbols state value
      -> catMon (f symbols) (enrichedCat init state)
    getConfig auto = prepend (initial auto) $ runMonoidObjectMorph (delta' auto)

    weight
      :: ( EnrichedCategory enrichedCat catMon tensor
         , AdjunctionGen f Identity catMon (MonoidObjectMorph catMon tensor)
         , MonoidObject (f symbols) catMon tensor
         )
      => AutomatonGen enrichedCat catMon tensor init symbols state value
      -> catMon (f symbols) (enrichedCat init value)
    weight auto = append (getConfig auto) $ final auto
  \end{minted}
\end{implementationBox}
Afin de faciliter l'utilisation de ces fonctions, on peut les spécialiser (en termes de signature) dans le cas particulier des arbres, par exemple.
\textbf{NB:} Dans le pseudo-code suivant est utilisée la fonction \texttt{packCast} du module \texttt{Enriched.Category.EnrichedCategory} permettant d'appliquer une conversion de type (classe de types \texttt{Castable} définie précédemment) au niveau d'une catégorie monoïdale, en définissant un morphisme du type \texttt{catMon (enrichedCat a b) (enrichedCat a c)} depuis deux types \texttt{b} et \texttt{c} satisfaisant la contrainte \texttt{Castable b c}.
\begin{implementationBox}[]{Calcul particulier du poids dans un RWTA}
  \begin{minted}[xleftmargin=1em, autogobble, fontsize=\footnotesize]{haskell}
    -- AutomataGen.TreeAutomata.BottomUp.BottomUpCompDet

    -- Le poids associé à un arbre n-aire est une fonction
    -- n-aire envoyant n variables sur un poids.
    weightAsScalarWithVar
      :: SeqRWTAWithVar weight var symbols state
      -> Graduation (FreeOperad symbols) n
      -> NFun n var weight
    weightAsScalarWithVar = runGradFun .: weightAsScalarWithVar'
     where
      weightAsScalarWithVar'
        :: SeqRWTAWithVar weight var symbols state
        -> Graduation (FreeOperad symbols) n
        -> Graduation (MultiFun var weight) n
      weightAsScalarWithVar' auto' = applyMorph (packCast Cat.. weight auto')

    -- Le poids associé à un arbre nullaire est une fonction
    -- 0-aire envoyant 0 variable sur un poids, c'est-à-dire un poids.
    weightAsScalar
      :: SeqRWTAWithVar weight var symbols state
      -> Graduation (FreeOperad symbols) Nat_0
      -> weight
    weightAsScalar = weightAsScalarWithVar

    -- Synonyme pour les automates Booléens.
    recognizesWithVar
      :: BottomUpCompDFTAWithVar var symbols state
      -> Graduation (FreeOperad symbols) n
      -> NFun n var Bool
    recognizesWithVar = weightAsScalarWithVar

    -- Synonyme pour les automates Booléens.
    recognizes
      :: BottomUpCompDFTAWithVar var symbols state
      -> Graduation (FreeOperad symbols) Nat_0
      -> Bool
    recognizes = weightAsScalar
  \end{minted}
\end{implementationBox}


Une autre utilisation de la fonction \texttt{append} est la modification possible des éléments généralisés initiaux et finaux.
Ainsi, la complémentation d'un automate d'arbres déterministe complet est aussi simple que dans le cas des automates sur la catégorie des types: il suffit d'inverser le poids de sortie.
\begin{implementationBox}[]{Complémentation d'un automate d'arbre déterministe complet}
  \begin{minted}[xleftmargin=1em, autogobble, fontsize=\footnotesize]{haskell}
    -- AutomataGen.TreeAutomata.BottomUp.BottomUpCompDet

    complement
      :: BottomUpCompDFTAWithVar var symbols state
      -> BottomUpCompDFTAWithVar var symbols state
    complement auto =
      auto { final = append (final auto) (gradedMorphFromFun not) }
  \end{minted}
\end{implementationBox}

\begin{remarqueBox}[]{Le type gradué \texttt{Symbol a}}
  Afin de simplifier la définition d'exemples, nous utiliserons dans la suite de ce document le type \texttt{Symbol a}, défini dans le module \texttt{Tools.GradSymbol} et permettant de définir simplement des symboles gradués depuis n'importe quel type.
  Sont également fournis des constructeurs de symboles et d'arbres, comme illustrés ci-dessous:
    \begin{minted}[xleftmargin=1em, autogobble, fontsize=\footnotesize]{haskell}
      -- Tools.GradSymbol

      -- Un symbole est déterminé par un Natural (son arité)
      -- et par une valeur de type a.
      data Symbol a where
        Symbol :: Natural n -> a -> Symbol a

      -- Le type Symbol a est gradué
      instance Graded (Symbol a) where
        -- Type support de la graduation
        data Graduation (Symbol a) n = GradSymb (Natural n) a

        -- Fonctions de conversion
        toGrad (Symbol nat a_) = GradSymb n a_
        fromGrad (GradSymb n a_) = Symbol n a_

      -- Construit un symbole nullaire
      zerothSymbol :: a -> Graduation (Symbol a) 'Z
      zerothSymbol = GradSymb singleton

      -- Construit un symbole unaire
      unarySymbol :: a -> Graduation (Symbol a) Nat_1
      unarySymbol = GradSymb one

      -- Construit un symbole binaire
      binarySymbol :: a -> Graduation (Symbol a) Nat_2
      binarySymbol = GradSymb two

      -- Construit un arbre nullaire
      tree0 :: Graduation a Nat_0 -> Graduation (FreeOperad a) 'Z
      tree0 s = GradOp s GradVNil

      -- Ajoute en racine d'un arbre un élément d'arité 1,
      -- préservant l'arité
      tree1
        :: Graded a
        => Graduation a Nat_1
        -> Graduation (FreeOperad a) n
        -> Graduation (FreeOperad a) n
      tree1 s tree = GradOp s $ mkSingle tree

      -- Ajoute une racine d'arité 2 au sommet de deux arbres,
      -- sommant les arités
      tree2
        :: Graded a
        => Graduation a Nat_2
        -> Graduation (FreeOperad a) n1
        -> Graduation (FreeOperad a) n2
        -> Graduation (FreeOperad a) (n1 + n2)
      tree2 s tree tree' = GradOp s $ mkTwo tree tree'
    \end{minted}
    Sont également fournies des fonctions utilitaires, comme le prédicat binaire \texttt{eqSymb} permettant de tester si deux symboles sont égaux.
\end{remarqueBox}
L'exemple suivant (Pseudo-Code~\ref{codeRWTA})\footnote{Code exécutable dans l'interpréteur \texttt{ghci} \emph{via} la commande \texttt{stack ghci src/HDRExample/HDRExampleBUCompDet.hs}}, s'appuyant sur les notations précédentes, montre comment définir le type des arbres dont les n{\oe}uds sont des chaînes de caractères, et comment les interpréter comme arbre syntaxique d'opérations arithmétiques modulaires simples \emph{via} un automate d'arbres.
Pour cela, une fois un entier \(n\) fixé (pour la congruence modulaire), il suffit de définir un automate d'arbres (Bottom-Up) avec
\begin{itemize}
  \item un état \(\mathrm{Just}\ k\) pour chaque entier \(k\) entre \(0\) et \(n - 1\) et l'état \(\mathrm{Nothing}\);
  \item une transition nullaire envoyant chaque chaîne numérique représentant un entier \(m\) sur \(\mathrm{Just}(m \mod n)\), et les autres chaînes sur \(\mathrm{Nothing}\);
  \item des transitions pour chaque opération arithmétique choisie, dont les arités correspondent;
  \item des transitions vers \(\mathrm{Nothing}\) dans les autres cas;
  \item comme poids final associé à chaque état lui-même si l'on souhaite évaluer l'expression, ou bien un Booléen dans le choix de la reconnaissance (par exemple, un test de parité);
  \item un poids initial arbitrairement choisi pour les variables, le cas échéant.
\end{itemize}
\begin{implementationBox}[label={codeRWTA}]{Calcul particulier du poids dans un RWTA}
  \begin{minted}[xleftmargin=1em, autogobble, fontsize=\footnotesize]{haskell}
    -- HDRExample.HDRExampleBUCompDet

    -- Composition d'une fonction unaire et d'une fonction binaire
    (.:) :: (b -> c) -> (a1 -> a2 -> b) -> a1 -> a2 -> c
    (.:) = (.) . (.)

    -- Symbole unaire défini par la chaine "-"
    minusString :: Graduation (Symbol String) Nat_1
    minusString = unarySymbol "-"

    sumString, binMinusString, prodString :: Graduation (Symbol String) Nat_2
    -- Symbole binaire défini par la chaine "+"
    sumString = binarySymbol "+"
    -- Symbole binaire défini par la chaine "-"
    binMinusString = binarySymbol "-"
    -- Symbole binaire défini par la chaine "*"
    prodString = binarySymbol "*"

    -- Définition des types des états de l'automate.
    -- L'état Nothing permet de gérer les erreurs,
    -- par exemple les échecs de conversion
    -- des chaines en entier.
    type State = Maybe Int

    -- Fonction de transition de l'automate;
    -- la définition principale est celle de la fonction aux'.
    -- Interprète des chaines de caractères en fonctions arithmétiques
    -- modulaires, où la borne est donnée en paramètre.
    delta_ :: Int -> GradedMorph (Symbol String) (MultiFun State State)
    delta_ n = GradedMorph aux
     where
      aux x = aux' (graduation' x) x
      -- Fonction associant à des symboles
      -- des fonctions n-aires sur les états
      aux'
        :: Natural n
        -> Graduation (Symbol String) n
        -> Graduation (MultiFun State State) n
      -- Pour un symbole d'arité 0, tentative de conversion
      -- de la chaine en entier puis calcul modulaire.
      aux' Zero (GradSymb _ s) = GradFun $ (`mod` n) <$> readMaybe s
      -- Pour le symbole d'arité 1 minusString, tentative d'inversion du signe
      -- de l'état puis calcul modulaire.
      aux' (Succ Zero) x
        | eqSymb x minusString = GradFun $ fmap ((`mod` n) . negate)
      aux' (Succ (Succ Zero)) x
        -- Pour le symbole d'arité 2 sumString,
        -- tentative de somme des deux états
        -- puis calcul modulaire.
        | eqSymb x sumString = GradFun $ liftA2 $ (`mod` n) .: (+)
        -- Pour le symbole d'arité 2 sumString,
        -- tentative de soustraction des deux états
        -- puis calcul modulaire.
        | eqSymb x binMinusString = GradFun $ liftA2 $ (`mod` n) .: (-)
        -- Pour le symbole d'arité 2 prodString,
        -- tentative de soustraction des deux états
        -- puis calcul modulaire.
        | eqSymb x prodString = GradFun $ liftA2 $ (`mod` n) .: (*)
      -- Dans tous les autres cas, quelque soit l'arité,
      -- renvoie de Nothing.
      aux' m _ = convertFromVectMultiFun m $ const Nothing

    -- Définition du type de variables;
    -- ici arbitrairement 3 variables.
    data Vars = X1 | X2 | X3

    -- Fonction d'initialisation des variables
    initFun :: Vars -> State
    initFun X1 = Just 65
    initFun X2 = Just 9
    initFun X3 = Nothing

    -- Fonction de poids: chaque état a pour
    -- poids lui même
    finalWeight :: State -> Maybe Int
    finalWeight = id

    -- Fonction de finalité:
    -- un état est final s'il est pair
    isFinal :: State -> Bool
    isFinal = maybe False even

    -- Définition d'un RWTA où le poids est Just p si
    -- l'arbre est l'arbre syntaxique d'une expression arithmétique
    -- valide, Nothing sinon.
    rwta :: Int -> SeqRWTAWithVar (Maybe Int) Vars (Symbol String) State
    rwta n = packRWTA initFun (delta_ n) finalWeight

    -- Définition d'un automate (sans variable) reconnaissant un arbre s'il
    -- est l'arbre syntaxique d'une expression arithmétique
    -- valide dont le résultat est pair.
    auto :: Int -> BottomUpCompDFTA (Symbol String) State
    auto n = pack (delta_ n) isFinal

    -- Définition d'un automate reconnaissant un arbre s'il
    -- est l'arbre syntaxique d'une expression arithmétique
    -- valide dont le résultat est pair.
    autoVar :: Int -> BottomUpCompDFTAWithVar Vars (Symbol String) State
    autoVar n = packVar initFun (delta_ n) isFinal

    -- Arbre syntaxique nullaire de l'expression valide -(513 + 838) * 37
    t :: Graduation (FreeOperad (Symbol String)) Nat_0
    t =
      tree2
          prodString
          ( tree1 minusString
          $ tree2 sumString (tree0 $ zerothSymbol "513")
          $ tree0
          $ zerothSymbol "838"
          )
        $ tree0
        $ zerothSymbol "37"

    -- Arbre syntaxique nullaire de l'expression invalide -(KO + 838) * 37
    t_KO :: Graduation (FreeOperad (Symbol String)) 'Z
    t_KO =
      tree2
          prodString
          ( tree1 minusString
          $ tree2 sumString (tree0 $ zerothSymbol "KO")
          $ tree0
          $ zerothSymbol "838"
          )
        $ tree0
        $ zerothSymbol "37"

    -- Arbre syntaxique binaire de l'expression -(Eps + Eps) * 37
    t' :: Graduation (FreeOperad (Symbol String)) Nat_2
    t' =
      tree2 prodString (tree1 minusString $ tree2 sumString GradEps GradEps)
        $ tree0
        $ zerothSymbol "37"

    -- Affichage du résultat du calcul des poids
    testDFTA :: IO ()
    testDFTA = do
      putStrLn $ mconcat ["t_KO = ", show t_KO]
      putStrLn $ mconcat ["t    = ", show t]
      putStrLn $ mconcat ["t'   = ", show t']

      putStrLn $ mconcat ["X1 -> ", show (initFun X1)]
      putStrLn $ mconcat ["X2 -> ", show (initFun X2)]
      putStrLn $ mconcat ["X3 -> ", show (initFun X3)]

      let n = 19

      putStrLn "\nPonderations of t_KO"
      print $ auto n `recognizes` t_KO
      print $ autoVar n `recognizes` t_KO
      print $ rwta n `weightAsScalar` t_KO
      print $ complement (auto n) `recognizes` t_KO

      putStrLn "\nPonderations of t"
      print $ auto n `recognizes` t
      print $ autoVar n `recognizes` t
      print $ rwta n `weightAsScalar` t
      print $ complement (auto n) `recognizes` t

      putStrLn "\nPonderations of t'(X1, X2)"
      print $ recognizesWithVar (autoVar n) t' X1 X2
      print $ weightAsScalarWithVar (rwta n) t' X1 X2
      print $ recognizesWithVar (complement $ autoVar n) t' X1 X2

      putStrLn "\nPonderations of t'(X2, X2)"
      print $ recognizesWithVar (autoVar n) t' X2 X2
      print $ weightAsScalarWithVar (rwta n) t' X2 X2
      print $ recognizesWithVar (complement $ autoVar n) t' X2 X2

      putStrLn "\nPonderations of t'(X1, X3)"
      print $ recognizesWithVar (autoVar n) t' X1 X3
      print $ weightAsScalarWithVar (rwta n) t' X1 X3
      print $ recognizesWithVar (complement $ autoVar n) t' X1 X3

  \end{minted}
\end{implementationBox}


L'exemple précédent présente une méthode d'analyse d'arbre où les étapes intermédiaires sont représentés pas les états de l'automate.
Ainsi, si le calcul ne s'effectue plus d'une façon modulaire, le nombre d'états nécessaire est arbitrairement grand.
Comme dans le cas des automates de mots, il est possible de pondérer des arbres \emph{via} des structures algébriques externes aux états de l'automate, mais encodées à l'intérieur même des morphismes des catégories enrichies.
Si les structures utilisées étaient le semi-anneau et le semi-module dans le cas des mots, celles utilisées dans le cas des arbres peuvent être définies à partir d'opérades particulières pour former des structures algébriques appelées \firstocc{monoïdes multi-opérateurs} dans la littérature (voir par exemple~\cite{FMV09}).
\begin{remarqueBox}[]{Monoïde multi-opérateurs}
  Classiquement, un \firstocc{monoïde multi-opérateurs}{monoïde!multi-opérateurs} est un monoïde commutatif \((M, \oplus, 0)\) muni d'un sous-ensemble de \(\bigcup_{n} M^n \rightarrow M\), ses opérateurs.
\end{remarqueBox}
Les opérateurs d'un monoïde multi-opérateurs peuvent être utilisés pour pondérer les transitions d'un automate d'arbre (Bottom-Up), une transition étiquetée par un symbole \(n\)-aire étant pondérée par un opérateur \(n\)-aire.

Pour représenter cela, nous considérerons, pour deux types \texttt{A} et \texttt{B} et une opérade \texttt{O} munie d'une somme graduée, les fonctions envoyant \texttt{A\textsuperscript{n}} sur les fonctions (partielles) de \texttt{B} vers les éléments \(n\)-aires de \texttt{O}; un monoïde multi-opérateurs sera donc le cas où l'opérade choisie est celle des fonctions sur un monoïde, la somme de deux fonctions de même arité étant définie comme la somme de leurs images.

\begin{remarqueBox}[]{Opérade plus des fonctions sur un monoïde}
  Une opérade munie d'une somme graduée est appelée \firstocc{opérade plus}{opérade!plus} dans la suite de ce document.
  La classe de type correspondante est définie dans le module \texttt{Algebra.Structures.Operad} comme suit:
  \begin{minted}[xleftmargin=1em, autogobble, fontsize=\footnotesize]{haskell}
    class Operad o => OperadPlus o where
      (<++>) :: Graduation o n -> Graduation o n -> Graduation o n
  \end{minted}
  Si \texttt{m} est un type monoïdal, les fonctions \(n\)-aires sur \texttt{m}, pour un \(n\) fixé, forment un monoïde, implanté dans le module \texttt{Algebra.Functions.GradedFunctions} comme suit:
  \begin{minted}[xleftmargin=1em, autogobble, fontsize=\footnotesize]{haskell}
  instance (Semigroup b, Singleton n)
    => Semigroup (Graduation (MultiFun a b) n) where
      -- L'image d'une somme de fonctions est la somme de leurs images
      x <> y = convertFromVectMultiFun
        $ \xs -> convertToVectMultiFun x xs <> convertToVectMultiFun y xs

  instance (Monoid b, Singleton n)
    => Monoid (Graduation (MultiFun a b) n) where
      -- La fonction élément neutre est la fonction renvoyant
      -- l'élément neutre du monoïde, quelles que soient les
      -- entrées
      mempty = convertFromVectMultiFun $ const mempty

  instance Semigroup m => OperadPlus (MultiFun m m) where
    -- La somme de l'opérade plus des fonctions sur un semigroupe
    -- est la somme définie ci-avant
    (<++>) = (<>)
  \end{minted}
\end{remarqueBox}
Les automates d'arbres pondérés par des fonctions d'opérades plus sont définis sur la catégorie enrichie dont les morphismes-objets sont constitués des fonctions envoyant un vecteur de n états sur une combinaison de couple (élément n-aire de l'opérade, état).
Cette catégorie enrichie, étendant la notion de transduction (envoyant un symbole n-aire sur une combinaison de valeurs n-aires) est implantée dans le module \texttt{Algebra.Functions.GradedTransduction}.
\begin{remarqueBox}[]{Les automates pondérés par des opérades}
  De tels automates permettent, par exemple, d'associer à un arbre nullaire sa hauteur ou sa largeur (alphabétique).
  Si un arbre n'est pas nullaire, son poids est alors une fonction.
  On retrouvera ces automates implantés dans le module \texttt{AutomataGen.TreeAutomata.BottomUp.BottomUpMultiOpMon}, contenant des constructeurs usuels (\texttt{packVar}) et des fonctions utilitaires pour convertir les morphismes poids en scalaire (\texttt{scalarWeight}, cas des arbres nullaires) ou fonctions (\texttt{weightWithVarAsFun}, cas général).
\end{remarqueBox}
L'exemple suivant (Pseudo-Code~\ref{codeOperadPlus})\footnote{exécutable \emph{via} la commande \texttt{stack ghci src/HDRExample/BUMultiOpMonoid.hs}} montre comment définir un automate d'arbre associant à un arbre sur le type \texttt{Char} la multiplication de sa hauteur par sa largeur alphabétique\footnote{nombre de ses n{\oe}uds et feuilles};
les variables acceptées sont des chaînes de caractères, considérées alors comme des arbres de hauteur 1 mais de largeur alphabétique correspondant à leur longueur.
Un tel automate peut être défini comme suit:
\begin{itemize}
  \item trois états: un pour la largeur (\texttt{Just WidthState}), un pour la hauteur (\texttt{Just HeightState}), un pour les combinaisons indésirables de transitions (\texttt{Nothing});
  \item pour chaque symbole nullaire, une transition vers les états pour le calcul de la largeur et de la hauteur avec un poids de \(1\);
  \item pour les symboles non-nullaires d'arité \(k\), une transition envoyant les vecteurs uniformes de valeurs (\texttt{Just WidthState}) sur (\texttt{Just WidthState}) (resp. (\texttt{Just HeightState}) sur (\texttt{Just HeightState})) de poids la fonction associant à \(k\) valeurs leur somme augmentée de \(1\) (resp.\ la fonction associant à \(k\) valeurs leur maximum augmenté de \(1\));
  \item en poids final la fonction identité;
  \item en configuration initiale pour les variables (chaînes de caractères) les fonctions constantes renvoyant leurs longueurs (pour l'état largeur) ou \(1\) (pour la hauteur).
\end{itemize}
\begin{implementationBox}[label={codeOperadPlus}]{Un automate sur une opérade plus}
  \begin{minted}[xleftmargin=1em, autogobble, fontsize=\footnotesize]{haskell}
    -- HDRExample.BUMultiOpMonoid

    -- Définition des états de l'automate, pour calculer
    -- la hauteur et la largeur
    data StateType = HeightState | WidthState

    -- L'automate a trois états: un pour calculer la hauteur
    -- (Just HeightState), un pour la largeur (Just WidthState),
    -- un pour les transitions non définies (Nothing).
    type State = Maybe StateType

    -- Le support est le monoïde multiplicatif des entiers
    type Val = Product Int

    -- Les variables sont des chaines de caractères
    type Variables = String

    -- Les symboles de l'arbres sont des caractères gradués.
    type Inputs = Symbol Char

    -- Fonction de transition associant à un symbole n-aire une transformation
    -- d'un vecteur de longueur n en une combinaison
    -- (fonction n-aire, état de l'automate)
    delta_ :: GradedMorph Inputs (Transduction (MultiFun Val Val) State State)
    delta_ = GradedMorph aux
     where
      aux x = aux' (graduation' x) x
      aux'
        :: Natural n
        -> Graduation Inputs n
        -> Graduation (Transduction (MultiFun Val Val) State State) n
      -- Pour les feuilles (arité 0), les éléments de l'opérade sont
      -- des entiers.
      aux' Zero _ = GradTrans $ const $ fromList
        [(Just HeightState, GradFun 1), (Just WidthState, GradFun 1)]
      -- Dans le cas des symboles (n + 1)-aires, cela dépend du contenu du
      -- vecteur d'états.
      aux' n@(Succ _) _ = GradTrans $ \vs -> if
        -- si tous les états sont HeighState, on utilise la fonction
        -- classique du calcul de la hauteur (maximum des paramètres
        -- + 1)
        | all (== Just HeightState) vs -> fromList
          [(Just HeightState, convertFromVectMultiFun n $ (1 +) . maximum)]
        -- si tous les états sont WidthState, on utilise la fonction
        -- classique du calcul de la largeur (somme des paramètres
        -- + 1)
        | all (== Just WidthState) vs -> fromList
          [(Just WidthState, convertFromVectMultiFun n $ (1 +) . sum)]
        -- sinon on se déplace dans l'état puits avec un poids neutre
        | otherwise -> fromList
          [(Nothing, convertFromVectMultiFun n $ const mempty)]

    -- La fonction de finalité laisse la fonction poids inchangée.
    finalWeight :: State -> Graduation (MultiFun Val Val) Nat_1
    finalWeight _ = GradFun id

    -- On associe aux variables des fonctions de poids constantes,
    -- 1 pour la hauteur, la longueur pour la largeur.
    initialWeight
      :: Variables -> HashMap State (Graduation (MultiFun Val Val) Nat_1)
    initialWeight s = Map.fromList
      [ (Just HeightState, GradFun $ const 1)
      , (Just WidthState , GradFun $ const $ genericLength s)
      ]

    -- Construction de l'automate
    auto :: BottomUpMultiOpMonWithVar Variables (MultiFun Val Val) Inputs State
    auto = TA.packVar initialWeight delta_ finalWeight

    -- a1 = g(a, f(b))
    -- h = 3
    -- w = 4
    a1 :: Graduation (FreeOperad Inputs) Nat_0
    a1 = tree2 g (tree0 a) (tree1 f (tree0 b))

    -- a2 = g(g(a, f(b)), h(g(a, f(b))))
    -- h = 5
    -- w = 10
    a2 :: Graduation (FreeOperad Inputs) Nat_0
    a2 = tree2 g a1 (tree1 h a1)

    -- a3 = g(g(g(a, f(b)), h(g(a, f(b)))), GradEps)
    -- h = 6
    -- w (s) = 11 + |s|
    a3 :: Graduation (FreeOperad Inputs) Nat_1
    a3 = tree2 g a2 GradEps

    -- Affichage des résultats pour les trois arbres définis ci-avant
    testBUMOM :: IO ()
    testBUMOM = do
      print (cast $ scalarWeight auto a1 :: Int) -- affiche 12
      print (cast $ scalarWeight auto a2 :: Int) -- affiche 50
      let test = "operade plus"
      print (cast $ weightWithVarAsFun auto a3 test 0 :: Int) -- affiche 138
  \end{minted}
\end{implementationBox}


\section{Automates généralisés et monades}

La définition de catégories enrichies \emph{ad hoc} n'est pas la seule façon de construire des automates de la littérature classique en utilisant des notions catégoriques.

Comme dans le cas des automates de mots, il est certain que les monades ont un intérêt particulier de par leur pouvoir d'abstraction.
Le cas des automates d'arbres Top-Down est particulièrement intéressant et remarquable de par le fait que le non-déterminisme, contrairement au cas des mots (et au cas des automates d'arbres Bottom-Up que nous traiterons plus tard), permet d'augmenter le pouvoir de représentation en termes de langage.
Il est en effet bien connu qu'il existe des langages reconnaissables par automates Top-Down non déterministes qui ne le sont pas par automates Top-Down déterministes.
De plus, les fonctions de transitions des automates Top-Down non-déterministes, des fonctions graduées envoyant un symbole \(n\)-aire sur un morphisme de \(Q\) (l'objet état) vers \(2^{(Q^n)}\), ne peuvent pas être vu comme des morphismes d'une catégorie enrichie \emph{type Kleisli}, qui seraient plutôt basés sur un morphisme de \(Q\) (l'objet état) vers \({(2^Q)}^n\).

La représentation de ces automates se suffit ainsi de l'extension du type \texttt{DualFunN a b} des fonctions entre \texttt{a} et \texttt{Vector n b} (vecteurs de longueurs \texttt{n} sur \texttt{b}) (Pseudo-Code~\ref{enrichedCatDualMultiFun}) en \texttt{DualFunNF f a b} des fonctions entre \texttt{a} et \texttt{f (Vector n b)}, le cas initial des automates Top-Down déterministes étant isomorphe au cas où \texttt{t} est le foncteur identité.
Ce type est implanté dans le module \texttt{Algebra.Functions} \texttt{.GradedFunctionsTopDown}, où on retrouvera également sa définition en tant que catégorie enrichie sur la catégorie monoïdale des types gradués lorsque \texttt{f} est une monade de la catégorie des types (\texttt{HashSet}, \texttt{Maybe}, \texttt{FreeSemimodule k}, \texttt{Identity}, \emph{etc.}).
Son utilisation permet alors de définir des automates Top-Down classiques avec ou sans variables (implantés comme des cas particuliers d'automates généralisés dans le module \texttt{AutomataGen.TreeAutomata.TopDown.TopDownMonadicAutomata}), en retrouvant, comme dans le cas des mots, le non-déterminisme par la monade \texttt{HashSet},
le déterminisme non-complet par la monade \texttt{Maybe}, des multiplicités dans un semi-anneau par la monade \texttt{FreeSemimodule k}, ou le déterminisme complet par la monade \texttt{Identity}.

L'exemple suivant (Pseudo-Code~\ref{codeExAutTopDownMult}), dont la définition par le constructeur d'automate \texttt{packVar} nécessite
\begin{itemize}
  \item une fonction associant à chaque état une combinaison linéaire de variables;
  \item une fonction graduée de transition associant à chaque symbole gradué \texttt{n}-aire une fonction associant un état à une combinaison linéaire de vecteur de longueur \texttt{n};
  \item une combinaison linéaire d'état pour configuration initiale,
\end{itemize}
 présente l'utilisation d'un automate Top-Down à multiplicités permettant de calculer le nombre d'occurrences d'un arbre (non nécessairement nullaire) dans un arbre nullaire donné.
 Plus particulièrement ici, le nombre d'occurrences d'un arbre \(t\) dans un arbre nullaire \(t'\) est calculé par l'automate défini comme suit:
 \begin{itemize}
   \item les états sont les sous-arbres (nullaires) de \(t'\);
   \item pour tout état de la forme \(f(t_1,\ldots,t_k)\), il y a une transition depuis cet état vers les états \((t_1,\ldots,t_k)\) par le symbole \(f\);
   \item chaque état a pour poids initial son nombre d'occurrences dans \(t'\);
   \item tout état a un poids final de \(1\).
 \end{itemize}
 Cet exemple est exécutable \emph{via} la commande \texttt{stack ghci src/HDRExample/HDRTDWFTA.hs}.
 \begin{implementationBox}[label={codeExAutTopDownMult}]{Un automate Top-Down à multiplicités}
   \begin{minted}[xleftmargin=1em, autogobble, fontsize=\footnotesize]{haskell}
     -- HDRExample.HDRTDWFTA

     -- Symboles de l'alphabet gradué
     type Inputs = Symbol Char

     -- États de l'automate
     type State = Graduation (FreeOperad Inputs) Nat_0

     -- Morphisme de transition
     delta_ :: GradedMorph Inputs (DualFunNF (FreeSemimodule Int) State State)
     delta_ = GradedMorph $
       -- Par la lettre f1,
       \f1 -> DualGradFunF $
         -- depuis l'état f2(t1, ..., tk) avec ts = (t1, ..., tk)
         \(GradOp f2 ts) ->
           -- si f1 = f2
           if eqSymb f1 f2
           -- la destination est le vecteur ts pour un poids de 1
           then monome (toVector ts) 1
           -- sinon pas de destination
           else empty

     -- Les états initiaux sont les sous-arbres nullaires de l'arbre
     -- paramètre, avec une multiplicité correspondant aux occurrences.
     inits :: State -> (FreeSemimodule Int) State
     inits = fromListWith (+) . aux
      where
       aux :: State -> [(State, Int)]
       aux t'@(GradOp f_ ts) = (t', 1) : foldMap aux (toVector ts)

     -- On considère le singleton () pour variables.
     -- Chaque état est envoyé sur () avec un poids de 1.
     finals :: State -> FreeSemimodule Int ()
     finals _ = monome () 1

     -- Définition de l'automate
     auto :: State -> TopDownWFTAWithVar () Int Inputs State
     auto t = packVar finals (inits t) delta_

     -- a1 = g(a, f(b))
     a1 :: Graduation (FreeOperad Inputs) Nat_0
     a1 = tree2 g (tree0 a) (tree1 f (tree0 b))

     -- a2 = g(g(a, f(b)), h(g(a, f(b))))
     a2 :: Graduation (FreeOperad Inputs) Nat_0
     a2 = tree2 g a1 (tree1 h a1)

     -- a3 = g(g(g(a, f(b)), h(g(a, f(b)))), g(a, f(b)))
     a3 :: Graduation (FreeOperad Inputs) Nat_0
     a3 = tree2 g a2 a1

     -- a4 = g(Eps, f(Eps))
     a4 :: Graduation (FreeOperad Inputs) Nat_2
     a4 = tree2 g GradEps (tree1 f GradEps)

     -- a5 = g(Eps, Eps)
     a5 :: Graduation (FreeOperad Inputs) Nat_2
     a5 = tree2 g GradEps GradEps

     testTDWFTA :: IO ()
     testTDWFTA = do
       -- affichage de la multiplicité de a3 dans a3 : 1
       print (weightWithVarAsScalar (auto a3) a3 :: Int)

      -- affichage de la multiplicité de a4 dans a3 : 3
       print (weightWithVarAsScalar (auto a3) a4 :: Int)

      -- affichage de la multiplicité de a5 dans a3 : 5
       print (weightWithVarAsScalar (auto a3) a5 :: Int)
   \end{minted}
 \end{implementationBox}
Le cas des automates d'arbres Bottom-Up non-déterministes est quant à lui un bon candidat pour la représentation de catégories enrichies \emph{type Kleisli}.
En effet, l'extension de la fonction de transition \(\delta \) présentée page~\pageref{section Def Aut Arbres} transforme la fonction de transition de signature \(Q^n\times\Sigma_n \rightarrow 2^Q\) en \({(2^Q)}^n\times\Sigma_n \rightarrow 2^Q\), transformation relativement proche de la promotion monadique.
On peut ainsi s'intéresser à la définition d'une catégorie enrichie dont les morphismes-objets entre deux objets \texttt{a} et \texttt{b} sont les morphismes-objets entre les objets \texttt{a} et \texttt{m b} d'une catégorie enrichie.
Pour cela, commençons par étendre la notion de foncteur à celle de foncteur enrichi.

Un \firstocc{foncteur enrichi}{foncteur!enrichi} \( F \) entre deux catégories
 \( \mathcal{C} \) et \( \mathcal{D} \) enrichies sur la même catégorie monoïdale \( \mathfrak{M}=(\mathcal{M},\otimes,I,\alpha,\lambda,\rho) \) est défini par:
\begin{itemize}
 \item une association de chaque objet \( A \) de \( \mathcal{C} \) à un objet \( F(A) \) de \( \mathcal{D} \),
 \item un morphisme \( F_{A,B} \) de \( \mathrm{Hom}_{\mathcal{M}}(\mathrm{Hom}_{\mathcal{C}}(A,B), \mathrm{Hom}_{\mathcal{D}}(F(A),F(B))) \)
\end{itemize}
tels que les diagrammes suivants commutent:

\hfill
\begin{tikzpicture}[baseline=(current  bounding  box.center)]
 \matrix (m) [matrix of math nodes,row sep=3em,column sep=6em,minimum width=3em]{
   \mathrm{Hom}_{\mathcal{C}}(B,C) \otimes \mathrm{Hom}_{\mathcal{C}}(A,B) & \mathrm{Hom}_{\mathcal{C}}(A,C)\\
   \mathrm{Hom}_{\mathcal{D}}(F(B),F(C)) \otimes \mathrm{Hom}_{\mathcal{D}}(F(A),F(B)) & \mathrm{Hom}_{\mathcal{D}}(F(A),F(C))\rlap{,}\\
 };
 \path[-stealth]
   (m-1-1)
     edge node [above] {\( \circ_{\mathcal{C},(A,B,C)} \)} (m-1-2)
     edge node [left] {\( F_{B,C}\otimes F_{A,B} \)} (m-2-1)
   (m-1-2)
     edge node [right] {\( F(A,C) \)} (m-2-2)
   (m-2-1)
     edge node [below] {\( \circ_{\mathcal{C},(F(A),F(B),F(C))} \)} (m-2-2)
     ;
\end{tikzpicture}
\hfill
\hfill

\hfill
\begin{tikzpicture}[baseline=(current  bounding  box.center)]
 \matrix (m) [matrix of math nodes,row sep=3em,column sep=4em,minimum width=3em]{
   \mathrm{Hom}_{\mathcal{C}}(A,A) & & \mathrm{Hom}_{\mathcal{D}}(F(A),F(A))\\
   & I\rlap{.} & \\
 };
 \path[-stealth]
   (m-1-1)
     edge node [above] {\( F_{A,A} \)} (m-1-3)
   (m-2-2)
     edge node [below left] {\( \mathrm{Id}_A \)} (m-1-1)
     edge node [below right] {\( \mathrm{Id}_{F(A)} \)} (m-1-3)
     ;
\end{tikzpicture}
\hfill
\hfill


\noindent En Haskell, il ne s'agira que d'un morphisme d'une catégorie monoïdale entre deux morphismes-objets\footnote{pour une bonne compatibilité, il faudrait vérifier les propriétés fondamentales des catégories enrichies.}.
\begin{implementationBox}[]{Les foncteurs enrichis}
\begin{minted}[xleftmargin=1em, autogobble, fontsize=\footnotesize]{haskell}
  -- Enriched.Category.EnrichedFunctor

  -- Un constructeur de types f peut être un foncteur entre deux catégories
  -- enrichies enrichedCat et enrichedCat' sur une catégorie monoïdale
  -- catMon si est définie la fonction fmap ci-dessous.
  class (EnrichedCategory enrichedCat catMon tensor,
    EnrichedCategory enrichedCat' catMon tensor)
      => EnrichedFunctor f enrichedCat enrichedCat' catMon tensor where

    -- Promotion fonctorielle : existence d'un morphisme dans catMon
    -- envoyant tout morphisme-objet enrichedCat a b sur
    -- (enrichedCat' (f a) (f b)).
    -- Cette promotion doit être compatible avec l'identité
    -- et l'associativité de la composition.
    fmap :: catMon (enrichedCat a b) (enrichedCat' (f a) (f b))
\end{minted}
\end{implementationBox}
\noindent Par exemple, on peut montrer que tout foncteur est équivalent à un foncteur enrichi sur la catégorie monoïdale des types.
\begin{implementationBox}[]{Les foncteurs enrichis de la catégorie monoïdale des types}
\begin{minted}[xleftmargin=1em, autogobble, fontsize=\footnotesize]{haskell}
  -- Category.OfStructures.CategoryOfTypes

  instance (FunctorGen f cat cat', CategoryGen cat, CategoryGen cat')
    => EnrichedFunctor f (EnrichedTypeCat cat) (EnrichedTypeCat cat') (->) Prod where
    fmap = EnrichedTypeCat . fmap . runEnrichedCat
\end{minted}
\end{implementationBox}
Le cas des fonctions \(n\)-aires est également intéressant: le rôle d'un endofoncteur \texttt{f} enrichi sur la catégorie enrichie de ces fonctions serait alors de promouvoir une fonction de \texttt{a\textsuperscript{n} -> b} en une fonction de \texttt{(f a)\textsuperscript{n} -> f b}.
Pour cela, nous nous restreindrons dans ce document au cas où \texttt{f} est une monade\footnote{On pourrait également s'intéresser aux \href{http://hackage.haskell.org/package/base-4.12.0.0/docs/Control-Applicative.html}{foncteurs applicatifs} (foncteurs monoïdaux particuliers) de la bibliothèque de base Haskell;
mais la manipulation de ces foncteurs pour promouvoir des fonctions \(n\)-aires peut rendre son utilisation complexe.}.
Ainsi, toute monade \texttt{m} de la catégorie des types permet de définir un endofoncteur sur la catégorie enrichie des fonctions \(n\)-aires.
Pour cela, il suffit de montrer comment envoyer un vecteur de \(n\) termes de type \texttt{m a} sur un terme \texttt{m b} pour une fonction donnée envoyant un vecteur de \(n\) termes de type \texttt{a} sur un terme de type \texttt{b}.
\begin{implementationBox}[]{Les foncteurs enrichis de la catégorie enrichies des fonctions \(n\)-aires}
\begin{minted}[xleftmargin=1em, autogobble, fontsize=\footnotesize]{haskell}
  -- Tools.Vector

  -- Définition du type de données des vecteurs de longueur n
  data Vector n a where
    -- VNil est le vecteur de longueur 0
    VNil  :: Vector 'Z a
    -- VCons concatène un élément de type a et
    -- un vecteur de longueur n pour former
    -- un vecteur de longueur n
    VCons :: a -> Vector n a -> Vector ('S n) a

  -- Promotion monadique d'une fonction vectorielle
  promoteFunctVectFun
    :: MonadGen m (->)
    => (Vector n a -> b)
    -> Vector n (m a)
    -> m b
  promoteFunctVectFun f VNil = return $ f VNil
  promoteFunctVectFun f (VCons mx mxs) =
    M.bind (\x -> promoteFunctVectFun (f . VCons x) mxs) mx

  -- Algebra.Functions.GradedFunctions

  -- Si m est une monade de la catégorie des types, m est
  -- un endofoncteur enrichi sur la catégorie des fonctions
  -- n-aires
  instance MonadGen m (->)
    => EnrichedFunctor m MultiFun MultiFun GradedMorph TensorProd where
  fmap = GradedMorph promoteGradMultiFun
   where
    promoteGradMultiFun f =
      convertFromVectMultiFun $ promoteFunctVectFun $ convertToVectMultiFun f
\end{minted}
\end{implementationBox}
L'enrichissement des monades peut se faire grâce à la notion d'élément généralisé; si les monades sont définies par deux morphismes \texttt{return} et \texttt{join} de la catégorie sur laquelle elles existent, une monade enrichie est définie
par deux éléments généralisés ayant le même rôle (à savoir la promotion monadique et la réduction de niveau monadique)\footnote{comme pour l'enrichissement des foncteurs, pour une bonne compatibilité, il faudrait vérifier les propriétés fondamentales des catégories enrichies, comme par exemple définies dans~\cite{Dub06} (chapitre 2, \emph{V-monads}).}, éléments à partir desquels les outils monadiques classiques peuvent être définis.
\textbf{NB\@: Dans le pseudo-code suivant, et plus généralement dans le cadre des catégories enrichies, deux compositions coexistent: celle des catégories (plus précisément celle de la catégorie monoïdale), fonction composant deux morphismes, et celle de la catégorie enrichie, morphisme de la catégorie monoïdale.
Pour lever les ambiguïtés, la composition de la catégorie enrichie sera préfixée par \texttt{EC.} et celle de la catégorie monoïdale par \texttt{Cat.}.
Il en sera de même pour l'identité le cas échéant.}
\begin{implementationBox}[]{Les monades enrichies}
\begin{minted}[xleftmargin=1em, autogobble, fontsize=\footnotesize]{haskell}
  -- Enriched.Category.EnrichedMonad

  -- Une monade enrichie est un endofoncteur enrichi particulier.
  class (EnrichedFunctor m enrichedCat enrichedCat catMon tensor)
    => EnrichedMonad m enrichedCat catMon tensor where

    -- Élément généralisé "choisissant" dans un morphisme-objet
    -- une "transformation" promouvant une valeur au niveau
    -- monadique
    return :: catMon (Unit catMon tensor) (enrichedCat a (m a))

    -- Élément généralisé "choisissant" dans un morphisme-objet
    -- une "transformation" réduisant une valeur d'un niveau
    -- monadique
    join :: catMon (Unit catMon tensor) (enrichedCat (m (m a)) (m a))

  -- Promotion monadique d'un morphisme objet
  bind
    :: ( EnrichedFunctor f enrichedCat enrichedCat catMon tensor
       , EnrichedMonad f enrichedCat catMon tensor
       )
    => catMon (enrichedCat a (f b)) (enrichedCat (f a) (f b))
  bind = append fmap join

  -- Composition de Kleisli enrichie.
  (<=<)
    :: (EnrichedMonad m enrichedCat catMon tensor, Bifunctor tensor catMon)
    => catMon
         (tensor (enrichedCat b (m c), enrichedCat a (m b)))
         (enrichedCat a (m c))
  (<=<) = (EC..) Cat.. combine bind Cat.id
   where
   combine
     :: (Bifunctor tensor catMon, _)
     => catMon (enrichedCat b (m c)) (enrichedCat (m b) (m c))
     -> catMon (enrichedCat a (m b)) (enrichedCat a (m b))
     -> catMon
          (tensor (enrichedCat b (m c), enrichedCat a (m b)))
          (tensor (enrichedCat (m b) (m c), enrichedCat a (m b)))
   combine m1 m2 = F.fmap (Cat.Prod m1 m2)
\end{minted}
\end{implementationBox}
Par exemple,
\begin{itemize}
\item toute monade définit une monade enrichie sur la catégorie monoïdale des types;
\item toute monade sur la catégorie des types définit une monade enrichie sur la catégorie enrichie des fonctions \(n\)-aires.
\end{itemize}
\begin{implementationBox}[]{Exemples de monades enrichies}
\begin{minted}[xleftmargin=1em, autogobble, fontsize=\footnotesize]{haskell}
  -- Category.OfStructures.CategoryOfTypes

  instance MonadGen m cat
    => EnrichedMonad m (EnrichedTypeCat cat) (->) Prod where
    return () = EnrichedTypeCat return
    join () = EnrichedTypeCat join

  -- Algebra.Functions.GradedFunctions

  instance MGen.MonadGen m (->)
    => EnrichedMonad m MultiFun GradedMorph TensorProd where
    return = GradedMorph $ \GradUn -> GradFun return
    join   = GradedMorph $ \GradUn -> GradFun join
\end{minted}
\end{implementationBox}
Afin de pouvoir définir la catégorie enrichie de Kleisli associée à une monade, il nous faut créer le type des morphismes-objets de cette catégorie, à savoir des morphismes-objets entre des objets \texttt{a} et \texttt{m b}.
Cependant, puisque nous manipulerons ces morphismes-objets à travers des morphismes d'une catégorie monoïdale, qui ne sont pas nécessairement des fonctions, il nous faut trouver un moyen de promouvoir les constructeurs de types et les fonctions de projection au niveau de la catégorie monoïdale.
Ainsi, nous nous restreindrons alors aux catégories monoïdales \texttt{catMon} satisfaisant les trois propriétés suivantes:
\begin{itemize}
\item le constructeur de type \texttt{Identity} est un foncteur depuis la catégorie des types vers la catégorie \texttt{catMon}, permettant d'envoyer toute fonction (y compris les constructeurs de types) de signature \texttt{a -> b} sur un morphisme de \texttt{catMon (Identity a) (Identity b)};
\item le constructeur de type \texttt{Identity} est une monade de la catégorie \texttt{catMon}, permettant d'exhiber le morphisme \texttt{return}, de signature \texttt{catMon a (Identity a)};
\item le constructeur de type \texttt{Identity} est une comonade\footnote{monade de la catégorie duale, caractérisée notamment par la fonction \texttt{extract} de signature duale de \texttt{return}} de la catégorie \texttt{catMon}, permettant d'exhiber le morphisme \texttt{extract}, de signature \texttt{catMon (Identity a) a}.
\end{itemize}


Si les deux dernières propriétés permettent de transformer un morphisme de signature \texttt{catMon}\\
\texttt{(Identity a) (Identity b)} en \texttt{catMon a b}, l'ajout de la première permet de transformer toute fonction de signature \texttt{a -> b} en un morphisme \texttt{catMon a b}:
\begin{implementationBox}[]{Conversion fonctions / morphismes}
\begin{minted}[xleftmargin=1em, autogobble, fontsize=\footnotesize]{haskell}
  -- Category.MonadGen

  -- Élimination du constructeur de types Identity dans un morphisme
  -- d'une catégorie
  removeId
    :: (MonadGen Identity cat, CoMonadGen Identity cat, CategoryGen cat)
    => cat (Identity a) (Identity b)
    -> cat a b
  removeId f = extract . f . return

  -- Promotion d'une fonction en un morphisme
  catPromote
    :: ( FunctorGen Identity (->) cat
       , MonadGen Identity cat
       , CoMonadGen Identity cat
       )
    => (a -> b)
    -> cat a b
  catPromote = removeId . fmap

  -- Compose une fonction promue à un morphisme
  catApply
    :: ( FunctorGen Identity (->) cat
       , CategoryGen cat
       , MonadGen Identity cat
       , CoMonadGen Identity cat
       )
    => (b -> c)
    -> cat a b
    -> cat a c
  catApply f m = catPromote f . m
\end{minted}
\end{implementationBox}
À l'aide de ces outils, on peut alors définir un constructeur de types pour les morphismes-objets de la catégorie enrichie de Kleisli associée à une monade enrichie.
Comme dans le cas des catégories de Kleisli (voir Pseudo-Code~\ref{CodeCatKleisli}), les éléments définissant la classe, à savoir les valeurs \texttt{id} et \texttt{(.)}, seront obtenus par la combinaison du constructeur de types \texttt{Kleisli} avec (respectivement) \texttt{return} et \texttt{(<=<)}.

\begin{implementationBox}[]{Catégorie enrichie de Kleisli associée à une monade enrichie}
\begin{minted}[xleftmargin=1em, autogobble, fontsize=\footnotesize]{haskell}
  -- Enriched.Category.Kleisli

  -- Définition du constructeur de type KleisliEnrichedCat, contenant une
  -- valeur de type enrichedCat a (m b), et de la fonction de projection
  -- runKleisli renvoyant cette valeur.
  data KleisliEnrichedCat m enrichedCat catMon tensor a b where
    Kleisli ::
      {runKleisli :: enrichedCat a (m b)}
        -> KleisliEnrichedCat m enrichedCat catMon  tensor a b

  -- Promotion du constructeur de type au niveau de la catégorie monoïdale.
  packKleisli ::
    (FunctorGen Identity (->) catMon, MonadGen Identity catMon, CoMonadGen Identity catMon)
      => catMon (enrichedCat a (m b)) (KleisliEnrichedCat m enrichedCat catMon tensor a b)
  packKleisli = catPromote Kleisli

  -- Promotion de la fonction de projection de type au niveau
  -- de la catégorie monoïdale.
  unpackKleisli ::
    (FunctorGen Identity (->) catMon
    , MonadGen Identity catMon
    , CoMonadGen Identity catMon
    )
    =>  catMon (KleisliEnrichedCat m enrichedCat catMon tensor a b) (enrichedCat a (m b))
  unpackKleisli = catPromote runKleisli

  -- Promotion de la fonction de projection de type au niveau
  -- des produits tensoriels de la catégorie monoïdale.
  unpackProdKleisli ::
    ( FunctorGen Identity (->) catMon
    , MonadGen Identity catMon
    , CoMonadGen Identity catMon
    , Bifunctor tensor catMon)
      => catMon
        (tensor
          (KleisliEnrichedCat m enrichedCat catMon tensor b c
          , KleisliEnrichedCat m enrichedCat catMon tensor a b))
        (tensor (enrichedCat b (m c), enrichedCat a (m b)))
  unpackProdKleisli = fmap $ Prod unpackKleisli unpackKleisli

  -- Définition de la catégorie enrichie de Kleisli associée
  -- à la monade enrichie m
  instance
    ( FunctorGen Identity (->) catMon
    , MonadGen Identity catMon
    , EnrichedMonad m enrichedCat catMon tensor
    , CoMonadGen Identity catMon
    , Monoidal catMon tensor
    )
      => EnrichedCategory (KleisliEnrichedCat m enrichedCat catMon tensor) catMon tensor
        where

    id  = catApply Kleisli return
    (.) = packKleisli Cat.. (<=<) Cat.. unpackProdKleisli
\end{minted}
\end{implementationBox}
Dans le cas des automates de mots généralisés, on peut encore montrer que les automates généralisés de Kleisli sont équivalents aux automates (catégoriques) de Kleisli.
\begin{implementationBox}[]{Équivalence entre automates de mots de Kleisli et automates généralisés de mots de Kleisli}
\begin{minted}[xleftmargin=1em, autogobble, fontsize=\footnotesize]{haskell}
  -- AutomataGen.WordAutomata.WordAut

  convertToClassicalKleisliAutomaton
    :: WordAut
         (KleisliEnrichedCat m (EnrichedTypeCat cat) (->) Prod)
         init
         symbols
         state
         value
    -> Automaton (KleisliCat m cat) init symbols state value
  convertToClassicalKleisliAutomaton (AG.Auto i d f) = A.Auto
    (Kleisli $ runEnrichedCat $ runKleisli $ i ())
    (Kleisli . runEnrichedCat . runKleisli . d)
    (Kleisli $ runEnrichedCat $ runKleisli $ f ())

  convertFromClassicalKleisliAutomaton
    :: (MonadGen m cat)
    => Automaton (KleisliCat m cat) init symbols state value
    -> WordAut
         (KleisliEnrichedCat m (EnrichedTypeCat cat) (->) Prod)
         init
         symbols
         state
         value
  convertFromClassicalKleisliAutomaton (A.Auto i d f) = AG.Auto
    (\() -> KECat.Kleisli $ EnrichedTypeCat $ KCat.runKleisli i)
    (KECat.Kleisli . EnrichedTypeCat . KCat.runKleisli . d)
    (\() -> KECat.Kleisli $ EnrichedTypeCat $ KCat.runKleisli f)
\end{minted}
\end{implementationBox}


Pour les automates d'arbres Bottom-Up, il suffit également de considérer des synonymes de types pour retrouver les automates monadiques classiques (déterministes non complets, non-déterministes, \emph{etc.})
\begin{implementationBox}[]{Automates d'arbres Bottom-Up monadiques}
  \begin{minted}[xleftmargin=1em, autogobble, fontsize=\footnotesize]{haskell}
    -- AutomataGen.TreeAutomata.BottomUp.KleisliTreeAutomata

    type KleisliBottomUpTreeAutWithVar m var symbols state
      = TreeAut
          (KleisliEnrichedCat m MultiFun GradedMorph TensorProd)
          var
          symbols
          state
          ()

    type KleisliBottomUpTreeAut m symbols state
      = KleisliBottomUpTreeAutWithVar m Void symbols state

    type BottomUpWFTAWithVar var k symbols state
      = KleisliBottomUpTreeAutWithVar (FreeSemimodule k) var symbols state
    type BottomUpWFTA k symbols state
      = KleisliBottomUpTreeAut (FreeSemimodule k) symbols state

    type BottomUpNFTAWithVar var symbols state
      = KleisliBottomUpTreeAutWithVar HashSet var symbols state
    type BottomUpNFTA symbols state = KleisliBottomUpTreeAut HashSet symbols state

    type BottomUpDFTAWithVar symbols state var
      = KleisliBottomUpTreeAutWithVar Maybe var symbols state
    type BottomUpDFTA symbols state = KleisliBottomUpTreeAut Maybe symbols state

    -- Construction d'un automate d'arbres Bottom-Up de Kleisli
    -- avec variables.
    packVar
      :: ( Graded symbols
         , Castable (m weight) k
         , EnrichedMonad m MultiFun GradedMorph TensorProd
         )
      => (var -> m state)
      -> GradedMorph symbols (MultiFun state (m state))
      -> (state -> k)
      -> TreeAut
           (KleisliEnrichedCat m MultiFun GradedMorph TensorProd)
           var
           symbols
           state
           weight
    packVar i delta_ finalWeights = Auto i_ transit end
     where
      i_      = packKleisli . gradedMorphFromFun i
      transit = packKleisli . delta_
      end     = packKleisli . gradedMorphFromFun (castInv . finalWeights)

    -- Construction d'un automate d'arbres Bottom-Up de Kleisli
    -- sans variable.
    pack
      :: ( Graded symbols
         , Castable (m weight) k
         , EnrichedMonad m MultiFun GradedMorph TensorProd
         )
      => GradedMorph symbols (MultiFun state (m state))
      -> (state -> k)
      -> TreeAut
           (KleisliEnrichedCat m MultiFun GradedMorph TensorProd)
           Void
           symbols
           state
           weight
    pack = packVar absurd

    -- Fonction auxiliaire pour calculer le poids associé
    -- à un arbre non nécessairement nullaire
    weightWithVar
      :: (Graded symbols, EnrichedMonad m MultiFun GradedMorph TensorProd)
      => TreeAut
           (KleisliEnrichedCat m MultiFun GradedMorph TensorProd)
           var
           symbols
           state
           weight
      -> Graduation (FreeOperad symbols) n
      -> NFun n var (m weight)
    weightWithVar auto = runGradFun . applyMorph (unpackKleisli Cat.. weight auto)

    -- Fonction auxiliaire pour calculer le poids associé
    -- à un arbre nullaire
    weightAsScalar
      :: ( Graded symbols
         , Castable (m ()) k
         , EnrichedMonad m MultiFun GradedMorph TensorProd
         , _
         )
      => TreeAut
           (KleisliEnrichedCat m MultiFun GradedMorph TensorProd)
           Void
           symbols
           state
           ()
      -> Graduation (FreeOperad symbols) Nat_0
      -> k
    weightAsScalar auto =
      runGradFun . applyMorph (EC.packCast Cat.. unpackKleisli Cat.. weight auto)

    -- Fonction auxiliaire pour calculer le poids associé
    -- à un arbre nullaire dans le cas Booléeen
    recognizes
      :: ( Graded symbols
         , Castable (m ()) Bool
         , EnrichedMonad m MultiFun GradedMorph TensorProd
         )
      => TreeAut
           (KleisliEnrichedCat m MultiFun GradedMorph TensorProd)
           Void
           symbols
           state
           ()
      -> Graduation (FreeOperad symbols) Nat_0
      -> Bool
    recognizes = weightAsScalar
  \end{minted}
\end{implementationBox}


Pour conclure cette partie sur les automates généralisés, remarquons que pour toute catégorie enrichie de Kleisli, il existe un foncteur enrichi envoyant cette catégorie enrichie sur la catégorie dont elle est issue.
En effet, comme dans le cas des catégories de Kleisli, il est possible d'envoyer un morphisme-objet entre \texttt{a} et \texttt{b} de la catégorie enrichie de Kleisli (c'est-à-dire un morphisme-objet entre \texttt{a} et \texttt{m b} de la catégorie enrichie de départ) sur le morphisme-objet entre \texttt{m a} et \texttt{m b}, en appliquant la fonction \texttt{bind}.
Ce foncteur enrichi, appelé classiquement \firstocc{foncteur de Kleisli}{foncteur!de Kleisli}, permet d'implanter d'une façon unique les algorithmes de déterminisation ou de complétion, et cela indépendamment de la catégorie enrichie: dans tous les cas, on envoie un automate généralisé de Kleisli sur un automate généralisé.
\begin{implementationBox}[label={codeConversionGeneraliseeKleisli}]{Foncteur enrichi de Kleisli et conversions}
\begin{minted}[xleftmargin=1em, autogobble, fontsize=\footnotesize]{haskell}
  -- Enriched.Category.Kleisli

  instance
    (EnrichedCategory enrichedCat catMon tensor
    , EnrichedMonad m enrichedCat catMon tensor
    , MonadGen Identity catMon
    , CoMonadGen Identity catMon
    , FunctorGen Identity (->) catMon)
    => EnrichedFunctor m (KleisliEnrichedCat m enrichedCat catMon tensor) enrichedCat catMon tensor
      where

    fmap = bind . unpackKleisli

  -- AutomataGen.AutomatonGen

  -- Conversion générale entre automates généralisés.
  -- Retire un niveau de structure monadique dans la catégorie
  -- enrichie et ajoute ce niveau aux configurations de l'automate.
  unKleisli
    :: EnrichedMonad m enrichedCat catMon tensor
    => AutomatonGen
         (KleisliEnrichedCat m enrichedCat catMon tensor)
         catMon
         tensor
         init
         symbols
         state
         value
    -> AutomatonGen
         enrichedCat
         catMon
         tensor
         init
         symbols
         (m state)
         (m value)
  unKleisli (Auto i d f) = Auto i' d' f'
   where
    i' = unpackKleisli . i
    d' = fmap . d
    f' = fmap . f

  -- Identique à unKleisli, avec une modification
  -- du poids de sortie
  unKleisliWithCast
    :: ( EnrichedMonad m enrichedCat catMon tensor
       , Castable (m value) value'
       )
    => AutomatonGen
         (KleisliEnrichedCat m enrichedCat catMon tensor)
         catMon
         tensor
         init
         symbols
         state
         value
    -> AutomatonGen enrichedCat catMon tensor init symbols (m state) value'
  unKleisliWithCast a = Auto i d f'
   where
    Auto i d f = unKleisli a
    f'         = packCast . f

  -- AutomataGen.WordAutomata.WordAut

  determinise
    :: (Hashable state, Eq state)
    => WordAut
         (KleisliEnrichedCat HashSet (EnrichedTypeCat (->)) (->) Prod)
         init
         symbols
         state
         ()
    -> WordAut (EnrichedTypeCat (->)) init symbols (HashSet state) Bool
  determinise = unKleisliWithCast

  complete
    :: WordAut
         (KleisliEnrichedCat Maybe (EnrichedTypeCat (->)) (->) Prod)
         init
         symbols
         state
         ()
    -> WordAut (EnrichedTypeCat (->)) init symbols (Maybe state) Bool
  complete = unKleisliWithCast

  -- AutomataGen.TreeAutomata.BottomUp.KleisliTreeAutomata

  determinise
    :: (Hashable state, Eq state, Graded symbols)
    => BottomUpNFTA symbols state
    -> BottomUpCompDFTA symbols (HashSet state)
  determinise = unKleisliWithCast

  complete
    :: (Graded symbols)
    => BottomUpDFTA symbols state
    -> BottomUpCompDFTA symbols (Maybe state)
  complete = unKleisliWithCast
\end{minted}
\end{implementationBox}
\begin{remarqueBox}[]{Les fonctions \texttt{unKleisli} et \texttt{applyFunctor}}
La fonction de conversion \texttt{applyFunctor} définie pour les conversions des automates est en quelque sorte plus générale dans son rôle que la fonction \texttt{unKleisli};
la première permet de transformer des catégories de Kleisli, la seconde de dé\emph{Kleisli}fier une catégorie enrichie de Kleisli.
\end{remarqueBox}
Les deux exemples suivants illustrent l'utilisation de la même fonction \texttt{unKleisli} pour déterminiser des automates de mots et d'arbres.
Ils sont exécutables \emph{via} les commandes\\
\texttt{stack ghci src/HDRExample/ExampleWordGenDet.hs}\\
et \texttt{stack ghci src/HDRExample/ExampleBUNTA.hs}.


Le Pseudo-Code~\ref{codeDetAutGenMot} montre comment déterminiser l'automate défini au Pseudo-Code~\ref{pseudoCodeDetComp}, en utilisant les automates généralisés: l'automate classique est converti en automate généralisé, il est ensuite déterminisé par la fonction \texttt{determinise} (synonyme de la fonction \texttt{unKleisliWithCast}) définie au Pseudo-Code~\ref{codeConversionGeneraliseeKleisli}, puis le résultat est reconverti en automate classique.
Les transitions de ce dernier automate sont alors affichées.
\begin{implementationBox}[label={codeDetAutGenMot}]{Déterminisation d'un automate (généralisé) de mots}
\begin{minted}[xleftmargin=1em, autogobble, mathescape=true, fontsize=\footnotesize]{haskell}

  -- HDRExample.ExampleWordGenDet

  import Automata.DFAComp                 as DFA
  import Automata.KleisliAutomata         as NFA
  import AutomataGen.WordAutomata.WordAut as WA
  import HDRExample.KleisliAuto           as KA

  -- automate du $\text{Pseudo-Code \ref{pseudoCodeDetComp}}$
  k_auto :: NFA KA.Symbol Int
  k_auto = KA.auto 3

  -- conversion en automate non déterministe généralisé
  w_auto :: WA.NonDetWordAut KA.Symbol Int
  w_auto = WA.convertFromClassicalKleisliAutomaton k_auto

  -- déterminisation (en tant qu'automate généralisé)
  -- appelant le synonyme de unKleisli
  w_auto2 :: WA.CompDetWordAut KA.Symbol (HashSet Int)
  w_auto2 = WA.determinise w_auto

  -- conversion en un automate classique
  k_auto2 :: DFAComp KA.Symbol (HashSet Int)
  k_auto2 = WA.convertToClassicalAutomaton w_auto2

  -- Affichage des états et des transitions
  -- des automates k_auto
  -- et de son déterminisé (obtenu en passant
  -- par les automates généralisés)
  testWordAutGen :: IO ()
  testWordAutGen = do

    putStrLn "k_auto transitions: \n\t"
    let trans = NFA.getTransitions k_auto
    putStrLn $ toString trans

    putStrLn "k_auto2=Det(k_auto) transitions: \n\t"
    let trans2 = DFA.getTransitions k_auto2
    putStrLn $ toString trans2
\end{minted}
\end{implementationBox}

\begin{figure}[H]
\centerline{
  \begin{tikzpicture}[node distance=4cm,bend angle=30,transform shape,scale=1]
    \node[state] (q1) {$1$};
    \node[state, left of=q1, accepting] (q2) {$2$};
    \draw (q1) ++(0cm,-1cm) node {$a, b$}  edge[->] (q1);
    \path[->]
      (q1) edge[->,above] node {$f$} (q2)
      (q1) edge[loop,->,above left] node {$f$} ()
      (q2) edge[loop,->,above right] node {$h$} ()
    ;
    \draw (q1) ++(1.5cm,0cm)  edge[out=-90,in=-60,->] node[below right] {$g$} (q1) edge[shorten >=0pt, bend right] (q1) edge[shorten >=0pt, bend left] (q1);
  \end{tikzpicture}
}
\caption{Un automate Bottom-Up non-déterministe.}%
\label{figExBUTnonDet}
\end{figure}
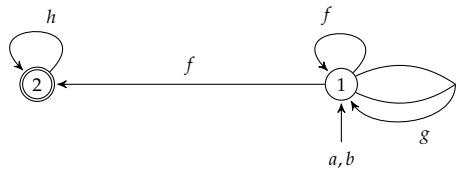

Le Pseudo-Code~\ref{codeDetGeneraliseeBotUpAut} montre comment définir et déterminiser l'automate de la Figure~\ref{figExBUTnonDet}
en appliquant un synonyme de la même fonction polymorphe \texttt{unKleisliWithCast}.

\begin{implementationBox}[label={codeDetGeneraliseeBotUpAut}]{Déterminisation d'un automate (généralisé) d'arbres}
\begin{minted}[xleftmargin=1em, autogobble, mathescape=true, fontsize=\footnotesize]{haskell}

  -- HDRExample.ExampleBUNTA

  -- synonyme de type
  type State = Int

  -- fonction de transition
  delta_ :: GradedMorph (Symbol Char) (MultiFun State (HashSet State))
  delta_ = GradedMorph aux
   where
    aux x = aux' (graduation' x) x
    aux'
      :: Natural n
      -> Graduation (Symbol Char) n
      -> Graduation (MultiFun State (HashSet State)) n
    aux' Zero x | eqSymb x a = GradFun $ singleton 1
                | eqSymb x b = GradFun $ singleton 1
    aux' (Succ Zero) x
      | eqSymb x f = GradFun $ \case
        1 -> fromList [1, 2]
        _ -> empty
      | eqSymb x h = GradFun $ \case
        2 -> fromList [2]
        _ -> empty
    aux' (Succ (Succ Zero)) x
      | eqSymb x g = GradFun $ \x_ y_ -> case (x_, y_) of
        (1, 1) -> singleton 1
        _      -> empty
    aux' n _ = convertFromVectMultiFun n $ const Set.empty

  -- états finaux
  isFinal :: Int -> Bool
  isFinal 2 = True
  isFinal _ = False

  -- définition de l'automate non déterministe
  auto :: BottomUpNFTA (Symbol Char) State
  auto = TA.pack delta_ isFinal

  -- déterminisation par la fonction (synonyme de)
  -- unKleisli
  auto2 :: BottomUpDFTA (Symbol Char) (HashSet State)
  auto2 = convertDet $ determinise auto

  -- affichage des états et des transitions de l'automate
  -- et de son déterminisé
  testBUNTA :: IO ()
  testBUNTA = do
    let arities = GCons a $ GCons b $ GCons h $ GCons f $ GCons g GNil

    putStrLn "Auto: "
    let (states, trans) =
          accessibleStatesAndTransitionsForWithInit Set.empty arities auto
    putStrLn $ "Transitions: " ++ toString trans
    putStrLn $ "States: " ++ toString states

    putStrLn "Det(Auto):"
    let (states2, trans2) =
          accessibleStatesAndTransitionsForWithInit Set.empty arities auto2
    putStrLn $ "Transitions: " ++ toString trans2
    putStrLn $ "States: " ++ toString states2
\end{minted}
\end{implementationBox}

\section{Conclusion}

Dans cette section, nous avons montré comment unifier les notions d'automates de mots et d'arbres \emph{via} l'utilisation des catégories enrichies, tout en conservant les possibilités d'agilité structurelle des catégories de Kleisli de la section précédente.
Pour cela, nous avons manipulé les catégories enrichies des types et des types gradués.

Bien évidemment, il est possible de définir d'autres types en entrée des automates généralisés: on pourrait s'intéresser aux automates de mots commutatifs, ou aux automates de PROs~\cite{LLMN19} (graphes particuliers, généralisation des arbres).
Il suffirait pour cela de changer la catégorie monoïdale sous-jacente des catégories enrichies à considérer: quelle serait par exemple la structure algébrique sous-jacente des monoïdes-objets d'une catégorie monoïdale sur les types bi-gradués?

La section suivante pose la question de l'extension de cette généralisation aux expressions; notamment, nous allons voir comment factoriser les structures d'expressions de mots et d'expressions d'arbres, ces structures n'utilisant pas, à première vue, des opérateurs de même nature.
Ainsi, nous répondrons à la question suivante.
Est-il possible, une fois une catégorie monoïdale fixée, d'en déduire un jeu d'opérateurs permettant de représenter le fonctionnement des automates généralisés associés?

\chapter{Expressions rationnelles}\label{chapRegExpCat}


Dans cette section, nous allons nous intéresser à la façon de décrire des expressions afin de généraliser la construction classique d'automates aux automates catégoriques.
Nous réaliserons cette généralisation en procédant en deux étapes.

Tout d'abord, nous allons montrer comment généraliser les constructions aux automates de Kleisli définis dans les sections~\ref{sec:automatesMotsCat} et~\ref{sec:automatesMotsKleisli}, c'est-à-dire non enrichie.
Pour cela, nous nous appuierons sur un type classique d'expressions à multiplicités (mais étendu à de nouveaux opérateurs) et nous étendrons d'une façon directe les constructions classiques de calcul de positions, de calcul de dérivées ou de calcul inductif en montrant comment factoriser les calculs au niveau monadique.

Nous montrerons ensuite, par un type d'expression nouveau, comment définir des expressions enrichies permettant de généraliser expressions de mots et d'arbres, et étendrons quelques constructions classiques sur celui-ci.

Nous conclurons alors ce document sur les pistes possibles offertes par cette définition.

\section{Définition et implantation des expressions de mots}

  Nous avons vu dans les chapitres précédents différentes modélisations des automates réalisées \emph{via} différentes monades.
  Afin de préserver la diversité des cas présentés, nous allons construire un type d'expressions assez général pour modéliser des opérations de natures variées, par exemple en autorisant toute fonction, non nécessairement unaire, pour combiner des poids, comme lors de la définition des automates alternants généralisés.

  Ainsi, dans cette section, nous considérerons des expressions définies, une fois un semi-anneau fixé, inductivement comme:
  \begin{enumerate}
    \item l'ensemble vide, associant à tout mot un poids nul,
    \item un caractère, associant à celui et uniquement à celui-ci un poids unitaire,
    \item le mot vide, associant à celui et uniquement à celui-ci un poids unitaire,
    \item un opérateur \(n\)-aire appliqué sur \(n\) expressions,
  \end{enumerate}
  où les opérateurs considérés seront les suivants:
  \begin{enumerate}
    \item la concaténation binaire,
    \item l'étoile unaire,
    \item la somme binaire,
    \item les multiplications unaires gauche et droite par un scalaire,
    \item la négation unaire et l'intersection binaire, pour les semi-anneaux Booléens,
    \item un opérateur générique de fonctions \(n\)-aires (par exemple pour construire des automates alternants généralisés utilisés dans le Pseudo-Code~\ref{codeGenAFA}).
  \end{enumerate}
  Afin de faciliter l'implantation, les particularités des opérateurs génériques seront définies à l'aide d'une famille de types dépendant de l'arité de l'opérateur et du type de poids.
  \begin{implementationBox}[]{Opérateurs des expressions de mots}
    \begin{minted}[xleftmargin=1em, autogobble, mathescape=true, fontsize=\footnotesize]{haskell}

      -- Expressions.WordExp.Operator

      -- Famille de types permettant de définir un opérateur
      -- générique en fonction de son arité.
      -- Un opérateur d'arité 1 est défini par un Booléen
      -- spécifiant si l'opérateur est post-fixe, ainsi que
      -- d'une fonction unaire.
      -- Un opérateur d'arité 2 est défini par un Booléen
      -- indiquant si l'opérateur est infixe, un second Booléen
      -- indiquant si l'opérateur est associatif à gauche ou
      -- à droite, un entier pour indice de priorité,
      -- ainsi que d'une fonction binaire.
      -- Pour les autres arités, seule la fonction est donnée.
      type family InfoOp n a where
        InfoOp Nat_1 a = (Bool, Fun Nat_1 a)
        InfoOp Nat_2 a = (Bool, Bool, Int, Fun Nat_2 a)
        InfoOp n a = Fun n a

      -- Type des opérateurs utilisés dans une expression.
      -- Oper a n est un opérateur n-aire sur une expression
      -- dont les poids sont de types a.
      data Oper a n  where
        -- Opérateur générique défini à partir
        Function ::
          String -- d'une chaîne pour sa description,
          -> Natural n -- d'une arité,
          -> InfoOp n a -- d'une information (définie ci-avant)
          -> Oper a n
        -- Concaténation, binaire
        Concat ::Oper a Nat_2
        -- Étoile d'itération, unaire
        Star ::Oper a Nat_1
        -- Somme, binaire
        Plus ::Oper a Nat_2
        -- Multiplication à gauche par un scalaire, unaire
        MultL ::a -> Oper a Nat_1
        -- Multiplication à droite par un scalaire, unaire
        MultR ::a -> Oper a Nat_1
        -- Complémentation Booléenne, unaire
        Not ::Oper Bool Nat_1
        -- Intersection Booléenne, binaire
        Inter ::Oper Bool Nat_2
    \end{minted}
  \end{implementationBox}
  Les opérateurs génériques permettent ainsi de définir simplement des opérateurs d'expressions génériques, comme le montre le Pseudo-Code suivant.
  \begin{implementationBox}[]{Opérateurs customisés}
    \begin{minted}[xleftmargin=1em, autogobble, mathescape=true, fontsize=\footnotesize]{haskell}

      -- Expressions.WordExp.Operator

      -- Opérateur binaire d'implication Booléenne
      implOper :: Oper Bool Nat_2
      implOper = Function "->" two (True, False, 4, GradFun (<=))

      -- Opérateur unaire d'exponentiation par un entier
      powerOper :: (Fractional a, Integral b, Show b) => b -> Oper a Nat_1
      powerOper n = Function ("^" ++ show n) one (True, GradFun (^^ n))

      -- Opérateur binaire de calcul de moyenne
      meanOper :: (Fractional a) => Oper a Nat_2
      meanOper =
        Function "moyenne" two (False, False, 0, GradFun (\x y -> (x + y) / 2))
    \end{minted}
  \end{implementationBox}
  À l'aide de ces opérateurs, les expressions sont définies comme suit.
  \begin{implementationBox}[label={codeExpressionsMots}]{Expressions de mots}
    \begin{minted}[xleftmargin=1em, autogobble, mathescape=true, fontsize=\footnotesize]{haskell}

      -- Expressions.WordExp.Expression

      -- Type de données (GADT) des expressions
      data Expression weight character where
        Epsilon :: Expression weight character
        Empty :: Expression weight character
        Symbol :: character -> Expression weight character
        Operator ::
          Oper weight n
          -> Vector n (Expression weight character)
          -> Expression weight character
    \end{minted}
  \end{implementationBox}

\section{Constructions d'automates de mots}

  Montrons alors comment factoriser certaines méthodes de construction d'automates connues pour obtenir à partir des expressions définies dans le Pseudo-Code~\ref{codeExpressionsMots} des automates de Kleisli.
  Plus précisément, nous allons montrer comment des mêmes expressions permettent d'obtenir des automates de Kleisli sur différentes monades,
  permettant ainsi d'interpréter une même expression de différentes façons possibles, à l'aide du polymorphisme et de la programmation au niveau des types (à l'aide des GADTs).

  Nous considérerons trois méthodes classiques de construction.
  La première, la méthode par induction, ne sera présentée que pour construire des automates non-déterministes ou à multiplicités, afin de présenter la notion de monade libre que l'on utilisera dans la section suivante.
  La deuxième, la méthode par positions, sera restreinte aux opérateurs \emph{compatibles} avec la notion de position (les expressions simples ou les multiplications par scalaires).
  Enfin, la dernière, la dérivation, sera la plus générale possible, afin d'essayer de tirer profit des propriétés de la monade ambiante.

  \begin{remarqueBox}[]{Les semianneaux étoilés}
    Il est important de noter que l'étoile de Kleene, lorsqu'elle s'applique sur des langages pondérés (ou séries), doit être accompagnée d'une opération similaire sur le semianneau des poids.
    Ainsi, dans la suite, nous considérerons le module Haskell \texttt{Data.Star} du package \href{http://hackage.haskell.org/package/semirings}{\texttt{semirings}}, définissant les \firstocc{semianneaux étoilés}{semianneau!étoilé}
    comme des semianneaux munis d'une opération unaire \({}^*\) satisfaisant l'équation suivante (voir~\cite{CLOZ04,LS05} pour d'autres définitions), pour tout élément \(x\) du semianneau:
    \begin{equation*}
      x^* = 1 + x^* \times x = 1 + x \times x^*.
    \end{equation*}
  \end{remarqueBox}

    \subsection{Induction et monade libre}\label{subsecConsIndMots}

    La première méthode que nous allons décrire est la construction inductive: nous montrerons dans la suite comment factoriser cette construction afin d'obtenir, uniquement par un processus de typage, des automates de Kleisli distincts (des automates non-déterministes ou à multiplicités) tout en conservant les bénéfices du polymorphisme pour les cas généraux.

    Remarquons que cette méthode inductive peut être interprétée à la façon d'un morphisme d'opérades, en considérant l'arbre syntaxique d'une expression.
    En effet l'idée de la construction est de:
    \begin{itemize}
      \item générer à partir de chaque atome (c'est-à-dire une feuille de l'arbre, ici le mot vide, les symboles ou l'ensemble vide) un automate \emph{reconnaissant unitairement} cet atome;
      \item associer à chaque opérateur \(n\)-aire une fonction \(n\)-aire sur l'ensemble des automates.
    \end{itemize}

    Les fonctions de somme et d'intersection Booléenne ont déjà été décrites dans la Remarque~\ref{RemUnionInterAuto} pour les automates non-déterministes et à multiplicités.

    La concaténation de deux automates de Kleisli peut également être factorisée en considérant que les monades d'ensemble et de semimodule libre, une fois appliquées, forment des semi-modules pour les types des poids de sorties des automates associés:
    \begin{itemize}
      \item tout ensemble est un semimodule Booléen (Pseudo-Code~\ref{ensembleSemimodBool});
      \item tout \(\textbb{K}\)-semimodule libre est (par essence) un \(\textbb{K}\)-semimodule (voir Pseudo-Code~\ref{semimodLibres}).
    \end{itemize}
    \begin{implementationBox}[label={ensembleSemimodBool}]{Monades et semimodules}
      \begin{minted}[xleftmargin=1em, autogobble, mathescape=true, fontsize=\footnotesize]{haskell}

        -- Algebra.Structures.Semimodule

        -- Les Booléens agissent sur les ensembles
        instance (Hashable a, Eq a) => Semimodule Bool (HashSet a) where
          -- True agit comme l'identité
          action True  e = e
          -- False annule l'ensemble et le vide
          action False _ = Set.empty

          actionR = flip action
      \end{minted}
    \end{implementationBox}
  Ainsi, la concaténation de deux automates se construit comme suit:
  \begin{enumerate}
    \item réalisation de la somme de deux automates en ne conservant comme configuration initiale que celle du premier automate;
    \item modification des morphismes de transitions en ajoutant comme destination aux états \(p\) du premier automate la configuration initiale du second après avoir fait agir dessus le poids de finalité  de \(p\);
    \item action, sur les poids de sortie des états du premier automate, du poids du mot vide du second automate.
  \end{enumerate}
  \begin{implementationBox}[]{Concaténation de deux automates de Kleisli}
    \begin{minted}[xleftmargin=1em, autogobble, mathescape=true, fontsize=\footnotesize]{haskell}

      -- Automata.InductionAutomatonFromExp

      concatGen
        :: ( MonadGen m (->)
           , Hashable state
           , Eq state
           , Eq state'
           , Hashable state'
           , Castable (m ()) weight
           , Semimodule weight (m ())
           , Semimodule weight (m (Sum (state, state')))
           , Semigroup (m (Sum (state, state')))
           , Castable (KleisliCat m (->) () ()) weight
           )
        => Automaton (KleisliCat m (->)) () symbols state ()
        -> Automaton (KleisliCat m (->)) () symbols state' ()
        -> Automaton (KleisliCat m (->)) () symbols (Sum (state, state')) ()
      concatGen a1@(Auto _ _ (Kleisli f1)) a2@(Auto _ _ (Kleisli f2)) = Auto i d' f'
       where
        -- réalisation de la somme des deux automates
        Auto i d _ = reduceInitialSum const $ binaryTransformationBySum a1 a2
        -- poids du mots vide dans a2
        epsilonWeight2 = weightValue a2 []
        -- nouveau poids de finalité pour les états de a1
        f'             = Kleisli $ \case
          Fst p -> f1 p `actionR` epsilonWeight2
          Snd q -> f2 q
        -- nouvelle transition pour les états de a1 :
        -- est ajoutée la configuration initiale de a2
        -- après action du poids de finalité de l'état
        -- de départ
        d' a = Kleisli $ \case
          p@(Fst p_) ->
            runKleisli (d a) p
              <> ((cast (f1 p_) :: weight) `action` fmap
                   Snd
                   (runKleisli initsByA ())
                 )
          q -> runKleisli (d a) q
          where initsByA = getConfig a2 [a]

    \end{minted}
  \end{implementationBox}
  L'étoile se construit d'une façon similaire:
  \begin{enumerate}
    \item ajout d'un état supplémentaire d'un poids final unitaire, qui devient configuration initiale après avoir fait agir dessus le poids du mot vide passé à l'étoile du semi-anneau étoilé; ses successeurs sont constitués par les successeurs de l'ancienne configuration initiale;
    \item modification des morphismes de transitions en ajoutant comme destination aux états \(p\) de l'automate initial la configuration initiale de cet automate après avoir fait agir dessus le poids de finalité de \(p\);
    \item action, sur les poids de sortie des états de l'automate initial, du poids étoilé du mot vide de cet automate.
  \end{enumerate}
  \begin{implementationBox}[]{Étoile d'un automate}
    \begin{minted}[xleftmargin=1em, autogobble, mathescape=true, fontsize=\footnotesize]{haskell}

      -- Automata.InductionAutomatonFromExp

      starGen
        :: ( Eq state
           , Hashable state
           , Castable (m ()) weight
           , Star weight
           , MonadGen m (->)
           , Semigroup (m (Maybe state))
           , Semimodule weight (m (Maybe state))
           , Castable (KleisliCat m (->) () ()) weight
           )
        => Automaton (KleisliCat m (->)) () symbols state ()
        -> Automaton (KleisliCat m (->)) () symbols (Maybe state) ()
      starGen aut@(Auto _ d (Kleisli f)) = Auto
        -- ajout d'un nouvel état constituant la configuration
        -- initiale de l'automate
        (Kleisli $ const $ action (star epsWeight) $ return Nothing)
        (Kleisli . d')
        (Kleisli finality)
       where
        -- les successeurs de la nouvelle configuration initiale
        -- sont constitués par les successeurs de l'ancienne
        -- configuration initiale
        d' a = maybe (initsBy a) succOf
         where
          -- sont ajoutés aux successeurs d'un état p de l'automate initial
          -- les successeurs de l'ancienne configuration initiale
          -- après action du poids final de p
          succOf p = ((cast (f p) :: weight) `action` initsBy a)
            <> fmap Just (runKleisli (d a) p)
        -- successeurs de l'ancienne configuration initiale
        initsBy a = fmap Just $ runKleisli (getConfig aut [a]) ()
        -- poids du mot vide dans l'automate initial
        epsWeight = weightValue aut []
        -- finalité (unitaire) de l'état ajouté
        finality Nothing  = castInv (one :: weight)
        -- finalité des "anciens" états, obtenus par action (à droite)
        -- du poids étoilé du mot vide
        finality (Just p) = actionR (f p) $ star epsWeight

    \end{minted}
  \end{implementationBox}
  La multiplication par un scalaire à gauche (resp. à droite) correspond à l'action à gauche (resp. à droite) de ce scalaire sur la configuration
   initiale (resp.\ sur les poids de sortie des états).
  L'intersection, déjà mentionnée précédemment, est réalisée à l'aide du produit tensoriel des types (c'est-à-dire le produit cartésien, Pseudo-Code~\ref{CodeProdTensTypes}) dans le cas des automates non-déterministes comme dans celui des automates à multiplicités.
  Enfin, le complémentaire d'un automate non-déterministe est obtenu en appliquant l'algorithme classique de déterminisation (passage à l'automate des parties), en inversant la finalité des états, puis en convertissant l'automate déterministe complet obtenu en automate non-déterministe.
  Dans le cas des automates à multiplicités Booléennes, on utilisera des conversions avec les automates non-déterministes.

  Intéressons-nous alors plus précisément à la conversion entre une expression et un automate.
  Pour une expression fixée, quelle serait la signature d'une fonction de conversion entre cette expression et un automate?
  Plus particulièrement, quel serait l'ensemble des états de l'automate produit?

  Une façon de faire serait de choisir un type entier par exemple, puis de renuméroter à chaque étape les états de l'automate obtenu.
  Une autre façon de faire est de remarquer que pour chaque opération de la construction inductive, un foncteur est appliqué sur les états:
  \begin{itemize}
    \item le foncteur (appelé \texttt{EndoSum} dans la suite) associant à un type \texttt{a} le type \texttt{Sum(a, a)} (Pseudo-Code~\ref{code:endoSumProd}) pour la somme,
    \item le foncteur \texttt{Maybe} pour l'étoile,
    \item le foncteur d'ensemble pour le complémentaire,
    \item le foncteur (appelé \texttt{EndoProd} dans la suite) associant à un type \texttt{a} le type \texttt{Prod(a, a)} (Pseudo-Code~\ref{code:endoSumProd}) pour le produit.
  \end{itemize}


\begin{implementationBox}[label={code:endoSumProd}]{Les foncteurs \texttt{EndoSum} et \texttt{EndoProd}}
  \begin{minted}[xleftmargin=1em, autogobble, mathescape=true, fontsize=\footnotesize]{haskell}

    -- Automata.InductionAutomatonFromExp

    data EndoSum a = EndoSum {runEndoSum :: Sum (a, a)}
      deriving Eq

    data EndoProd a = EndoProd {runEndoProd :: Prod (a, a)}
      deriving Eq

    instance FunctorGen EndoSum (->) (->) where
      fmap f (EndoSum (Fst a)) = EndoSum $ Fst $ f a
      fmap f (EndoSum (Snd a)) = EndoSum $ Snd $ f a

    instance FunctorGen EndoProd (->) (->) where
      fmap f (EndoProd (Product (a, b))) = EndoProd (Product (f a, f b))
  \end{minted}
\end{implementationBox}
En d'autres termes, les états d'un automate construit par la méthode inductive ont pour type un élément des combinaisons possibles d'applications de ces foncteurs; si on interprète les foncteurs comme des symboles, le type obtenu ressemble au monoïde libre engendré par ces symboles.
En effet, il s'agit d'une structure libre, et même d'un monoïde (objet) libre: une \firstocc{monade libre}{monade!libre}.

\begin{remarqueBox}[]{Les monades libres}
  Dans la catégorie des types, la monade libre associée à un (endo-) foncteur \texttt{F} correspond à un constructeur de types envoyant un type \texttt{a} sur un type isomorphe à \texttt{\(\sum_{n=0}\)F\textsuperscript{n}(a)}.
  Le fait que \(F\) soit un foncteur est suffisant pour produire une monade, comme le montre le code suivant.

    \begin{minted}[xleftmargin=1em,   autogobble, fontsize=\footnotesize]{haskell}
      -- Automata.InductionAutomatonFromExp

      -- Type de données des monades libres
      data FreeMonad f a = Free (f (FreeMonad f a)) | Pure a

      -- Si f est un foncteur de la catégorie des types,
      -- FreeMonad f l'est aussi
      instance FunctorGen f (->) (->) => FunctorGen (FreeMonad f) (->) (->) where

        fmap f (Pure x) = Pure $ f x
        fmap f (Free g) = Free $ fmap (fmap f) g

      -- Si f est un foncteur de la catégorie des types,
      -- FreeMonad f est une monade de la catégorie des types.
      instance (FunctorGen f (->) (->)) => MonadGen (FreeMonad f) (->) where
        return = Pure
        join (Pure a) = a
        join (Free g) = Free $ fmap join g
    \end{minted}
  Certaines monades connues sont des monades libres.
  Par exemple, en considérant le foncteur constant défini comme
  \begin{minted}[xleftmargin=1em,   autogobble, fontsize=\footnotesize]{haskell}
    data Const a = Const
  \end{minted}
  on peut montrer que la monade \texttt{Maybe} est (isomorphe à) la monade libre associé à \text{Const}.

  Une monade libre peut également être isomorphe à une structure de données.
  Par exemple, on peut montrer que la monade libre associée au foncteur \texttt{EndoProd} défini précédemment correspond aux arbres binaires valués uniquement aux feuilles.

  Cependant, ce n'est pas parce qu'une monade, une fois appliquée, produit un type isomorphe à une application d'une monade libre qu'elle est elle-même libre.
  Par exemple, considérons le foncteur (qui est même une monade)
  \begin{minted}[xleftmargin=1em,   autogobble, fontsize=\footnotesize]{haskell}
    data Writer s a =  Writer {runWriter :: (a, s)}
  \end{minted}
  On peut remarquer que les types \texttt{FreeMonad (Writer s) ()} et \texttt{[s]} sont isomorphes.
  Cependant, la monade liste n'est pas une monade libre\footnote{comme expliqué dans la documentation du module \href{http://hackage.haskell.org/package/free-5.1.3/docs/Control-Monad-Free.html}{Control.Monad.Free} du package \href{https://hackage.haskell.org/package/free}{Free}.}.
\end{remarqueBox}
Considérons alors le foncteur \texttt{StateType} défini comme suit:
\begin{implementationBox}[]{Le foncteur \texttt{StateType}}
  \begin{minted}[xleftmargin=1em, autogobble, mathescape=true, fontsize=\footnotesize]{haskell}

    -- Automata.InductionAutomatonFromExp

    data StateType a
      = MaybeState (Maybe a)
      | SetState (HashSet a)
      | SumState (EndoSum a)
      | ProdState (EndoProd a)
      deriving Eq

    instance FunctorGen StateType (->) (->) where
      type FunctorConstraint StateType (->) (->) val = (Hashable val, Eq val)

      fmap f (MaybeState m_a              ) = MaybeState $ fmap f m_a
      fmap f (SetState   as               ) = SetState $ fmap f as
      fmap f (SumState   (EndoSum (Fst a))) = SumState $ EndoSum $ Fst $ f a
      fmap f (SumState   (EndoSum (Snd a))) = SumState $ EndoSum $ Snd $ f a
      fmap f (ProdState  aa               ) = ProdState $ fmap f aa
  \end{minted}
\end{implementationBox}
La transformation d'une expression en un automate dont les états sont construits à partir de la monade libre \texttt{FreeMonad StateType} ne nécessite plus que quelques fonctions auxiliaires (définies dans le module \texttt{Automata.InductionAutomatonFromExp}) permettant de transformer les états des automates construits:
\begin{itemize}
  \item \texttt{packEndoAut} (resp. \texttt{packEndoProdAut}) est une fonction permettant de convertir les états d'un automate du type \texttt{Sum(a, a)} en \texttt{EndoSum a} (resp. \texttt{Prod(a, a)} en \texttt{EndoProd a}) en utilisant l'isomorphisme naturel;
  \item \texttt{packCombiFA} est une fonction permettant d'injecter un automate dont les états sont de type \texttt{f (FreeMonad StateType a)} (pour \texttt{f} un foncteur parmi \texttt{Maybe}, \texttt{HashSet}, \texttt{EndoProd} ou \texttt{EndoSum}) dans un automate dont les états sont de type \texttt{FreeMonad StateType a}.
\end{itemize}
La construction inductive peut alors s'exprimer comme suit:
\begin{itemize}
  \item l'automate associé au mot vide a pour configuration initiale l'état \texttt{Pure False}, qui a un poids de finalité unitaire; il n'y a pas de transitions;
  \begin{implementationBox}[]{La construction inductive associée au mot vide}
    \begin{minted}[xleftmargin=1em, autogobble, mathescape=true, fontsize=\footnotesize]{haskell}
      inductiveGen Epsilon = Just $ packKleisli
        (return $ Pure False)
        (const $ const mempty)
        finality
       where
        finality (Pure False) = one
        finality _            = zero
    \end{minted}
  \end{implementationBox}
  \item l'automate associé au vide n'a pas de transitions et tous ses états ont un poids de finalité nul;
    \begin{implementationBox}[]{La construction inductive associée au vide}
      \begin{minted}[xleftmargin=1em, autogobble, mathescape=true, fontsize=\footnotesize]{haskell}
        inductiveGen Empty =
          Just $ packKleisli mempty (const $ const mempty) (const zero)
      \end{minted}
    \end{implementationBox}
  \item l'automate associé à un symbole \texttt{a} a (au moins) 2 états du type \texttt{FreeMonad StateType Bool}: un état \texttt{Pure False}, initial et un état \texttt{Pure True}, seul état de finalité non-nulle, unitaire; il ne possède qu'une unique transition, envoyant \texttt{Pure False} sur \texttt{Pure True};
    \begin{implementationBox}[]{La construction inductive associée à un symbole}
      \begin{minted}[xleftmargin=1em, autogobble, mathescape=true, fontsize=\footnotesize]{haskell}
        inductiveGen (Symbol a) = Just
          $ packKleisli (return $ Pure False) delta_ finality
         where
          delta_ b (Pure False) | b == a = return $ Pure True
          delta_ _ _                     = mempty
          finality (Pure True) = one @weight
          finality _           = zero
      \end{minted}
    \end{implementationBox}
  \item la concaténation s'effectue \emph{via} les fonctions \texttt{concatGen}, \texttt{packCombiFA} et \texttt{packEndoProdAut}
    \begin{implementationBox}[]{La construction inductive associée à la concaténation}
      \begin{minted}[xleftmargin=1em, autogobble, mathescape=true, fontsize=\footnotesize]{haskell}
        inductiveGen (Operator Concat (VCons e1 (VCons e2 VNil))) = liftA2
          ((packCombiFA . packEndoAut) .: concatGen)
          (inductiveGen e1)
          (inductiveGen e2)
      \end{minted}
    \end{implementationBox}
  \item l'étoile s'effectue \emph{via} les fonctions \texttt{starGen} et \texttt{packCombiFA}
    \begin{implementationBox}[]{La construction inductive associée à l'étoile}
      \begin{minted}[xleftmargin=1em, autogobble, mathescape=true, fontsize=\footnotesize]{haskell}
        inductiveGen (Operator Star (VCons e VNil)) =
          fmap (packCombiFA . starGen) $ inductiveGen e
      \end{minted}
    \end{implementationBox}
  \item la somme s'effectue \emph{via} les fonctions \texttt{packCombiFA} et \texttt{packEndoAut}
    \begin{implementationBox}[]{La construction inductive associée à la somme}
      \begin{minted}[xleftmargin=1em, autogobble, mathescape=true, fontsize=\footnotesize]{haskell}
        inductiveGen (Operator Plus (VCons e1 (VCons e2 VNil))) = liftA2
          sumGenInd
          (inductiveGen e1)
          (inductiveGen e2)
         where
          sumGenInd =
            (packCombiFA . packEndoAut)
              .: (  (reduceInitialSum (<>) . reduceFinalSum id id)
                 .: binaryTransformationBySum
                 )
      \end{minted}
    \end{implementationBox}
  \item les opérateurs de multiplication par un scalaire \texttt{x} ne sont que des applications d'action sur les configurations initiales ou les poids de finalité
    \begin{implementationBox}[]{La construction inductive associée à la multiplication par un scalaire}
      \begin{minted}[xleftmargin=1em, autogobble, mathescape=true, fontsize=\footnotesize]{haskell}
        inductiveGen (Operator (MultL x) (VCons e VNil)) = fmap change $ inductiveGen e
          where change (Auto (Kleisli i) d f) = Auto (Kleisli $ action x . i) d f

        inductiveGen (Operator (MultR x) (VCons e VNil)) = fmap change $ inductiveGen e
          where change (Auto i d (Kleisli f)) = Auto i d (Kleisli $ flip actionR x . f)
      \end{minted}
    \end{implementationBox}
  \item l'intersection s'effectue \emph{via} les fonctions \texttt{packCombiFA} et \texttt{packEndoProdAut}
    \begin{implementationBox}[]{La construction inductive associée à l'intersection}
      \begin{minted}[xleftmargin=1em, autogobble, mathescape=true, fontsize=\footnotesize]{haskell}
        inductiveGen (Operator Inter (VCons e1 (VCons e2 VNil))) = liftA2
          intersectionGen
          (inductiveGen e1)
          (inductiveGen e2)
         where
          intersectionGen = (packCombiFA . packEndoProdAut)
            .: binaryTransformationByProductWith (const $ const ())
      \end{minted}
    \end{implementationBox}
    \item l'opération de complémentaire est une application directe des mécanismes explicités précédemment, et le cas des fonctions génériques est écarté (seul cas renvoyant \texttt{Nothing}).
\end{itemize}
Pour essayer cette construction, des exemples sont fournis dans les sources, implantés dans le module \texttt{HDRExample.WordExpInductive},  contenant les fonctions suivantes:
\begin{itemize}
  \item \texttt{printRandomBoolAutomaton} tire une expression rationnelle Booléenne au hasard, avec dix opérateurs, sur les symboles \texttt{a}, \texttt{b} ou \texttt{c}; cette expression est ensuite convertie en un automate non-déterministe,
  et s'affiche alors sa représentation \texttt{Dot}\footnote{Format de description de graphes, dont la représentation graphique peut-être obtenu par \href{http://www.graphviz.org/}{Graphviz}, ou en ligne comme \href{https://dreampuf.github.io/GraphvizOnline}{ici}.};
  \item \texttt{printRandomAutomatonWithMult} tire une expression rationnelle au hasard, avec dix opérateurs parmi les opérateurs simples (\emph{i.e.} somme, étoile et concaténation) et les multiplications par un scalaire compris entre 1 et 10, sur les symboles \texttt{a}, \texttt{b} ou \texttt{c};
  cette expression est ensuite convertie en un automate à multiplicités entières\footnote{ici est utilisé un semianneau étoilé pour les entiers, défini dans le module \texttt{Expressions.WordExp.Expression}}, et s'affiche alors sa représentation \texttt{Dot};
  \item \texttt{printRandomAutomata} tire une expression rationnelle au hasard, avec dix opérateurs parmi les opérateurs simples (\emph{i.e.} somme, étoile et concaténation) uniquement, sur les symboles \texttt{a}, \texttt{b} ou \texttt{c}; cette expression est ensuite convertie en deux automates (non-déterministe et à multiplicités entières), et s'affichent alors leurs représentations \texttt{Dot};
  \item \texttt{printAutomatonWithMult} lit une expression rationnelle sur l'entrée standard, sur les symboles \texttt{a}, \texttt{b} ou \texttt{c} utilisant la syntaxe suivante:
    \begin{itemize}
      \item \texttt{+} pour la somme;
      \item \texttt{.} pour la concaténation;
      \item \texttt{*} pour l'étoile;
      \item \texttt{[x]:} et \texttt{:[x]} pour les multiplications par le scalaire \texttt{x};
    \end{itemize}
    cette expression est ensuite convertie en un automate à multiplicités entières, et s'affiche alors sa représentation \texttt{Dot};
    \item \texttt{printBoolAutomaton} lit une expression rationnelle sur l'entrée standard, sur les symboles \texttt{a}, \texttt{b} ou \texttt{c} utilisant la syntaxe suivante:
      \begin{itemize}
        \item \texttt{+} pour la somme;
        \item \texttt{.} pour la concaténation;
        \item \texttt{*} pour l'étoile;
        \item \texttt{\textasciitilde} pour le complémentaire;
        \item \texttt{\&} pour l'intersection;
      \end{itemize}
      cette expression est ensuite convertie en un automate non déterministe, et s'affiche alors sa représentation \texttt{Dot};
\end{itemize}
Pour l'utiliser, une fois la commande \texttt{stack ghci src/HDRExample/WordExpInductive.hs} lancée, il suffit de taper par exemple \texttt{printAutomatonFrom} puis \texttt{(a+b+c)*.c.(a+b)\&\textasciitilde(c*.(a+b+c))} pour représenter l'automate inductif associé à l'expression \({(a+b+c)}^*c(a+b)\cap\neg(c^*(a+b+c))\).

\subsection{Automate de Kleisli des positions}

La deuxième méthode décrite dans ce document est la méthode des positions.
Comme dans le cas classique, nous ne l'appliquerons qu'à des expressions avec des opérateurs simples (somme, concaténation et étoile de Kleene) ou des multiplications par scalaire, les autres opérateurs n'étant pas nécessairement compatibles avec la notion de position.
Pour cela, montrons comment une généralisation simple et directe des formules connues~\cite{CF03,CLOZ04} permet de construire des automates de Kleisli des positions pour (au moins) les monades d'ensembles (\texttt{HashSet}), de semimodules libres (\texttt{FreeSemimodule}), d'expressions Booléennes (\texttt{BoolExpr}) ou d'expressions généralisées (\texttt{GenExpr}).

Comme dans le cas précédent de la construction inductive, remarquons que ces monades sont des semimodules pour certains types de valeurs pouvant pondérer les expressions, les ensembles et les semimodules libres ayant été traités précédemment.
Trivialement,
\begin{itemize}
  \item les expressions Booléennes forment un semimodule Booléen;
  \item plus généralement, les expressions généralisées sur un type \texttt{a} forment un semimodule sur \texttt{a}, si \texttt{a} est un semianneau.
\end{itemize}

La construction classique de l'automate des positions est composée de trois composantes que nous utiliserons également ici:
\begin{itemize}
  \item la définition des fonctions de position;
  \item la linéarisation;
  \item l'utilisation des fonctions de position sur une expression linéarisée pour construire l'automate des positions.
\end{itemize}

\subsubsection{Les fonctions de position}
Les fonctions de position sur une monade \texttt{m}, présentées dans la partie précédente pour le cas Booléen, peuvent alors être définies de la façon suivante pour une expression \texttt{e} avec des poids de type \texttt{p} et des symboles de types \texttt{c}:
\begin{itemize}
  \item \texttt{Null e} est de type \texttt{p}, et est interprété comme le poids du mot vide;
  \item \texttt{Pos e} est l'ensemble des symboles apparaissant dans l'expression (\emph{i.e.} \texttt{HashSet c});
  \item \texttt{First e} est un \emph{conteneur monadique} (\emph{i.e.} de type \texttt{m c}) des symboles commençant les mots de l'expression;
  \item \texttt{Last e} est une fonction associant à chaque symbole un poids de type \texttt{p}, correspondant au poids de finalité du symbole;
  \item \texttt{Follow e} est une fonction associant à chaque symbole un conteneur monadique (\emph{i.e.} de type \texttt{m c}) des symboles pouvant lui succéder.
\end{itemize}
Ces cinq fonctions, calculées par de la fonction \texttt{glushkovPosGen} du Pseudo-Code~\ref{codeGluFunSignature},
s'expriment inductivement comme suit.
\begin{implementationBox}[label={codeGluFunSignature}]{Le calcul inductif des fonctions de position}
  \begin{minted}[xleftmargin=1em, autogobble, mathescape=true, fontsize=\footnotesize]{haskell}
    -- Expressions.WordExp.GlushkovFunctions

    -- Renvoie un quintuplet des fonctions de Glushkov:
    -- (Null, Pos, First, Last, Follow)
    glushkovPosGen
      :: ( Star weight
         , Eq c
         , Hashable c
         , Semimodule weight (m c)
         , MonadGen m (->)
         )
      => Expression weight c
      -> Maybe (weight, HashSet c, m c, c -> weight, c -> m c)
  \end{minted}
\end{implementationBox}
Le mot vide est un des cas de base:
\begin{itemize}
  \item le poids du mot vide y est unitaire;
  \item il n'y a ni position ni position initiale;
  \item les positions y ont un poids de finalité nul;
  \item les positions n'y ont pas de successeur.
\end{itemize}
\begin{implementationBox}[]{Le calcul inductif des fonctions de position: le mot vide}
  \begin{minted}[xleftmargin=1em, autogobble, mathescape=true, fontsize=\footnotesize]{haskell}
    glushkovPosGen Epsilon = Just (one, mempty, mempty, const zero, const mempty)
  \end{minted}
\end{implementationBox}
Le cas du vide lui est relativement proche:
\begin{itemize}
  \item le poids du mot vide y est nul;
  \item il n'y a ni position ni position initiale;
  \item les positions y ont un poids de finalité nul;
  \item les positions n'y ont pas de successeur.
\end{itemize}
\begin{implementationBox}[]{Le calcul inductif des fonctions de position: le vide}
  \begin{minted}[xleftmargin=1em, autogobble, mathescape=true, fontsize=\footnotesize]{haskell}
    glushkovPosGen Empty = Just (zero, mempty, mempty, const zero, const mempty)
  \end{minted}
\end{implementationBox}
Enfin, le cas des symboles est le dernier des cas de bases:
\begin{itemize}
  \item le poids du mot vide y est nul;
  \item il n'y a qu'une position, qui est la seule initiale;
  \item le poids de finalité de cette position, unitaire, est le seul non-nul;
  \item les positions n'y ont pas de successeur.
\end{itemize}
\begin{implementationBox}[]{Le calcul inductif des fonctions de position: les symboles}
  \begin{minted}[xleftmargin=1em, autogobble, mathescape=true, fontsize=\footnotesize]{haskell}
  glushkovPosGen (Symbol c) = Just
    ( zero
    , return c
    , return c
    , \x -> if x == c then one else zero
    , const mempty
    )
  \end{minted}
\end{implementationBox}
Le cas de la somme de deux expressions \texttt{e1} et \texttt{e2} est relativement simple:
\begin{itemize}
  \item le poids du mot vide y correspond à la somme de ceux dans les deux expressions \texttt{e1} et \texttt{e2};
  \item les positions sont celles apparaissant dans \texttt{e1} ou \texttt{e2};
  \item les positions initiales sont la somme (monoïdale) des positions initiales\footnote{les conteneurs monadiques étant par hypothèse des semimodules, ils sont également des monoïdes} des deux expressions \texttt{e1} ou \texttt{e2};
  \item le poids de finalité des symboles sont les sommes (dans le semianneau des poids) des poids correspondant dans les deux expressions \texttt{e1} et \texttt{e2};
  \item les successeurs des positions s'obtiennent comme somme (des conteneurs monadiques) de ceux de \texttt{e1} et \texttt{e2}.
\end{itemize}
\begin{implementationBox}[]{Le calcul inductif des fonctions de position: la somme}
  \begin{minted}[xleftmargin=1em, autogobble, mathescape=true, fontsize=\footnotesize]{haskell}
  glushkovPosGen (Operator Plus (VCons e1 (VCons e2 VNil))) = do
    (null1, pos1, first1, last1, fol1) <- glushkovPosGen e1
    (null2, pos2, first2, last2, fol2) <- glushkovPosGen e2
    return
      ( null1 `plus` null2
      , pos1 <> pos2
      , first1 <> first2
      , \c -> last1 c `plus` last2 c
      , \c -> fol1 c <> fol2 c
      )
  \end{minted}
\end{implementationBox}
Le cas de l'étoile d'une expression \texttt{e} utilise le fait que les conteneurs monadiques considérés sont des semimodules:
\begin{itemize}
  \item le poids du mot vide y correspond à l'étoile de celui de \texttt{e};
  \item les positions sont celles apparaissant dans \texttt{e};
  \item les positions initiales sont celles de \texttt{e}, après avoir fait agir le poids étoilé du mot vide;
  \item le poids de finalité des symboles sont ceux de \texttt{e}, après avoir multiplié par le poids étoilé du mot vide à droite;
  \item les successeurs des positions s'obtiennent comme somme (des conteneurs monadiques) de ceux de \texttt{e} avec les positions initiales de \texttt{e}, après avoir fait agir le poids de finalité de la position ambiante multiplié par le poids étoilé de l'étoile.
\end{itemize}
\begin{implementationBox}[]{Le calcul inductif des fonctions de position: l'étoile}
  \begin{minted}[xleftmargin=1em, autogobble, mathescape=true, fontsize=\footnotesize]{haskell}
  glushkovPosGen (Operator Star (VCons e1 VNil)) = do
    (null1, pos1, first1, last1, fol1) <- glushkovPosGen e1
    let starEpsWeight = star null1
    return
      ( starEpsWeight
      , pos1
      , starEpsWeight `action` first1
      , \c -> last1 c `times` starEpsWeight
      , \c -> fol1 c <> ((last1 c `times` starEpsWeight) `action` first1)
      )
  \end{minted}
\end{implementationBox}
Le cas de la concaténation de deux expressions \texttt{e1} et \texttt{e2} utilise également le fait que les conteneurs monadiques considérés sont des semimodules:
\begin{itemize}
  \item le poids du mot vide y correspond au produit de ceux de \texttt{e1} et \texttt{e2};
  \item les positions sont celles apparaissant dans \texttt{e1} et \texttt{e2};
  \item les positions initiales sont celles de \texttt{e1}, ainsi que celle de \texttt{e2} après avoir fait agir le poids du mot vide dans \texttt{e1};
  \item le poids de finalité des symboles sont ceux de \texttt{e2}, ainsi que ceux de \texttt{e1} après avoir multiplié par le poids du mot vide de \texttt{e2} à droite;
  \item les successeurs d'une position \texttt{c} s'obtiennent comme somme (des conteneurs monadiques) de ceux de \texttt{e1} avec ceux de \texttt{e2}, mais aussi comme les éléments initiaux de \texttt{e2} après avoir fait agir le poids final de \texttt{c} dans \texttt{e1}.
\end{itemize}
\begin{implementationBox}[]{Le calcul inductif des fonctions de position: la concaténation}
  \begin{minted}[xleftmargin=1em, autogobble, mathescape=true, fontsize=\footnotesize]{haskell}
  glushkovPosGen (Operator Concat (VCons e1 (VCons e2 VNil))) = do
    (null1, pos1, first1, last1, fol1) <- glushkovPosGen e1
    (null2, pos2, first2, last2, fol2) <- glushkovPosGen e2
    return
      ( null1 `times` null2
      , pos1 <> pos2
      , first1 <> (null1 `action` first2)
      , \c -> last1 c `times` null2 `plus` last2 c
      , \c -> fol1 c <> (last1 c `action` first2) <> fol2 c
      )
  \end{minted}
\end{implementationBox}


%
Les cas des multiplications par scalaire ne sont que des actions à gauche ou à droite:
\begin{implementationBox}[]{Le calcul inductif des fonctions de position: les multiplications par des scalaires}
  \begin{minted}[xleftmargin=1em, autogobble, mathescape=true, fontsize=\footnotesize]{haskell}
    glushkovPosGen (Operator (MultL x) (VCons e1 VNil)) = do
      (null1, pos1, first1, last1, fol1) <- glushkovPosGen e1
      return (x `times` null1, pos1, x `action` first1, last1, fol1)

    glushkovPosGen (Operator (MultR x) (VCons e1 VNil)) = do
      (null1, pos1, first1, last1, fol1) <- glushkovPosGen e1
      return (null1 `times` x, pos1, first1, \c -> last1 c `times` x, fol1)
  \end{minted}
\end{implementationBox}
Les autres cas (complémentaire, intersection ou fonctions génériques) renvoient la valeur \texttt{Nothing}.

\subsubsection{La linéarisation à l'aide de la monade State}
Comme dans le cas des expressions classiques, il suffit de parcourir l'expression et d'indicer d'une façon croissante les symboles rencontrés.
Pour cela, nous utiliserons le type simple suivant:
\begin{equation*}
  \texttt{data Position index c = Pos index c}
\end{equation*}
L'opération de linéarisation est une parfaite candidate à l'utilisation de la monade \texttt{State} (définie dans le Pseudo-Code~\ref{codeStateMonad}).
\begin{remarqueBox}[label={remarqueMonadeState}]{La monade \texttt{State} du package \texttt{mtl}}
  Dans la suite, pour alléger le code, nous utiliserons l'implantation de la monade \texttt{State} du module \href{https://hackage.haskell.org/package/mtl-2.2.2/docs/Control-Monad-State-Lazy.html}{\texttt{Control.Monad.State.Lazy}} du package \texttt{mtl}, et plus particulièrement les fonctions suivantes:
  \begin{itemize}
    \item \texttt{state :: (s -> (a, s)) -> State s a } équivalent du constructeur de valeurs \texttt{State} de l'implantation précédente;
    \item \texttt{evalState :: State s a -> s -> a} permettant d'évaluer la fonction interne à l'aide d'une valeur.
  \end{itemize}
\end{remarqueBox}
En effet, pour linéariser, il suffit de définir une fonction prenant en paramètre l'expression à linéariser et l'indice de départ, et renvoyant l'expression linéarisée, ainsi que le successeur de l'indice maximal atteint, nécessaire pour linéariser le reste de l'expression lors des cas inductifs.
L'utilisation de la monade \texttt{State} permet de simplifier la syntaxe en réalisant la transmission de l'indice ambiant d'une façon naturelle.
\begin{implementationBox}[]{La linéarisation à l'aide de la monade \texttt{State}}
  \begin{minted}[xleftmargin=1em, autogobble, mathescape=true, fontsize=\footnotesize]{haskell}
    -- Expressions.WordExp.GlushkovFunctions

    -- Fonction de linéarisation prenant en paramètre l'indice de départ
    -- et l'expression à linéariser. La fonction construit tout d'abord une
    -- fonction auxiliaire (monade State) via la fonction linearize',
    -- puis l'évalue à l'aide de l'indice.
    linearizeFrom
      :: (Enum index) => index -> Expression a c -> Expression a (Position index c)
    linearizeFrom start expr = evalState (linearize' expr) start
      where
        -- Construit une fonction de linéarisation d'expressions
        -- à l'aide de la monade State.
        linearize'
          :: (Enum index)
          => Expression a c
          -> State index (Expression a (Position index c))
        -- Pour le mot vide et l'expression vide,
        -- l'expression et l'indice sont inchangés.
        linearize' Epsilon          = return Epsilon
        linearize' Empty            = return Empty
        -- Pour le cas d'un symbole, on renvoie le couple composé
        -- du symbole indicé par l'indice courant, ainsi que le
        -- successeur de cet indice (qui le remplace en tant
        --  qu'indice courant)
        linearize' (Symbol c      ) = state $ \n -> (Symbol (Pos n c), succ n)
        -- Dans le cas d'un opérateur, il suffit de linéariser le vecteur
        -- de ces opérandes à l'aide de la fonction linearizeOnVect
        -- définie ci-après
        linearize' (Operator op es) = fmap (Operator op) (linearizeOnVect es)
        -- Construit une fonction de linéarisation de vecteur
        -- d'expressions à l'aide de la monade State.
        linearizeOnVect
          :: (Enum index)
          => Vector n (Expression a c)
          -> State index (Vector n (Expression a (Position index c)))
        -- Si le vecteur est vide, pas de modification
        linearizeOnVect VNil = return VNil
        -- Sinon, on linéarise la première expression (ce qui modifie
        -- l'indice courant), et on recommence avec le reste du vecteur
        linearizeOnVect (VCons e es) =
          liftA2 VCons (linearize' e) (linearizeOnVect es)

    -- Fonction utilitaire pour linéariser une expression à partir de 1
    linearize :: (Enum index) => Expression a c -> Expression a (Position index c)
    linearize = linearizeFrom (toEnum 1)
  \end{minted}
\end{implementationBox}

\subsubsection{La construction de l'automate des positions}
Une fois une expression linéarisée, la construction de l'automate (monadique) des positions se fait de la façon (classique) suivante, à l'aide des fonctions de positions:
\begin{itemize}
  \item les états sont obtenus par le calcul de \texttt{Pos}, en y ajoutant un état initial supplémentaire;
  \item les poids finaux des états sont donnés par le calcul de \texttt{Last} et de \texttt{Null};
  \item les successeurs sont définis par le calcul de \texttt{Follow} et \texttt{First}.
\end{itemize}
\begin{implementationBox}[]{La construction de l'automate des positions}
  \begin{minted}[xleftmargin=1em, autogobble, mathescape=true, fontsize=\footnotesize]{haskell}
    -- Automata.KleisliAutomata

    -- type des états de l'automate des positions
    data GlushkovState index c = Init | P (Position index c)

    -- Construit l'automate des positions de l'expression e,
    -- et renvoie également la liste des symboles désindicés
    -- apparaissant dans l'expression de départ
    glushkovWithPos
      :: (Eq c
         , Hashable c
         , Star weight
         , Semimodule weight (m (Position Word c))
         , MonadGen m (->)
         , Castable (m weight') weight)
      => Expression weight c
      -> Maybe (Automaton (KleisliCat m (->)) () c (GlushkovState Word c) weight', [c])
    glushkovWithPos e = do
      -- linéarisation de e
      let e' = linearize e
      -- calcul des fonctions de positions
      (null_, pos_, first_, last_, follow_) <- glushkovPosGen e'
      let
        -- fonction permettant de filtrer les positions
        --  d'un conteneur monadique
        -- égales à un symbole donné une fois désindicées
        pos_of a =  bind (\c'@(Pos _ c) -> if a /= c then mempty else return c')
        -- calcul de la fonction de transition
        -- à partir de First et Follow
        delta_ p Init  = fmap P $ pos_of p first_
        delta_ q (P p) = fmap P $ pos_of q $ follow_ p
        -- poids de finalité des états, à partir de
        -- Null et Last
        final_ Init  = null_
        final_ (P p) = last_ p
        -- alphabet des symboles présents
        sigma = Set.toList $ Set.map (\(GF.Pos _ c) -> c) pos_
        in return (packKleisli (M.return Init) delta_ final_, sigma)
  \end{minted}
\end{implementationBox}
Pour essayer cette construction, des exemples sont fournis dans les sources, dans le module Haskell \texttt{HDRExample.WordExpGlushkov} contenant les fonctions suivantes:
\begin{itemize}
  \item \texttt{printRandomBoolAutomaton}, \texttt{printRandomAutomatonWithMult}, \texttt{printAutomatonWithMult} et \texttt{printBoolAutomaton} ont le même rôle que les fonctions homonymes du module précédent \texttt{HDRExample.WordExpInductive};
  \item les fonctions \texttt{performValidityTestWithMult} et \texttt{performRandomValidityTestWithMult} cons\-truisent trois automates (un automate à poids par la méthode inductive et deux automates des positions, à multiplicités et alternant généralisé) depuis une expression (tirée aléatoirement le cas échéant), puis génèrent des mots aléatoires de longueur comprises entre 0 et 10 et en nombre fixé lors de l'appel pour comparer les résultats de calcul des poids,
  \item la fonction \texttt{performValidityTestWithMultTrace} réalise un test équivalent sur deux automates des positions (à multiplicités et alternant généralisé) sur des mots aléatoires sont le nombre est fixé à l'appel en utilisant une construction mémoïsée et traçant les constructions des deux automates au fur et à mesure, permettant de ne construire que la partie utile (c'est-à-dire utilisée) de l'automate.
\end{itemize}
Pour utiliser ce module, une fois la commande \texttt{stack ghci src/HDRExample/WordExpGlushkov.hs} lancée, il suffit de taper par exemple \texttt{performValidityTestWithMultTrace 250} puis l'expression à multiplicités \texttt{([5]:a*+b*:[3]).c*:[2].(a+b+c)*} pour tester la validité des constructions depuis l'expression \((5a^*+b^*3)\cdot c^*2\cdot {(a+b+c)}^*\) sur \(250\) mots tirés aléatoirement.


\subsection{Dérivation}

La dernière construction que nous allons aborder dans cette section est la méthode par dérivation, qui est sans doute la plus flexible.
Pour cela, nous allons nous appuyer sur les méthodes classiques de dérivation (de Brzozowski~\cite{Brz64} et d'Antimirov~\cite{Ant96}),
mais également sur les extensions de la dérivation classique aux expressions étendues~\cite{CCM11b} ainsi que sur la notion de \emph{support de dérivation}~\cite{CCM14}.
De plus, nous allons ne considérer que des semianneaux étoilés pour la pondération.

L'idée est la suivante: si la dérivation partielle d'Antimirov étend la dérivation de Brzozowski en utilisant des ensembles,
on peut essayer de généraliser en utilisant d'autres conteneurs monadiques ayant de ``bonnes propriétés'' (pas nécessairement les mêmes pour tous les conteneurs d'ailleurs).

Parmi ces propriétés, on pourra remarquer le fait de pouvoir ``contracter''  un conteneur monadique d'expressions en une expression (c'est le rôle principal des supports de dérivation de~\cite{CCM14}):
par exemple, un ensemble d'expressions peut être interprété, lors de la dérivation partielle, comme une somme d'expressions.
Cette transformation va nous permettre d'augmenter le nombre des potentiels opérateurs utilisables dans les expressions tout en conservant la possibilité de construire un automate.

Nous allons alors nous intéresser à cinq monades pour lesquelles nous allons pouvoir dériver les expressions précédentes sous leur forme non-restreinte, y compris pour les opérateurs de fonctions génériques;
et comme suite logique, nous en déduirons la construction d'automate associée, qui est la même que dans les dérivations classiques:
les états sont des expressions, les transitions obtenues par dérivation, et la finalité par le poids du mot vide.

Remarquons que la finitude de ces automates n'est pas garantie;
cependant, grâce à l'évaluation paresseuse, cela ne posera pas de problèmes
et seule la partie utile de l'automate sera construite.

Ces cinq monades seront les suivantes:
\begin{itemize}
  \item la monade Identité pour la dérivation de Brzozowski;
  \item la monade des ensembles pour la dérivation d'Antimirov;
  \item la monade des semi-modules libres pour la dérivation partielle pondérée (en utilisant les formules de~\cite{LS05});
  \item la monade des expressions Booléennes pour construire des automates alternants;
  \item la monade des expressions généralisées pour construire des automates alternants généralisés.
\end{itemize}

Les propriétés suffisantes partagées par ces conteneurs contiennent les suivantes:
\begin{itemize}
  \item une fois appliqués, ces conteneurs sont des éléments d'un monoïde;
  \item une fois appliqués, ces conteneurs sont des semimodules pour le semianneau étoilé des poids, mais aussi sur celui des expressions\footnote{à quelques isomorphismes près}.
\end{itemize}

Ainsi, la dérivation monadique pour les atomes, pour les expressions à opérateurs simples ou pour les multiplications par scalaire se décrit comme suit.
On reconnaîtra les formules classiques~\cite{Ant96,LS05}, où le calcul inductif du poids du mot vide est nécessaire.
\begin{implementationBox}[label={codeDerivationMonadique}]{La dérivation monadique}
  \begin{minted}[xleftmargin=1em, autogobble, mathescape=true, fontsize=\footnotesize]{haskell}
    -- Expressions.WordExp.Expression

    -- Calcul du poids du mot vide dans l'expression.
    nullable :: Star a => Expression a c -> a
    nullable Epsilon    = one
    nullable Empty      = zero
    nullable (Symbol _) = zero
    nullable (Operator (Function _ n info) es) =
      convertToVectMultiFun (getFun n info) $ fmap nullable es
    nullable (Operator Concat (VCons e1 (VCons e2 VNil))) =
      nullable e1 `times` nullable e2
    nullable (Operator Star (VCons e VNil)) = star $ nullable e
    nullable (Operator Plus (VCons e1 (VCons e2 VNil))) =
      nullable e1 `plus` nullable e2
    nullable (Operator (MultL x) (VCons e VNil)) = x `times` nullable e
    nullable (Operator (MultR x) (VCons e VNil)) = nullable e `times` x
    nullable (Operator Not       (VCons e VNil)) = not $ nullable e
    nullable (Operator Inter (VCons e1 (VCons e2 VNil))) =
      nullable e1 && nullable e2

    -- Expressions.WordExp.Derivation

    -- Dérivation monadique d'une expression par un symbole
    derive :: c -> Expression a c -> m (Expression a c)

    -- la dérivation du vide ou du mot vide
    -- renvoie l'élément neutre du monoïde
    -- quel que soit le symbole
    derive _ Epsilon = mempty
    derive _ Empty   = mempty

    -- la dérivation d'un symbole renvoie
    -- la promotion monadique (return) du
    -- mot vide lorsqu'il y a correspondance,
    -- l'élément neutre du monoïde sinon
    derive c (Symbol c') | c == c'   = return epsilon
                         | otherwise = mempty

    -- la dérivation d'une somme est la somme
    -- monoïdale des dérivations
    derive c (Operator Plus (VCons e1 (VCons e2 VNil))) =
      derive c e1 <> derive c e2

    -- la dérivation d'une concaténation d'expressions e1 et e2
    -- est la somme monoïdale entre
    -- * la dérivation de e1 après action droite de e2
    -- * la dérivation de e2 après action gauche du poids
    --   du mot vide dans e1
    derive c (Operator Concat (VCons e1 (VCons e2 VNil))) =
      (derive c e1 `actionR` e2) <> (nullable e1 `action` derive c e2)

    -- la dérivation d'une étoile d'une expression, notée e*
    -- est la dérivation de e, sur laquelle on fait agir
    -- e* à droite, puis le poids étoilé du mot vide dans e
    -- à gauche
    derive c estar@(Operator Star (VCons e VNil)) =
      star (nullable e) `action` (derive c e `actionR` estar)

    -- la dérivation par un scalaire est l'action de ce scalaire
    -- sur la dérivation
    derive c (Operator (MultL x) (VCons e VNil)) = x `action` derive c e
    derive c (Operator (MultR x) (VCons e VNil)) = derive c e `actionR` x
  \end{minted}
\end{implementationBox}

Les opérateurs restants, ceux d'intersection et de complémentaire Booléens, ainsi que les fonctions génériques, peuvent être implantés d'au moins deux façons différentes:
\begin{itemize}
  \item le conteneur monadique peut supporter une opération \emph{encodant} une ou la totalité de ces opérations;
  \item dans le cas contraire, on peut convertir le conteneur en une expression, puis dériver cette dernière.
\end{itemize}
Ainsi, on peut définir une classe de types pour modéliser cette propriété, avec des opérations définies par défaut dans le cas où l'opération ne peut être implantée autrement que par conversion du conteneur en expression:
\begin{implementationBox}[]{La dérivation monadique: opérateurs étendus}
  \begin{minted}[xleftmargin=1em, autogobble, mathescape=true, fontsize=\footnotesize]{haskell}
    -- Expressions.WordExp.Derivation

    -- Définition d'une conversion d'un conteneur monadique
    -- en expression
    class Convertible a c m where
      convertTo :: m (Expression a c) -> Expression a c

    -- Définition des transormations associées aux opérateurs
    -- étendus pour la dérivation monadique d'une expression
    -- de type Expression a c sur la monade m
    class Derivable a c m where
      -- complémentaire d'un conteneur monadique
      negation :: m (Expression Bool c) ->  m (Expression Bool c)
      default negation ::
        Convertible a c m =>
          m (Expression a c) ->
            m (Expression a c)
      negation = return . neg . convertTo

      -- intersection de deux supports monadiques
      intersection ::
        m (Expression Bool c)
          -> m (Expression Bool c)
            -> m (Expression Bool c)
      default intersection ::
        (Convertible a c m) =>
          m (Expression a c) ->
            m (Expression a c) ->
              m (Expression a c)
      intersection es1 es2 = return $ inter' (convertTo es1) (convertTo es2)

      -- Application d'une fonction générique n-aire
      -- sur un vecteur de n expressions
      functionExpr ::
        String
          -> Natural n
            -> InfoOp n a
              -> Vector n (m (Expression a c)) -> m (Expression a c)
      default functionExpr ::
        (Convertible a c m) =>
          String ->
            Natural n ->
              InfoOp n a ->
                Vector n (m (Expression a c)) ->
                  m (Expression a c)
      functionExpr s n op = return . fun s n op . fmap convertTo
  \end{minted}
\end{implementationBox}
Certaines monades peuvent ``améliorer'' le traitement de ces opérations en intégrant des transformations associées;
par exemple:
\begin{itemize}
  \item la monade \texttt{BoolExpr} peut intégrer les opérateurs d'intersection et de complémentaire en associant leur équivalent au sein des expressions Booléennes, \texttt{(<\textasciitilde>)} et \texttt{(<\&{}\&>)}
  \item la monade \texttt{GenExpr} peut intégrer les opérateurs d'intersection, de complémentaire et de fonctions génériques en utilisant le constructeur \texttt{packFunction};
  \item les autres monades (ensemble, semimodule libre et identité) utilisent les formules par défaut définies précédemment.
\end{itemize}
Une fois ces opérations définies, la dérivation d'une expression de type \texttt{Expression a c} sur une monade \texttt{m} est complétée par les cas suivants:
\begin{implementationBox}[]{La dérivation monadique: opérateurs étendus}
  \begin{minted}[xleftmargin=1em, autogobble, mathescape=true, fontsize=\footnotesize]{haskell}
    -- Expressions.WordExp.Derivation

    -- la dérivation du complémentaire d'une expression
    -- est le complémentaire de la dérivation
    derive c (Operator Not (VCons e VNil)) = negation $ derive c e

    -- la dérivation de l'intersection de deux expressions
    -- est l'intersection des dérivations
    derive c (Operator Inter (VCons e1 (VCons e2 VNil))) =
      intersection (derive c e1) $ derive c e2

    -- la dérivation d'une fonction générique n-aire
    -- appliquée sur un vecteur de n expressions
    -- est l'application (monadique) de cette fonction
    -- sur le vecteur des dérivations
    derive c (Operator (Function s n info) es) =
      functionExpr s n info $ fmap (derive c) es
  \end{minted}
\end{implementationBox}
\noindent L'extension de la dérivation par un symbole à la dérivation par rapport à un mot, c'est-à-dire un élément du monoïde libre, se fait d'une façon similaire à l'extension de la fonction de transition d'un automate d'un symbole à un mot (Pseudo-code~\ref{codeDeltaExtension}):
grâce à une adjonction fonctorielle.
\begin{implementationBox}[]{La dérivation monadique: extension aux mots}
  \begin{minted}[xleftmargin=1em, autogobble, mathescape=true, fontsize=\footnotesize]{haskell}
    -- Expressions.WordExp.Derivation

    deriveByWord :: [c] -> Expression a c -> m (Expression a c)
    deriveByWord = runKleisli . promoteFun (Kleisli . derive)
  \end{minted}
\end{implementationBox}
\noindent La construction de l'automate associée à la dérivation peut alors se décrire en appliquant la méthode classique.
\begin{implementationBox}[]{La dérivation monadique: construction de l'automate}
  \begin{minted}[xleftmargin=1em, autogobble, mathescape=true, fontsize=\footnotesize]{haskell}
    -- Automata.KleisliAutomata

    -- Construction de l'automate associé à une dérivation, dont les états
    -- sont des expressions :
    -- * la configuration initiale correspond à l'expression de départ
    -- * le morphisme de transition est la dérivation
    -- * la pondération d'un état correspond au poids du mot vide dans
    --   celui-ci.
    derivationAutomaton ::
      Expression a c ->
        Automaton (KleisliCat m (->)) () c (Expression a c) a'
    derivationAutomaton e = packKleisli (return e) derive nullable
  \end{minted}
\end{implementationBox}
Pour essayer cette construction, des exemples sont fournis dans les sources dans le module\\
\texttt{HDRExample.WordExpDerivation}, contenant les fonctions suivantes:
\begin{itemize}
  \item \texttt{printRandomBoolAutomaton}, \texttt{printRandomAutomatonWithMult}, \texttt{printAutomatonWithMult} et \texttt{printBoolAutomaton} ont le même rôle que les fonctions homonymes du module précédent \texttt{HDRExample.WordExpInductive};
  \item les fonctions \texttt{performValidityTestWithMult} et \texttt{performRandomValidityTestWithMult} cons\-truisent trois automates (un automate à poids par la méthode inductive, un automate à poids obtenu par dérivation et un automates des positions alternant généralisé) depuis une expression (tirée aléatoirement le cas échéant),
  puis génèrent des mots aléatoires de longueur comprises entre 0 et 10 et en nombre fixé lors de l'appel pour comparer les résultats de calcul des poids,
  \item la fonction \texttt{performValidityTestWithMultTrace} réalise un test équivalent sur trois automates (deux automates, un à multiplicités et l'autre alternant généralisé, construits par dérivation, et un automate des positions alternant généralisé)
  en utilisant une construction mémoïsée et traçant les constructions des deux automates au fur et à mesure, permettant de ne construire que la partie utile (c'est-à-dire utilisée) de l'automate.
\end{itemize}
Pour l'utiliser, une fois la commande \texttt{stack ghci src/HDRExample/WordExpDerivation.hs} lancée, il suffit de taper par exemple \texttt{performValidityTestWithMultTrace 250} puis l'expression à multiplicités \texttt{([5]:a*+b*:[3]).c*:[2].(a+b+c)*} pour tester la validité des constructions depuis l'expression \(5a^*+b^*3\cdot c^*2\cdot {(a+b+c)}^*\) en comparant les poids de \(250\) mots tirés aléatoirement.

\section{Expressions enrichies}
Le cas des expressions rationnelles de mots peut servir de modèle à un type de données permettant de généraliser la représentation au niveau des catégories enrichies;
cette méthode nous permettra de fournir un type polymorphe permettant de factoriser les expressions de mots et d'arbres, ainsi que différentes méthodes de construction d'automates.

Pour cela, nous allons tout d'abord montrer comment définir des expressions simples (plus restreintes que dans la section précédente, c'est-à-dire avec moins d'opérateurs) et comment factoriser certaines informations sur ce type.

Nous montrerons ensuite comment transformer une telle expression en un automate enrichi selon trois méthodes: par la méthode des positions, par dérivation et par induction.
Chacune de ces méthodes permettra de calculer un automate enrichi de Kleisli (en toute généralité) afin de conserver la démarche précédente d'unification de construction.

\subsection{Définition et implantation générale}
Pour qu'une expression puisse représenter le mécanisme décrit par un automate enrichi, elle doit pouvoir être paramétrée par ce qui caractérise la base d'une catégorie enrichie: un produit tensoriel.
En effet, il faut pouvoir exprimer à travers les paramètres du type de données la différence entre une expression de mots et une expression d'arbres.

Ainsi, nous utiliserons les produits tensoriels pour décrire les atomes des expressions précédemment liés aux symboles.
Cependant, un produit tensoriel étant un bifoncteur, il doit s'exprimer à partir d'un couple d'objets (au niveau des catégories).
Si le premier est l'objet alphabet, le second sera alors, comme utilisé dans les automates enrichis, l'objet initial; on pourra ainsi l'utiliser comme une généralisation des variables des expressions d'arbres classiques de la littérature.

Mais plus que cela, cet objet initial va permettre également de définir les opérations de concaténation et d'itération, à partir de la substitution de variables.


Le type des expressions enrichies est ainsi paramétré par trois types:
\begin{itemize}
  \item un type pour les symboles;
  \item un type pour les variables;
  \item un produit tensoriel.
\end{itemize}

\begin{implementationBox}[]{Les expressions enrichies}
\begin{minted}[xleftmargin=1em, autogobble, mathescape=true, fontsize=\footnotesize]{haskell}
  -- Expressions.EnrichedExp.EnrichedExp

  data Expression tensor var symbols
    = Empty
    | Var var
    | Tensor (tensor (symbols, var))
    | Sum (Expression tensor var symbols) (Expression tensor var symbols)
    | Sub var (Expression tensor var symbols) (Expression tensor var symbols)
    | Star var (Expression tensor var symbols)
\end{minted}
\end{implementationBox}
Ce type peut être spécialisé pour retrouver les expressions classiques.
\begin{implementationBox}[]{Les expressions de mots et d'arbres depuis les expressions enrichies}
\begin{minted}[xleftmargin=1em, autogobble, mathescape=true, fontsize=\footnotesize]{haskell}
  -- Expressions.EnrichedExp.WordExp

  -- Les expressions de mots sont obtenus en utilisant le produit
  -- tensoriel de la catégories des types (le produit cartésien)
  -- et en utilisant une unique variable, ()
  type WordExp symbols = Expression Prod () symbols

  -- Expressions.EnrichedExp.TreeExp

  -- Les expressions d'arbres sont obtenus en utilisant le produit
  -- tensoriel de la catégories des types gradués
  type TreeExp symbols var = Expression TensorProd var symbols
\end{minted}
\end{implementationBox}
\noindent Les variables ont, comme nous venons de l'annoncer, un rôle à jouer au niveau de l'objet initial des automates.
Autrement dit, les variables vont apparaître ``au début'' de la lecture.
Ainsi, en considérant une combinaison de deux expressions, la sémantique de l'opération de substitution ne sera pas exactement la sémantique usuelle: une expression de la forme \(E_1 \cdot_v E_2\) réalisera la substitution de la variable \(v\) dans les ``parcours'' de \(E_2\) par un ``parcours'' de \(E_1\).
Par exemple, en termes de représentation, on pourrait imaginer les arbres représentés orientés horizontalement, les feuilles à gauche et les racines à droite.
\begin{remarqueBox}[]{Substitution des expressions enrichies et arbres}
Remarquons que ``la direction'' de l'opération de concaténation pour les expressions de mots et celle de substitution pour les expressions d'arbres ne sont pas les mêmes.
Il suffit alors de considérer une fonction de construction alternative pour rétablir l'ordre classique des expressions d'arbres par exemple.
  \begin{minted}[xleftmargin=1em, autogobble, mathescape=true, fontsize=\footnotesize]{haskell}
    -- Expressions.EnrichedExp.TreeExp

    concatVar
      :: var -> TreeExp symbols var -> TreeExp symbols var -> TreeExp symbols var
    concatVar v = flip $ Sub v
      where
      -- flip est une fonction de base prédéfinie en Haskell
      flip f x y = f y x
  \end{minted}
Autre différence, ce ne sont pas les symboles nullaires qui paramètrent le produit de substitution ou l'étoile.
Le pouvoir de représentation n'est pas changé, puisqu'il suffit de remplacer un symbole nullaire par une variable, voire d'inclure ces symboles nullaires dans le type des variables en utilisant un type somme par exemple.
\end{remarqueBox}
Le module \texttt{Expressions.EnrichedExp.EnrichedExp} contient également la définition de plusieurs fonctions outils permettant de manipuler les expressions en toute généralité, basées sur des propriétés partagées quel que soit le produit tensoriel considéré, telles que les réductions ACI (Associativité --- Commutativité --- Idempotence)
de la somme (\texttt{reduceACISum}), le calcul des multiplicités de la décomposition d'une expression somme (\texttt{getWeightedSum}) ou le renversé d'une expression (\texttt{reverseExp}).
C'est aussi le cas des fonctions qui ne tiennent pas compte des atomes symboliques (c'est-à-dire les atomes utilisant les produits tensoriels).

Parmi ces fonctions, les généralisations de la fonction déterminant le poids du mot vide dans une expression (utilisée dans les méthodes de construction par position et par dérivation).
Même si nous ne l'avons pas encore explicité, le mot vide est bien présent en tant que sous-expression des expressions enrichies spécialisées aux mots: il s'agit de la variable \texttt{()}.
La fonction \texttt{nullable} (Pseudo-Code~\ref{codeDerivationMonadique}) peut être étendue comme suit, où les seules différences proviennent de la modification de la concaténation en substitution,
nécessitant de promouvoir cette substitution au niveau monadique, \emph{via} la fonction \texttt{(=<<)}.
\begin{implementationBox}[label={pseudoCodeNullableVarVariables}]{Les calculs de nullabilité pour les expressions enrichies}
\begin{minted}[xleftmargin=1em, autogobble, mathescape=true, fontsize=\footnotesize]{haskell}
  -- Expressions.EnrichedExp.EnrichedExp

  -- Poids d'une variable lors d'un "parcours" dans l'expression
  -- (en fonction de l'interprétation du produit tensoriel)
  nullableVar ::
    (Eq var, Star weight) =>
      var ->
        Expression tensor var symbols ->
          weight
  nullableVar _ (Tensor _) = zero
  nullableVar _ Empty      = zero
  nullableVar v (Var v') | v == v'   = one
                         | otherwise = zero
  nullableVar v (Sum e1 e2) = nullableVar v e1 `plus` nullableVar v e2
  nullableVar v (Sub v' e1 e2)
    | v == v' = nullableVar v e1 `times` nullableVar v e2
    | otherwise = nullableVar v e2 `plus` (nullableVar v e1 `times` nullableVar v' e2)
  nullableVar v (Star v' e)
    | v == v'   = star $ nullableVar v e
    | otherwise = nullableVar v e `times` star (nullableVar v e)

  -- Calcul d'un conteneur monadique contenant les variables
  -- apparaissant lors d'un parcours de l'expression
  -- (en fonction de l'interprétation du produit tensoriel)
  variables
    :: ( Eq var
       , MonadGen m (->)
       , Monoid (m var)
       , Star weight
       , Semimodule weight (m var)
       )
    => Expression tensor var symbols
    -> m var
  variables Empty         = mempty
  variables (Var    v   ) = return v
  variables (Tensor _   ) = mempty
  variables (Sum e1 e2  ) = variables e1 <> variables e2
  variables (Sub v e1 e2) = substIfEq =<< variables e2
   where
    substIfEq v' | v == v'   = variables e1
                 | otherwise = return v'
  variables (Star v e) =
    (M.return v <> variables e) `actionR` (star $ nullableVar v e)
\end{minted}
\end{implementationBox}
D'autres fonctions nécessitent de considérer les spécificités des produits tensoriels pour être calculés (par exemple \emph{via} la définition de classe de types).
C'est le cas de la génération aléatoire d'expressions (\texttt{randomExpression}), mais aussi des factorisations/généralisations des fonctions de position \texttt{Last} (pour les mots) et \texttt{Racine} ou de la linéarisation\footnote{utilisant la monade \texttt{State} (voir Remarque~\ref{remarqueMonadeState})}.
\begin{implementationBox}[]{Poids des symboles de fin de parcours d'une expression enrichie}
\begin{minted}[xleftmargin=1em, autogobble, mathescape=true, fontsize=\footnotesize]{haskell}
  -- Expressions.EnrichedExp.EnrichedExp

  -- Classe de types permettant d'abstraire l'extraction (monadique)
  -- de symbole d'un produit tensoriel
  class ExtractFinal tensor where
    extractFinal :: tensor(symbols, var) -> m symbols

    extractFinalWeight ::
      (Eq symbols, Semiring weight) =>
        tensor(symbols, var) -> symbols -> weight

  -- Calcul d'un conteneur monadique des symboles "terminant un parcours"
  -- de l'expression (en fonction de l'interprétation du produit tensoriel)
  final
    ::( Eq var
       , Eq symbols
       , Monoid (m symbols)
       , Semimodule weight (m symbols)
       , Star weight
       )
    => Expression tensor var symbols
    -> m symbols
  final Empty       = mempty
  final (Var    _ ) = mempty
  final (Tensor t ) = extractFinal t
  final (Sum e1 e2) = final e1 <> final e2
  final (Sub v e1 e2) =
    (final e1 `actionR` (nullableVar v e2)) <> final  e2
  final (Star v e) = final e `actionR` star (nullableVar v e)

  -- Calcul du "poids final" d'un symbole
  -- de l'expression (en fonction de l'interprétation du produit tensoriel)
  finalWeight
    :: (Eq var, Eq symbols, ExtractFinal tensor, Star weight)
    => symbols
    -> Expression tensor var symbols
    -> weight
  finalWeight _ Empty       = zero
  finalWeight _ (Var    _ ) = zero
  finalWeight f (Tensor t ) = extractFinalWeight t f
  finalWeight s (Sum e1 e2) = finalWeight s e1 `plus` finalWeight s e2
  finalWeight s (Sub v e1 e2) =
    (finalWeight s e1 `times` nullableVar v e2) `plus` finalWeight s e2
  finalWeight s (Star v e) = finalWeight s e `times` star (nullableVar v e)

  -- Expressions.EnrichedExp.WordExp

  instance ExtractFinal Prod where

    extractFinal (Product (f, _)) = return f

    extractFinalWeight (Product (f, _)) g | f == g    = one
                                          | otherwise = zero

  -- Expressions.EnrichedExp.TreeExp
  instance ExtractFinal TensorProd where
    extractFinal (TensorProd f _) = return $ fromGrad f

    extractFinalWeight (TensorProd f _) g | fromGrad f == g = one
                                          | otherwise       = zero
\end{minted}
\end{implementationBox}
\begin{implementationBox}[]{Linéarisation d'une expression enrichie}
\begin{minted}[xleftmargin=1em, autogobble, mathescape=true, fontsize=\footnotesize]{haskell}
  -- Expressions.EnrichedExp.EnrichedExp

  -- Classe de type permettant d'abstraire la linéarisation d'un produit tensoriel
  class TensorLin tensor where
    tensorLin ::
      Enum index =>
        tensor (symbols, var) ->
          State index (tensor (Position index symbols, var))

  -- Linéarisation d'une expression, pour un produit
  -- tensoriel "linéarisable".
  linearize
    :: (Enum index, TensorLin tensor)
    => index
    -> Expression tensor var symbols
    -> Expression tensor var (Position index symbols)
  linearize start expr = evalState (linearize' expr) start
   where
    linearize'
      :: (Enum index, TensorLin tensor)
      => Expression tensor var symbols
      -> State index (Expression tensor var (Position index symbols))
    linearize' Empty       = return Empty
    linearize' (Var    v ) = return $ Var v
    linearize' (Tensor t ) = fmap Tensor $ tensorLin t
    linearize' (Sum e1 e2  ) = liftA2 Sum (linearize' e1) (linearize' e2)
    linearize' (Sub v e1 e2) = liftA2 (Sub v) (linearize' e1) (linearize' e2)
    linearize' (Star v e) = fmap Star v $ linearize' e

  -- Expressions.EnrichedExp.WordExp
  instance TensorLin Prod where
    tensorLin (Product (f, v)) =
      state $ \n -> (Product (Pos n f, v), succ n)

  -- Expressions.EnrichedExp.TreeExp
  instance TensorLin TensorProd where
    tensorLin (TensorProd f vs) =
      state $ \n -> (TensorProd (GradPos n f) vs, succ n)
\end{minted}
\end{implementationBox}
\noindent Remarquons qu'ici tous les symboles sont indicés, y compris les symboles nullaires dans le cas des expressions d'arbres.
Seules les variables ne sont pas indicées.


\subsection{Constructions d'automates enrichis}
Les fonctions précédentes décrites ne sont pas suffisantes pour construire des automates enrichis depuis une expression enrichie.
Ainsi, en fonction des méthodes de construction, nous allons ajouter d'autres outils afin de finaliser le calcul de l'automate associé.

\subsubsection{L'automate des positions}

L'automate des positions que nous allons décrire s'inspire de la version de Laugerotte \emph{et al.}~\cite{LOSZ13} pour les arbres, permettant de construire un automate Top-Down en calculant pour chacune des positions un vecteur de prédécesseurs
(c'est-à-dire un vecteur de successeurs si l'expression est lue de droite à gauche)\footnote{la version classique, utilisant la fonction \texttt{follow}, est implantée pour les mots dans le module \texttt{Expressions.EnrichedExp.WordExp}, en conservant l'extension aux monades}.
Si la version proposée n'est pas factorisée pour les arbres et les mots, la proximité syntaxique des fonctions mises en {\oe}uvre la laisse entrevoir\footnote{Cette factorisation sera discutée dans les perspectives.}.

L'idée est la suivante: nous allons construire un automate enrichi
\begin{itemize}
\item où la configuration initiale est obtenue à partir des variables,
\item où le poids final est obtenu par les fonctions \texttt{final} et \texttt{variables} précédentes,
\item où les transitions sont obtenues à partir de la fonction de calculs des prédécesseurs (variables ou symboles).
\end{itemize}
Ainsi, avant de décrire la construction de l'automate, commençons par expliciter le calcul des prédécesseurs.

Si un prédécesseur d'un symbole est un atome (un symbole ou l'unique variable \texttt{()}) pour les mots, un prédécesseur d'un symbole \(n\)-aire dans une expression d'arbres est un vecteur de longueur \(n\) d'atomes.
De plus, pour conserver la généricité de la construction, la fonction renverra un conteneur monadique de ces combinaisons d'atomes.

Ainsi,
\begin{implementationBox}[label={pseudoCodeSigPredec}]{Signatures des fonctions \texttt{predec}}
\begin{minted}[xleftmargin=1em, autogobble, mathescape=true, fontsize=\footnotesize]{haskell}
  -- Expressions.EnrichedExp.WordExp

  -- Calcul d'un conteneur monadique des prédécesseurs d'un
  -- symbole d'une expression de mots
  predec
    :: ( Semigroup (m (Either () symbols))
       , Eq symbols
       , MonadGen m (->)
       , Semigroup (m ())
       , Star weight
       , Semimodule weight (m symbols)
       , Semimodule weight (m ())
       , Semimodule weight (m (Either () symbols))
       )
    => symbols
    -> WordExp symbols
    -> m (Either () symbols)

  -- Expressions.EnrichedExp.TreeExp

  -- Calcul d'un conteneur monadique des prédécesseurs d'un
  -- symbole n-aire d'une expression d'arbres
  predec
    :: ( Graded symbols
       , Singleton n
       , Eq var
       , Eq symbols
       , MonadGen m (->)
       , Monoid (m (Vector n (Either var symbols)))
       , Semigroup (m var)
       , Semimodule weight (m symbols)
       , Semimodule weight (m var)
       , Semimodule weight (m (Either var symbols))
       , Star weight
       )
    => Graduation symbols n
    -> TreeExp symbols var
    -> m (Vector n (Either var symbols))
\end{minted}
\end{implementationBox}
\noindent Le cas du vide, des variables et de la somme sont classiques, identiques et triviaux dans chacun des cas:
\begin{implementationBox}[]{Fonctions \texttt{predec}: cas triviaux}
\begin{minted}[xleftmargin=1em, autogobble, mathescape=true, fontsize=\footnotesize]{haskell}
  -- Expressions.EnrichedExp.WordExp
  -- Expressions.EnrichedExp.TreeExp

  predec _ Empty   = mempty
  predec _ (Var _) = mempty
  predec g (Sum e1 e2  ) = predec g e1 <> predec g e2
\end{minted}
\end{implementationBox}
\noindent Le cas de la substitution et de l'itération est un peu plus complexe.
Pour une expression \(E_1 \cdot_v E_2\), les prédécesseurs d'un symbole \(c\) sont les prédécesseurs de \(c\) dans \(E_1\) et les prédécesseurs de \(c\) dans \(E_2\), conteneur dans lequel on substitue les occurrences de la variable \(v\) par les éléments finaux (symboles et variables) de \(E_1\).
L'opération de substitution d'occurrences est légèrement différent entre les mots et les arbres, de par la présence d'une ``couche'' de vecteurs supplémentaires.
Le cas de l'itération est similaire.
\begin{implementationBox}[]{Fonctions \texttt{predec}: itération et substitution}
\begin{minted}[xleftmargin=1em, autogobble, mathescape=true, fontsize=\footnotesize]{haskell}
  -- Expressions.EnrichedExp.WordExp

  -- Substitution de la variable () par une combinaison monadique de
  -- symboles et variables dans une combinaison monadique
  substitute
    :: (MonadGen m (->))
    => ()
    -> m (Either () symbols)
    -> m (Either () symbols)
    -> m (Either () symbols)
  substitute _ toAdd = (=<<) aux
   where
    aux (Left _) = toAdd
    aux other    = return other

  -- Expressions.EnrichedExp.TreeExp

  -- Substitution d'une variable par une combinaison monadique de
  -- vecteurs de symboles et variables dans une combinaison monadique
  -- de vecteurs
  substitute
    :: (Eq var, MonadGen m (->))
    => var
    -> m (Either var symbols)
    -> m (Vector n (Either var symbols))
    -> m (Vector n (Either var symbols))
  substitute v toAdd = (=<<) aux
   where
    aux vect = mSequence $ fmap aux' vect
    aux' (Left v') | v == v' = toAdd
    aux' other               = return other


  -- Expressions.EnrichedExp.WordExp
  -- Expressions.EnrichedExp.TreeExp
  predec g (Sub v e1 e2) = predec g e1 <> substitute
    v
    (fmap Right (final e1) <> fmap Left (variables e1))
    (predec g e2)

  predec g (Star v e) = substitute
    v
    (         (  fmap Right (final e)
              <> fmap Left (return v <> variables e)
              )
    `actionR` star (nullableVar v e :: weight)
    )
    (predec g e)
\end{minted}
\end{implementationBox}
\noindent Les cas des symboles sont quant à eux par essence même différents, de par la présence des produits tensoriels.
\begin{implementationBox}[]{Fonctions \texttt{predec}: produits tensoriels}
\begin{minted}[xleftmargin=1em, autogobble, mathescape=true, fontsize=\footnotesize]{haskell}
  -- Expressions.EnrichedExp.WordExp

  predec g (Tensor (Product (f, ()))) | f == g    = return $ Left ()
                                      | otherwise = mempty

  -- Expressions.EnrichedExp.TreeExp

  predec g (Tensor (TensorProd f vs))
      | f == g = return $ fmap Left $ toUngradedVector vs
      | otherwise = mempty

\end{minted}
\end{implementationBox}
\noindent L'automate des positions s'obtient alors directement dans le cas des arbres, selon la construction classique\footnote{Remarquons que dans le code source proposé, la factorisation ne peut se faire aussi directement au niveau monadique pour des problèmes (techniques) de factorisation de contraintes quantifiées}:
\begin{implementationBox}[]{L'automate enrichi des positions: cas des arbres}
\begin{minted}[xleftmargin=1em, autogobble, mathescape=true, fontsize=\footnotesize]{haskell}
  -- Expressions.EnrichedExp.TreeExp

  treePosition e_ = packVar
    (either return (const mempty))
    (fmap Right (final e) <> fmap Left (variables e))
    (GradedMorph $ \x -> DualGradFunF $ \case
      Right (Pos n g) | fromGrad x == g -> predec (GradPos n x) e
      _ -> mempty
    )
    where e = linearize 1 e_

\end{minted}
\end{implementationBox}
Dans le cas des mots, pour conserver l'utilisation de la fonction \texttt{predec} sans autres calculs, il suffit de retourner l'expression avant linéarisation:
\begin{implementationBox}[]{L'automate enrichi des positions: cas des mots}
\begin{minted}[xleftmargin=1em, autogobble, mathescape=true, fontsize=\footnotesize]{haskell}
  -- Expressions.EnrichedExp.WordExp

  predecPositionFA e = packKleisli
    (fmap Right (final e') <> fmap Left (variables e))
    (\x -> \case
      Right (Pos n g) | x == g -> predec (Pos n g) e'
      _                        -> mempty
    )
    (\case
      Left () -> one
      _       -> zero
    )
    where e' = linearize 1 $ reverseExp e
\end{minted}
\end{implementationBox}
\noindent Il est aussi possible de calculer un automate sans renverser l'expression, mais en ``renversant'' la fonction \texttt{predec}:
\begin{implementationBox}[]{L'automate enrichi des positions: cas alternatif des mots}
\begin{minted}[xleftmargin=1em, autogobble, mathescape=true, fontsize=\footnotesize]{haskell}
  -- Expressions.EnrichedExp.WordExp

  predecPositionFA' e = packKleisli
    (  succsOf (Left ())
    <> ((nullableVar () e' :: weight) `action` return (Left ()))
    )
    (\x -> \case
      p@(Right p'@(Pos _ p'')) | p'' == x ->
        ((finalWeight p' e' :: weight) `action` return (Left ())) <> succsOf p
      _ -> mempty
    )
    (\case
      Left () -> one
      _       -> zero
    )
   where
    e' = linearize 1 e
    symbolPrecList =
      fmap (\p -> (p, predec p e')) $ Set.toList $ getSymbols e'
    succsOf p = foldMap (aux p) symbolPrecList
    aux p (q, predsOfQ) =
      (\q' -> if q' == p then return (Right q) else mempty) =<< predsOfQ
\end{minted}
\end{implementationBox}


Le module \texttt{HDRExample.HDRExampleEnrichedPosition} contient des exemples de constructions.
Une fois \texttt{stack ghci src/HDRExample/HDRExampleEnrichedPosition.hs} lancé dans le terminal, il est possible d'utiliser les fonctions suivantes:
\begin{itemize}
\item la fonction \texttt{printWordPositionAut} attend la saisie d'une expression rationnelle de mots (saisie \emph{via} les opérateurs \texttt{+}, \texttt{*} et \texttt{.}, et les symboles \texttt{a}, \texttt{b} et \texttt{c}) avant d'appliquer les constructions de l'automate des positions non-déterministe et à multiplicités sur cette expression avant d'afficher le résultat au format dot;
\item la fonction \texttt{printTreePositionAut} calcule une expression d'arbre aléatoire ne reconnaissant que des arbres nullaires, ayant pour symboles nullaires \texttt{a}, \texttt{b} et \texttt{c}, unaires \texttt{f} et \texttt{h}, binaires \texttt{g} et pour variables \texttt{()}, puis affiche son automate des positions non-déterministes au format Dot.
\end{itemize}

\subsubsection{L'automate enrichi des dérivées}

Comme dans le cas précédent de la construction par la méthode des positions, la dérivation, même si elle n'est pas factorisée au niveau des expressions enrichies, est syntaxiquement proche entre les spécialisations des expressions enrichies (mots et arbres).
Ainsi, montrons comment calculer des automates enrichis (en conservant la généralisation monadique) en utilisant la dérivation.
Remarquons que comme dans le cas précédent, pour conserver la proximité syntaxique entre dérivation (Top-Down, adaptée de~\cite{KM11} et étendue aux monades) des arbres et des mots, nous appliquerons la dérivation des mots par la droite, et inverserons l'expression lors de la construction.
\begin{implementationBox}[]{La dérivation d'expressions enrichies: signatures}
\begin{minted}[xleftmargin=1em, autogobble, mathescape=true, fontsize=\footnotesize]{haskell}
  -- Expressions.EnrichedExp.WordExp

  enrichedDerivation
    :: ( Eq symbols
       , Semimodule weight (m (WordExp symbols))
       , MonadGen m (->)
       , Star weight
       )
    => symbols
    -> WordExp symbols
    -> m (WordExp symbols)

  -- Expressions.EnrichedExp.TreeExp

  enrichedDerivation
    :: ( Eq symbols
       , Graded symbols
       , Eq var
       , Semimodule weight (m (Vector n (TreeExp symbols var)))
       , MonadGen m (->)
       , Star weight
       )
    => Graduation symbols n
    -> TreeExp symbols var
    -> m (Vector n (TreeExp symbols var))
\end{minted}
\end{implementationBox}
D'une façon équivalente au calcul des prédécesseurs (Pseudo-Code~\ref{pseudoCodeSigPredec}), le calcul inductif est, \emph{mutatis mutandis}, le même entre les deux implantations (arbres et mots), situation proche d'une factorisation au niveau des expressions enrichies.
\begin{implementationBox}[]{La dérivation d'expressions enrichies: implantation}
\begin{minted}[xleftmargin=1em, autogobble, mathescape=true, fontsize=\footnotesize]{haskell}
  -- Expressions.EnrichedExp.WordExp
  -- Expressions.EnrichedExp.TreeExp

  -- Les cas de l'expression vide, d'une variable et de la somme
  -- sont identiques.
  enrichedDerivation _ Empty   = mempty
  enrichedDerivation _ (Var _) = mempty
  enrichedDerivation f (Sum e1 e2) =
    enrichedDerivation f e1 <> enrichedDerivation f e2

  -- La substitution et l'itération nécessitent une fonction auxiliaire dont
  -- le rôle est de concaténer (produit de substitution) une expression
  -- à un conteneur monadique

  -- Expressions.EnrichedExp.WordExp
  concatExpr
    :: (FunctorGen f (->) cat')
    => var
    -> Expression tensor var symbols
    -> cat'
         (f (Expression tensor var symbols))
         (f (Expression tensor var symbols))
  concatExpr v e1 = fmap (Sub v e1)

  -- Expressions.EnrichedExp.TreeExp
  concatExpr
    :: (FunctorGen f (->) cat')
    => var
    -> Expression tensor var symbols
    -> cat'
         (f (Vector n (Expression tensor var symbols)))
         (f (Vector n (Expression tensor var symbols)))
  concatExpr v e1 = fmap (fmap (Sub v e1))

  -- Expressions.EnrichedExp.WordExp
  -- Expressions.EnrichedExp.TreeExp

  -- La substitution et l'itération sont identiques,
  -- entre mots et arbres, modulo l'utilisation de
  -- la fonction concatExpr
  enrichedDerivation f (Sub v e1 e2) =
    ((nullableVar v e2 :: weight) `action` enrichedDerivation f e1)
      <> concatExpr v e1 (enrichedDerivation f e2)

  enrichedDerivation f e'@(Star v e) =
    (star (nullableVar v e) :: weight)
      `action` concatExpr v e' (enrichedDerivation f e)

  -- Le cas des produits tensoriels est le seul spécifique,
  -- dépendant totalement du produit tensoriel.

  -- Expressions.EnrichedExp.WordExp
  enrichedDerivation y (Tensor (Product (x, v)))
    -- ci dessous, v == (), et ainsi
    -- Var v == epsilon
    | x == y    = return $ Var v
    | otherwise = mempty

  -- Expressions.EnrichedExp.TreeExp
  enrichedDerivation f (Tensor (TensorProd g vars))
    | f == g = return (fmap Var (toUngradedVector vars) )
    | otherwise = mempty
\end{minted}
\end{implementationBox}
\noindent Les automates monadiques enrichis des dérivées\footnote{Comme dans le cas de la méthode des positions, le code source fourni n'est pas factorisé, même si relativement identique, entre monades pour le cas des arbres, dû à un problème d'expression de contraintes.}
sont ainsi calculés à partir de cette dérivation, mais aussi des fonctions \texttt{variables} ou \texttt{nullableVar} (Pseudo-Code~\ref{pseudoCodeNullableVarVariables}).
\begin{implementationBox}[]{La dérivation d'expressions enrichies: construction d'automates}
\begin{minted}[xleftmargin=1em, autogobble, mathescape=true, fontsize=\footnotesize]{haskell}
  -- Expressions.EnrichedExp.WordExp

  -- Automate (enrichi, monadique, de mots) des dérivées
  rightDerivationAutomaton e =
    packKleisli
      -- l'état initial est construit depuis l'expression inversée
      (return $ reverseExp e)
      -- le morphisme de transition est calculé depuis la dérivation
      enrichedDerivation
      -- la finalité est la traversabilité
      (nullableVar ())

  -- Expressions.EnrichedExp.TreeExp

  -- Automate (enrichi, monadique, d'arbres) des dérivées
  treeDerivation e = packVar
    -- les états finaux sont les variables finales
    variables
    -- l'état initial est l'expression
    (return e)
    -- le morphisme de transition est construit depuis
    -- la dérivation enrichie
    (GradedMorph $ \x -> DualGradFunF $ enrichedDerivation x)
\end{minted}
\end{implementationBox}
Même si ces constructions ne sont pas factorisées, dû à des contraintes techniques, leur proximité syntaxique est encore une fois à remarquer, laissant présager d'une amélioration dans la suite de ce projet.

Le module \texttt{HDRExample.HDRExampleEnrichedDerivation} contient des exemples de constructions.
Une fois \texttt{stack ghci src/HDRExample/HDRExampleEnrichedDerivation.hs} lancé dans le terminal, il est possible d'utiliser les fonctions suivantes:
\begin{itemize}
\item la fonction \texttt{printWordDerivationAut} attend la saisie d'une expression rationnelle de mots (saisie \emph{via} les opérateurs \texttt{+}, \texttt{*} et \texttt{.}, et les symboles \texttt{a}, \texttt{b} et \texttt{c}) avant d'appliquer les constructions de l'automate des dérivées non-déterministe et à multiplicités sur cette expression avant d'afficher le résultat au format Dot;
\item la fonction \texttt{printTreeDerivationAut} calcule une expression d'arbre aléatoire ne reconnaissant que des arbres nullaires, ayant pour symboles nullaires \texttt{a}, \texttt{b} et \texttt{c}, unaires \texttt{f} et \texttt{h}, binaires \texttt{g} et pour variables \texttt{()}, puis affiche son automate des dérivées non-déterministe au format Dot.
\end{itemize}


\subsubsection{La construction inductive}
La dernière méthode de construction que nous allons décrire dans ce document est la méthode inductive, déjà décrite dans la Section~\ref{subsecConsIndMots} pour le cas des mots.
Cette fois, la construction sera factorisée au niveau des expressions enrichies pour construire des automates de mots ou des automates d'arbres Bottom-Up; en effet, les classes de types seront suffisantes afin d'exprimer les différences entre les différents enrichissements des catégories.
Si la construction enrichie s'inspire de la version précédente sur les mots (notamment en utilisant une technique similaire à la construction de la monade libre), les algorithmes mis en place seront légèrement différents afin de faciliter l'écriture au niveau des expressions enrichies.

Nous commencerons tout d'abord à expliquer comment réaliser les opérations encodées par les opérateurs et les atomes d'expressions en automates enrichis, puis nous déclarerons un type général permettant d'encoder (à la manière des monades libres).
Pour cela, nous présenterons des propriétés communes à certaines catégories enrichies permettant de construire d'une façon similaire des automates enrichis, tout en conservant la généralisation monadique obtenue précédemment afin de construire depuis une même expression des automates non-déterministes, à multiplicités, \emph{etc.}

Les deux constructions les plus simples sont le cas de l'expression vide et de l'atome variable.
Pour ces deux constructions, il suffira de considérer des catégories monoïdales pour lesquelles les morphismes entre l'unité (monoïdale) et les objets morphismes de la catégorie enrichie (c'est-à-dire les éléments généralisés) sont isomorphes avec une catégorie de Kleisli.
\begin{implementationBox}[]{La construction inductive: Interprétation des morphismes en fonctions de Kleisli}
\begin{minted}[xleftmargin=1em, autogobble, mathescape=true, fontsize=\footnotesize]{haskell}
  -- AutomataGen.Inductive

  -- Classe de types permettant d'expliciter le lien entre
  -- morphismes (de la catégorie monoïdale) et morphisme
  -- de Kleisli
  class AsFunction enrichedCat catMon tensor where

    -- Famille de types permettant d'"extraire" la monade
    type ExtractM enrichedCat catMon tensor :: * -> *

    -- Transformation d'un morphisme depuis l'unité monoïdale
    -- en une fonction de Kleisli
    toFun ::
      catMon (Unit catMon tensor) (enrichedCat a b)
        -> a -> ExtractM enrichedCat catMon tensor b

    -- Transformation d'une fonction de Kleisli en morphisme
    -- depuis l'unité monoïdale
    fromFun ::
      (a -> ExtractM enrichedCat catMon tensor b)
        -> catMon (Unit catMon tensor) (enrichedCat a b)

    -- Injection d'une fonction constante dans un morphisme
    constCatMon :: ExtractM enrichedCat catMon tensor b -> catMon x (enrichedCat a b)

  -- Définition pour la catégorie enrichie de Kleisli des types
  -- associée à une monade enrichie m
  instance AsFunction (KleisliEnrichedCat m (EnrichedTypeCat (->)) (->) Prod) (->) Prod where
    type ExtractM (KleisliEnrichedCat m (EnrichedTypeCat (->)) (->) Prod) (->) Prod = m

    toFun c = runEnrichedCat $ runKleisli $ c ()
    fromFun c () = Kleisli $ EnrichedTypeCat c

    constCatMon mb _ = Kleisli $ EnrichedTypeCat $ const mb

  -- Définition pour la catégorie enrichie de Kleisli des types gradués
  -- associée à une monade enrichie m
  instance AsFunction
    (KleisliEnrichedCat m MultiFun GradedMorph TensorProd) GradedMorph TensorProd
    where

    type ExtractM (KleisliEnrichedCat m MultiFun GradedMorph TensorProd) GradedMorph TensorProd = m

    toFun (GradedMorph c) = runGradFun $ runGradK $ c GradUn
    fromFun c = GradedMorph $ \GradUn -> GradKleisli $ GradFun c

    constCatMon mb =
      GradedMorph $
        \x -> GradKleisli $ convertFromVectMultiFun (graduation' x) (const mb)
\end{minted}
\end{implementationBox}
À l'aide de cette classe de types, la construction de l'automate enrichi ne reconnaissant rien est simple: tous les morphismes sont obtenus depuis la constante nulle.
\begin{implementationBox}[]{La construction inductive: l'automate vide}
\begin{minted}[xleftmargin=1em, autogobble, mathescape=true, fontsize=\footnotesize]{haskell}
  -- AutomataGen.Inductive

  autoEmpty
    :: ( EnrichedCategory enrichedCat catMon tensor
       , Monoid (ExtractM enrichedCat catMon tensor state)
       , AsFunction enrichedCat catMon tensor
       , Semiring altValue
       , Castable (ExtractM enrichedCat catMon tensor value) altValue
       )
    => AutomatonGen enrichedCat catMon tensor init symbols state value
  autoEmpty =
    Auto
      -- le morphisme initial est obtenu depuis la constante nulle
      (constCatMon mempty)
      -- le morphisme de transitions est obtenu depuis la constante nulle
      (constCatMon mempty)
      -- le morphisme initial est obtenu depuis le poids nul
      (constCatMon $ castInv (zero :: altValue))
\end{minted}
\end{implementationBox}
L'automate associé à une variable \texttt{var} n'a pas de transition (c'est à dire que son morphisme de transition est obtenu depuis la constante nulle), sa configuration initiale est la promotion monadique des variables, et son morphisme de finalité est obtenu depuis la fonction associant l'unité à la variable \texttt{var}, la constante nulle aux autres.
\begin{implementationBox}[]{La construction inductive: l'automate associé à une variable}
\begin{minted}[xleftmargin=1em, autogobble, mathescape=true, fontsize=\footnotesize]{haskell}
  -- AutomataGen.Inductive

  fromVar
    :: ( Eq var
       , EnrichedCategory enrichedCat catMon tensor
       , MonadGen (ExtractM enrichedCat catMon tensor) (->)
       , AsFunction enrichedCat catMon tensor
       , Castable (ExtractM enrichedCat catMon tensor value) altValue
       , Semiring altValue
       , Monoid (ExtractM enrichedCat catMon tensor var)
       )
    => var
    -> AutomatonGen enrichedCat catMon tensor var symbols var value
  fromVar v = Auto
    -- le morphisme initial est basé sur la promotion monadique
    -- des variables
    (fromFun return)
    -- le morphisme de transition est obtenu depuis la constante nulle
    (constCatMon mempty)
    -- le morphisme de finalité est obtenu en transformant en élément
    -- généralisé la fonction associant à v le poids unité,
    -- le poids nul aux autres variables
    (fromFun $ \v' -> if v == v'
      then castInv (one :: altValue)
      else castInv (zero :: altValue)
    )
\end{minted}
\end{implementationBox}
Le cas de l'itération se suffit également de la classe de types définie précédemment.
Cependant, contrairement à la méthode inductive précédente sur les mots, nous décomposerons ici l'opération d'itération des expressions en deux étapes: l'itération non-nulle suivie de l'ajout de variable.
L'itération positive par une variable \texttt{v} s'obtient alors d'une façon classique:
\begin{itemize}
\item le morphisme initial s'obtient en combinant l'ancien morphisme initial avec le poids étoilé de \texttt{v} le cas échéant;
\item le morphisme de transitions se compose de l'ancien dans lequel est ajoutée à toute ``destination'' la ``configuration initiale'' après action du poids de finalité de cette destination;
\item le morphisme de finalité est l'ancien morphisme combiné avec l'action du poids étoilé de \texttt{v}.
\end{itemize}
Afin de combiner morphismes et éléments généralisés, nous utiliserons une nouvelle fois la fonction \texttt{append} du module \texttt{Enriched.Category.EnrichedCategory} (Pseudo-Code~\ref{codeCombiElemtGen}).
\begin{implementationBox}[]{La construction inductive: l'itération positive}
\begin{minted}[xleftmargin=1em, autogobble, mathescape=true, fontsize=\footnotesize]{haskell}
  -- AutomataGen.Inductive

  autoPositiveStar
    ::( Semiring altValue
       , AsFunction enrichedCat catMon tensor
       , Castable (ExtractM enrichedCat catMon tensor value) altValue
       , Semimodule altValue (ExtractM enrichedCat catMon tensor state)
       , Semimodule altValue (ExtractM enrichedCat catMon tensor value)
       , MonadGen (ExtractM enrichedCat catMon tensor) (->)
       )
    => init
    -> AutomatonGen enrichedCat catMon tensor init symbols state value
    -> AutomatonGen enrichedCat catMon tensor init symbols state value
  autoPositiveStar var (Auto i d f) = Auto initPositiveStar
                                           deltaPositiveStar
                                           finalPositiveStar
   where
    -- poids de la variable var dans l'automate de départ
    epsWeight        = toFun (append i f) var
    -- fonction de Kleisli associée au morphisme initial
    i'               = toFun i
    -- fonction de Kleisli associée au morphisme final
    f'               = toFun f
    -- le nouveau morphisme initial s'obtient en convertissant
    -- en élément généralisé la fonction faisant agir
    -- le poids étoilé de la variable var sur l'ancienne configuration
    -- initiale le cas échéant
    initPositiveStar = fromFun $ \var' -> if var == var'
      then (star $ cast epsWeight :: altValue) `action` i' var
      else i' var'
    -- le nouveau morphisme final s'obtient en transformant en élément
    -- généralisé la fonction faisant agir le poids étoilé de la variable
    -- sur le poids de finalité de chaque état
    finalPositiveStar = fromFun
      $ \p -> f' p `actionR` (star $ cast epsWeight :: altValue)
    -- le nouveau morphisme de transition s'obtient en combinant l'ancien
    -- morphisme de transition avec l'élément généralisé obtenu depuis la
    -- fonction ajoutant à chaque état-destination p la configuration
    -- initiale associée à la variable var après action du poids de p
    deltaPositiveStar = append d $ fromFun $ \p ->
      return p <> ((cast $ f' p :: altValue) `action` i' var)
\end{minted}
\end{implementationBox}
La somme de deux automates enrichis se suffit quant à elle de l'adjonction de deux classes de types permettant
\begin{itemize}
\item de simuler au niveau des catégories monoïdales le dédoublement de l'entrée d'un automate (par un morphisme qu'on pourrait presque qualifié de \emph{diagonal}) par le foncteur \texttt{Prod} (Pseudo-Code~\ref{CodeProdTensTypes}),
\item d'utiliser le foncteur \texttt{Sum} (Pseudo-Code~\ref{CodeProdTensTypes}) au sein d'une catégorie enrichie pour combiner les résultats obtenus par le dédoublement précédent.
\end{itemize}
\begin{implementationBox}[]{La construction inductive: le dédoublement et le regroupement pour la somme}
\begin{minted}[xleftmargin=1em, autogobble, mathescape=true, fontsize=\footnotesize]{haskell}
  -- AutomataGen.Inductive

  -- Définition d'un morphisme presque "diagonal"
  class Diagonalizable catMon where
   diag :: catMon x (Prod(x, x))

  -- Cas de la catégorie monoïdale des types
  instance Diagonalizable (->) where
    diag x = Product (x, x)

  -- Déclaration des couples d'éléments gradués (valable pour
  -- des éléments de mêmes graduation uniquement), nécessaire
  -- à la définition de Diagonalizable pour les morphismes
  -- gradués
  instance (Graded (First a), Graded (Second a)) => Graded (Prod a) where

    fromGrad (GradProd x1 x2) = Product (fromGrad x1, fromGrad x2)
    toGrad (Product (x1, x2)) = GradProd (toGrad x1) (toGrad x2)

    data Graduation (Prod a) n =
      GradProd (Graduation (First a) n) (Graduation (Second a) n)

  -- La nouvelle graduation permet de définir Prod comme foncteur
  -- depuis la catégorie produit des types gradués
  instance FunctorGen Prod (ProdCat GradedMorph GradedMorph) GradedMorph where
    fmap (Prod (GradedMorph f) (GradedMorph g)) =
     GradedMorph $ \(GradProd x y) -> GradProd (f x) (g y)

  -- Cas de la catégorie monoïdale des types gradués
  instance Diagonalizable GradedMorph where
    diag = GradedMorph $ \x -> GradProd x x

  -- Extension de la fonction 'Data.Either.either' au
  -- niveau monoïdal, envoyant un produit de morphismes
  -- sur une somme
  class AsEither catMon enrichedCat where

   eitherFromProd ::
     catMon
       (Prod (enrichedCat a c, enrichedCat b c))
         (enrichedCat (Sum (a, b)) c)

  -- Définition pour la catégorie enrichie de Kleisli des types
  -- associée à une monade enrichie m
  instance
    AsEither (->) (KleisliEnrichedCat m (EnrichedTypeCat (->)) (->) Prod)
    where

    eitherFromProd (Product (Kleisli f1, Kleisli f2)) =
      let f1' = runEnrichedCat f1
          f2' = runEnrichedCat f2
      in  Kleisli $ EnrichedTypeCat $ packSumFunctions f1' f2'

  -- Définition pour la catégorie enrichie de Kleisli des types gradués
  -- associée à une monade enrichie m
  instance
    AsEither
      GradedMorph (KleisliEnrichedCat m MultiFun GradedMorph TensorProd)
    where

   eitherFromProd =
     GradedMorph $
       \(GradProd (GradKleisli f1) (GradKleisli f2)) ->
         GradKleisli $ GradFun $
           convertFromVectMultiFun' (graduation' f1) (aux f1 f2)
     where
       aux (GradFun f1) (GradFun f2) VNil = f1 <> f2
       aux f1 f2 ys@(VCons _ _) = case vectToSum ys of
         Nothing       -> mempty
         Just (Fst xs) -> convertToVectMultiFun f1 xs
         Just (Snd xs) -> convertToVectMultiFun f2 xs
       -- fonction issue du module Tools.Vector
       -- transformant un vecteur d'éléments Sum(a, b)
       -- en
       -- * un vecteur de b si tous les éléments sont de type b
       -- * un vecteur de a si tous les éléments sont de type a
       -- * Nothing sinon
       vectToSum
         :: Vector ( 'S n) (Sum (a, b))
         -> Maybe (Sum (Vector ( 'S n) a, Vector ( 'S n) b))
       vectToSum (VCons (Fst x) xs) = case xs of
         VNil      -> Just $ Fst $ VCons x VNil
         VCons _ _ -> case vectToSum xs of
           Just (Fst xs_) -> Just $ Fst $ VCons x xs_
           _              -> Nothing
       vectToSum (VCons (Snd x) xs) = case xs of
         VNil      -> Just $ Snd $ VCons x VNil
         VCons _ _ -> case vectToSum xs of
           Just (Snd xs_) -> Just $ Snd $ VCons x xs_
           _              -> Nothing
\end{minted}
\end{implementationBox}


À l'aide de ces outils, la somme de deux automates enrichis se décrit, d'une façon relativement classique, comme suit:
\begin{itemize}
\item le nouveau morphisme initial est la somme des morphismes initiaux interprétés comme fonctions de Kleisli;
\item le nouveau morphisme de transition s'obtient en dupliquant l'entrée par le morphisme diagonal, en appliquant à chacune de ces composantes le morphisme de transitions des anciens automates, puis en combinant le résultat par le morphisme \texttt{eitherFromProd}, chacun de ces morphismes la catégorie monoïdale pouvant être concaténé par la composition de la catégorie;
\item le nouveau morphisme final est obtenu en appliquant le premier morphisme sur les états du premier automate, le second morphisme aux états du second automate.
\end{itemize}
\begin{implementationBox}[]{La construction inductive: le cas de la somme}
\begin{minted}[xleftmargin=1em, autogobble, mathescape=true, fontsize=\footnotesize]{haskell}

  -- AutomataGen.Inductive

  autoSum
    :: ( FunctorGen (ExtractM enrichedCat catMon tensor) (->) (->)
       , AsFunction enrichedCat catMon tensor
       , Semigroup (ExtractM enrichedCat catMon tensor (Sum (state, state')))
       , Diagonalizable catMon
       , AsEither catMon enrichedCat
       , Bifunctor Prod catMon
       , MonadGen (ExtractM enrichedCat catMon tensor) (->)
       )
    => AutomatonGen enrichedCat catMon tensor init symbols state value
    -> AutomatonGen enrichedCat catMon tensor init symbols state' value
    -> AutomatonGen
         enrichedCat
         catMon
         tensor
         init
         symbols
         (Sum (state, state'))
         value
  autoSum (Auto i1 d1 f1) (Auto i2 d2 f2) = Auto initSum deltaSum finalSum
   where
    -- fonctions de Kleisli associées aux morphismes initiaux
    i1' = toFun i1
    i2' = toFun i2
    -- fonctions de Kleisli associées aux morphismes finaux
    f1' = toFun f1
    f2' = toFun f2
    -- promotions des morphismes de transitions au niveau
    -- du foncteur Sum
    d1' = append d1 $ fromFun (return . Fst)
    d2' = append d2 $ fromFun (return . Snd)
    -- le nouveau morphisme initial est la combinaison des deux anciens
    initSum =
      fromFun $ \var -> fmap Fst (i1' var) <> fmap Snd (i2' var)
    -- le nouveau morphisme final s'obtient en combinant les deux anciens
    finalSum = fromFun $ packSumFunctions f1' f2'
    -- le nouveau morphisme de transitions s'obtient par composition
    -- du morphisme diagonal, des promotions des deux anciens morphismes de
    -- transition, puis en combinant les résultats
    deltaSum = eitherFromProd . fmap (Prod d1' d2') . diag

    -- Category.OfStructures.CategoryOfTypes

    -- Injection de deux fonctions dans une 'Sum'
    packSumFunctions :: (a -> c) -> (b -> c) -> Sum (a, b) -> c
    packSumFunctions f _ (Fst a) = f a
    packSumFunctions _ g (Snd b) = g b
\end{minted}
\end{implementationBox}

La substitution par une variable \texttt{var} se fait également d'une façon classique, en ajoutant des transitions entre les deux automates de la façon suivante:
\begin{itemize}
\item le morphisme initial est identique à la somme pour les variables différentes de \texttt{var}, sinon correspond à la combinaison de la configuration initiale associée à \texttt{var} du premier automate combinée à celle du second, après action du poids de la variable \texttt{var} dans le premier automate;
\item le morphisme de transition s'obtient tout d'abord en construisant le même morphisme que la somme, puis en ajoutant aux états destinations des transitions du premier automate la configuration initiale associée à \texttt{var} du second automate, après action du poids final de l'état dans le premier automate;
\item le morphisme final s'obtient en ne conservant que celui du second automate.
\end{itemize}
\begin{implementationBox}[]{La construction inductive: le cas de la substitution}
\begin{minted}[xleftmargin=1em, autogobble, mathescape=true, fontsize=\footnotesize]{haskell}

  -- AutomataGen.Inductive

  autoCat
    :: ( Eq init
       , Diagonalizable catMon
       , AsEither catMon enrichedCat
       , Bifunctor Prod catMon
       , EnrichedCategory enrichedCat catMon tensor
       , AsFunction enrichedCat catMon tensor
       , MonadGen (ExtractM enrichedCat catMon tensor) (->)
       , Castable (ExtractM enrichedCat catMon tensor value) altValue
       , Semimodule altValue (ExtractM enrichedCat catMon tensor value)
       , Semimodule
           altValue
           (ExtractM enrichedCat catMon tensor (Sum (state, state')))
       )
    => init
    -> AutomatonGen enrichedCat catMon tensor init symbols state value
    -> AutomatonGen enrichedCat catMon tensor init symbols state' value
    -> AutomatonGen
         enrichedCat
         catMon
         tensor
         init
         symbols
         (Sum (state, state'))
         value
  autoCat var (Auto i1 d1 f1) (Auto i2 d2 f2) = Auto
    initConcat
    (append deltaSum (fromFun aux))
    finalConcat
   where
    -- promotions des morphismes de transitions au niveau
    -- du foncteur Sum
    d1'      = append d1 $ fromFun (return . Fst)
    d2'      = append d2 $ fromFun (return . Snd)
    -- identique à la construction de la somme
    deltaSum = eitherFromProd . fmap (Prod d1' d2') . diag
    -- ajout de l'action des poids finaux sur les états
    -- destinations de transitions
    aux q'@(Fst q) =
      return q' <> ((cast (f1' q) :: altValue) `action` fmap Snd (i'2 var))
    aux q' = return q'
    -- poids de var dans le premier automate
    emptyWeight1 = toFun (append i1 f1) var
    -- fonctions de Kleisli associées aux morphismes initiaux
    i'1          = toFun i1
    i'2          = toFun i2
    -- fonctions de Kleisli associées aux morphismes finaux
    f1'          = toFun f1
    f2'          = toFun f2
    -- nouveau morphisme initial, calculé en fonction
    -- de la variable considérée pour la configuration
    -- initiale
    initConcat   = fromFun $ \var' -> if var == var'
      then fmap Fst (i'1 var)
        <> action (cast emptyWeight1 :: altValue) (fmap Snd (i'2 var))
      else fmap Fst (i'1 var') <> fmap Snd (i'2 var')
    -- nouveu morphisme final, obtenu depuis le morphisme final
    -- du second automate
    finalConcat = fromFun $ packSumFunctions (const mempty) f2'
\end{minted}
\end{implementationBox}
Le cas restant est le cas le plus spécifique, puisqu'il repose sur le produit tensoriel entre symboles et variables: l'idée générale est de construire un automate dont les états sont des variables et des symboles, de considérer les variables comme configurations initiales,
les symboles comme des configurations finales (unitaires), et d'ajouter une transition entre les variables et symboles dont le produit tensoriel est composé (en toute généralité).
Pour représenter cette spécificité, nous allons décrire une type de classes permettant de transformer un produit tensoriel en un morphisme (de la catégorie monoïdale), ce qui nous servira de transition.
\begin{implementationBox}[]{La construction inductive: extraction d'une transition d'un produit tensoriel}
\begin{minted}[xleftmargin=1em, autogobble, mathescape=true, fontsize=\footnotesize]{haskell}

  -- AutomataGen.Inductive

  -- Classe de type représentant la conversion d'un produit tensoriel
  -- en un morphisme, utilisé pour décrire une transition
  class AddTrans enrichedCat catMon tensor where

    toTrans :: tensor(a, b) -> catMon a (enrichedCat b a)

  -- Cas de la catégorie monoïdale des types
  instance AddTrans (KleisliEnrichedCat m (EnrichedTypeCat (->)) (->) Prod) (->) Prod where

    toTrans prod@(Product (x, _)) y =
      Kleisli $ EnrichedTypeCat $ \v' ->
        if Product (y, v') == prod then return x else mempty

  -- Cas de la catégorie monoïdale des types gradués
  instance
    AddTrans (KleisliEnrichedCat m MultiFun GradedMorph TensorProd)
      GradedMorph TensorProd
    where

    toTrans (TensorProd x ys) = GradedMorph $ \y ->
      GradKleisli $ convertFromVectMultiFun (graduation' y) $ \ys' ->
        if fromGrad x == fromGrad y && vectEq ys' (toUngradedVector ys)
          then return (fromGrad x)
          else mempty
\end{minted}
\end{implementationBox}
%


La considération de cette unique transition pour construire l'automate intervient comme suit.
\begin{implementationBox}[]{La construction inductive: le cas du produit tensoriel}
\begin{minted}[xleftmargin=1em, autogobble, mathescape=true, fontsize=\footnotesize]{haskell}

  -- AutomataGen.Inductive

  fromTensor
    :: ( Castable (ExtractM enrichedCat catMon tensor value) altValue
       , Semiring altValue
       , AddTrans enrichedCat catMon tensor
       , Eq var
       , Eq symbols
       , EnrichedCategory enrichedCat catMon tensor
       , MonadGen (ExtractM enrichedCat catMon tensor) (->)
       , AsFunction enrichedCat catMon tensor
       , Bifunctor Prod catMon
       , Monoid (ExtractM enrichedCat catMon tensor (Sum (var, symbols)))
       , AsEither catMon enrichedCat
       , Diagonalizable catMon
       )
    => tensor (symbols, var)
    -> AutomatonGen
         enrichedCat
         catMon
         tensor
         var
         symbols
         (Sum (var, symbols))
         value
  fromTensor tensor = Auto
    -- configuration initiale associée à chaque variable
    (fromFun $ return . Fst)
    -- transition obtenu depuis l'extraction tensorielle
    (eitherFromProd
      . fmap
        (Prod
          ( append (toTrans tensor) $ fromFun (return . Snd) )
          (constCatMon mempty)
        )
      . diag
    )
    -- morphisme de finalité:
    -- seuls les symboles ont un poids non-nul, unitaire
    (fromFun$ \case
      Snd _ -> castInv (one :: altValue)
      _     -> castInv (zero :: altValue)
    )
\end{minted}
\end{implementationBox}
Le passage au calcul de l'automate depuis une expression enrichie nécessite la prise en compte du problème soulevé lors de la définition de la méthode inductive sur les expressions de mots: ne connaissant pas à l'avance le nombre d'opérations ni leur nature à appliquer inductivement, comment exprimer le type des états de l'automate?
Nous appliquerons ici une méthode équivalente à la construction de la monade libre engendré par un foncteur, mais sans déclarer tous les outils pour simplifier.
Tout d'abord, remarquons que
\begin{itemize}
\item l'automate vide est construit sur n'importe quel type d'états;
\item l'automate associé à une variable a pour états des variables;
\item l'itération positive ne modifie pas les états de l'automate;
\item l'automate associé à un produit tensoriel a pour état des symboles et des variables, \emph{via} le constructeur de types \texttt{Sum};
\item la somme et la substitution ajoutent un niveau de foncteur somme au niveau du type des états (passant de \texttt{state} et \texttt{state'} à \texttt{Sum(state, state')}).
\end{itemize}
Ainsi, les états d'un automate inductif sont obtenus par application successive du foncteur \texttt{Sum} depuis les types \texttt{var} et \texttt{symbols}, représentable par le type suivant:
\begin{implementationBox}[]{La construction inductive: le type des états}
\begin{minted}[xleftmargin=1em, autogobble, mathescape=true, fontsize=\footnotesize]{haskell}

  -- AutomataGen.Inductive

  -- Les états sont:
  data StateType var symbols
   -- soit des variables
   = Var var
   -- soit des sommes variables et symboles
   | Sum (Sum (var, symbols))
   -- soit des sommes récursives de ce type
   | SumRec (Sum(StateType var symbols, StateType var symbols))
\end{minted}
\end{implementationBox}
En plus de la définition de ce type, le module \texttt{AutomataGen.Inductive} contient la définition de fonctions utilitaires permettant de projeter les constructions précédentes:
\begin{itemize}
\item \texttt{varToStateType} transforme un automate dont les états sont de types \texttt{var} en un automate dont les états sont de type \texttt{StateType var symbols};
\item \texttt{sumVarSymbToStateType} transforme un automate dont les états sont de types \texttt{Sum(var, symbols)} en un automate dont les états sont de type \texttt{StateType var symbols};
\item \texttt{sumToStateType} transforme un automate dont les états sont de types \texttt{Sum(StateType var symbols, StateType var symbols)} en un automate dont les états sont de type \texttt{StateType var symbols}
\end{itemize}
par des processus similaires à ceux mis en {\oe}uvre par la technique de la monade libre précédente.
À l'aide de ces outils, la construction de l'automate enrichi associé à une expression enrichie se déclare sans surprise comme suit.

\begin{implementationBox}[]{La construction inductive}
\begin{minted}[xleftmargin=1em, autogobble, mathescape=true, fontsize=\footnotesize]{haskell}

  -- AutomataGen.Inductive

  fromExpr
    :: ( Eq symbols
       , Eq var
       , AsFunction enrichedCat catMon tensor
       , Semiring altValue
       , Semimodule
           altValue
           (ExtractM enrichedCat catMon tensor (StateType var symbols))
       , Star altValue
       , Monoid
           ( ExtractM
               enrichedCat
               catMon
               tensor
               (Sum (StateType var symbols, StateType var symbols))
           )
       , Diagonalizable catMon
       , AsEither catMon enrichedCat
       , FunctorGen (ExtractM enrichedCat catMon tensor) (->) (->)
       , FunctorGen Prod (ProdCat catMon catMon) catMon
       , Monoid (ExtractM enrichedCat catMon tensor var)
       , Monoid (ExtractM enrichedCat catMon tensor (Sum (var, symbols)))
       , MonadGen (ExtractM enrichedCat catMon tensor) (->)
       , EnrichedCategory enrichedCat catMon tensor
       , Castable (ExtractM enrichedCat catMon tensor value) altValue
       , Semimodule altValue (ExtractM enrichedCat catMon tensor value)
       , Semimodule
           altValue
           ( ExtractM
               enrichedCat
               catMon
               tensor
               (Sum (StateType var symbols, StateType var symbols))
           )
       , AddTrans enrichedCat catMon tensor
       )
    => Expression tensor var symbols
    -> AutomatonGen
         enrichedCat
         catMon
         tensor
         var
         symbols
         (StateType var symbols)
         value
  fromExpr Empty   = varToStateType autoEmpty
  fromExpr (Var v) = varToStateType $ fromVar v
  fromExpr (Tensor t) = sumVarSymbToStateType $ fromTensor t
  fromExpr (Sum e1 e2) =
    sumToStateType $ autoSum (fromExpr e1) (fromExpr e2)
  fromExpr (Sub v e1 e2) =
    sumToStateType $ autoCat v (fromExpr e1) (fromExpr e2)
  fromExpr (Star v e) = sumToStateType $ autoSum
    (autoPositiveStar v $ fromExpr e)
    (fromExpr (Var v))
\end{minted}
\end{implementationBox}
Le module \texttt{HDRExample.HDRExampleEnrichedInductive} contient des exemples de constructions.
Une fois \texttt{stack ghci src/HDRExample/HDRExampleEnrichedInductive.hs} lancé dans le terminal, il est possible d'utiliser les fonctions suivantes:
\begin{itemize}
\item la fonction \texttt{printInductiveAut} attend la saisie d'une expression rationnelle de mots (saisie \emph{via} les opérateurs \texttt{+}, \texttt{*} et \texttt{.}, et les symboles \texttt{a}, \texttt{b} et \texttt{c}) avant d'appliquer les constructions de l'automate inductif non-déterministe et à multiplicités sur cette expression avant d'afficher le résultat au format dot;
\item la fonction \texttt{printInductiveTreeAut} calcule une expression d'arbre aléatoire ne reconnaissant que des arbres nullaires, ayant pour symboles nullaires \texttt{a}, \texttt{b} et \texttt{c}, unaires \texttt{f} et \texttt{h}, binaires \texttt{g} et pour variables \texttt{()}, puis affiche son automate Bottom-Up inductif non-déterministe au format dot.
\end{itemize}
Afin de tester la validité des constructions, le module \texttt{HDRExample.HDRExampleEnrichedTests} propose deux fonctions permettant de comparer les résultats des différentes méthodes.
Ainsi, une fois \texttt{stack ghci src/HDRExample/HDRExampleEnrichedTests.hs} lancé dans le terminal, il est possible d'utiliser les fonctions suivantes:
\begin{itemize}
\item la fonction \texttt{testWordConstructions} génère une (courte) expression rationnelle aléatoire, puis construit un automate à multiplicités (non enrichi) selon la méthode proposée à la Section~\ref{subsecConsIndMots}, et 8 automates enrichis
(Glushkov enrichi à multiplicités, inductif enrichi à multiplicités, deux Glushkov inversés à multiplicités, enrichi à multiplicités des dérivées gauches et droites, enrichi alternant généralisé  des dérivées gauches et droites) puis compare les résultats de ces automates pour 100 mots aléatoires;
\item la fonction \texttt{testTreeConstructions} génère une (courte) expression enrichie aléatoire d'arbres, puis construit les automates Top-Down à multiplicités par dérivation et par la méthode des positions, ainsi que l'automate Bottom-Up inductif à multiplicités associé à cette expression, puis compare les résultats de ces automates pour 100 arbres aléatoires.
\end{itemize}

\section{Conclusion et perspectives}

Dans cette partie, nous avons proposé une ébauche de définition d'expressions rationnelles pouvant utiliser les différentes notions présentées dans les parties précédentes.
Même si l'implantation proposée n'est pas optimale, elle permet d'entrevoir le potentiel de factorisation, mais aussi de permettre de définir des machines à l'aide de monades bien connues de la communauté de la programmation fonctionnelle.

Cette proposition de structure permet également de soulever le rôle des variables dans les automates d'arbres et de proposer une utilisation de celles-ci dans d'autres structures monoïdales;
par exemple, il est possible de définir des expressions de mots sur un autre ensemble de variables que l'ensemble \(\mathbb{1}\).
Dans ce cas, le rôle d'une variable pourrait être, par exemple, de ``sélectionner'' une machine particulière d'un réseau de machines.

Le choix du produit tensoriel est également important et peut permettre de changer avec finesse le comportement des expressions.
On pourrait imaginer qu'en fonction du choix du produit tensoriel, il soit possible d'obtenir des expressions commutatives par exemple.

Nous avons pu également constater que l'implantation proposée ne permet pas une factorisation totale des constructions, bien que leur très forte proximité syntaxique ne laisse pas de doute sur la possibilité d'y arriver.
Pour cela, il peut être intéressant de redéfinir les fonctions de construction, que ce soit les fonctions de position ou la dérivation non plus comme des fonctions (c'est-à-dire des morphismes de la catégorie des types) mais comme des morphismes d'autres catégories monoïdales enrichissant la catégorie enrichie ambiante.
Par exemple, on pourrait imaginer une signature de la forme \texttt{catMon symbols (enrichedCat expression expression)} pour un morphisme de dérivation.
Le problème de la définition d'une construction inductive se pose alors: comment définir un tel morphisme par induction sur la structure d'expression rationnelle, définie comme une structure de données, et ainsi ``vivant'' dans la catégorie des types.
Une piste serait alors de s'intéresser à la notion d'opérade dans d'autres catégories que la catégorie des types: l'idée serait de donner les cas de bases sur les constructeurs de types, ou sur les opérateurs formant la base de l'opérade isomorphe aux expressions,
puis étendre ces cas (par induction ``automatique'') par un mécanisme semblable à \texttt{promoteFunToOperad}.

La dérivation Bottom-Up d'expression d'arbres, quant à elle, pourrait s'étendre en prenant en compte une dérivation par un vecteur de variables; la dérivation serait alors paramétrée également par la définition du produit tensoriel.

Une autre piste serait de s'intéresser à une extension de la notion d'\texttt{Arrows}~\cite{Hug00} (voir également~\cite{Atk11} pour les liens possibles avec l'enrichissement) sur les catégories enrichies, ce qui semblerait être relativement proche des classes de types utilisées pour la construction de l'automate généralisé par induction.



\newpage


\chapter*{Conclusion et perspectives de recherche}\label{chapterConclu}
\addcontentsline{toc}{part}{Conclusion et perspectives}

\begin{flushright}
  \emph{Une sortie, c'est une entrée qu'on prend dans l'autre sens.}\\
  Boris Vian
\end{flushright}

Nous avons ainsi montré comment implanter d'une façon relativement factorisée les notions d'automates d'arbres et de mots en les généralisant au niveau des catégories enrichies.
Nous avons également proposé une ébauche de définition concernant les expressions rationnelles.

Ces propositions de représentation laissent entrevoir de futures extensions à d'autres monoïdes objets de catégories monoïdales afin de généraliser les éléments de bases à pondérer.
On pourrait également se poser la question de l'extension de la pondération non plus à des éléments, mais à des éléments généralisés.

Concernant l'implantation, nous avons indiqué plusieurs fois que celle-ci n'avait pas été produite dans un but d'utilisation intensive.
Il s'agit avant tout d'une preuve de concept qui doit être alors remise à plat depuis sa base afin de tirer profit au maximum des outils déjà disponibles dans la communauté Haskell.
Par exemple, le genre \texttt{Nat} utilisé pour représenter les entiers au niveau des types (pour paramétrer la longueur des vecteurs ou l'arité des éléments gradués par exemple) est une construction simpliste pouvant être optimisée (il existe d'ailleurs de très bons modules pour cela dans la communauté Haskell).
Un autre exemple d'optimisation possible est la résolution d'une grande perte d'efficacité étant due à la notion de preuve au niveau des types (comme les preuves d'égalité de types de genre \texttt{Nat}, telles que la preuve récursive de la commutativité de la somme).
Les types sont ``oubliés'' après la compilation, et il faut alors utiliser des mécanismes, parfois lourds, pour les récupérer afin de s'assurer que tout est correct en terme d'équivalence de types pour réaliser une programmation au niveau des types correcte.
Cependant, la correction n'est nécessaire que pour guider le compilateur dans l'inférence d'égalité (au sens fort) de types.
En d'autres termes, on peut se passer d'un grand nombre de ces morceaux de code à l'exécution car ces calculs ne servent qu'à indiquer au compilateur qu'il peut réaliser des opérations prévues sur des types sur d'autres types en toute sécurité.

Une fois ces modifications faites, c'est-à-dire une fois avoir mis en place un mécanisme permettant de réaliser des tests forts au niveau du typage tout en étant capable de les ``désactiver'',
on pourrait renforcer encore plus la sécurité du code en encodant au niveau des types les propriétés mathématiques des structures algébriques mises en {\oe}uvre.
Ainsi, un monoïde ne se satisferait plus d'une fonction binaire et d'un élément particulier, mais on pourrait lui ajouter une preuve de l'associativité et de l'unitarité de la fonction sous-jacente.
On pourrait alors restreindre les éléments contenus dans le constructeur \texttt{MonoidMorph} aux morphismes de monoïdes.

Une autre modification importante pour l'implantation pourrait être le fait de généraliser la classe de types des catégories à d'autres genres (c'est-à-dire d'autres ``types de types'') que le genre \texttt{*} d'Haskell.
On pourrait ainsi définir une catégorie dont les objets seraient des types du genre \texttt{Nat}, support qui serait nécessaire à encoder proprement la notion de \emph{PRO}~\cite{LLMN19}, puis peut-être les reconnaisseurs associés.


\printbibliography%

\printindex

\addcontentsline{toc}{part}{Index}

\appendix
\newpage

\part{Annexes}

\chapter{Automates de mots}
Dans ce chapitre sont regroupés des exemples supplémentaires d'utilisation des différentes notions présentées dans ce document n'ayant pas nécessairement leur place dans le fil conducteur de la discussion.

Nous y trouverons des extensions de constructions simplistes (\(k\)-DFAs), des généralisations d'algorithmes (combinaisons booléennes quelconques de DFAs), des utilisations exotiques (produit parallèle d'automates) ou des réinterprétations de constructions débouchant sur des extensions de modélisations à la fois des implantations des automates mais également de construction classiques en Haskell (automates séquentiels et monade \texttt{Writer}).

\section{Automates à multiplicités séquentiels et monade \texttt{Writer}}\label{sec::annexe:autMultWriter}
Nous avons vu dans la Section~\ref{subsecAutMult} comment encoder les automates à multiplicités comme des automates de Kleisli de la monade associée à la construction des semimodules libres.
Ces automates nécessitent la définition de la somme et du produit des semianneaux; le produit lors d'un parcours pour combiner les poids des transitions successives pendant la traversée sur un chemin, et la somme pour combiner les poids des différents chemins.

Les automates pour lesquels il n'existe au plus qu'un chemin par mot sont dits \firstocc{séquentiels}{automates!séquentiels}, l'équivalent du qualificatif \emph{déterministe} pour des poids non Booléens.
Remarquons que si un automate est séquentiel, la somme du semianneau n'est plus nécessaire; un monoïde est alors suffisant.
Ainsi, en lieu d'un semimodule libre dont les éléments sont définis comme des sommes (formelles) de couples, un produit cartésien est suffisant.

Pour réaliser cela, il est possible d'utiliser le constructeur de type \texttt{Writer} du module\\
\texttt{Control.Monad.Writer} du package \texttt{mtl}, dont la définition est la suivante\footnote{en réalité, à un isomorphisme près.}.

\begin{implementationBox}[]{Le constructeur Writer}
  \begin{minted}[xleftmargin=1em,   autogobble, fontsize=\footnotesize]{haskell}

    -- Constructeur binaire de types Writer
    data Writer w a = Writer {runWriter :: (a, w)}

    -- Fonctions permettant de récupérer la seconde composante du couple
    -- contenu dans un Writer w a
    execWriter :: Writer w a -> w
    execWriter m = snd (runWriter m)

    -- Permet de transformer une valeur de type w
    -- en un élément pondérant un Writer
    tell :: w -> Writer w ()
    tell w = Writer ((), w)
  \end{minted}
\end{implementationBox}
\noindent Ce constructeur de type binaire, une fois complété par un type monoïdal, permet de définir une monade.
\begin{implementationBox}[]{Monade Writer}
  \begin{minted}[xleftmargin=1em,   autogobble, fontsize=\footnotesize]{haskell}

    -- module HDRExample.WriterAutParProd

    -- Le constructeur Writer w est un foncteur de
    -- la catégorie des types
    instance FunctorGen (Writer w) (->) (->) where
      fmap f (Writer (a, w)) = Writer (f a, w)

    -- Le foncteur Writer w est une monade si w est un monoïde
    instance Monoid w => MonadGen (Writer w) (->) where
      join (Writer (Writer (a, w), w'))  = Writer (a, w <> w')
      return a = Writer (a, mempty)
  \end{minted}
\end{implementationBox}
Un automate de Kleisli associé à cette monade est exactement un automate séquentiel.
L'exemple suivant construit un automate séquentiel à partir d'un prédicat \texttt{test} sur des symboles; plus précisément, il associe à un mot \texttt{w}, mot sur un type \texttt{Symbol}, le couple \texttt{(c, w')} où
\begin{itemize}
  \item \texttt{c} est l'entier de \texttt{Symbol}s de \texttt{w} satisfaisant le prédicat \texttt{test};
  \item \texttt{w'} est le sous-mot de \texttt{w} constitué des \texttt{Symbol}s satisfaisant \texttt{test}.
\end{itemize}
Pour cela, il suffit d'utiliser comme structure de pondération le monoïde défini comme le produit tensoriel des monoïdes suivants: le monoïde additif des entiers (utilisant le constructeur de type unaire \texttt{Sum} en Haskell) et le monoïde libre des \texttt{Symbols} (les listes en Haskell).
\begin{remarqueBox}[]{Couple de monoïdes}
  En Haskell, un couple de monoïde est automatiquement une instance de la classe de types monoïde.
    \begin{minted}[xleftmargin=1em,   autogobble, fontsize=\footnotesize]{haskell}

      instance (Semigroup w, Semigroup w') => Semigroup (w, w')
        (x, y) <> (x, y') = (x <> y, x' <> y')

      instance (Monoid w, Monoid w') => Monoid (w, w')
        mempty = (mempty, mempty)
    \end{minted}
    Remarquons que cette implantation correspond exactement au produit tensoriel direct de monoïdes.
\end{remarqueBox}

\begin{implementationBox}[label={codeExAutSequentiel}]{Exemple d'automate séquentiel}
  \begin{minted}[xleftmargin=1em,   autogobble, fontsize=\footnotesize]{haskell}

    -- module HDRExample.WriterAutParProd

    -- Définition de l'équivalence des valeurs ((), w) et w
    instance Monoid n => Castable (Writer n ()) n where
      cast    = execWriter
      castInv = tell

    -- Synonyme pour le type des états de l'automate (un seul état)
    type StateAuto = ()

    -- Automate comptant le nombre de symboles d'un mot w vérifiant le prédicat
    -- donné en paramètre et renvoyant le sous mot de w formé des symboles
    -- vérifiant le prédicat
    autoCount
      :: (Symbol -> Bool)
      -> Automaton
           (KleisliCat (Writer (Sum Int, [Symbol])) (->))
           ()
           Symbol
           StateAuto
           ()
    autoCount test = packKleisli initial_b delta_b final_b
     where
      -- l'état initial est l'(unique) état () avec pour poids le couple
      -- (0, [])
      initial_b = writer ((), (Sum 0, []))
      -- Après lecture de x, si ce symbole vérifie le prédicat,
      -- on ajoute 1 et on concatène x;
      -- sinon, on ajoute 0 et on concatène la chaine vide
      delta_b x () | test x    = Writer ((), (Sum 1, [x]))
                   | otherwise = Writer ((), (Sum 0, []))
      final_b :: () -> (Sum Int, [Symbol])
      -- Le poids final de l'état () est neutre.
      final_b () = (Sum 0, [])
  \end{minted}
\end{implementationBox}
Tout comme la monade \texttt{State} (voir Section~\ref{subsecMonadState}), la monade \texttt{Writer} admet des \emph{transformers}, permettant de combiner toute monade avec elle.
Le type \texttt{WriterT m w a}, pour \texttt{m} une monade et \texttt{w} un monoïde, est isomorphe au type \texttt{m(a, w)}.
Naturellement, on peut alors y associer des automates de Kleisli.
Remarquons alors qu'en choisissant la monade des listes, les valeurs sont des listes de couples, relativement proches de la structure de semimodules libres; malheureusement, il n'est pas possible en Haskell d'exprimer la monade des semimodules libres depuis un \emph{transformer} de la monade \texttt{Writer}, la définition de base des monades y étant restreintes (d'où notre extension de départ).

Cependant, cette proximité soulève des questions intéressantes, prenant un sens concret en termes catégoriques.
Remarquons donc que:
\begin{itemize}
  \item la monade \texttt{Writer} est isomorphe à un produit tensoriel, celui de la catégorie des types (le produit cartésien) entre un monoïde et un type;
  \item la monade \texttt{FreeSemimodule} est isomorphe à un produit tensoriel, celui des monoïdes commutatifs, entre un semianneau et un monoïde commutatif.
\end{itemize}
\begin{remarqueBox}{Le produit tensoriel de monoïdes commutatifs}
  Le produit de deux monoïdes commutatifs \( \mathcal{M}=(M,+,0) \) et \( \mathcal{M}'=(M',+',0') \), est le monoïde dont l'ensemble sous-jacent est l'ensemble des sommes formelles commutatives d'éléments de \(M\times M'\), quotientée par la plus petite congruence générée par les trois équations suivantes:
  \begin{align*}
    (x,z) \oplus (x,t) &= (x,z +' t), &
    (x,z) \oplus (y,z) &= (x+y,z), &
    (0,z) = (0,0') &= (x,0').
  \end{align*}
\end{remarqueBox}
Dans les deux cas, une des composantes est un monoïde objet:
\begin{itemize}
  \item pour la monade \texttt{Writer}, un monoïde est un monoïde objet de la catégorie des types;
  \item pour la monade \texttt{FreeSemimodule}, un semianneau est un monoïde objet de la catégorie des monoïdes commutatifs (la multiplication du monoïde objet est la multiplication du semianneau).
\end{itemize}
Le produit tensoriel par un monoïde objet permet de généraliser la notion de module d'un semianneau.
Un \( \mathfrak{M} \)-\firstocc{module}~\cite{Br14, MacLane71}, pour un monoïde objet \( \mathfrak{M}=
  (M,\mu,\nu) \) d'une catégorie monoïdale \( (\mathcal{C},\otimes,I,\alpha,\lambda,\rho) \)
  est un couple \( (A, \mapsto) \) avec \( A \) un objet de  \( \mathcal{C} \) et \( \mapsto \)
  un morphisme de \( \mathrm{Hom}_{\mathcal{C}}(M\otimes A, A) \) faisant commuter les diagrammes suivants:

\hfill
\begin{minipage}{0.7\linewidth}
  \begin{equation}\label{eq:module}
    \begin{tikzpicture}[baseline=(current  bounding  box.center)]
      \matrix (m) [matrix of math nodes,row sep=3em,column sep=4em,minimum width=3em]{
        M \otimes M \otimes A & M \otimes (M \otimes A) & M\otimes A \\
        M \otimes A & & A\rlap{,}\\
      };
      \path[-stealth]
        (m-1-1)
          edge node [above] {\( \alpha_{M,M,A} \)} (m-1-2)
          edge node [left] {\( \mu \otimes \mathrm{Id}_A \)} (m-2-1)
        (m-1-2)
          edge node [above] {\( \mathrm{Id}_M \otimes \mapsto \)} (m-1-3)
        (m-1-3)
          edge node [right] {\( \mapsto \)} (m-2-3)
        (m-2-1)
          edge node [below] {\( \mapsto \)} (m-2-3)
          ;
    \end{tikzpicture}
  \end{equation}
\end{minipage}
\hfill\hfill

\hfill
\begin{minipage}{0.7\linewidth}
  \begin{equation}\label{eq:moduleIdent}
    \begin{tikzpicture}[baseline=(current  bounding  box.center)]
      \matrix (m) [matrix of math nodes,row sep=3em,column sep=4em,minimum width=3em]{
        I\otimes A & & M \otimes A\\
        & A\rlap{.} & \\
      };
      \path[-stealth]
        (m-1-1)
          edge node [below left] {\( \lambda_A \)} (m-2-2)
          edge node [above] {\( \nu \otimes\mathrm{Id}_A \)} (m-1-3)
        (m-1-3)
          edge node [below right] {\( \mapsto \)} (m-2-2)
          ;
    \end{tikzpicture}
  \end{equation}
\end{minipage}
\hfill\hfill

Pour deux \( \mathfrak{M} \)-modules \( (A,\mapsto) \) et \( (A',\mapsto') \), un \firstocc{morphisme de module} est un morphisme \( f \) de \( \mathcal{C} \) faisant commuter les diagrammes suivants:

\hfill
\begin{minipage}{0.7\linewidth}
  \begin{equation}\label{eq:mod-morph}
    \begin{tikzpicture}[baseline=(current  bounding  box.center)]
      \matrix (m) [matrix of math nodes,row sep=3em,column sep=4em,minimum width=3em]{
       M\otimes A & A \\
        M \otimes A' & A'\rlap{.}\\
      };
      \path[-stealth]
        (m-1-1)
          edge node [above] {\(  \mapsto \)} (m-1-2)
          edge node [left] {\( \mathrm{Id}_M\otimes f \)} (m-2-1)
        (m-1-2)
          edge node [right] {\( f \)} (m-2-2)
        (m-2-1)
          edge node [below] {\( \mapsto' \)} (m-2-2)
          ;
    \end{tikzpicture}
  \end{equation}
\end{minipage}
\hfill\hfill

De plus, le foncteur envoyant le \( \mathfrak{M} \)-module \( (A,\mapsto) \)
  depuis la catégorie \(\mathrm{Mod}(\mathfrak{M}) \) des \( \mathfrak{M} \)-modules sur \( A \)
  dans la catégorie \( \mathcal{C} \), c'est-à-dire le foncteur d'oubli
  \( \mathrm{Forget} \), admet un adjoint à gauche \( \mathrm{Free}_{\mathfrak{M}} \)~\cite{MacLane71}
  envoyant un objet \( X \) de \( \mathcal{C} \) sur le \( \mathfrak{M} \)-module \( (M\otimes X,
  (\mu\otimes\mathrm{Id}_X)\circ\alpha^{-1}_{M,M,X}) \), le
  \( \mathfrak{M} \)-\firstocc{module libre généré par} \( X \); autrement dit,
\begin{equation}\label{eq:mod-adjunction}
  \mathrm{Hom}_{\mathcal{C}}(X,A) \sim \mathrm{Hom}_{\mathcal{\mathrm{Mod}
    (\mathfrak{M})}}(\mathrm{Free}_{\mathfrak{M}}(X),(A,\mapsto)).
\end{equation}
Cette adjonction peut être définie par l'unité \( \eta_X
  =(\nu \otimes \mathrm{Id}_X) \circ \lambda_X^{-1} \) envoyant \( X \) sur \( M\otimes X \)
  et la co-unité \( \varepsilon_{(\mathrm{Free}_{\mathfrak{M}}(A))} = \mapsto \)
  envoyant \( (M\otimes A, (\mu\otimes\mathrm{Id}_A)\circ\alpha^{-1}_{M,M,A}) \)
  sur \( (A,\mapsto) \).
En effet, \( \mapsto \) est un morphisme de module car satisfait l'Équation~\eqref{eq:mod-morph}:


\hfill
\begin{tikzpicture}[baseline=(current  bounding  box.center)]
  \matrix (m) [matrix of math nodes,row sep=3em,column sep=9em,minimum width=3em]{
   M\otimes (M\otimes A) & M\otimes A \\
     M\otimes A & A\rlap{,}\\
  };
  \path[-stealth]
    (m-1-1)
      edge node [above] {\(  (\mu\otimes\mathrm{Id}_A)\circ\alpha^{-1}_{M,M,A} \)} (m-1-2)
      edge node [left] {\( \mathrm{Id}_M\otimes \mapsto \)} (m-2-1)
    (m-1-2)
      edge node [right] {\( \mapsto \)} (m-2-2)
    (m-2-1)
      edge node [below] {\( \mapsto \)} (m-2-2)
      ;
\end{tikzpicture}
\hfill\hfill

\noindent en conséquence de la forme modifiée suivante de l'Équation~\eqref{eq:module}:

\hfill
\begin{tikzpicture}[baseline=(current  bounding  box.center)]
  \matrix (m) [matrix of math nodes,row sep=3em,column sep=6em,minimum width=3em]{
    M \otimes M \otimes A & M \otimes (M \otimes A) & M\otimes A \\
    M \otimes A & & A\rlap{.}\\
  };
  \path[-stealth]
    (m-1-1)
      edge node [left] {\( \mu \otimes \mathrm{Id}_A \)} (m-2-1)
    (m-1-2)
      edge node [above] {\( \alpha^{-1}_{M,M,A} \)} (m-1-1)
      edge node [above] {\( \mathrm{Id}_M \otimes \mapsto \)} (m-1-3)
    (m-1-3)
      edge node [right] {\( \mapsto \)} (m-2-3)
    (m-2-1)
      edge node [below] {\( \mapsto \)} (m-2-3)
      ;
\end{tikzpicture}
\hfill
\hfill

En conséquences directes,
\begin{itemize}
  \item les automates à multiplicités peuvent être exprimés comme automates de Kleisli sur les monades des modules libres, généralisant la monade des semimodules libres;
  \item la monade \texttt{Writer} pourrait être généralisée autrement que par les \emph{transformers}, en définissant un type monadique
    \begin{minted}[xleftmargin=1em,   autogobble, fontsize=\footnotesize]{haskell}

      data WriterTens tensor w a = Writer {runWriter :: tensor (a, w)}
    \end{minted}
    où \texttt{tensor} est un produit tensoriel d'une catégorie monoïdale et \texttt{m} un monoïde objet de cette même catégorie monoïdale;
   \item la monade \texttt{State} pourrait être généralisée similairement.
\end{itemize}
Un exemple d'utilité d'une monade à base de module serait le cas d'une fonctionnalité de \emph{logs} (utilisation possible de la monade \texttt{Writer}) arborescents générés facilement à partir des opérades, les monoïdes objets de la catégorie monoïdale des types gradués (voir Pseudo-Code~\ref{CodeOperades} et Section~\ref{secOperadesReturn}).

\section{Produits parallèles d'automates de types différents}\label{sec::annexeProdParAut}
Le produit parallèle d'automate est utilisé pour calculer, par exemple, l'union ou l'intersection de deux automates déterministes (voir Remarque~\ref{RemUnionInterAuto}).
Dans sa définition, rien n'empêche de combiner deux automates de types différents.
Par exemple, le code suivant montre comment combiner un automate alternant (construit similairement à celui du Psudo-Code~\ref{codeAFA}) et un automate séquentiel à multiplicités (celui du Pseudo-Code~\ref{codeExAutSequentiel}) en un automate utilisant les deux parallèlement.
\begin{implementationBox}[]{Produit parallèle d'automates}
  \begin{minted}[xleftmargin=1em,   autogobble, fontsize=\footnotesize]{haskell}
    -- module HDRExample.WriterAutParProd

    -- Synonymes de types
    type Symbol = Char
    type State = Maybe Char

    -- Automate reconnaissant les mots contenant tous les symboles
    -- d'une liste donnée en paramètre
    autoForAll :: [Symbol] -> AFA Symbol State
    autoForAll l = packKleisli (initial_a l) delta_a final_a
     where
      initial_a = foldr1 (<&&>) . fmap (packVar . Just)
      delta_a x (Just y) | x == y    = packVar Nothing
                         | otherwise = packVar $ Just y
      delta_a _ Nothing = packVar Nothing
      final_a = isNothing

    -- fonction de retrait des accents
    removeAccents :: String -> String
    removeAccents =
      T.unpack . T.filter (not . property Diacritic) . normalize NFD . T.pack

    -- teste si un caractère est une voyelle
    isVowel :: Char -> Bool
    isVowel c = fmap toLower (removeAccents [c]) `isInfixOf` "aeiouy"

    -- instance facilitant l'extraction des poids de sortie de
    -- l'automate autoForAll
    instance (Castable (cat a c) e, Castable (cat' b d) f)
      => Castable (ProdCat cat cat' (a, b) (c, d)) (e, f) where
        cast (Prod f g) = (cast f, cast g)
        castInv (e, f) = Prod (castInv e) (castInv f)

    -- Ouvre un fichier texte, puis pour chacun de ses mots (suite
    -- de caractères suivies d'un séparateur d'espacement),
    -- affiche ce mot et la chaine de ses voyelles
    -- si les deux conditions suivantes sont respectées:
    -- * le mot contient toutes les lettres de la chaine symbols,
    -- * son nombre de voyelles est paire
    testAFAFromFile :: String -> String -> IO ()
    testAFAFromFile file symbols =
      let auto = autoForAll symbols `parallelProduct` autoCount isVowel
          aux w = do
            let (reco, (p, res)) =
                  auto `weightValue` w :: (Bool, (Sum Int, String))
            when (reco && getSum p `mod` 2 == 0)
                 (putStrLn $ mconcat ["- ", w, ": ", res])
      in  do
            f <- readFile file
            let ws = words f
            mapM_ aux ws
  \end{minted}
\end{implementationBox}
Cet exemple peut être utilisé pour filtrer les mots contenus dans le fichier
\texttt{file} contenant toutes les occurrences des lettres d'une chaîne \texttt{chaine} et un nombre pair de voyelles en lançant l'exécutable défini dans le fichier \texttt{MainParProd.hs} par la commande
\texttt{stack exec AnnexeExAFA file chaine}.
Ainsi, la commande \texttt{stack exec AnnexeExAFA liste\_francais.txt abcd} affichera la liste des mots contenus dans le fichier \texttt{liste\_francais.txt} (à peu près 22000 mots) contenant les quatre lettres \texttt{a, b, c, d} et ayant un nombre pair de voyelles.

\section{Utilisation des vecteurs}

Les vecteurs, c'est-à-dire les listes de longueur fixée, peuvent être utilisés pour représenter des constructions classiques, algorithmiques ou structurelles, de la théorie des automates.
Par exemple, il est possible d'exprimer la combinaison générale booléenne d'automates déterministes, sans utiliser des combinaisons des opérations d'union, d'intersection ou de complémentaire (voir Remarque~\ref{RemUnionInterAuto}).

\begin{implementationBox}[label={CodeCombiGenBoolDFAs}]{Combinaison Booléenne d'automates}
  \begin{minted}[xleftmargin=1em,  autogobble, fontsize=\footnotesize]{haskell}
    -- module Automata.DFAComp

    -- Soit f une fonction Booléenne n-aire et a1, ..., an
    -- n DFAs.
    -- boolCombi n f a1 ... an calcule la combinaison booléenne
    -- associée à f des automates a1 ... an.
    boolCombiOfCompDFA
      :: Natural ( 'S n)
      -> NFun ( 'S n) Bool Bool
      -> DFAComp symbol state
      -> NFun
           n
           (DFAComp symbol state)
           (DFAComp symbol (Vector ( 'S n) state))
    boolCombiOfCompDFA (Succ n) f aut = convertFromVectMultiFun' n $ \as ->
      let autos = VCons aut as
      in
        pack
          (fmap (\(Auto i _ _) -> i ()) autos)
          (\a ps -> combine (fmap (`delta` a) autos) ps)
          (convertToVectMultiFun (packFun (Succ n) f) . combine (fmap final autos))
  \end{minted}
\end{implementationBox}
\noindent Ces mêmes vecteurs permettent de définir des représentations des \(k\)-DFAs, des automates déterministes ayant \(k\) états initiaux.
\begin{implementationBox}[]{KDFAs}
  \begin{minted}[xleftmargin=1em,  autogobble, fontsize=\footnotesize]{haskell}
    -- module Automata.DFAComp

    toKDFA
      :: Vector k state -> DFAComp symbol state -> DFAComp symbol (Vector k state)
    toKDFA initial (Auto _ d f) =
      pack initial (\a ps -> fmap (d a) ps) (or . fmap f)
  \end{minted}
\end{implementationBox}
\noindent Les constructions précédentes sont manipulables comme suit.
\begin{implementationBox}[]{Exemple DFA complets}
  \begin{minted}[xleftmargin=1em,  autogobble, fontsize=\footnotesize]{haskell}
    -- module HDRExample.DFAComp2

    -- fonction auxiliaire de composition d'une fonction unaire
    --  avec une fonction binaire.
    (.:) :: (c -> d) -> (a -> b -> c) -> a -> b -> d
    (.:) = (.) . (.)

    -- Fonction de transition modulaire; si p est une valeur
    -- numérique entière, envoie p sur (p + 1) modulo n
    deltaMod :: Integral state => state -> symbol -> state -> state
    deltaMod n _ p = (p + 1) `mod` n

    -- a1 reconnait les mots de longueurs multiples de 3
    -- a2 reconnait les mots de longueurs multiples de 7
    -- a3 reconnait les mots de longueurs multiples de 11
    a1, a2, a3 :: DFAComp symbol Int
    a1 = pack 0 (deltaMod 3) (== 0)
    a2 = pack 0 (deltaMod 7) (== 0)
    a3 = pack 0 (deltaMod 11) (== 0)

    -- a4 reconnait les mots reconnus à la fois par
    -- a1, a2 et a3
    a4 :: DFAComp symbol (Vector Nat_3 Int)
    a4 = boolCombiOfCompDFA three ((&&) .: (&&)) a1 a2 a3

    -- a4 reconnait les mots reconnus par a1, a2
    -- ou a3
    a5 :: DFAComp symbol (Vector Nat_3 Int)
    a5 = boolCombiOfCompDFA three ((||) .: (||)) a1 a2 a3

    -- a6 correspond à l'automate a3 dans lequel les états 2 et 5 sont
    -- initiaux
    a6 :: DFAComp symbol (Vector Nat_2 Int)
    a6 = toKDFA (2 <+> mkSingle 5) a3

    -- Teste si le mot w est reconnu par les automates a1, ..., a6
    test :: Show symbols => [symbols] -> IO ()
    test w = do
      putStrLn $ mconcat ["a1 recognizes ", show w, ": ", show (a1 `recognizes` w)]
      putStrLn $ mconcat ["a2 recognizes ", show w, ": ", show (a2 `recognizes` w)]
      putStrLn $ mconcat ["a3 recognizes ", show w, ": ", show (a3 `recognizes` w)]
      putStrLn $ mconcat ["a4 recognizes ", show w, ": ", show (a4 `recognizes` w)]
      putStrLn $ mconcat ["a5 recognizes ", show w, ": ", show (a5 `recognizes` w)]
      putStrLn $ mconcat ["a6 recognizes ", show w, ": ", show (a6 `recognizes` w)]
  \end{minted}
\end{implementationBox}

\end{document}